\documentclass[oneside,12pt]{Latex/Classes/PhDthesisPSnPDF}

\usepackage{lineno}
\usepackage{amsbsy}
\usepackage{xspace}
\usepackage{wtmmPkg}
\usepackage{natbib}
\usepackage{multirow}
\usepackage{paralist}
\usepackage{fancyhdr} 
\pdfoutput=1

\newcommand{\arcsec}{^{\prime\prime}}
\DeclareRobustCommand{\ion}[2]{%
\relax\ifmmode
\ifx\testbx\f@series
{\mathbf{#1\,\mathsc{#2}}}\else
{\mathrm{#1\,\mathsc{#2}}}\fi
\else\textup{#1\,{\mdseries\textsc{#2}}}%
\fi}

\newcommand{\sube}{ {_{e}} }
\newcommand{\subj}{ {_{j}} }
\newcommand{\subi}{ {_{i}} }
\newcommand{\ji}{ {_{j,i}} }

\newcommand{\rlam}{ {$R(\lambda)$} }

\newcommand{\rsun}{R$_{\odot}$}

\newcommand{\aap}{    {\it Astronomy \& Astrophysics}}

\newcommand{\apj}{    {\it Astrophysical Journal}}

\ifpdf  
    \pdfcatalog { /PageMode (/UseOutlines)
                  /OpenAction (fitbh)  }
\fi

\title{EUV and X-ray Spectroscopy of the Active Sun}

\ifpdf
  \author{\href{mailto:claireraftery@gmail.com}{Claire L. Raftery}}
  \collegeordept{\href{http://www.physics.tcd.ie}{School of Physics}}
  \university{\href{http://www.tcd.ie}{University of Dublin, Trinity College}}

  \crest{
  \vspace{1cm}
  \includegraphics[width=0.3\textwidth]{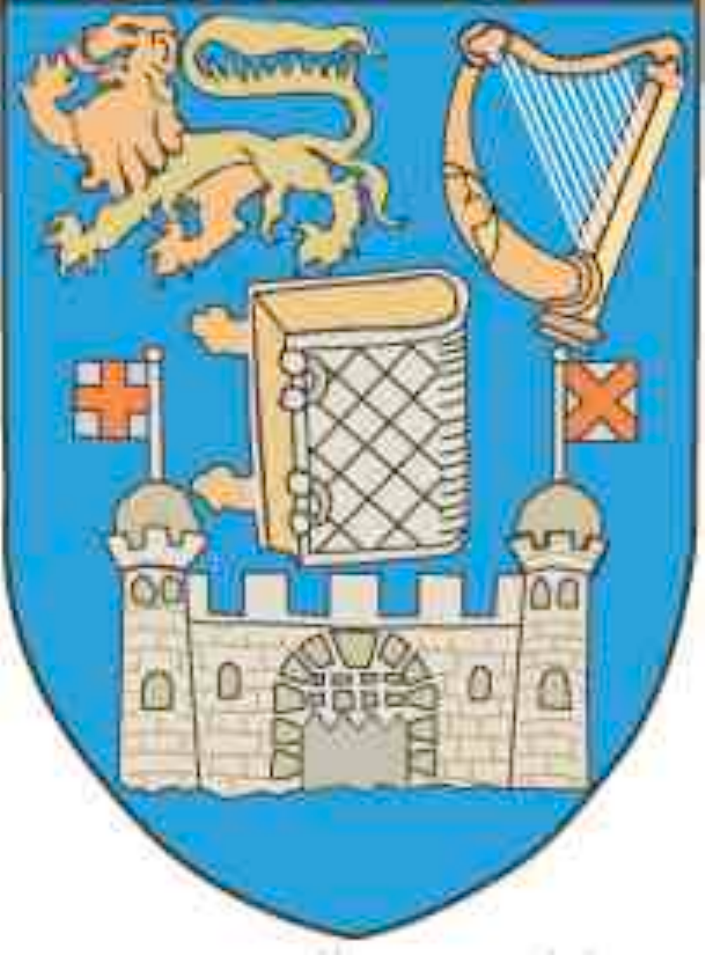}
   \vspace{1cm}
    }
  
\else
  \author{\\ Claire L. Raftery}

  \collegeordept{School of Physics}
  \university{University of Dublin, Trinity College}
  \crest{\includegraphics[width=6cm]{TCD_Crest.pdf}}
\fi
\degree{Philosophi\ae Doctor (PhD)}
\degreedate{2009 December}

\usepackage{setspace}
\begin{document}


\renewcommand\baselinestretch{1.2}
\baselineskip=18pt plus1pt


\maketitle  

%








\frontmatter


\begin{declaration}        

I, Claire L. Raftery, hereby certify that I am the sole author of this thesis and that all the work presented in it, unless otherwise referenced, is entirely my own. I also declare that this work has not been submitted, in whole or in part, to any other university or college for any degree or other qualification. 
\vspace{5mm}

This thesis work was conducted from October 2006 to December 2009 under the supervision of Dr. Peter T. Gallagher at Trinity College, University of Dublin.

\vspace{5mm}
In submitting this thesis to the University of Dublin I agree that the University Library may lend or copy the thesis upon request. 
\vspace{5mm}

\textbf{Name:} Claire L. Raftery	
 \begin{center}
 \textbf{Signature:}  ........................................		\textbf{Date:}  ..............
\end{center}

\end{declaration}



\begin{abstracts}        

This thesis strives to improve our understanding of solar activity, specifically the behaviour of solar flares and coronal mass ejections. An investigation into the hydrodynamic evolution of a confined solar flare was carried out using RHESSI, CDS, GOES and TRACE. Evidence for pre-flare heating, explosive and gentle chromospheric evaporation and loop draining were observed in the data. The observations were compared to a 0-D hydrodynamic model, EBTEL, to aid interpretation. This led to the conclusion that the flare was not heated purely by non-thermal beam heating as previously believed, but also required direct heating of the plasma. An observational investigation in to the initiation mechanism of a coronal mass ejection and eruptive flare was then carried out, again utilising observations from a wide range of spacecraft: MESSENGER/SAX, RHESSI, EUVI, Cor1 and Cor2. Observations provided evidence of CME triggering by internal tether-cutting and not by breakout reconnection. A comparison of the confined and eruptive flares suggests that while they have different characteristics, timescales and topologies, these two phenomena are the result of the same fundamental processes. Finally, an investigation into the sensitivity of EUV imaging telescopes was carried out. This study established a new technique for calculating the sensitivity of EUV imagers to plasmas of different temperatures for four different types of plasma: coronal hole, quiet sun, active region and solar flare. This was carried out for six instruments: Proba-2/SWAP, TRACE, SOHO/EIT, STEREO A/EUVI, STEREO B/EUVI and SDO/AIA. The importance of considering the multi-thermal nature of these instruments was then put into the context of investigating explosive solar activity.

\end{abstracts}



\begin{dedication} 

For Mam and Dad, \\my motivation and inspiration.

\end{dedication}


\begin{acknowledgements}      
I must begin by thanking my supervisor Dr. Peter Gallagher. Without his encouragement and guidance, this thesis would never have happened. His unwavering and contagious enthusiasm and passion for his subject is a constant source of motivation for me. Thank you for your endless patience. 

I would also like to thank ``...all the Queen's men...'', not only for regularly putting Humpty Dumpty back together again but also for being unending fountains of knowledge and patience. Ryan Milligan, Shaun Bloomfield and James McAteer, thank you for putting up with me. To Ryan, for your guidance, friendship and help in getting me to NASA. I would have been lost without you. To Shaun, for never being too busy to discuss a new theory or problem. To James, for always having the answer (and for letting me take over your house for the summer!). I would also like to thank Dr. Graham Harper for taking the time to proof read my thesis. I really appreciate your time and guidance. Thanks to Brian Espey for setting me off on the Astro path and to the staff of the physics department (especially John and Jemmer) for all the help along the way. Thanks are also due to the rest of the SWAP team and to Dan Seaton, David Berghmans and Anik De Groof in particular: I am looking forward to seeing what SWAP has to offer! 

To Paul Conlon and Jason Byrne: the other two of the first three. If you did nothing else guys, you kept me laughing!! It will be a long time before I forget your antics. To the next three: Larisza Krista, Shane Maloney and David Long, the other two: Paul ``Higgo'' Higgins and Sophie Murray and the last one: Joe Roche. Thanks for the hilarity, pints and morning tea (Dave you will have to find a replacement for me I'm afraid). 

It goes without saying that a big thanks goes to the people at Goddard, especially to Brian Dennis for having faith in me and to Richard Schwartz, Kim Tolbert and Andy Gopie for their endless RHESSI knowledge. A big thank you to James Klimchuk for picking me out of a crowd and trusting me with his code. Thanks are also due to Dominic Zarro for his tireless efforts in getting my finances sorted. My visits to DC wouldn't have been the same without Alex Young, Jack Ireland, Emilie Drobnes and Mike Marsh. I will be back for pho soon.

Thank you to my best friends Laura and Leah. For listening to my moaning and sobbing and for reminding me of the good things in life. I will always appreciate it. Thanks are especially due to Hazel. You have a unique ability to make me laugh when there is nothing worth laughing at. Thank you for kicking my butt when it needed kicking and passing the snorkel when I was sinking. 

Finally a lifetime of gratitude is owed to my family. To my parents Dominic and Catherine. The unwavering support that you show me is the reason I am who I am. You gave me roots and you gave me wings and encouraged me to follow the wind where it took me. I will be forever grateful for you both. And I cannot forget my ``big'' brothers, David and Richard. You are the comic relief in my life and it wouldn't be the same without you. 

Please note: Many images in this online version are of lower resolution in order to reduce filesize.

\end{acknowledgements}


\chapter{List of Publications}
\label{chapter:publications}

\textbf{\Large{\underline{Refereed}}}
\begin{enumerate}

 \item \textbf{Raftery, C. L.}, Gallagher, P. T. \& Bloomfield, D. S. (2010)\\
 ``Temperature response of EUV imagers'', \\ \aap, submitted
 
 \item \textbf{Raftery, C. L.}, Gallagher, P. T., McAteer, R. T. J, Lin, C. H. \& Delahunt, G . (2010)\\
 ``The flare-CME connection: A study of the physical relationship between a solar flare and an associated CME'', \\\apj, in review

 \item \textbf{Raftery, C. L.}, Gallagher, P. T., Milligan, R. O. \& Klimchuk, J. A. (2009)\\
``Multi-wavelength observations and modelling of a canonical solar flare'', \\ \aap, 494, 1127

\item Lin, C.-H, Gallagher, P. T. \& \textbf{Raftery, C., L.} (2009)\\
``Investigating the driving mechanisms of coronal mass ejections'', 
\\ \aap, in review

 \item Adamakis, S., \textbf{Raftery, C. L.}, Walsh, R.~W. \& Gallagher, P. T., (2009)\\
 ``A Bayesian approach to comparing theoretic models to observational data: A case study from solar flare physics'', 
 \\ \apj, in prep

\end{enumerate}


\setcounter{secnumdepth}{3} 
\setcounter{tocdepth}{3}    
\tableofcontents            


\listoffigures	

\listoftables  


\mainmatter



\doublespacing
\pagestyle{fancy}


\chapter{Introduction}

\ifpdf
    \graphicspath{{1_introduction/figures/PNG/}{1_introduction/figures/PDF/}{1_introduction/figures/}}
\else
    \graphicspath{{1_introduction/figures/EPS/}{1_introduction/figures/}}
\fi


\hrule height 1mm
\vspace{0.5mm}
\hrule height 0.4mm 
\noindent 
\\ {\it This Chapter introduces the fundamental physics and concepts that are discussed in this thesis. This begins with a general introduction to the Sun and the various layers of the solar atmosphere and is followed by a detailed discussion of the solar corona and solar flares. The physical conditions required for flares are presented, including an introduction to concepts such as magnetohydrodynamics, magnetic reconnection and chromospheric evaporation. Finally, a discussion of coronal mass ejections and their connection to solar flares is presented.  }
\\ 
\hrule height 0.4mm
\vspace{0.5mm}
\hrule height 1mm 

\newpage

\section{The dynamic Sun}
\label{int_intro}

The Sun has piqued the interest of humankind for thousands of years. Records from the Stone Age show that even then, the behaviour of our nearest star was noted and utilised. Newgrange, for example, is a megalithic passage tomb located in Co. Meath, Ireland. Built in $\sim$3,200 BC it is believed to be 200 years older than Stonehenge, making it the oldest surviving ``solar observatory'' in the world. While not an observatory in the modern sense, the construction of this place of worship required a detailed understanding of the behaviour of the Sun and its motion relative to the Earth. At dawn on the days surrounding the winter solstice, the tomb is illuminated by light from the Sun. This happens as a result of a near perfectly aligned roof box.

Much has changed since Neolithic times. Today solar observatories utilise cutting edge technology to make high quality observations of the Sun. These high resolution, high cadence data have revolutionised not only solar physics, but stellar astronomy, geophysics and planetary sciences to name but a few. With recent advances in space technology, the dynamic nature of the Sun, and its atmosphere in particular, is only now beginning to come to light. Explosive releases of energy, charged particles and the solar wind all have an impact on the Earth. This, of course, is in stark contrast to the image of the Sun most people are familiar with. While it is widely understood that the Sun is the source of energy and heat for the Earth, few people understand the detrimental impact ``space weather'' can have on Earth. From loss of communications satellites to electricity grid failures, even to the interruption of long-haul flights, the full understanding and accurate prediction of solar storms is essential to the future well being of this planet's inhabitants. 

Although many details of solar behaviour remain elusive, the general understanding of the Sun and its atmosphere has improved significantly in the past number of years. Revolutionary spacecraft both old and new such as the Solar Maximum Mission \citep[SMM;][]{SMM}, the Solar and Heliospheric Observatory \citep[SOHO;][]{SOHO} and the Reuven Ramaty High Energy Solar Spectroscopic Imager \citep[RHESSI;][]{Lin02} are the cornerstones of a vast fleet of solar observatories. Between ground- and space-based observatories, there are instruments designed to study almost every aspect of solar activity. The solar corona is the focus of much investigation and instruments such as the Transition Region and Coronal Explorer \citep[TRACE;][]{Handy99} and the Coronal Diagnostic Spectrometer \citep[CDS;][]{Harrison95} on board SOHO are dedicated to the study of the upper layers of the solar atmosphere. 

\begin{figure}[!t]
\centering
\includegraphics[width=0.9\textwidth, trim =170 0 0 0, clip = true]{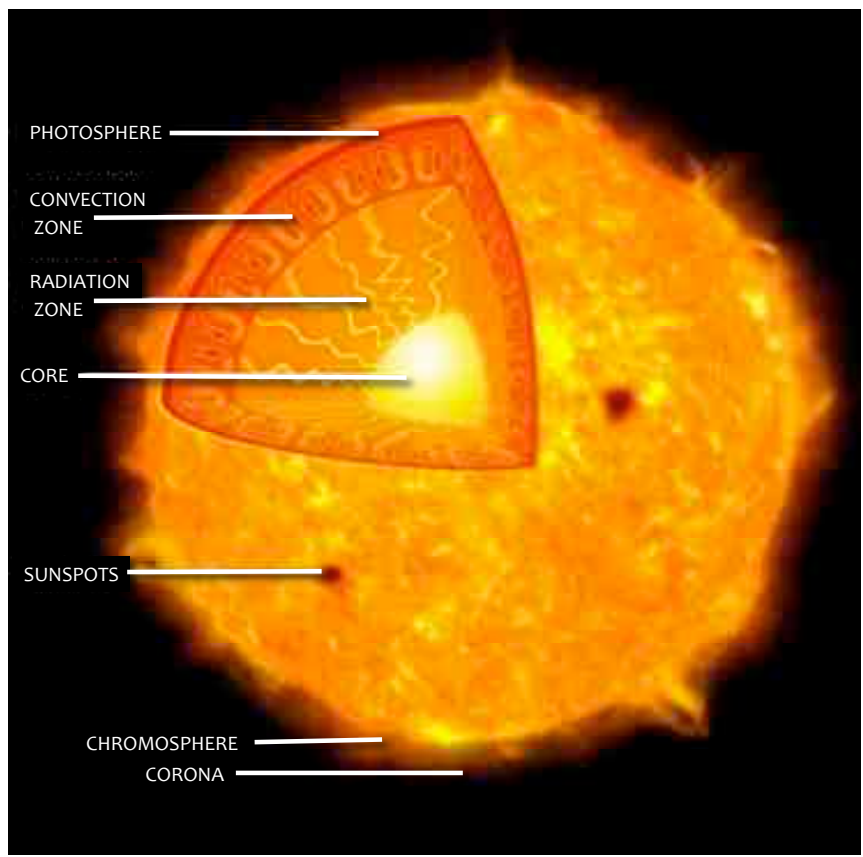} 
\caption{Cartoon showing the layers of the Sun from the core to the outer atmosphere. }
\label{fig:core}
\end{figure}

\begin{figure}[!t]
\centering
\includegraphics[width=\textwidth, trim =0 100 0 0, clip = true]{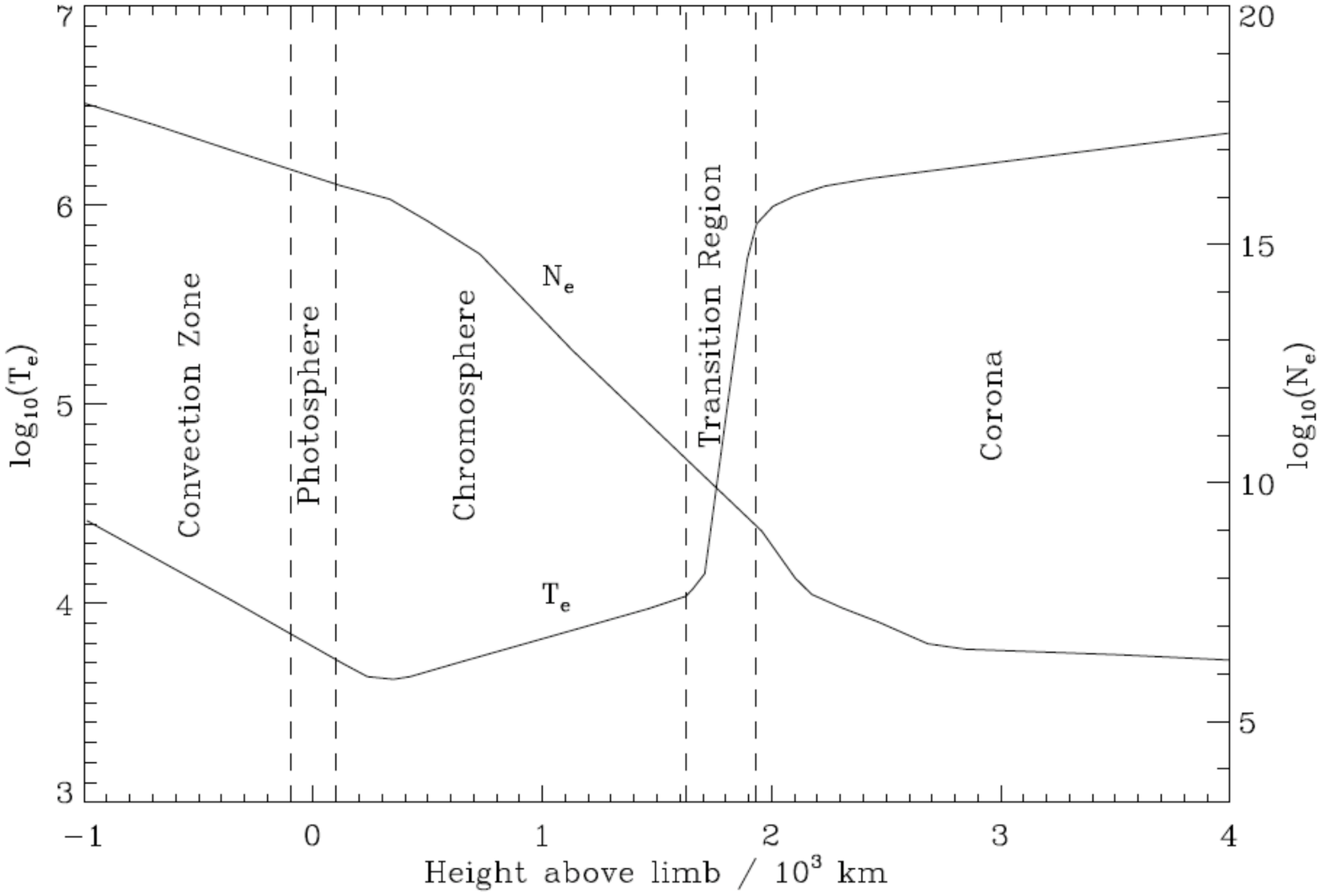} 
\caption{Temperature (T$_e$) and density (N$_e$) of the solar atmosphere as a function of height \citep{gallagher_thesis, Gabriel82}.}
\label{fig:solar_mod}
\end{figure}

The Sun's atmosphere, the layers of which are shown in Figure \ref{fig:core}, is defined to be the part of the Sun that lies above the visible surface, or photosphere. It can be divided into four regions based on their differing physical properties. Thermodynamic properties such as temperature and density are highly sensitive to height above the photosphere, as shown in Figure \ref{fig:solar_mod}. This can have a significant impact on the composition and characteristics of the plasma in each of the layers. The density and temperature gradients also affect a parameter known as the plasma $\mathrm \beta$. It is given by the ratio of the gas to magnetic pressure:
\begin{equation}
\beta = \frac{P_{gas}}{P_{B}} = \frac{n_ek_BT}{B^2/8\pi}
\label{eqn:beta}
\end{equation}
where $P_{gas}$ is gas pressure, $P_B$ is magnetic pressure, $B$ is magnetic field strength, $n_e$ is electron density, $k_B$ is Boltzmann's constant and $T$ is temperature. The density change though the atmosphere ($\Delta n\approx10^{11}$~cm$^{-3}$) is greater than that of temperature ($\Delta T\approx10^3$~K) or the square of the magnetic field strength ($\Delta B^2\approx10^2$~G). Therefore between the photosphere and the corona the $\beta$ value drops from $\sim10$ to $\sim10^{-1}$ \citep{Aschwanden_book, Gary01}. 
\begin{figure}[!t]
\centering
\includegraphics[width=\textwidth, trim =45 200 40 40, clip = true]{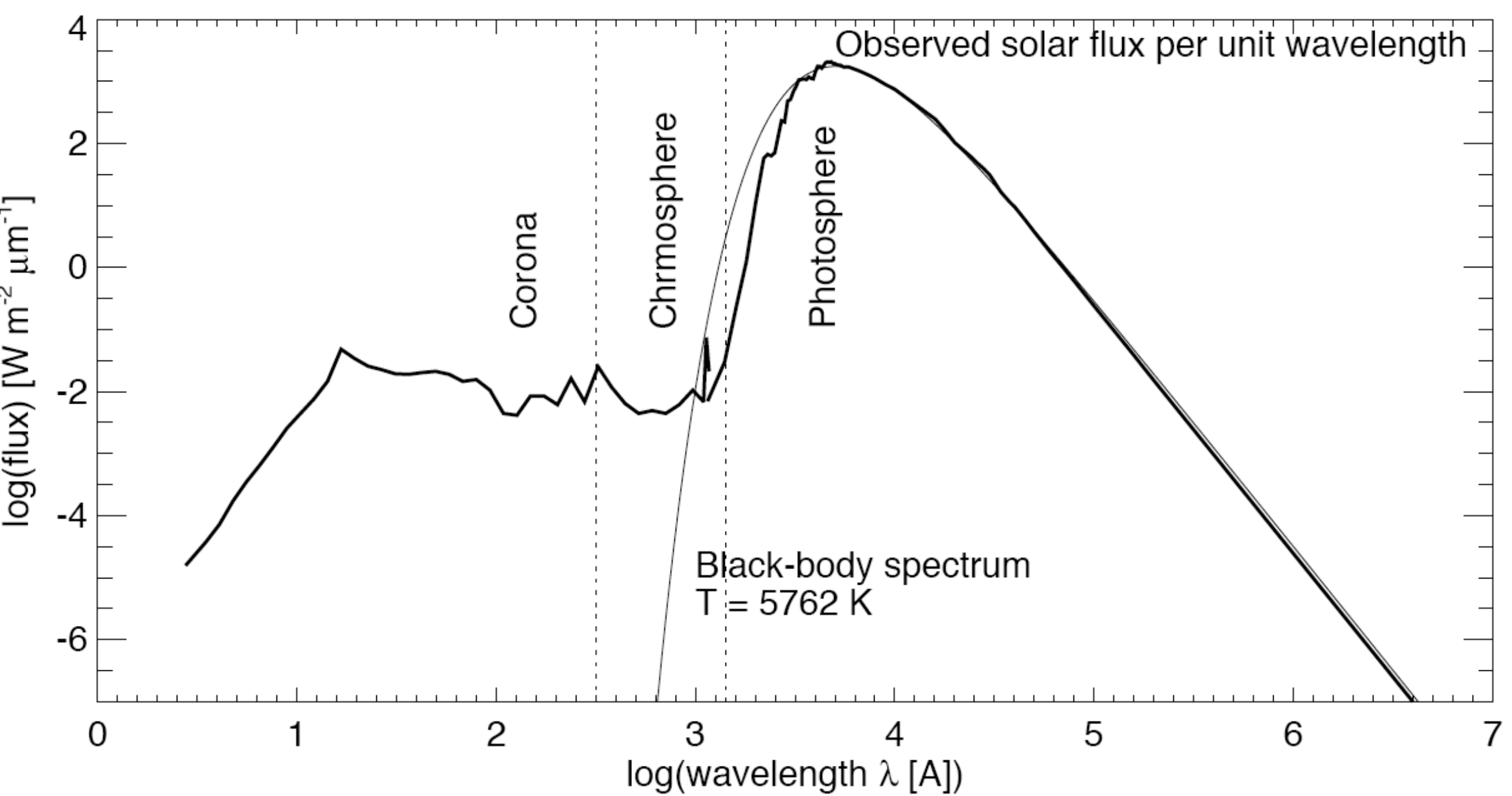} 
\caption{The observed solar flux through the various layers of the atmosphere. The photospheric flux compares very well to a black-body spectrum at a temperature of 5,762~K \citep{Aschwanden_book}.  }
\label{fig:bb}
\end{figure}
\begin{figure}[!t]
\centering
\includegraphics[width=0.75\textwidth, trim =0 100 0 0, clip = true]{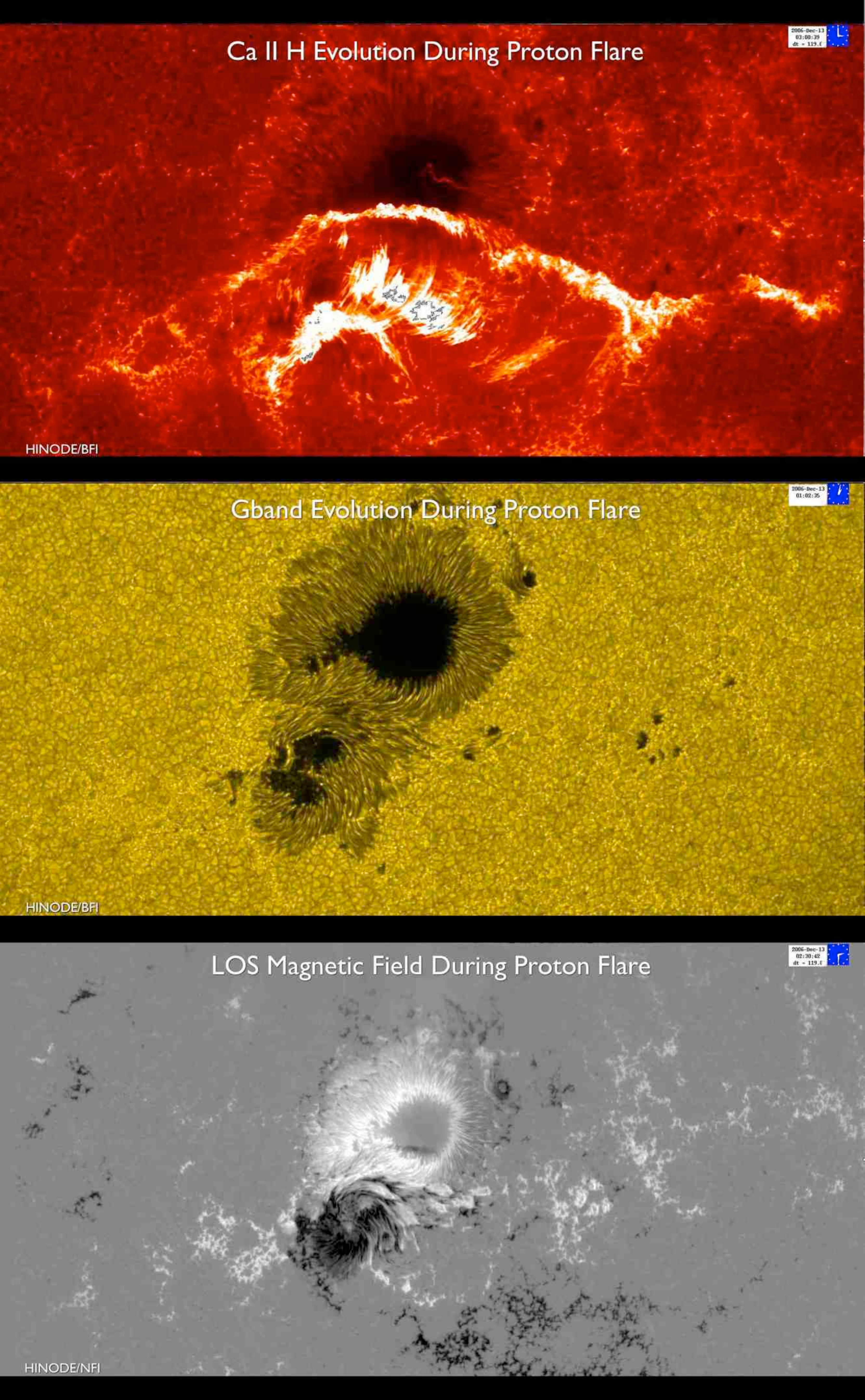} 
\caption{Three different views of two interacting sunspots that resulted in a solar flare, observed with Hinode/SOT. Top panel shows the chromospheric \ion{Ca}{ii} line, the middle frame shows G-band (optical) emission and the bottom panel shows the magnetogram image of both sunspots with their opposite polarities (black vs. white).}
\label{fig:sunspot}
\end{figure}

\subsection{The photosphere}
The photosphere is a cool, dense region and is the optically thick layer of the atmosphere that is seen when viewing the Sun in optical continuum. The visible spectrum of the Sun is in excellent agreement with a black body radiator with an effective temperature of $\sim$5,800~K (Figure \ref{fig:bb}) and has a density of $\sim$10$^{17}$ cm$^{-3}$. The photosphere is opaque and emits a continuous spectrum which is crossed by Fraunhofer lines. Granulation is a feature of the photosphere. This is the term used to describe the photospheric manifestation of the large convective motion that occurs in the convection zone beneath the photosphere. The continuous shifting of plasma by convective motions acts to tangle and stress the magnetic field, increasing its non-potential energy. Another interesting feature found in the photosphere are sunspots. These are large concentrations of magnetic flux which appear as dark regions in white light and continuum images. Their dark appearance is as a result of the suppression of convection in those areas, resulting in temperatures of 3,000-4,000~K which is much cooler than the surrounding photosphere. Sunspots can often be divided into two parts - the central, dark umbra, where the magnetic field is approximately normal to the photosphere and the surrounding, lighter penumbra, where the magnetic fields are more inclined. Figure \ref{fig:sunspot} shows a composite image of two interacting sunspots taken with the Solar Optical Telescope (SOT) on board Hinode. The top panel shows the chromospheric \ion{Ca}{ii} line. In this image you can see the dark umbra of the northern sunspot and chromospheric granulation patterns. The flare can be seen as the bright elongated feature stretching laterally between the sunspots. The middle frame shows G-band (optical) emission. Both sunspots are clearly visible, along with granulation patterns. Note there is no indication of a flare occurring in this image. The bottom panel shows the magnetogram image of both sunspots of opposite polarities (black vs. white). The vertical nature of the field in the umbra is clear here as there are no magnetic field measurements made in this region. The penumbra are showing strong magnetic field strengths. Large magnetic field gradients exist in the mixed polarity region in the center of this image making it the location of the solar flare observed in the top panel.

\begin{figure}[!t]
\centering
\includegraphics[width=\textwidth, trim =0 0 0 0, clip = true]{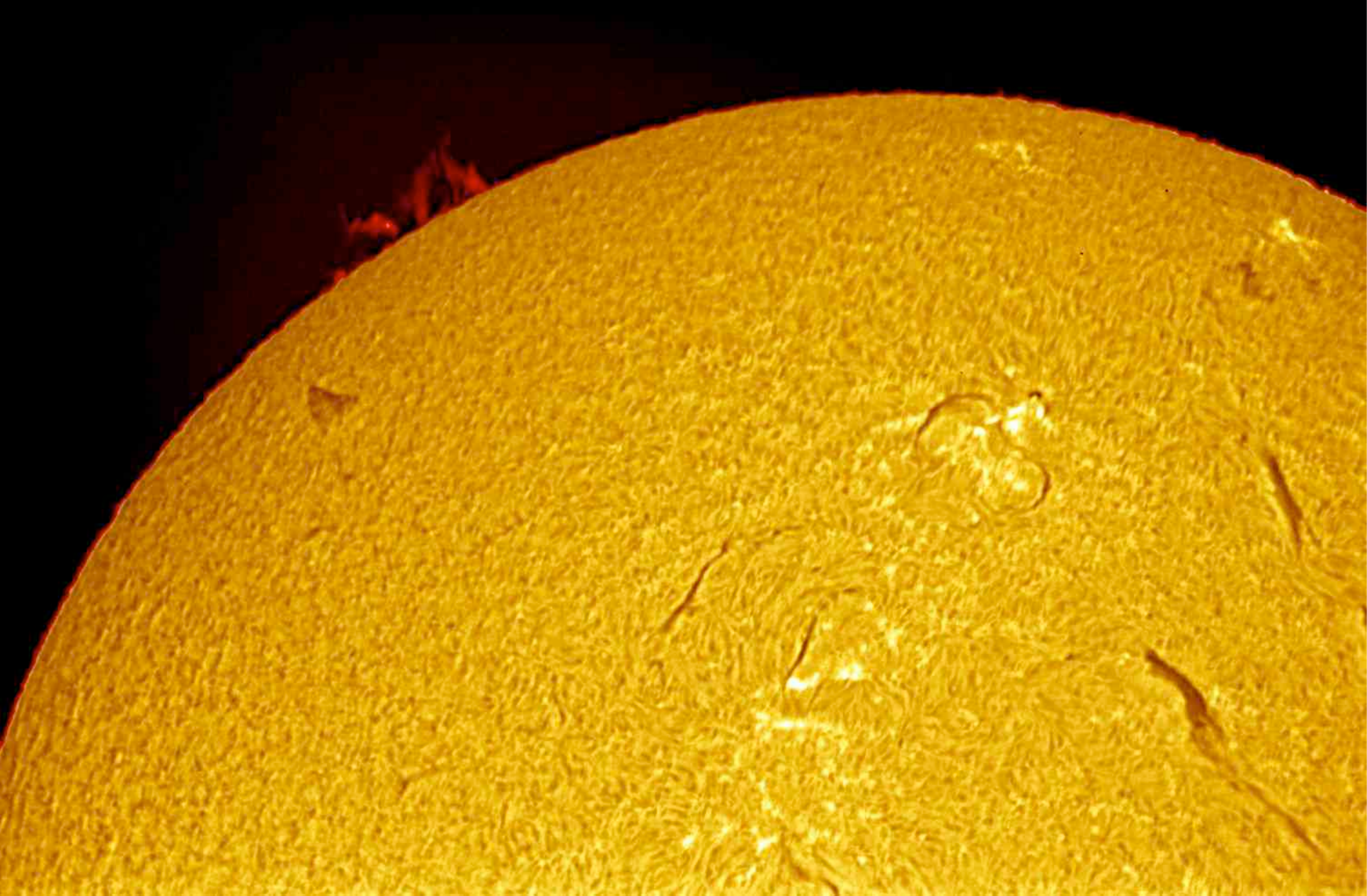} 
\caption{H$\alpha$ image of the Sun showing a prominence (top left on the limb) and filaments (dark regions on disk). Credit: Jack Newton. }
\label{fig:prom}
\end{figure}
\begin{figure}[!t]
\centering
\includegraphics[width=\textwidth, trim =0 0 0 0, clip = true]{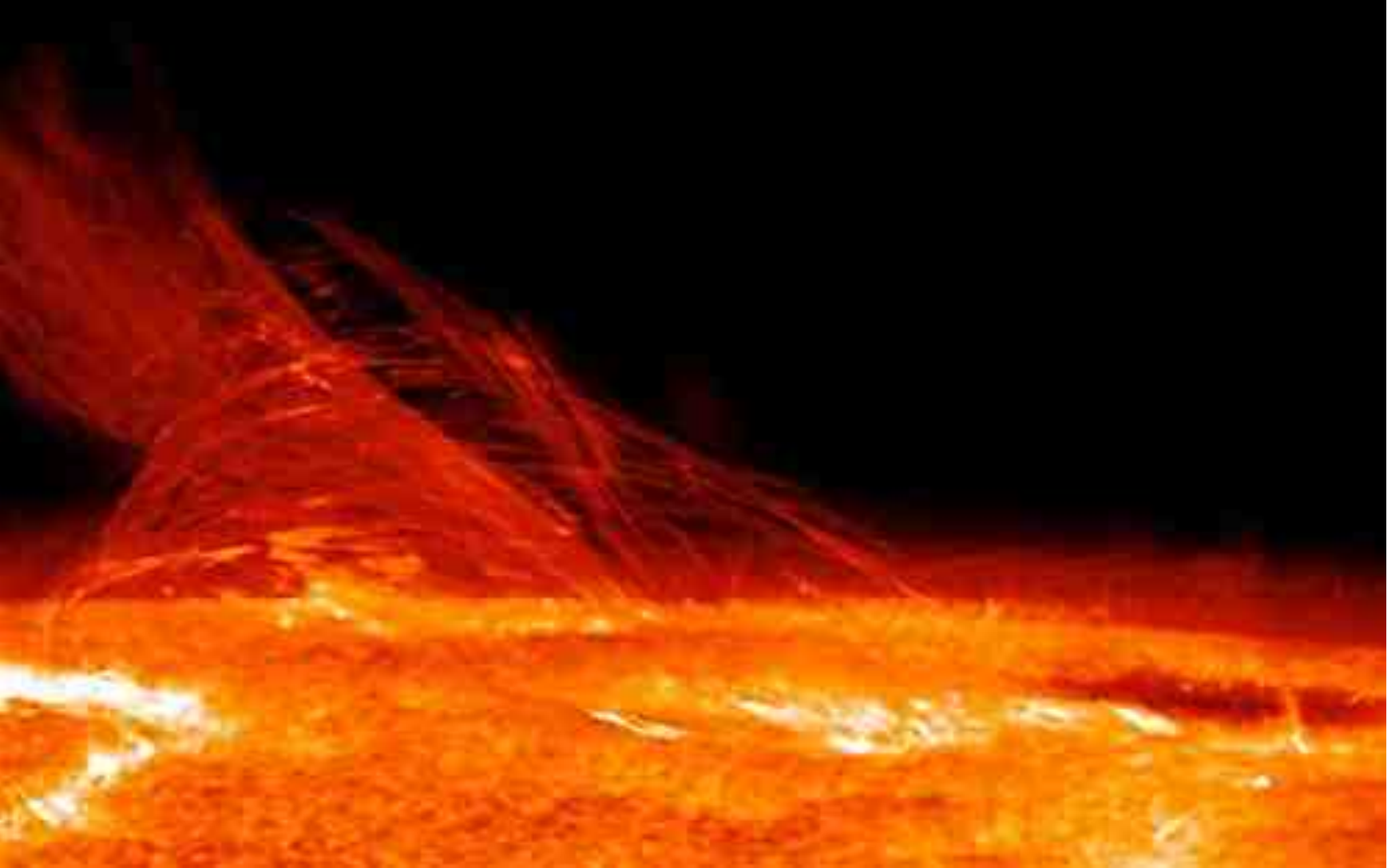} 
\caption{Image of the structure of the solar chromosphere taken with Hinode/SOT. }
\label{fig:chromo}
\end{figure}
\subsection{The chromosphere}
It is clear from Figure \ref{fig:sunspot} that the outline of sunspots are not as well defined as height is increased from the photosphere into the chromosphere. At densities of $\sim$10$^{15}$~cm$^{-3}$ and temperatures of $\sim$10$^4$~K, the chromosphere is dominated by absorption lines and continuum emission. The strong H$\alpha$ line shows bright regions called \emph{plage} in the vicinity of active regions and sunspots. Another interesting feature of note in the chromosphere are \emph{filaments}. These are regions of cool plasma suspended above the photosphere. They are seen as long, dark structures on disk in Figure \ref{fig:prom} and as arcade-like features on the limb, where they are known as \emph{prominences}. It is believed that the chromosphere is heated by some combination of conduction of heat from the hotter transition region and by the deposition of energy by waves. It is believed that acoustic waves are generated by turbulent motions in the photosphere which then form shocks as they propagate upwards through the chromosphere. With a plasma $\beta$ of $\sim$1, this highly ordered layer has very well defined structures, as Figure \ref{fig:chromo} clearly demonstrates \citep{Carlsson97, Depont04} . 
\subsection{The transition region}

\begin{figure}[!t]
\centering
\includegraphics[width=0.9\textwidth, trim =0 0 0 0, clip = true]{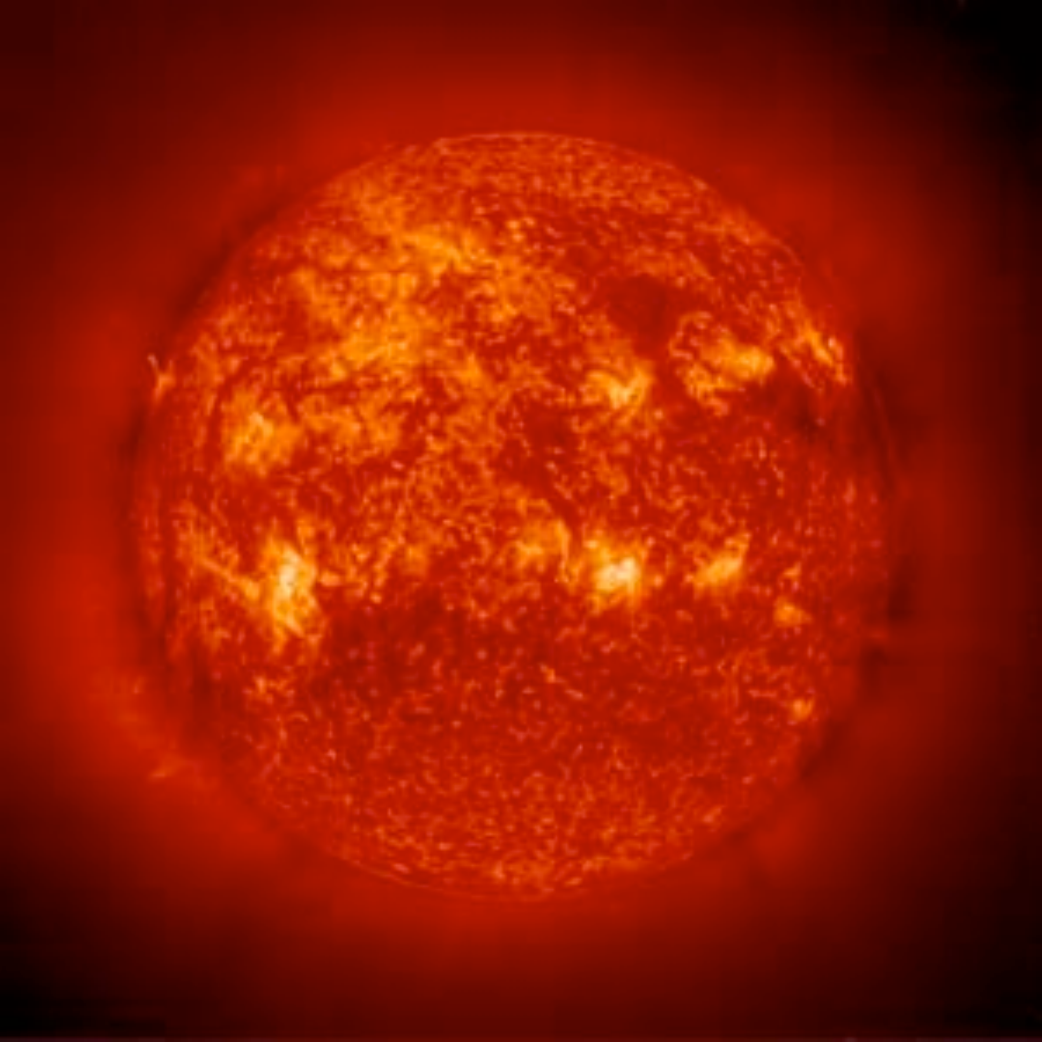} 
\caption{A SOHO/EIT image of the Sun's transition region observed in the 304~\AA\ (\ion{He}{ii}) passband. }
\label{fig:304}
\end{figure}

Above the chromosphere, there is a very narrow layer only a few hundred km thick known as the transition region \citep{Mariska_book, Gallagher98}. The transition region is a highly dynamic interface between the chromosphere and the corona \citep{Dowdy86, Feldman83, Feldman87, Gallagher99}. Unlike the smooth temperature and density profiles found in the chromosphere, there are very steep gradients across the transition region, with values changing from $\sim$10$^4 - $10$^6$~K and $\sim$10$^{15} - $10$^9$~cm$^{-3}$ across a height of the order $10^2$~km high. Conduction $(F_c)$ is highly sensitive to both the temperature $(T)$ and the temperature gradient as it scales as $F_c = \kappa_0 T^{5/2} \nabla T$ where $\kappa_0$ is the coefficient of thermal conductivity (see Equation \ref{eqn:spitzer} for further details). The steep temperature gradient across the transition region means conduction is very efficient at transferring heat energy from the hot corona downwards, heating the upper layers of the transition region as it does so. As a result, the upper transition region emits strongly in the ultraviolet (UV) and extreme ultraviolet (EUV) portion of the spectrum. The transition region appears brightest in areas called \emph{active regions} (Figure \ref{fig:304}). These are the EUV manifestations of the dense magnetic field regions observed as sunspots in the photosphere.  

\subsection{The corona}
\begin{figure}[!t]
\centering
\includegraphics[width=0.9\textwidth, trim =0 0 0 0, clip = true]{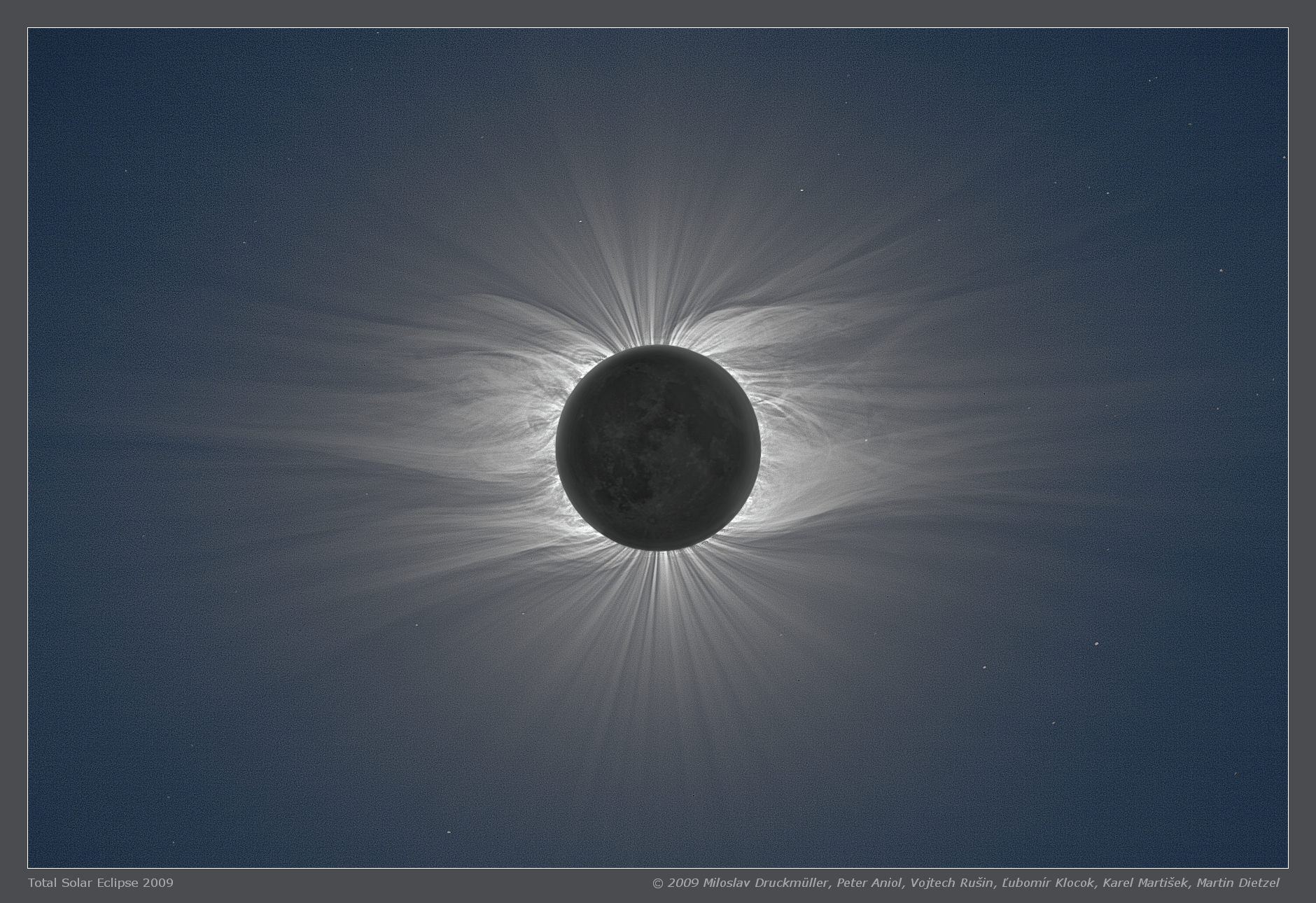} 
\caption{ The visible solar corona up to 6 \rsun\ taken during the 2009 solar eclipse. The linear open field of polar coronal holes are clear, along with the closed loops of helmet streamers closer to the equator \citep{corona}. }
\label{fig:eclipse}
\end{figure}

As we increase in height into the corona, increased temperatures of $\gtrsim$10$^6$~K and greatly reduced densities of $\sim$10$^9$~cm$^{-3}$ compared to the photosphere mean the corona emits strongly in EUV and in X-rays, particularly in active regions where large intricate loop systems are present. The complexity of these loop systems along with large magnetic field gradients make them the most likely place for solar flares to occur \citep{Gallagher02, Conlon08, Mcateer09}. The tenuous, ambient emission in the visible corona however, can be very difficult to observe due to the stark intensity contrast between it and photospheric emission. Until the development of coronagraphs, the corona could only be imaged in visible light during a solar eclipse. Figure \ref{fig:eclipse} shows an image of the extended corona taken during the 2009 solar eclipse with the moon blocking out emission from the solar disk, making it possible to image the visible corona in striking detail. The linear features in this image result from emitting plasma flowing along magnetic field lines. 

\begin{figure}[!t]
\centering
\includegraphics[width=0.7\textwidth, trim =0 0 15 0, clip = true]{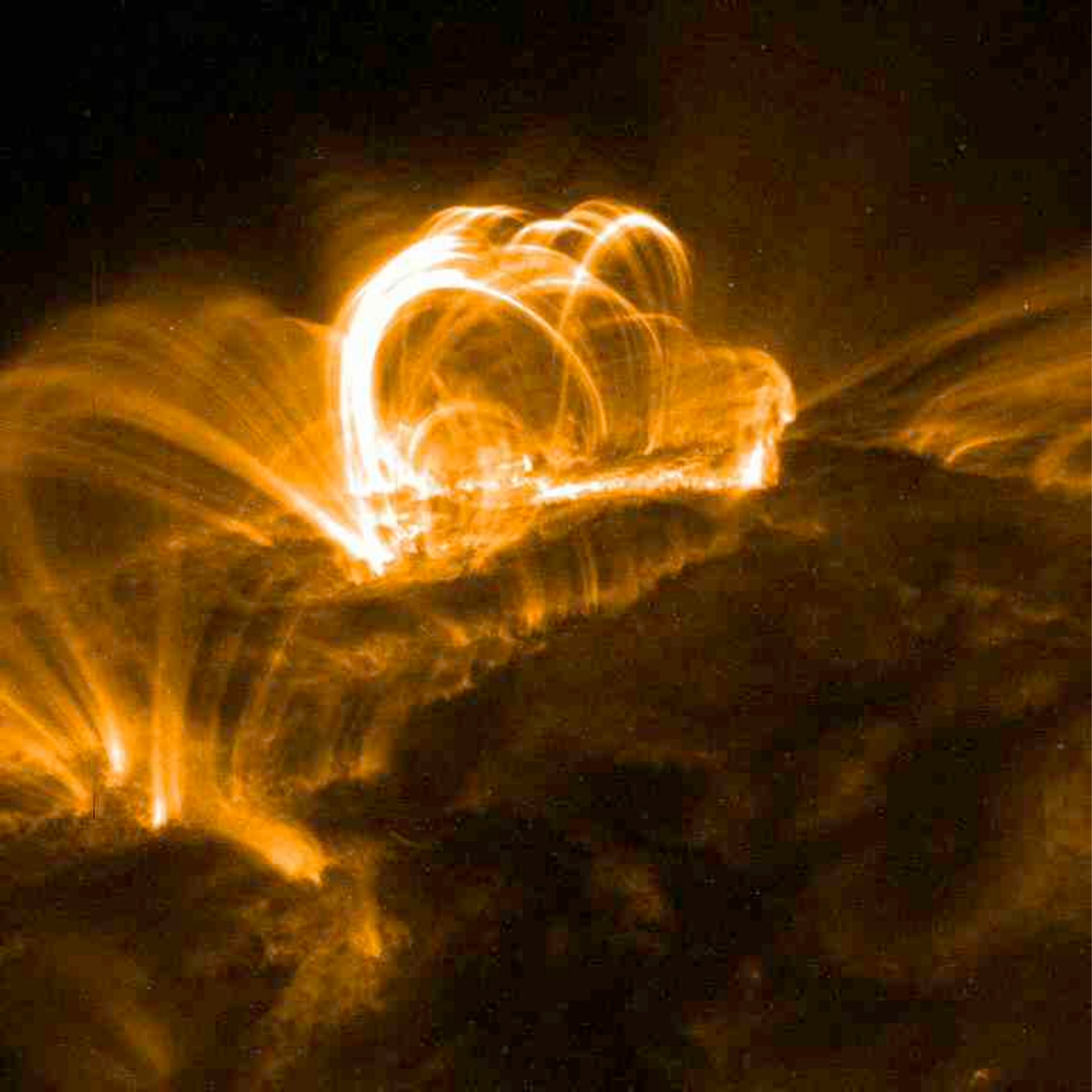} 
\caption{Extreme ultraviolet image of a set of flaring loops taken with TRACE.} 
\label{fig:solar_flare}
\end{figure}
There are two distinct features that can be seen in Figure \ref{fig:eclipse}. At the poles there are regions of open field called \emph{coronal holes}. These features exist permanently at the poles and intermittently closer to the  equator. Open magnetic field lines stretching between the photosphere and interplanetary space have a large pressure gradient along them. This drives plasma out of the solar atmosphere into space in what is known as the \emph{solar wind} \citep{solar_wind}. As a result, these regions tend to have a low density \citep[$\sim10^7$~cm$^{-3}$;][]{Wilhelm06}, showing up as dark regions in EUV images. 

Closer to the equator in Figure \ref{fig:eclipse}, streamers and helmet streamers can be seen. These large, trans-equatorial loop systems have coronal plasma trapped along the field lines. High in the atmosphere however, the weakening field is dragged into a cusp shape by the solar wind. Helmet streamers are often found above active regions and prominences. Active regions are areas of enhanced magnetic activity. These tend to form in bands north and south of the equator that migrate towards the equator as the solar cycle progresses and are frequently associated with sunspots. The increased magnetic stresses in these regions result in large, impulsive releases of energy known as solar flares (\S \ref{sect:flares}). Flares are often associated with coronal mass ejections \citep[\S \ref{sect:CMEs};][]{Tousey73}. These are the ejections of plasma, energy and magnetic field into interplanetary space. The remainder of the solar disk is what is known as \emph{quiet sun}. Despite the name, these areas are far from quiet. It is believed that small scale activity is occurring constantly in the quiet sun by way of micro- and nano-flares \citep{Parker83, Gallagher99, Klimchuk01, Schmelz09}.

\begin{figure}[!t]
\centering
\includegraphics[width=\textwidth, trim =40 40 40 40, clip = true]{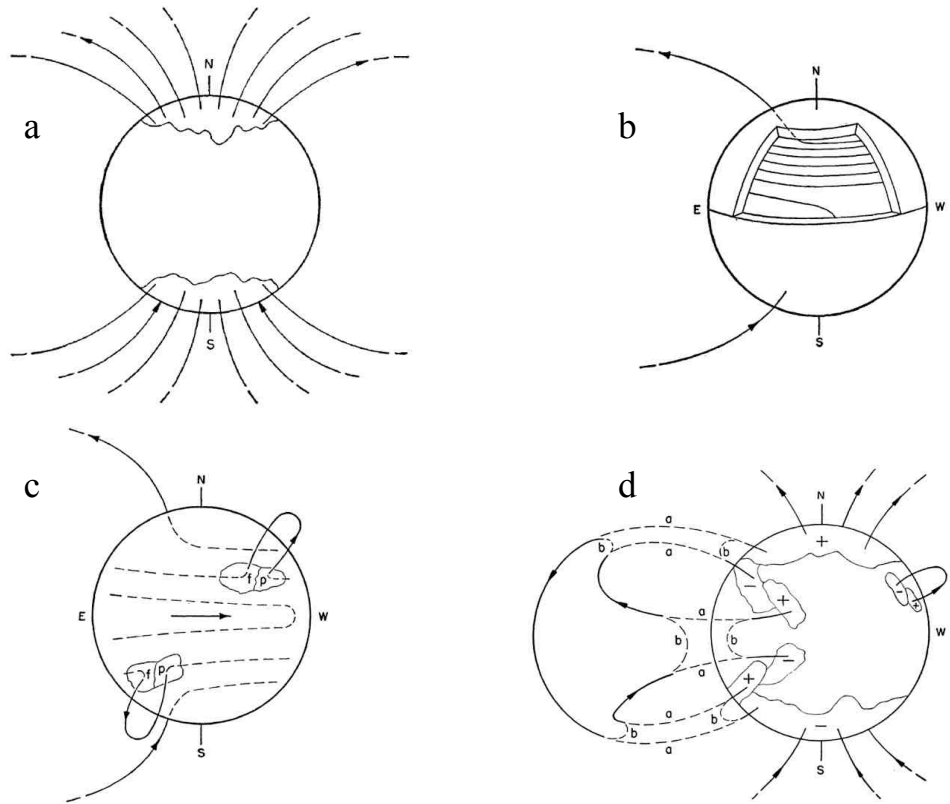} 
\caption{The solar dynamo \citep[$\alpha \Omega$ effect, adapted from; ][]{Babcock61}. Panel (a) shows the standard bipolar poloidal field of the Sun. Panel (b) shows differential rotation resulting in the wrapping of field lines around the Sun, leading to a toroidal topology ($\alpha$ effect). Panel (c) shows the increasing internal pressure inside flux tubes results in them rising through the photosphere ($\Omega$ effect). Panel (d) shows that magnetic reconnection with the bipolar field and with neighbouring magnetic loop systems results in a gradual cancelling of the toroidal field and a return to a poloidal topology. }
\label{fig:diff_rot}
\end{figure}

Active regions, and therefore solar flares, exist as a result of the emergence of areas of concentrated magnetic field through the photosphere. This occurs directly as a result of differential rotation \citep[see][for review]{Hoyng90}. However, the details of the solar dynamo remain elusive. The consensus within the community is that the solar cycle can be explained by the $\alpha\Omega$ effect \citep[Figure \ref{fig:diff_rot};][]{Babcock61}. The basis for this is the MHD mean field dynamo equation:
\begin{equation}
\frac{\partial \langle B \rangle }{\partial t} = \nabla \times \left( \langle v \rangle \times \langle B \rangle + \langle v'  \times B' \rangle - \eta \nabla \times \langle B \rangle\right)
\label{eqn:ind}
\end{equation}
where $B$ and $v$ are the net magnetic field strength and flow velocities of the large scale mean components and $B'$ and $v'$ refer to the small-scale turbulent motions \citep{Charbonneau05}. $\eta$ is the magnetic diffusivity of the Sun. This essentially refers to the viscosity of the fluid. At the tachocline \citep[base of the convection zone;][]{Spiegel92}, two characteristics affect the magnetic field. Firstly, $\eta$ is known to be small \citep{Charbonneau05} and so the cross product (advection) terms in Equation \ref{eqn:ind} dominate. Low diffusion means the magnetic field is ``frozen-in'' to the plasma and will move with both the large-scale, global plasma flows and the small turbulent motions. Secondly, above the tachocline, the convection zone rotates differentially (i.e. the Sun no longer rotates as a solid body above the tachocline). This forces the magnetic field to deviate from its initial poloidal state (Figure \ref{fig:diff_rot}a) and become wrapped up, generating a toroidal field (Figure \ref{fig:diff_rot}b). As regions of high magnetic field density develop, the internal pressure of the magnetic flux tube begins to increase until the pressure gradient is sufficient to cause the magnetic flux tube to rise. The flux tube protrudes through the photosphere and can appear as a sunspot (Figure \ref{fig:diff_rot}c). Regions where flux emergence is prominent can become \emph{active regions}. As flux emerges through the photosphere, it is twisted and therefore contains non-potential energy. In an attempt to return to a force-free state, excess free energy can be released in solar flares and coronal mass ejections (Figure \ref{fig:diff_rot}d). 

\section{Solar Flares}
\label{sect:flares}
\begin{figure}[!t]
\centering
\includegraphics[width=\textwidth, trim =0 00 0 0, clip = true]{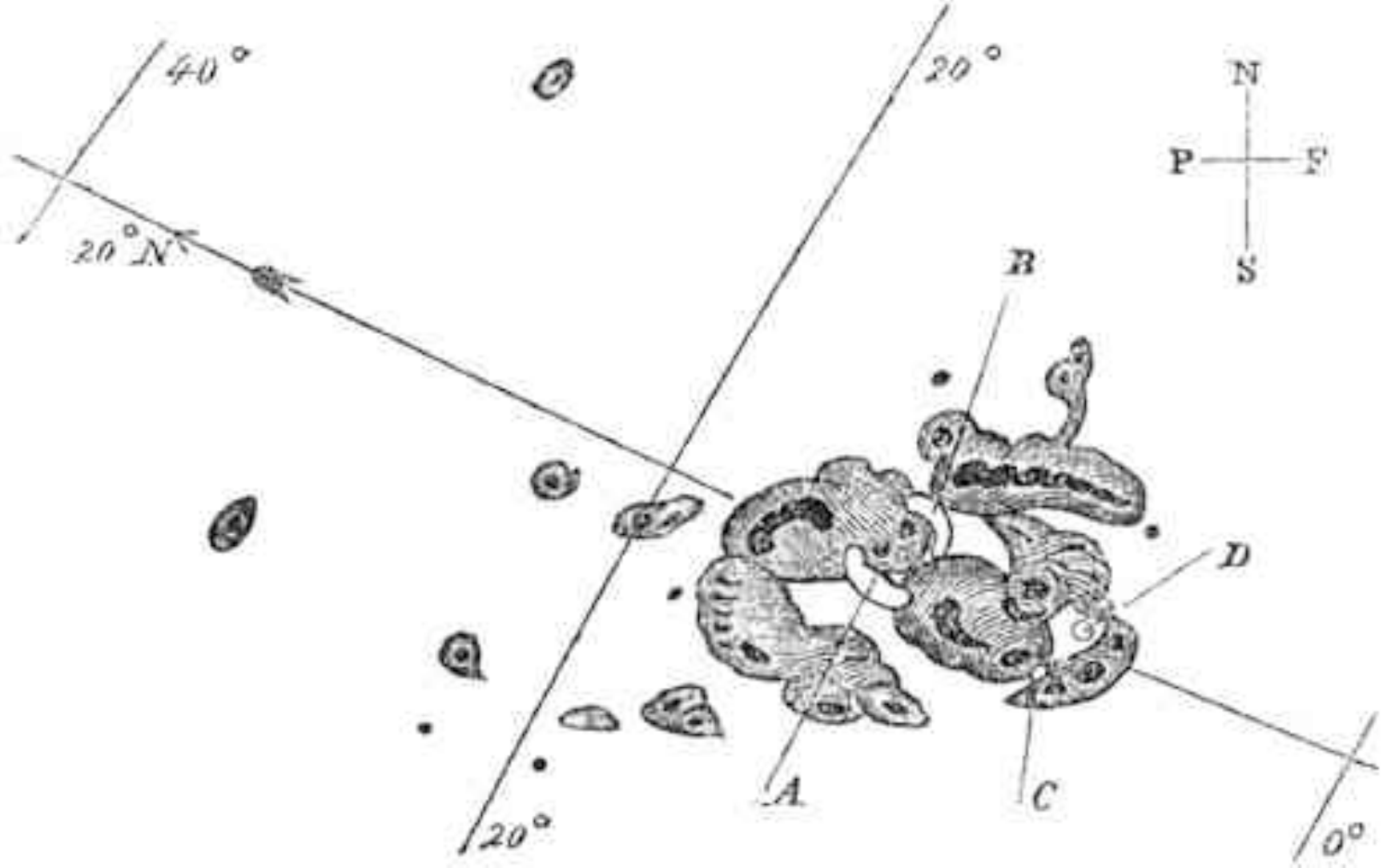} 
\caption{Diagram sketched by \citet{Carrington1859}. This is the first recorded image of a solar flare, indicated by the white regions marked $A$ and $B$.   }
\label{fig:carrington}
\end{figure}

\begin{figure}[!t]
\centering
\includegraphics[width=\textwidth, trim =0 00 0 0, clip = true]{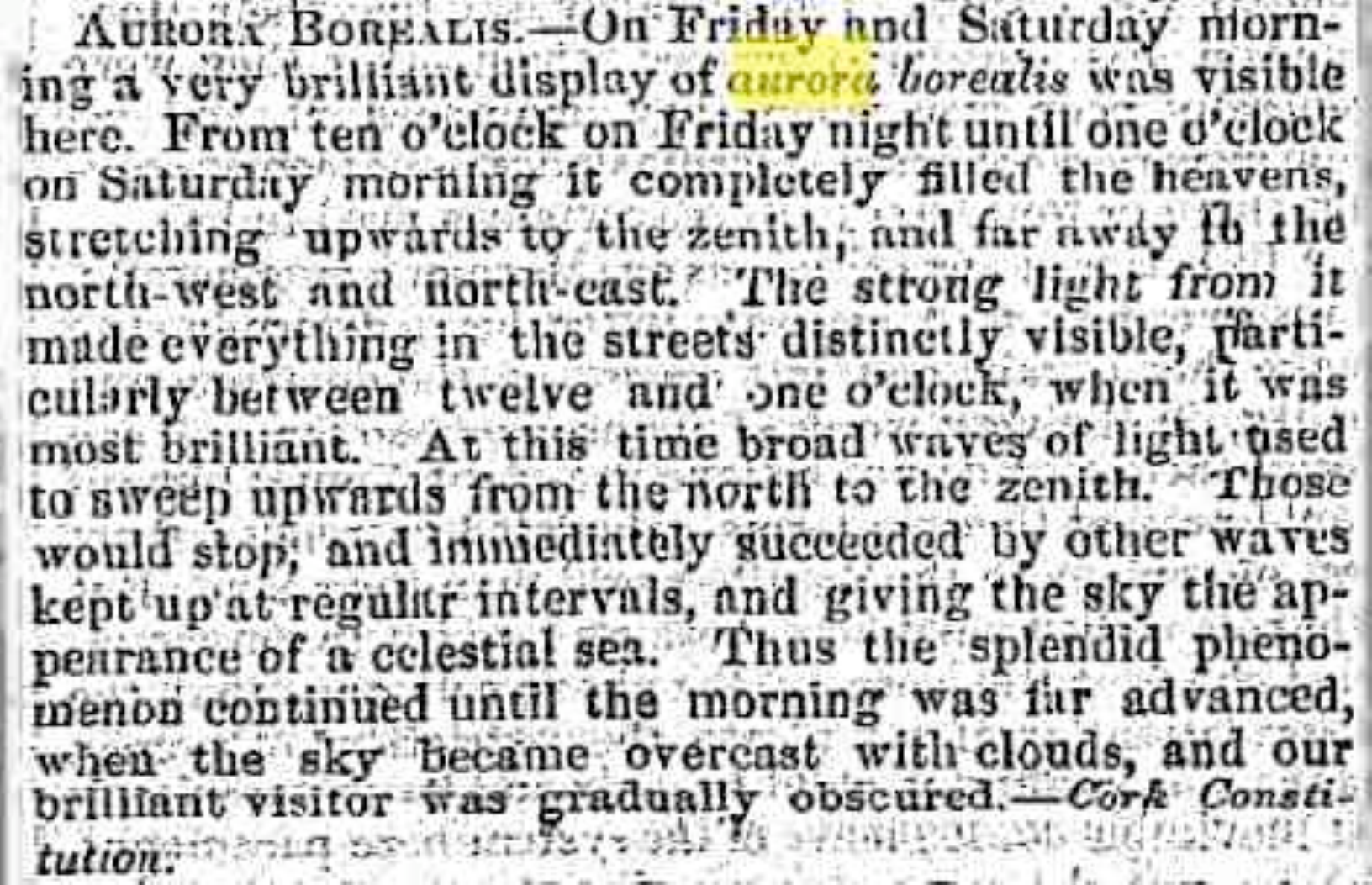} 
\caption{Irish Times article published on 8-Sept-1859 describing auroral activity in Cork, Ireland following the 1859 flare (courtesy of Peter T. Gallager).  }
\label{fig:times}
\end{figure}

The study of solar flares began more than 150 years ago. Contrary to general knowledge, the first recording of the Sun-Earth connection was published by an Irishman, Colonel Edward \citet{Sabine1852}. Sabine hypothesised that the number of sunspots was connected with the level of auroral activity observed at Earth: ``...it is quite conceivable that affections of the gaseous envelope of the Sun, or causes occasioning these affections, may give rise to sensible \emph{magnetical} effects at the surface of our planet, without producing sensible \emph{thermic} effects.'' The first image of a solar flare was published 7 years later by \citet{Carrington1859}, shown in Figure \ref{fig:carrington}. The magnetic storm associated with this flare (now known to be a CME) was recorded all over the world. Articles were published in the \emph{Irish Times} detailing the visibility  of the aurora as far as Cork in the south of Ireland at 51$^{\circ}$ (Figure \ref{fig:times}). Thus began the inquest into the cause of these transient and very intense brightenings (flares) and their ``affection'' to the Earth (CMEs). 

The mid-20th century brought the technology to record these brightenings using rockets and balloon flights. The first hard X-ray emission from a flare was recorded by \citet{Balloon_hxr}. The development of orbiting satellites in the latter half of the century resulted in some ground breaking telescopes. Observatories of note include Skylab and SMM which were vital in furthering the understanding of the solar flare phenomenon \citep{Sturrock80, Strong99_smm}. The Japanese Solar-A mission, later named Yohkoh \citep{Yohkoh}, was designed to study solar flares in the keV - GeV range. One of the most noteworthy revelations of the Yohkoh mission was the observation of what became known as the \emph{Masuda Flare} \citep{Masuda94}. This paper was revolutionary as it proclaimed the reconnection region above the solar flare as the location of particle acceleration. While Yohkoh was concentrating on the soft and hard X-ray emission (SXR and HXR respectively), SOHO was broadening horizons with its suite of twelve instruments. While the primary focus of the SOHO mission was not the investigation of solar flares, its diverse set of instruments has ensured it played its part in the understanding of their behaviour. From measurements of the magnetic field out to white light observations of coronal mass ejections, this observatory continues to facilitate the study of flares over much of their temperature ranges, length scales and wavelengths. While SOHO was unique in its vast range of instruments, RHESSI was revolutionary in a different sense. The primary goal of this mission is the investigation of particle acceleration and energy release during solar flares. RHESSI's unique ability to simultaneously produce high-resolution images and spectra in the X- to $\gamma$- ray regime is as a result of a combination of finely tuned rotating grids and cooled germanium detectors (see \S \ref{sect:rhessi} for further details). This instrument has facilitated the investigation of not only particle acceleration, but also the location of reconnection regions \citep[e.g.][]{Krucker2009} and the motion of this reconnection region with the evolution of flares \citep[e.g.][]{Grigis05}. It also facilitated a statistical study of 25,000 microflares \citep{Hannah1, Hannah2}. 
\begin{figure}[!t]
\centering
\includegraphics[width=\textwidth, trim =0 0 5 0, clip = true]{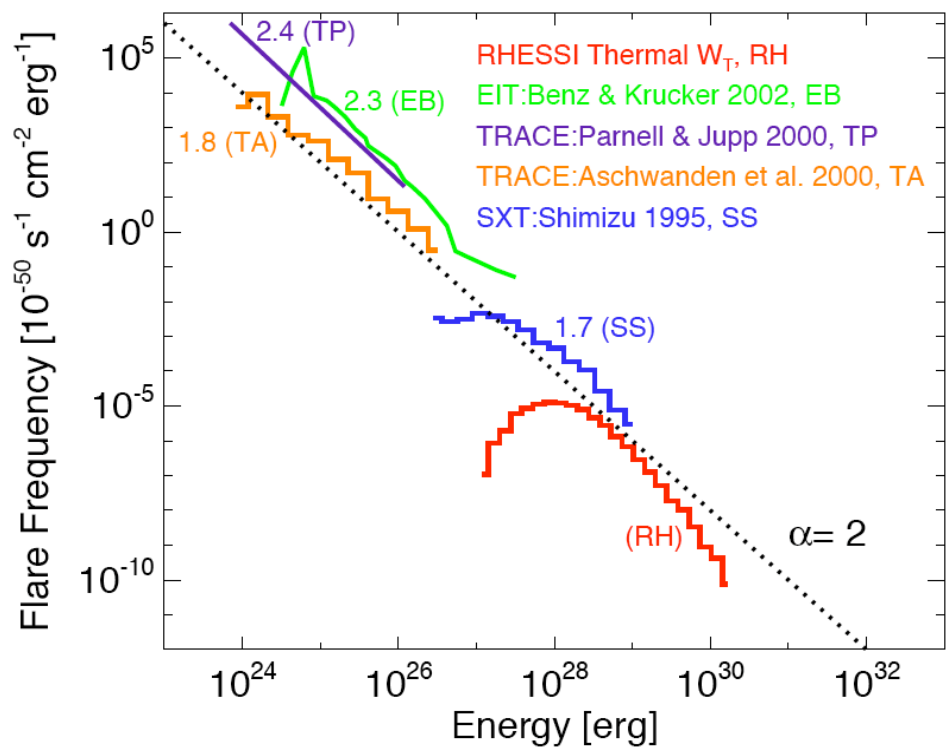} 
\caption{Frequency distribution of the thermal energy of microflares (RH), nanoflares (TA, TP, EB) and active region brightenings (SS). They have been fitted with a power-law function of index 2 (dotted line). \citep{Hannah1}. }
\label{fig:flare_freq}
\end{figure}

The aforementioned investigation by \citet{Hannah1} reveals one of the many important reasons for studying solar flares. The frequency distribution of flares, shown in Figure \ref{fig:flare_freq}, shows that, for the most part, the frequency distribution of flare energy follows a power-law function of index $2.0\pm0.4$. This is important as it is believed that this distribution of energies may be what is driving the heating of the corona. It is believed that the energy of the many nano flares that occur may be sufficient to supply enough energy to the corona to achieve the temperatures we observe. The critical value is the power-law index. If the index is less than 2 then the energy in micro- and nano-flares is insufficient. However if it is $\ge$2 then ``minor'' flares have the potential to be the heating mechanism for the solar corona. The magnitude of flares are classified according to the maximum flux observed by the GOES Satellites. A flare is assigned a class based on a logarithmic scaling shown in Table \ref{table:goes_class}. The highest classification is ``X-class'' with a flux of 10$^{-4}$~W~m$^{-2}$. For each dex from 10$^{-4}$ to 10$^{-8}$~W~m$^{-2}$, the classes $X, M, C, B$ and $A$ are assigned. A secondary classification is used to indicate the level within each class. E.g. an M3.2 flare has a maximum GOES flux of 3.2$\times10^{-5}$~W~m$^{-2}$. See \S \ref{sect:instr_goes} for further details. 

\begin{table}[!t] 
\caption{GOES classification scheme for solar flares.} 

\centering 
\begin{tabular}{c c} 
\hline\hline

GOES class & Minimum flux\\
			& 	[$Watts~m^{-2}]$  	\\
\hline
X			&		10$^{-4}$		\\ 
M			&		10$^{-5}$	 	\\ 
C			&		10$^{-6}$		\\ 
B			&		10$^{-7}$		\\ 
A			&		10$^{-8}$		\\ 
 \hline
\hline
     \end{tabular} 

\label{table:goes_class} 
\end{table}

\begin{figure}[!t]
\centering
\includegraphics[width=0.9\textwidth, trim =25 50 25 50, clip = true]{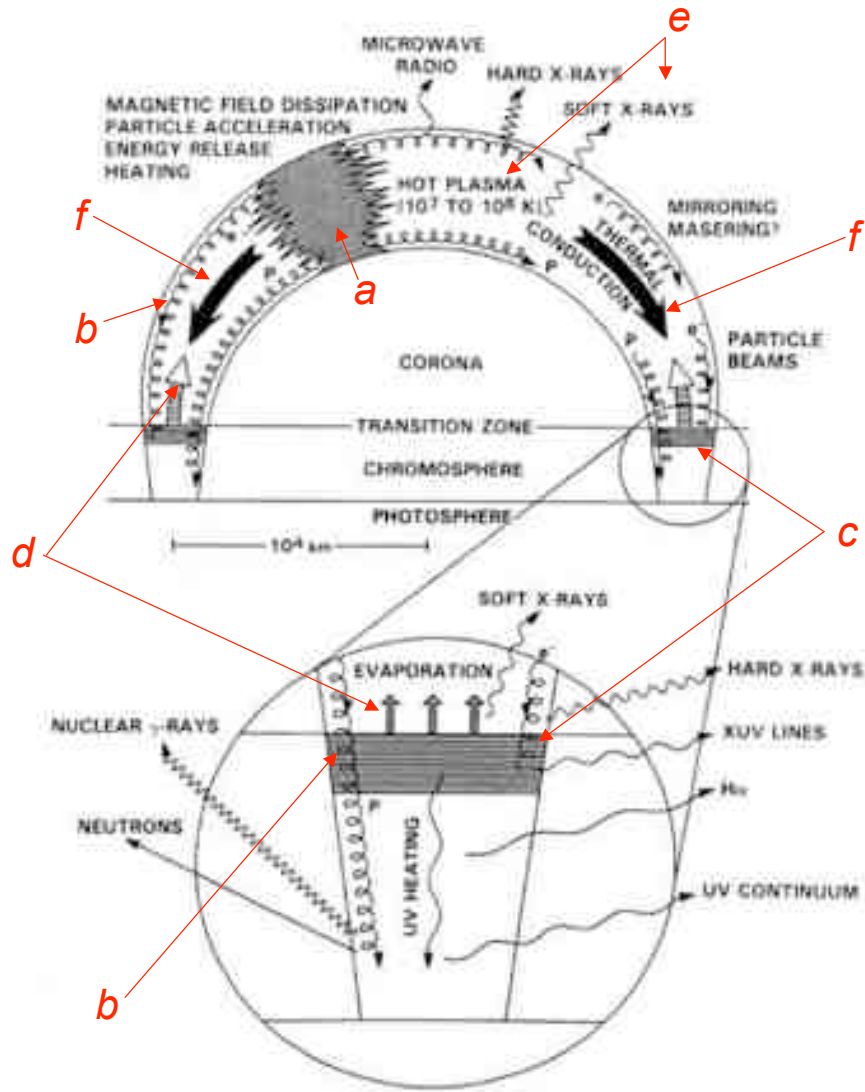} 
\caption{Schematic diagram showing the processes involved in the standard model for solar flares that has been adapted from \citet{Dennis89}. The main processes in the evolution of a solar flare are labelled. $(a)$ is the energy deposition region, $(b)$ shows the path of propagating electrons towards the dense chromosphere $(c)$. Heating of the chromosphere results in chromospheric evaporation $(d)$ which fills the loops with hot, SXR emitting plasma $(e)$. Direct heating of the looptop can result in large pressure gradients that drive conductive fronts $(f)$. }
\label{fig:flare_mod}
\end{figure}

Solar flares are highly complex events that span temperature ranges over four orders of magnitude ($10^4 - 10^7$~K) and energy ranges over six orders of magnitude ($\sim$keV - GeV). These impulsive bursts of energy are some of the most powerful events in the solar system, releasing up to $10^{32}~$ergs (10$^{25}$~J) in tens of minutes \citep{Emslie04}. There is generally considered to be two classifications of solar flares: compact and eruptive. In compact flares little or no loss of material occurs, while in eruptive flares, there is generally an associated coronal mass ejection that carries away plasma and magnetic field from the system. During all types of flares, there exists two phases: the impulsive phase during which plasma is heated to high temperatures and the decay phase during which the flare cools back to equilibrium \citep{Dennis89}. 

Figure \ref{fig:flare_mod} gives an overview of the processes involved in the \emph{standard model} for solar flares. During the impulsive (or rise) phase of a flare, energy is deposited into the loop (marked $a$ on Figure \ref{fig:flare_mod}). It is generally accepted that this is driven by magnetic reconnection (\S \ref{sect:MHD}) and follows the thick target model of \citet{Brown71}. During magnetic reconnection, energy stored in the loops (e.g. as twist) is released and is used to accelerate coronal particles. These particles propagate down magnetic field lines towards the chromosphere (Figure \ref{fig:flare_mod}$b$). The sudden increase in density at the chromosphere (Figure \ref{fig:flare_mod}$c$) results in the beam particles interacting with ambient particles in the chromosphere through coulomb collisions, resulting in the emission of Bremsstrahlung HXR radiation. The energy transferred by the beam particles is absorbed by the chromosphere and, where possible, radiated away. However, should the rate of energy deposition be too great for the chromosphere to efficiently radiate, pressure gradients can build up in the plasma. This results in the expansion of the plasma into the loop (Figure \ref{fig:flare_mod}$d$), filling the loop with hot, SXR and EUV emitting plasma (Figure \ref{fig:flare_mod}$e$) in a process known as chromospheric evaporation. 

\citet{Fisher84, Fisher85, Fisher85b, Fisher85c} were among the first to study chromospheric evaporation in detail from a theoretical perspective. By running simulations to replicate the behaviour of the chromosphere to different fluxes of non-thermal beams of electrons, they established a threshold for the flux of non-thermal electrons required to drive explosive chromospheric evaporation. It was found that fluxes of non-thermal particles of less than 10$^{10}$~ergs~cm$^{-2}$~s$^{-1}$ do not generate sufficient pressure gradients to drive explosive evaporation and the velocities observed were of less than 20~km~s$^{-1}$. This is what is now classified as \emph{gentle chromospheric evaporation}. Gentle evaporation can also be driven by pressure gradients that result from the direct heating of the looptop, driving conduction fronts towards the chromosphere \citep[Figure \ref{fig:flare_mod}$f$;][]{Antiochos78}. This is believed to occur early in the decay phase when conduction is most efficient due to high plasma temperatures. \emph{Explosive chromospheric evaporation} results from fluxes of greater than $\sim$3$\times10^{10}$~ergs~cm$^{-2}$~s$^{-1}$. This drives upflows of hundreds of km~s$^{-1}$ due to the large pressure difference between the heated material and the tenuous corona. Low velocity downflows (tens of km~s$^{-1}$) predicted by their models were later observed by \citet{Zarro89}. Although the velocity of the downflows are orders of magnitude smaller than the upflows, the components and densities of the chromosphere and corona are such that momentum within the system is conserved \citep{Canfield_nat, Teriaca06}. 
 \begin{figure}[!t]
      \includegraphics[width=0.7\textwidth, trim =-60 30 60 50, clip = true]{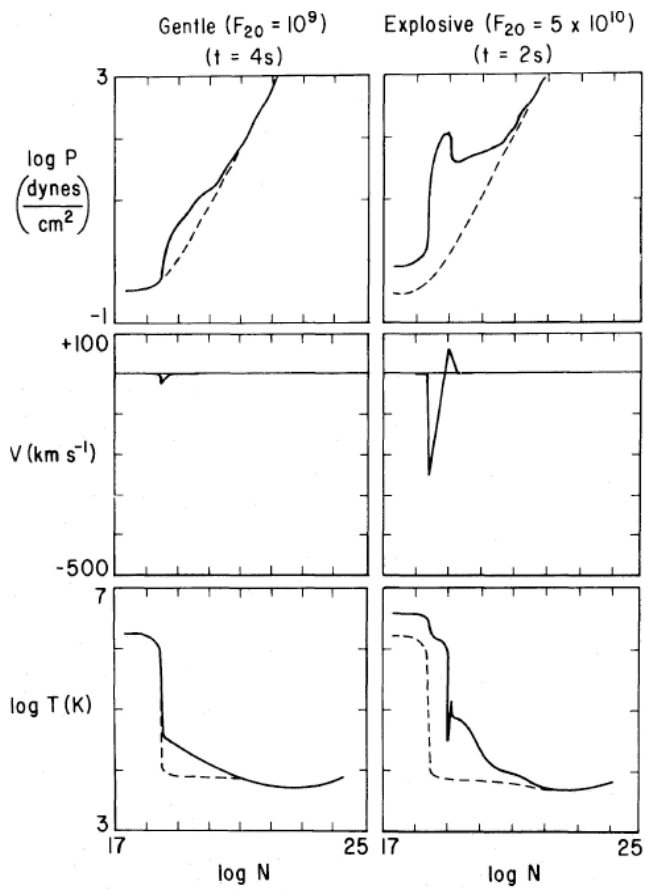}
\caption{Simulation results from \citet{Fisher85} showing temperature ($T$), number density ($n$), pressure ($P$) and velocity ($v$) as a function of column depth measured from the loop apex ($N$) at significant times in the simulations (i.e. 2 and 4 seconds after the electron beam is imposed on the system). The flux of non-thermal electrons at energies less than 20~keV (F$_{20}$) are shown for the explosive (right) and gentle (left) chromospheric evaporation profiles and preflare profiles (dashed curves). Positive velocities correspond to downflows (redshifts) and negative velocities correspond to upflows (blueshifts).}
   \label{fig:fisher}
\end{figure}

The velocities expected from chromospheric evaporation were calculated in \citet{Fisher84}. It begins with the equation of motion in one dimension:
\begin{equation}
\rho\frac{Dv}{Dt} = -\nabla P
\label{eqn:motion}
\end{equation}
where $v$ is velocity, $P$ is pressure and 
\begin{equation}
\frac{D}{Dt} = \left(\frac{\partial}{\partial t} + v\frac{\partial }{\partial z} \right)
\label{eqn:conv}
\end{equation}
is the so-called convective derivative in one dimension. This takes account of the rate of change of the velocity as it moves in space ($z$) and time ($t$). Since the mass density, $\rho$ can be expressed as the product of the mean mass per hydrogen nucleus, $m$ and number density of hydrogen nucleii, $n$, Equation \ref{eqn:motion} can be rewritten as
\begin{equation}
m\left(\frac{\partial v}{\partial t} + v\frac{\partial v}{\partial z}\right) = -\frac{1}{n}\frac{\partial P}{\partial z}
\label{eqn:fisher2}
\end{equation}
If we ignore the time derivative, assuming a constant velocity evolution at any given height in the loop. Assuming that the coronal protons and electrons have the same temperature, the pressure can be defined as $P_{tot} = P_{ion}+P_{electron}$ and since it can be assumed that $P_{ion} = P_{electron} = nk_BT$ in the corona, we can write $P_{tot} = 2nk_BT$ \citep{Kivelson_book, Antiochos76, Krall98}. Therefore, Equation \ref{eqn:fisher2} reduces to:
\begin{equation}
mv\frac{\partial v}{\partial z} = -2k_BT\frac{1}{n}\frac{\partial n}{\partial z}
\end{equation}
Integrating between the chromosphere $(ch)$ and the front of the expanding material in the corona $(co)$, we obtain:
\begin{equation}
\frac{m(v_{co}-v_{ch})^2}{2} = 2k_BT\left[ln(n_{ch}) - ln(n_{co})\right]
\end{equation}
If we assume that once the flare occurs, velocities in the chromosphere are negligible, rearranging gives us 
\begin{equation}
v_{co} = \left[\frac{4kT}{m}ln\left(\frac{n_{ch}}{n_{co}}\right)\right]^{1/2}
\end{equation}
This can be written in terms of the sound speed $c_s = \left(\gamma k_BT/m\right)^{1/2}$ where $\gamma$ is the heat capacity ratio. 
 \begin{equation}
 v_{co} = \left[\left(\frac{6}{5}\right) ln\left(\frac{n_{ch}}{n_{co}}\right)\right]^{1/2}c_s
\end{equation}
 The plasma velocity in the corona works out to be approximately twice the coronal sound speed. Changing the density ratio ($n_{ch}/n{co}$) from e.g. $10^2$ to $10^3$ has little effect on this with $v_{co}$ changing from $\sim2.35c_s$ to $\sim2.88c_s$. \citet{Milligan_explosive} recorded upflow velocities of $230\pm38$~km~s$^{-1}$ and simultaneous downflow velocities of $36\pm16$ and $43\pm22$~km~s$^{-1}$ at chromospheric and transition region temperatures respectively. \citet{Raftery09} observed maximum upflows of $95\pm10$~km~s$^{-1}$ and simultaneous downflows of $23\pm10$~km~s$^{-1}$. Considering the model adopts a constant velocity approach, thus ignoring any ``start-up'' time required to accelerate the plasma from rest, this result is quite reasonable. 
 
The rate at which the loop fills as a result of chromospheric evaporation was found to be closely correlated to the HXR flux \citep{Neupert68}. This effect, named after its discoverer relates the flux ($F$) of hard and soft X-rays as follows:
\begin{equation}
\frac{dF_{SXR}}{dt} \propto F_{HXR}
\end{equation}
This means that the derivative of an SXR light curve can be used to approximate the size and duration of an associated HXR burst. This is believed to stem from the time difference between the heating of chromospheric plasma during chromospheric evaporation and the filling of the loop with evaporated plasma. It takes a finite time for the heated plasma to rise into the loop and reach a sufficient density to begin to emit in SXRs. This suggests that the energetic electrons responsible for the HXR flux are also responsible for chromospheric heating of the flare loop. In reality, multiple-loop systems and other heating mechanisms result in slight deviations from this relationship \citep[e.g.][]{Dennis_Zarro93}. However, it can be a very useful technique for estimating the duration of a HXR burst. The flux of HXRs are often determined using observations from the RHESSI spacecraft. This spacecraft however, is subject to regular night-time passes, or \emph{Earth occultations} due to its low Earth orbit. As the satellite orbits the Earth, it can pass into the night-side of the Sun-Earth line. RHESSI also passes through the South Atlantic Anomaly (SAA). This region is where the Earth's inner Van Allen belt is at its closest. Since the Van Allen belt is aligned with the planet's magnetic axis and not its rotational axis, the belt is closest to the Earth over the south Atlantic ocean. As a satellite passes through this region, it will be exposed to strong radiation, contaminating any observations it is taking. The frequency and duration of the Earth occultations and SAA passes vary throughout the year. However, with a 90 minute orbit and up to 20 minutes of interference, they can still have a dramatic effect on the number of events that are observed unhindered. However, SXR observations from the GOES satellite in 0.5-4~\AA\ and 1-8~\AA\ range are taken every 3 seconds with no interruptions. Therefore, when HXR observations from RHESSI are contaminated, the HXR burst can be approximated by the derivative of the GOES SXR lightcurve. 
\begin{figure}[!t]
\centering
\includegraphics[width=\textwidth, trim =0 0 0 0, clip = true]{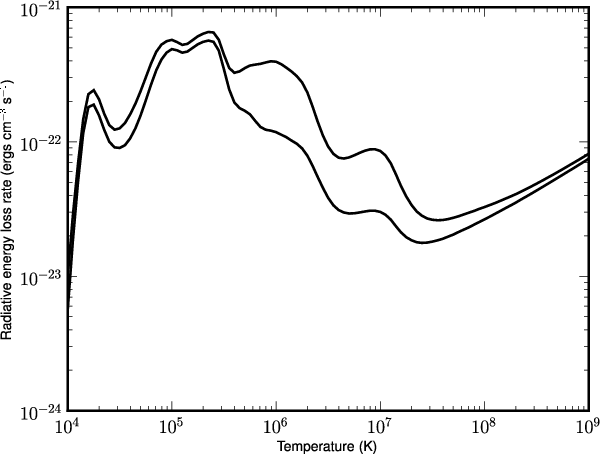} 
\caption{The radiative loss function $(E_R(T))$ as a function of temperature for coronal abundances (upper curve) and photospheric abundances (lower curve) \citep{Dere09}.  }
\label{fig:rtv}
\end{figure}

The peak of the SXR flux often occurs simultaneously with the peak of emission measure. This is not surprising, since the emission will have its highest intensity when the loop is at its densest. This occurs just after the time of maximum temperature. Like the Neupert effect, this delay is as a result of the loop filling time. Around the time of maximum temperature/emission measure conduction is found to be the dominant heat transfer process. Spitzer conductivity \citep{Spitzer56} is widely accepted as the form for conduction in solar plasmas:
\begin{equation}
\label{eqn:conduction}
F_{C} = \kappa_0T^{5/2}\nabla T
\label{eqn:spitzer}
\end{equation}
where $\kappa_0$ is the Spitzer coefficient. Spitzer conductivity is not the most suitable form for solar flares, considering it was devised for non-magnetised plasma in equilibrium. However, in the absence of a more appropriate relation, it is widely accepted within the solar physics community. During the very early decay phase while temperatures are still high, conduction is very efficient in redistributing heat throughout a system, although it does not actually remove heat from the system. If for example, the corona is studied as an isolated system, it can be heated by the conduction of heat into the corona and cooled by the removal of heat to the transition region. During the early decay phase while temperatures are high and the temperature gradient is steep, conduction is very efficient. It is the primary mechanism for heat loss in the corona for the first $\sim$10$^2$~seconds. As the loop thermalises and temperatures begin to fall, its efficiency is reduced. As the temperature approaches $\sim10^5$~K, the efficiency of heat loss by radiation is maximised (Figure \ref{fig:rtv}). The radiative loss rate is given by:
\begin{equation}
\label{eqn:radiative}
\frac{\partial E_{R}}{\partial t} = n_e^2\Lambda(T)
\label{eqn:rad}
\end{equation}
where $\Lambda(T)$ is the optically thin radiative loss function, as shown in Figure \ref{fig:rtv}, making radiation an effective cooling mechanism for $\sim10^3$~seconds in the mid- to late decay phase \citep{Culhane1970, Raftery09}. 

These processes describe the standard model for solar flares. While the standard model is widely accepted, it is also known to include some serious flaws, such as the \emph{number problem}. It has been shown that the number of electrons required to produce the Bremsstrahlung emission by the thick target model is approximately $10^{32} - 10^{37}$ electrons per second \citep{Holman03}. With 10$^9$~cm$^{-3}$ as an average electron density in the corona, a cubic volume of $10^{23}$~cm$^3$ is required to be evacuated into the chromosphere every second. Considering reconnection is believed to take place across length scales of the order of meters (see \S \ref{sect:MHD} for further details), and an average coronal loop is $\sim10^{10}$~cm long, this proves to be a significant problem in the standard thick-target model. The volume of the corona surrounding the entire loop would have to be accelerated and replenished every second. The concept of \emph{return currents} \citep[e.g.][]{Benz_review} has been presented as a possible mechanism of returning accelerated electrons back to the corona. \citet{Fletcher_hudson} have also presented a possible solution to the number problem by moving the site of particle acceleration to the chromosphere, where the number density of electrons is significantly higher. 

\subsection{Magnetohydrodynamics}
\label{sect:MHD}
Magnetic reconnection is widely believed to be the driving force behind solar flares. The interaction of solar magnetic field with plasma can be understood using the equations of magnetohydrodynamics (MHD). These set of equations describe the behaviour of the electric field \textbf{E}, the magnetic field \textbf{B}, current density \textbf{j} and plasma velocity $\textbf{v}$. 

Gauss' law relates the distribution of electric charge $\rho_q$ to the resulting electric field by way of the permittivity of free space, $\epsilon_0$:
\begin{equation}
\label{eqn:max4}
\nabla \cdot \textbf{E} = \frac{\rho_q}{\epsilon_0} 
\end{equation}

Faraday's law of electromagnetic induction states that changing a magnetic field in time will induce an electric field:
\begin{equation}
\label{eqn:max3}
\nabla \times \textbf{E} = -\frac{\partial \textbf{B}}{\partial t}
\end{equation}

Gauss' law for magnetism states that there are no magnetic monopoles:
\begin{equation}
\label{eqn:max2}
\nabla \cdot \textbf{B} = 0
\end{equation}

Amp\`ere's law states that either a current or a time varying electric field will produce a magnetic field. 
\begin{equation}
\label{eqn:max1}
\nabla \times \textbf{B} = \frac{1}{c^2}\frac{\partial \textbf{E}}{\partial t} + \mu_0 \textbf{j}
\end{equation}
where $\mu_0$ is the permeability of free space and is related to the speed of light by $c^2 = 1/\mu_0\epsilon_0$. In addition to Maxwell's equations (Equations \ref{eqn:max4} to \ref{eqn:max1}), Ohm's law states that the current density is related to both the electric field of the plasma and the motion of the plasma at velocity $\textbf{v}$ relative to the magnetic field. 
\begin{equation}
\label{eqn:ohm}
\textbf{j} = \sigma \left( \textbf{E} + \textbf{v} \times \textbf{B} \right)
\end{equation}
where $\sigma$ is electrical conductivity. 

MHD also incorporates the  equations of fluid dynamics for a plasma with density $\rho$ and pressure $P$. The equation of motion for a parcel of fluid states that the rate of change of fluid velocity is governed by the pressure gradients:
\begin{equation}
\label{eqn:fluid1}
\rho\frac{D\textbf{v}}{Dt} = - \nabla P - \rho g + (\textbf{j} \times \textbf{B})
\end{equation}
The mass continuity equation states that the rate at which mass enters a system is equal to the rate at which mass leaves the system:
\begin{equation}
\label{eqn:fluid2}
\frac{D\rho}{Dt} + \rho \nabla \cdot \textbf{v} = 0 
\end{equation}
The energy equation describes the rate of change of energy ($E$) in a plasma as a result of various heat sources and sinks, such as the heating rate ($Q$), the divergence of the conductive flux ($F_c$) and the radiative loss rate ($E_r$). 
\begin{equation}
\frac{\partial E}{\partial t} = Q - \nabla \cdot F_c - E_r
\label{eqn:mhd_energy}
\end{equation}

\begin{figure}
\centering
\includegraphics[width=1\textwidth, trim =0 0 0 0, clip = true]{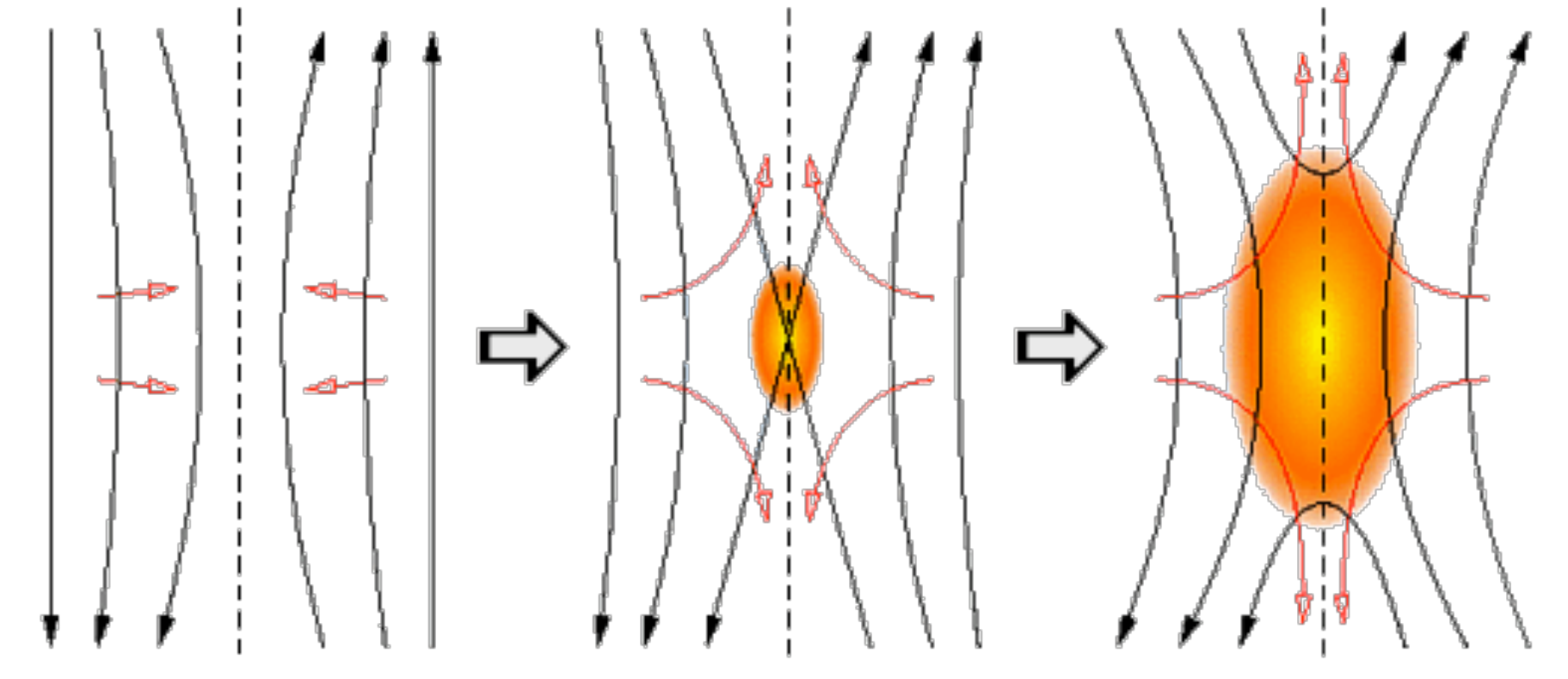} 
\caption{Schematic diagram of magnetic reconnection. Field lines approach the diffusion region where they reconnect. The resulting topology is energetically more stable.}
\label{fig:rxn}
\end{figure}

Assuming non-relativistic velocities, Equation \ref{eqn:max1} can be approximated as:
\begin{equation}
\label{eqn:max1_aprox}
\nabla \times \textbf{B} = \mu_0 \textbf{j}
\end{equation}
Substituting Equations \ref{eqn:ohm} and \ref{eqn:max1_aprox} into Equation \ref{eqn:max3}, gives:
\begin{equation}
\label{eqn:before_eta}
\frac{\partial \textbf{B}}{\partial t} = \nabla \times \left( \textbf{v} \times \textbf{B} - \frac{1}{\sigma \mu_0} \nabla \times \textbf{B} \right)
\end{equation}
Utilising the identity:
\begin{equation}
\label{eqn:iden}
\nabla \times \left( \nabla \times \textbf{B} \right) = \nabla \left( \nabla \cdot \textbf{B} \right) - \left( \nabla \cdot \nabla \right) \textbf{B}, 
\end{equation}
and recalling Equation \ref{eqn:max2}, Equation \ref{eqn:before_eta} can be rewritten as what is known as the induction equation:
\begin{equation}
\label{eqn:induction}
\frac{\partial \textbf{B}}{\partial t} = \nabla \times \left( \textbf{v} \times \textbf{B} \right) + \eta \nabla^{2}\textbf{B} 
\end{equation}
where magnetic diffusivity is given by $\eta = 1/\sigma \mu_0$. The two terms on the right hand side of Equation \ref{eqn:induction} describe advection and diffusion respectively. The ratio of these terms is known as the magnetic Reynolds number, $R_{m}$:
\begin{equation}
\label{eqn:rm}
R_{m} = \frac{ \nabla \times \left( \textbf{v} \times \textbf{B} \right) }{\eta \nabla^{2}\textbf{B}}  \approx \frac{v L}{\eta} 
\end{equation}
For a perfectly conducting plasma, i.e. $ \sigma \to \infty$ and $\eta \to 0$, the change in the magnetic field is completely dominated by advection and the field is carried along by the flowing plasma. Thus Equation \ref{eqn:induction} can be approximated by:
\begin{equation}
\label{eqn:advection}
\frac{\partial \textbf{B}}{\partial t} \approx \nabla \times \left( \textbf{v} \times \textbf{B} \right) 
\end{equation} 
and the fields are said to be ``frozen-in'' to the plasma. Although the plasma is not perfectly conducting, in most astrophysical plasmas, $R_{m} \gg 1$. For example, in the solar corona, $\eta \approx 1$~m$^{2}$~s$^{-1}$, $L \approx 10^{5}$~m, and $v \approx 10^{3}$~m~s$^{-1}$ gives a value of $R_{m} \approx 10^{8}$. The primary exception to this is in the presence of large magnetic field gradients. Under these circumstances, the changing field can be approximated as:
\begin{equation}
\label{eqn:diffusion}
\frac{\partial \textbf{B}}{\partial t} \approx  \eta \nabla^{2}\textbf{B} 
\end{equation}
Here, $R_{m} \ll 1$ and diffusion dominates over advection, allowing the magnetic field to slip through the plasma. This is one of the conditions required for magnetic reconnection. A simple schematic of the processes involved in magnetic reconnection is shown in Figure \ref{fig:rxn}. When regions of opposite polarity magnetic flux are in close proximity to each other, large magnetic field gradients will be established as the value of the field goes from positive in one region, to zero at the neutral line to negative at the opposing flux. The large gradients in the magnetic field result in $R_m \ll 1$. This allows the magnetic field to diffuse through the plasma and reconnect, resulting in an energetically more favourable topology. In doing so, the plasma is ejected perpendicular to the in-flowing fields which creates a drop in pressure. This in turn pulls more plasma and magnetic field into the diffusion region, repeating the process.

Since Equation \ref{eqn:induction} can be approximated as Equation \ref{eqn:diffusion} in the diffusion region, the timescales can be estimated as 
\begin{equation}
\tau_{rxn} \approx \frac{L^2}{\eta}
\label{eqn:rxn_time}
\end{equation}
The timescales for reconnection are known to be on the order of seconds. Therefore, to satisfy Equation \ref{eqn:rxn_time}, there is a requirement for very short length scales (on the order of meters). The reconnection model presented by Sweet and Parker \citep{Sweet58, Parker57} utilised a thin current sheet with length of the order of a coronal loop along which reconnection can take place. With $L\sim1$~Mm, the reconnection rate is of the order of $10^{20}$~seconds, or roughly a billion years! Since reconnection on the Sun requires timescales on the order of seconds, \citet{Petschek64} proposed a system where the diffusion region is of the order of meters, a fraction of a scale length. This means that direct observational evidence of magnetic reconnection is not yet possible. The low density of the diffusion region, combined with its small length scale is beyond the capability of current instruments. However, indirect measurements of this phenomenon are widespread in literature, from solar jets \citep{Bain09} to magnetospheres \citep{Slavin09} and in laboratory plasmas \citep[e.g.][]{Cothran03}.

\subsection{Hydrodynamic modelling}
While the use of MHD is essential to the understanding of the behaviour of field and plasma, to quantitatively solve the equations of MHD in 3 dimensions is a very complex task that requires a huge amount of computing power. As such, approximations are often made to these equations in 0- and 1- dimensional codes. Even the computation load of 1-D codes is not trivial, however the use of 0-D hydrodynamic simulations are very quick ($\sim$seconds on a personal computer versus hours to weeks on a cluster for 1-D). The term ``0-D'' stems from the fact that these models sacrifice spatial resolution in favour of efficiency by assuming that any energy in a coronal loop undergoing magnetic reconnection is uniformly distributed in the corona. This is a reasonable representation since temperature, density and pressure are approximately uniform along the magnetic field, with the exception of the steep gradients in the transition region. It has been shown that these codes are very robust in their responses and compare very well to more detailed 1-D codes \citep[see e.g.][for details of comparison]{Klim07}.

In general, hydrodynamic models begin with the 1-D time dependent version of the energy equation, given in Equation \ref{eqn:mhd_energy}.
\begin{equation}
\frac{\partial E}{\partial t} = Q - \frac{\partial}{\partial s} \left(\kappa_0 T^{\frac{5}{2}}\frac{dT}{ds}\right) - n_e^2\Lambda(T)
\label{eqn:basic_hydro}
\end{equation}
This equation states that the rate of change in energy of a system $(E)$ is balanced by the heating rate $(Q)$, and the loss rates by conduction radiation (two right hand terms which are defined in Equations \ref{eqn:spitzer} and \ref{eqn:rad}).

\subsubsection{The Cargill model}
\label{sect:cargill}
The simple model presented by \citet{Cargill93, Cargill94} is an effective method for making a first approximation for the cooling timescale of a flare. Beginning with the basic energy equation \ref{eqn:basic_hydro} and assuming velocities generated by evaporation are much smaller than the sound speed, we can substitute for the different terms. Energy is given by $3/2nk_BT$, conduction is approximated from the Spitzer formula given in Equation \ref{eqn:spitzer} as $2/7 \kappa_0T^{7/2}/L^2$. 
\begin{equation}
\frac{3}{2}\frac{\partial \left[nk_BT\right]}{\partial t} = Q - \frac{2}{7}\kappa_0\frac{T^{7/2}}{L^2} -n^2\Lambda(T)
\label{eqn:cargill1}
\end{equation}

The temperature evolution of a system can be obtained from Equation \ref{eqn:cargill1}. Following work laid out in \citet{Antiochos78} and \citet{Antiochos80}, the assumption is made that at any given time, cooling is done by only a single mechanism, i.e. in the early decay phase radiation is ignored and in the late decay phase, conduction is ignored. For example, the conductive cooling time can be calculated by setting $Q$ and the radiative loss rate to 0. This leaves
\begin{equation}
\frac{3}{2}\frac{\partial \left[nk_BT\right]}{\partial t} = - \frac{2}{7}\kappa_0\frac{T^{7/2}}{L^2}
\end{equation}
Rearranging this equation to give
\begin{equation}
\frac{3nk_B}{2} \int_{T_{0}}^T T^{-7/2} dT = \frac{-2}{7} \frac{\kappa_0}{L^2} \int_{t_{0}}^t dt
\end{equation}
we can integrate to get the temperature evolution for the conductive phase of a flare:
\begin{equation}
T(t) = T_0\left(1+\frac{t}{\tau_{c_0}}\right)^{-2/7}
\label{eqn:tc}
\end{equation}
The temperature evolution for the radiation phase is obtained in a similar fashion and is given by
\begin{equation}
T(t) = T_0\left(1-\frac{3}{2}\frac{t}{\tau_{r_0}}\right)
\label{eqn:tr}
\end{equation}
In Equations \ref{eqn:tc} and \ref{eqn:tr}, the parameters $\tau_c$ and $\tau_r$ correspond to the conductive and radiative cooling timescales:
\begin{equation}
\label{equation:con_cool_time}
\tau_{c} = \frac{3nk_BT}{\kappa_0T^{7/2}/L^2} 
\end{equation}
and 
\begin{equation}
\label{equation:rad_cool_time}
\tau_{r} = \frac{3nk_BT}{n^2\Lambda(T)} 
\end{equation}
respectively. 

\begin{figure}
\centering
\includegraphics[width=0.6\textwidth, trim =40 40 260 40, clip = true]{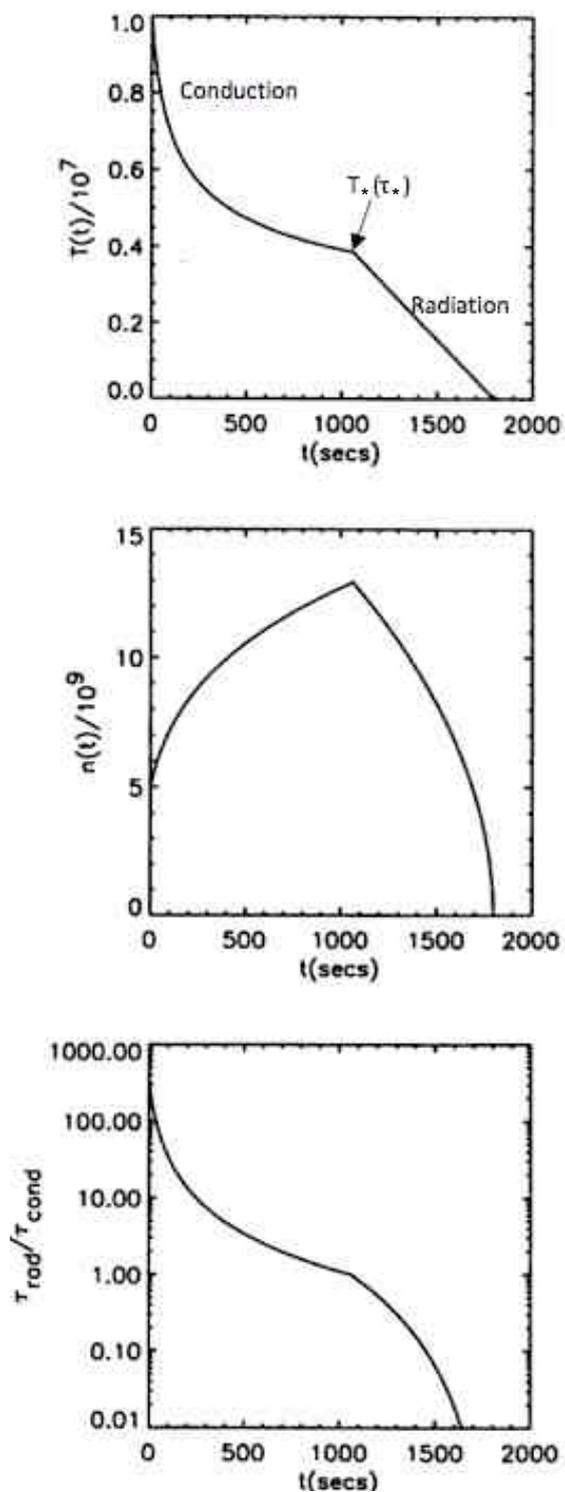} 
\caption{Sample plots from \citet{Cargill94} showing the cooling profile of a flare of density $5\times10^9$~cm$^{-3}$, initial temperature of $10^7$~K and a half length of $2.5\times10^9$~cm. The top two panels show the temperature and density evolution of the flare. The cutoff time and temperature ($\tau_{*}, T_{*}$) are shown in the top panel. The bottom panel shows the instantaneous radiative and conductive cooling times.  }
\label{fig:car_fig}
\end{figure}

The time and temperature at which the cooling mechanism dominance changes, $\tau_{*}$ and $T_{*}$ respectively, can then be expressed as: 
\begin{equation}
\label{equation:crit_time}
\tau_{*} = \tau_{c_{0}} \left[ \left( \frac{\tau_{r_{0}}} {\tau_{c_{0}}}   \right)^{7/12} -  1 \right],
\end{equation}
and
\begin{equation}
\label{equation:crit_temp}     
T_{*} = T_{0}\left( \frac{\tau_{r} }{\tau_{c}}\right)^{-1/6}
\end{equation}
$T(t)$ is shown in the top panel of Figure \ref{fig:car_fig} with $T_{*}(\tau_{*})$ highlighted as the break between conductive and radiative cooling. Note that this occurs in conjunction with the peak density (middle panel, Figure \ref{fig:car_fig}). The bottom panel shows the ratio of $\tau_r/\tau_c$. The radiative cooling time is significantly longer than the conductive cooling time during the first $\sim1000$ seconds of the flare. After this time, the radiative cooling time becomes shorter than the conductive timescale, making radiation the dominant cooling mechanism. 

One assumption of the Cargill model is that of independent cooling mechanisms. It is assumed that while conduction is efficiently removing heat from the system, radiation is negligible and is therefore ignored, and vice versa. This is a reasonable thing to assume at the beginning and end of the decay phase. However, there is a time when both conduction and radiation have significant contribution to the removal of energy from the system. Therefore, during this time, a single loss mechanism will underestimate the energy removed from the system and potentially result in a higher temperature or longer cooling time.

\subsubsection{Enthalpy Based Thermal Evolution of Loops (EBTEL)}
\label{sect:ebtel}
The Enthalpy Based Thermal Evolution of Loops model \citep[EBTEL;][]{Klim07} accounts for both conductive and radiative losses throughout the lifetime of the flare, thus eliminating some of the limitations of the Cargill model. EBTEL takes explicit account of the important role of enthalpy in the energetics of evolving loops. The basic assumption of the EBTEL model is that $n, P$ and $T$ in the corona can be represented by spatial averages since they generally vary by less than a factor of 3 through the corona. The base of the corona is defined to be the point at which conduction switches from a heating term in the transition region to a cooling term in the corona. This is based on the categorisation of the enthalpy flux. As the heating rate increases during, say, a flare, the chromosphere is unable to radiate all the absorbed energy, and so plasma rises into the corona. Therefore, an excess in heat flux can be associated with the impulsive phase of a flare. Conversely, when the heating rate decreases, the coronal temperature begins to fall, creating pressure gradients in the loop and causing plasma to flow towards the footpoints. Thus a deficit of heat flux is associated with plasma cooling. 

By incorporating the enthalpy $H$ of the system, we take account of the work done on and by the system:
\begin{equation}
\frac{\partial H}{\partial t} = Q -\frac{\partial}{\partial s}\left( \kappa_0 T^{5/2} \frac{\partial T}{\partial s} \right) - n_e^2\Lambda(T)
\label{eqn:eb1}
\end{equation}
where $Q$ is the combined direct and non-thermal heating rate, $s$ is distance along the loop, $\kappa_0$ is the Spitzer conductivity coefficient, $T$ is the temperature, $n_e$ is the electron density and $\Lambda(T)$ is the radiative loss function. Considering the enthalpy can be written the sum of the internal energy ($E$) and the work done ($PV$), we can write
\begin{equation}
\frac{\partial E}{\partial t} + v\frac{\partial [E + P]}{\partial s} = Q - \frac{\partial}{\partial s}\left( \kappa_0 T^{5/2} \frac{\partial T}{\partial s}  \right) - n_e^2\Lambda(T)
\label{eqn:eb2}
\end{equation}
Substituting for the thermal and kinetic energies:
\begin{equation}
\label{equation:energy}
E = \frac{3}{2}P + \frac{1}{2}\rho v^{2}
\end{equation}
we can write 
\begin{eqnarray}
\frac{3}{2}\frac{\partial P}{\partial t}+ \frac{1}{2}\frac{\partial \left(\rho v^2\right)}{\partial t} &=& Q - \frac{\partial \left(Pv\right)}{\partial s} - \frac{3}{2}\frac{\partial Pv}{\partial s} - \frac{1}{2} \frac{\partial \left(\rho v^2\right)v}{\partial s} \nonumber \\
& &-\frac{\partial}{\partial s}\left( \kappa_0 T^{5/2} \frac{\partial T}{\partial s} \right)  - n_e^2\Lambda(T)
\end{eqnarray}
Assuming a subsonic flow, the kinetic terms can be neglected giving
\begin{equation}
\frac{3}{2}\frac{\partial P}{\partial t} = Q - \frac{5}{2}\frac{\partial Pv}{\partial s} - \frac{\partial}{\partial s}\left( \kappa_0 T^{5/2} \frac{\partial T}{\partial s}  \right)  - n_e^2\Lambda(T)
\label{eqn:eb_final}
\end{equation}

EBTEL makes the assumption that the energy equation can be solved independently for the corona and the transition region. If we designate the subscript ``$0$'' to define the base of the corona (see above for definition) and assume a coronal half length of $L$ and a transition region half length of $l$, integrating Equation \ref{eqn:eb_final} over these regions gives: 
\begin{equation}
\frac{3}{2}L \frac{\partial \bar{P}}{\partial t} \approx \bar{Q}L + \frac{5}{2}P_0v_0 + F_{c_{0}} - F_{r_{c}}
\label{eqn:eb_co}
\end{equation}
\begin{equation}
\frac{3}{2}l \frac{\partial \bar{P}}{\partial t} \approx \bar{Q}l - \frac{5}{2}P_0v_0 - F_{c_{0}} - F_{r_{tr}}
\label{eqn:eb_tr}
\end{equation}
for the corona and transition region respectively where the overbar refers to spatially averaged values along the appropriate region. $F_r$ refers to the radiative cooling rate per unit area and $F_c$ to the conductive flux. Note the difference in signs for the conduction term. This implies that heat conducted \emph{out of} the corona is conducted \emph{into} the transition region. It is assumed that the enthalpy flux is negligible at the base of the transition region (i.e. the majority of the heat flux energy is distributed through the upper layers, heating each consecutive layer). Since $l\partial \bar{P}_{tr}/\partial t \ll L\partial \bar{P}_{c}/\partial t$, it is also assumed that pressure in the transition region is constant and we can write Equation \ref{eqn:eb_tr} as
\begin{equation}
\frac{5}{2}P_0v_0 = -F_{c_{0}} - F_{r_{tr}}
\label{eqn:eb_tr2}
\end{equation}
Equation \ref{eqn:eb_tr2} directly describes the directionality of the heat flux. When $|F_{c_{0}}| > F_{r_{tr}}$ then transition region radiation is insufficient to remove the heat and it is conducted into the corona. If, however $|F_{c_{0}}| < F_{r_{tr}}$, then the transition region is radiating efficiently enough to draw heat from the corona. This, combined with Equation \ref{eqn:eb_co} leads to 
\begin{equation}
\frac{d\bar{P}}{dt} \approx \frac{2}{3}\left[ \bar{Q} - \frac{1}{L}\left(F_{r_{c}}+F_{r_{tr}}\right)\right]
\label{eqn:eb_p_pre}
\end{equation}
One important assumption made when solving Equation \ref{eqn:eb_p_pre} is 
\begin{equation}
\frac{F_{r_{tr}}}{F_{r_{c}}} = c_{rad}
\label{eqn:eb_const}
\end{equation}
at all times, where $c_{rad}$ is a constant of the system. This is a reasonable approximation during the decay phase of the flare since the long decay timescales mean the system does not deviate far from static equilibrium at any given time. However, during the impulsive phase this is not necessarily valid. The quickly changing density of the system can invalidate this approximation. Despite comparisons to 1-D hydro models that suggest that this ratio does not have a significant effect on the resulting parameters \citep{Klim07}, an investigation carried out by \citet{Adamakis08} revealed that the resulting output parameters are in fact sensitive to this ratio. 

Since $F_{r_{c}} = \bar{n}_e^2\Lambda(\bar{T})L$ and from Equation \ref{eqn:eb_const}, $F_{r_{tr}}$ can be written in terms of $F_{r_{c}}$, the coronal pressure can be written as:
\begin{equation}
\frac{d\bar{P}}{dt} \approx \frac{2}{3}\left[ \bar{Q} - \frac{\bar{n}_e^2\Lambda(\bar{T})L}{L}\left(1+c_{rad}\right)\right]
\label{eqn:eb_p}
\end{equation}

The conservation of mass requires that the enthalpy flux through the transition region be approximately constant. Since the total mass contained within a length $L$ will change with the evaporation of material into the corona, this must equal the electron flux from the transition region to the corona (i.e. through the base of the corona), we can write:
\begin{equation}
\frac{\partial \left( \bar{n}L\right)}{\partial t} = nv \approx J_0
\end{equation}
Combining this with Equation \ref{eqn:eb_tr2} and considering $P=2nk_BT$, we can describe the rate of change of density as 
\begin{equation}
\frac{\partial \bar{n}}{\partial t} = \frac{1}{5k_BT_0L} F_{r_{tr}}(c_{rad}+1)
\label{eqn:eb_n}
\end{equation}

Finally, the temperature evolution can be obtained from the ideal gas law as:
\begin{equation}
\frac{d\bar{T}}{dt} \approx \bar{T} \left( \frac{1}{\bar{P}} \frac{d\bar{P}}{dt} - \frac{1}{\bar{n}} \frac{d\bar{n}}{dt} \right)
\label{eqn:eb_t}
\end{equation}
The evolution of the average coronal pressure, density and temperature can therefore be obtained as a function of time from Equations \ref{eqn:eb_p}, \ref{eqn:eb_n} and \ref{eqn:eb_t}. 

\begin{figure}
\centering
\includegraphics[width=\textwidth, trim =35 40 40 30, clip = true]{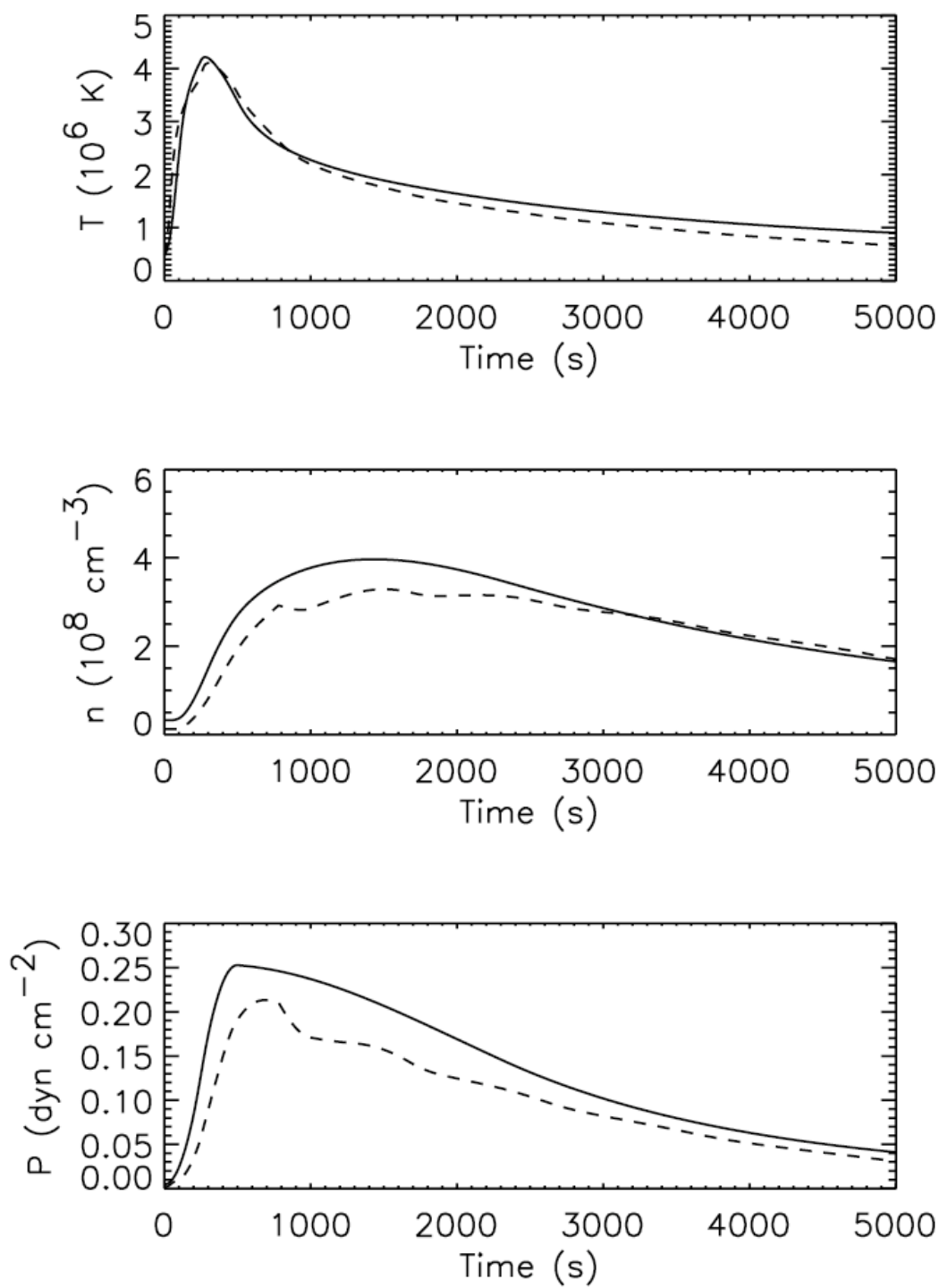} 
\caption{Sample plots from \citet{Klim07} showing the coronal averaged temperature, density and pressure of an impulsively heated nanoflare. The solid lines are the EBTEL simulation while the dashed lines are from a more complex 1-D simulation for comparison \citep{Antiochos91}. }
\label{fig:klim_fig}
\end{figure}

Figure \ref{fig:klim_fig} shows the time evolution of the three parameters: temperature, density and pressure of a nanoflare heated impulsively for 500 seconds. The comparison of the EBTEL simulation (solid line) to a 1-D simulation (dashed line) shows very good agreement between the two models, despite the significantly simpler approach taken with EBTEL. It is also interesting to note that unlike Figure \ref{fig:car_fig}, there are no discontinuities in the EBTEL evolution. This is a significant improvement on the Cargill model. The efficiency of combining losses by conduction and radiation throughout the flare are evident from the sharp reduction in temperature in the first $\sim$600 seconds of the decay phase. Note the change in slope of the temperature evolution around the time of the density maximum. As the temperature falls, conduction is no longer as significant. Simultaneously, the density peaks and radiation becomes the dominant loss mechanism.  

The basic model scenario is such that the corona is heated primarily by conduction fronts (direct heating). However, it is also possible to use EBTEL to approximate the combined effects of direct heating and heating by a non-thermal electron beam (non-thermal heating). The inclusion of the non-thermal particle heating is not thorough but is sufficient to give a reasonable estimate of the effect of a beam. It is assumed that any accelerated particles originate from within the system, specifically the corona. Thus, the number density of the entire loop does not change. It is also assumed that the accelerated particles can stream freely and not interact with any particles until they reach the chromosphere. The final, and most concerning approximation, is that all energy found in the beam is used to evaporate chromospheric plasma upwards into the loop. This is a concern because it is well known that propagating particles result in the expansion of plasma downwards as well as upwards \citep[e.g.][]{Fisher84, Milligan_explosive, Raftery09}. Chromospheric emissions are another result of accelerated particles interacting with the chromosphere, albeit a tiny fraction ($10^{-5}$) of their energy \citep{Dennis_nugget}.

The EBTEL model can accommodate any combination of direct and non-thermal heating functions, though it cannot accommodate purely non-thermal heating. Following the procedure above, the average temperature, density and pressure in the corona as a function of time are calculated. A secondary, but important result is the evolution of the conductive and radiative losses in the coronal portion of the loop, allowing the user assess to the efficiency of the cooling mechanisms for the duration of the event. Other model options, including the differential emission measure of the loop may also be calculated. However, the inclusion of a non-thermal beam results in ambiguous transition region differential emission measure values, as the dependence on deposition depth are not considered.

\section{Eruptive flares and CMEs}
\label{sect:CMEs}
\begin{figure}
\centering
\includegraphics[width=0.7\textwidth, trim =150 100 300 250, clip = true]{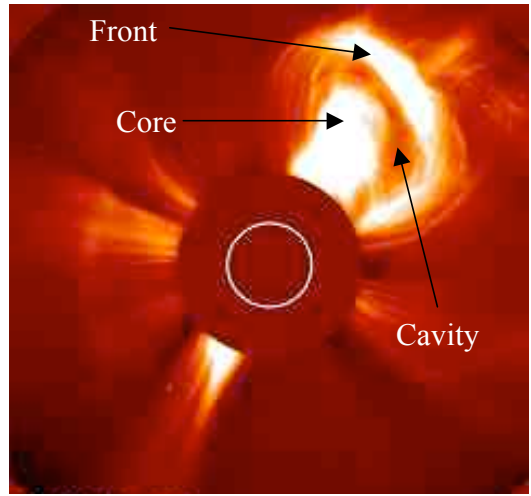} 
\caption{A coronagraph image of a typical CME. The bright front, dark cavity and bright core are highlighted. }
\label{fig:cme}
\end{figure}

Solar flares are well known to be associated with coronal mass ejections (CMEs). CMEs are ejections of material, magnetic field and energy from the solar corona. They are generally bulbous structures threaded with magnetic field that grow radially as they propagate away from the Sun. CMEs occur across many different length- and time-scales. CMEs appear on ever increasing lengthscales as the propagate away from the Sun. They are most frequently observed using white light emission by imaging the Thompson scattered light of the K-corona (photospheric emission scattered off free electrons inside the CME). They generally have a three part structure: a bright leading edge, a dark sparse cavity and a bright dense core. The three components are highlighted in Figure \ref{fig:cme}.

The association of CMEs with solar flares is widely known but not well understood. \citet{Gosling93} postulated that geomagnetic storms are produced by CMEs and not, as previously believed, by solar flares. This declaration, dubbed ``The Solar Flare Myth'', led to the misunderstanding that solar flares were not an important aspect of solar physics research as they had no effect on life on Earth. This belief divided the community and was contested on numerous occasions (e.g. \citeauthor{Hudson95} \citeyear{Hudson95}; \citeauthor{Reames95} \citeyear{Reames95}; \citeauthor{Svestka01} \citeyear{Svestka01}). The significance of solar flares has since been restored and was summarised nicely by  \citet{Svestka01}: ``It is misleading to claim that flares are not important in solar-terrestrial relations. Although they do not cause the CME phenomenon that propagates from the Sun eventually hitting the Earth, they are excellent indicators of coronal storms and actually indicate the strongest, fastest and most important storms.'' Since then, the ideas connecting flares and CMEs have left the ``cause and effect'' paradigm, and it is now widely believed that eruptive flares and CMEs both result from the same driving mechanism \citep{Zhang2001}. 

\begin{figure}
\centering
\includegraphics[width=0.8\textwidth, trim =45 40 40 75, clip = true]{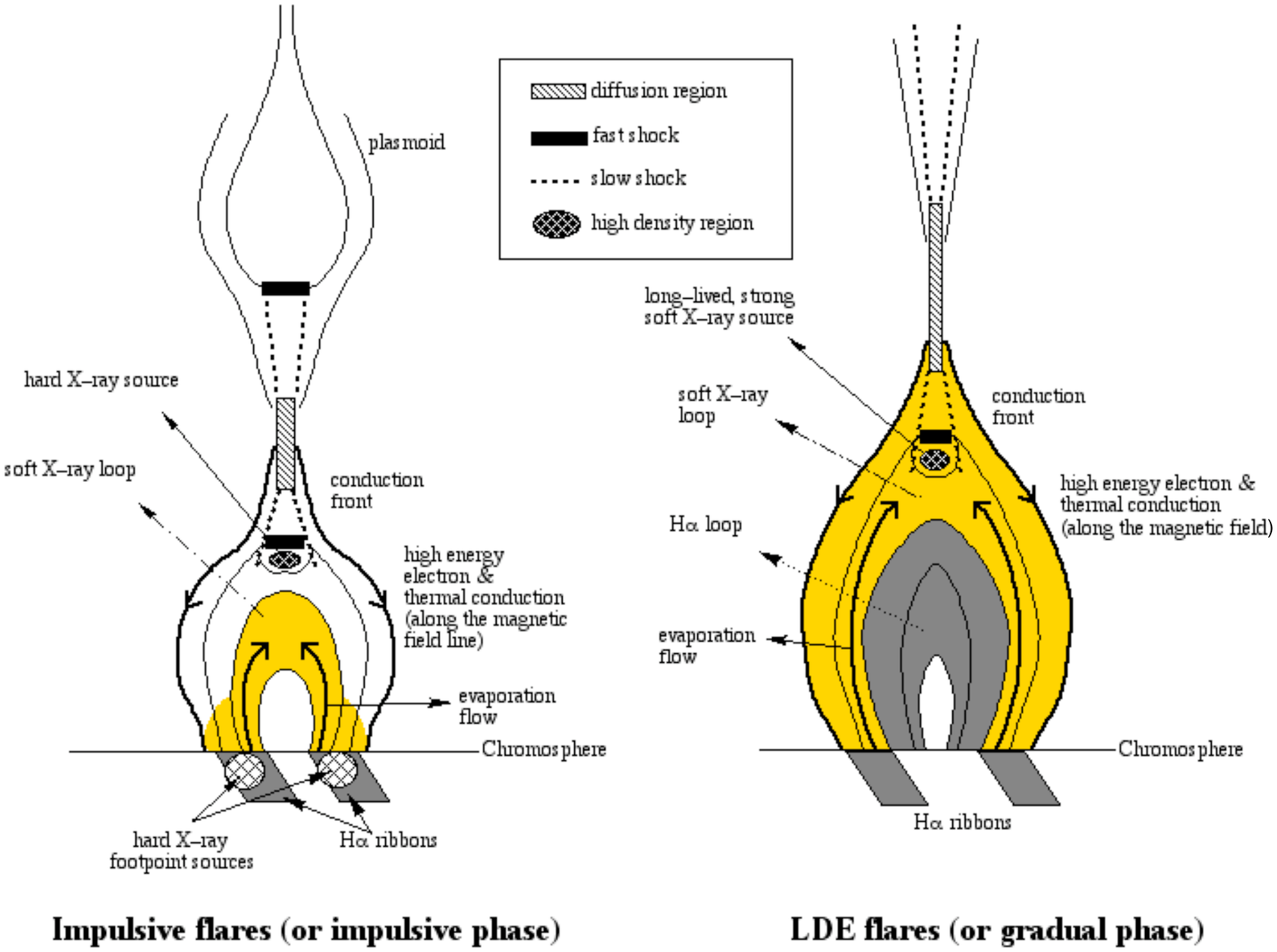} 
\caption{The CSHKP model, adapted from \citet{Magara96}. }
\label{fig:cshkp}
\end{figure}

The combined work of \citet{Carmichael64}, \citet{Sturrock66}, \citet{Hirayama74} and \citet{Kopp76}, known as the CSHKP model, began the unification of the flare-CME theories. This model began with the proposition that low lying coronal loops and open overlying field lines could create an upside-down Y-type magnetic topology (Figure \ref{fig:cshkp}, left panel). Following this, shearing motions may trigger a tearing instability near the neutral line located above the loop (within the diffusion region). This can result in magnetic reconnection and the acceleration of particles which propagate to lower altitudes and result in a solar flare beneath the reconnection region. The sling shot effect of the reconnected fields can result in shocks perpendicular to the incoming field. Accelerated particles can cause HXR footpoints and chromospheric evaporation of hot material into the loop to create SXR emitting loops beneath the reconnection region as with confined flares. Thin target emission is believed to be the cause of short lived HXR sources that form at the top of the loop/base of the reconnection region and may be the signature of the reconnection site \citep[e.g.][]{Krucker2009}. The reconnection and reorganisation of the field results in the next set of field lines being drawn into the diffusion region to be reconnected. Thus a run-away process begins. With each set of reconnecting field lines, the CME is freed some more: as more field reconnects there is less overlying field preventing the propagation of the CME and it can therefore rise faster. This in turn results in faster reconnection and an explosive burst of acceleration occurs. This acceleration burst has been found to be closely associated with the hard X-ray profile of the underlying eruptive flare \citep{Temmer08}. As successive field lines are reconnected, the post flare loops appear to rise as they are heated into the passbands of our instruments. Thus, as the CME rises, the post flare arcade is also observed to rise beneath the CME.

\subsection{CME initiation}
\label{sect:cme_models}
There have been many models put forward to try to simulate the behaviour of eruptive flares and CMEs. A successful model must be able to reproduce the observations as closely as possible. However, with ever evolving and improving observations, it is difficult to simulate every aspect of the system. This is especially difficult considering the large energy and length scales involved. There have been many reviews on this topic. For example \citet{Forbes_cme06} contains an overview of the entire research field, including CME initiation, propagation, structure, modelling and shock formation. The review by \citet{Klimchuk_cme01} contains a description of CME initiation models using illustrations analogous to simple mechanical systems such as springs, pulleys and bombs. While this is instructive, it can also be misleading in how simplified the models appear to be. The author believes that \citet{Moore06} strikes a good balance between conceptual progression and physical description. \citet{Moore06} describe three independent methods of initiating a CME: internal tether cutting, external tether cutting and ideal MHD instability. 

\begin{figure}
\centering
\includegraphics[width=\textwidth, trim =20 00 15 0, clip = true]{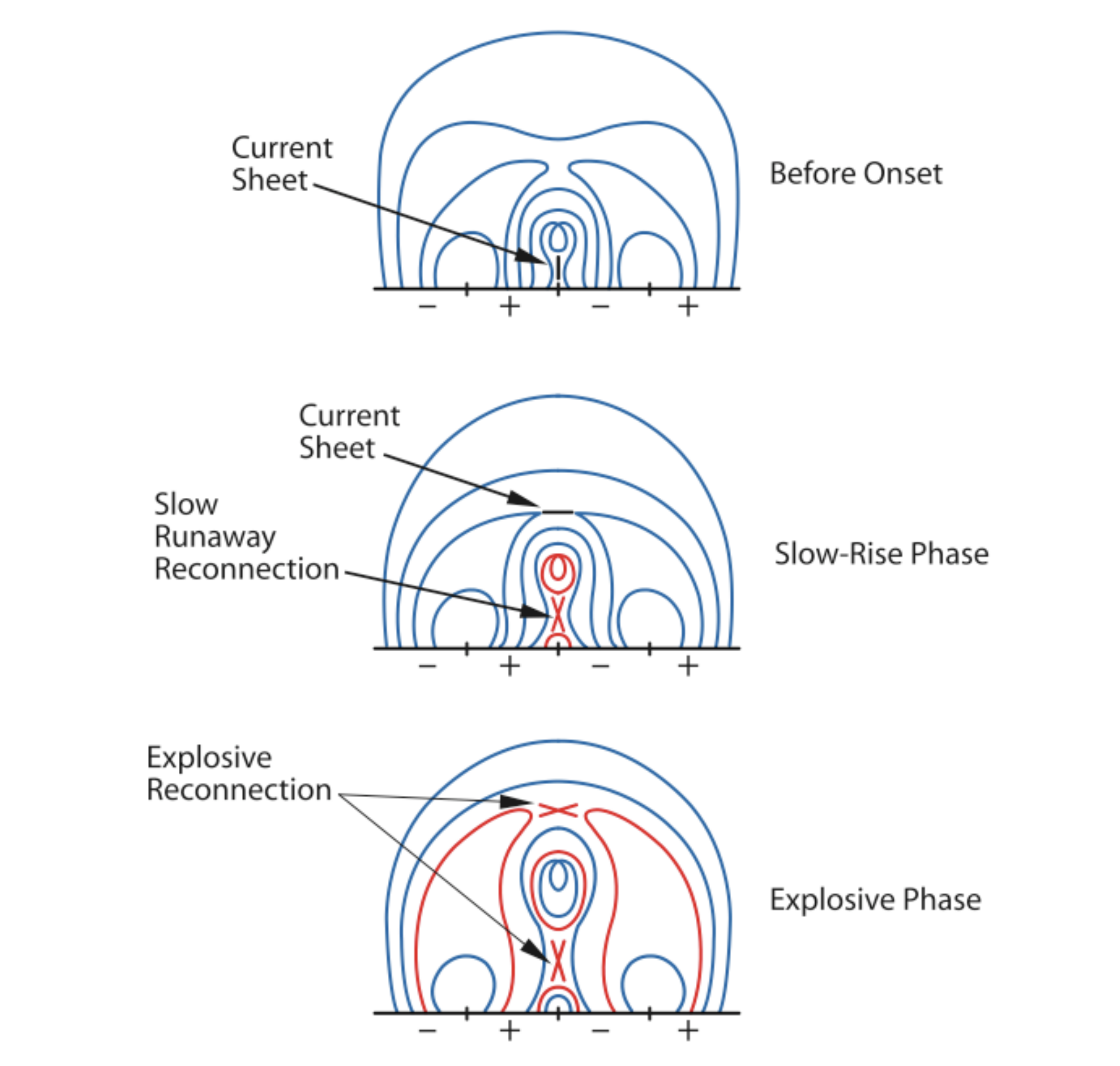} 
\caption{Schematic diagram describing internal tether cutting, from \citet{Moore06}. The filament is shown as the innermost loop. It is tethered by a central arcade and has neighbouring sidelobe arcades on each side. The entire system is then covered by overlying field lines. }
\label{fig:moore_int}
\end{figure}

In the internal tether-cutting case the initial topology is as shown in the top panel of Figure \ref{fig:moore_int}. This 2-D sketch implies a central sheared core (represented by the innermost loop of the quadrupole), tethered by a central arcade. The presence of neighbouring arcades is also a possibility but not a necessity for this model. Before eruption, the central arcade is in force free equilibrium and no current sheet exists between it and the overlying field. However, a current sheet does exist between the legs of the arcade as a result of their slow shearing due to photospheric motions. When this current sheet becomes sufficiently thin enough to allow reconnection across it (as in Figure \ref{fig:moore_int} middle panel), a ``run-away tether-cutting'' process begins. The field lines above the reconnection site are now no longer tethered to the photosphere and begin to erupt upwards while the field lines beneath the reconnection site become the site of a solar flare. As the reconnection progresses, the plasmoid above the reconnection site slowly erupts upwards and compresses the null point, forming a current sheet with the overlying field. This will result in explosive breakout reconnection above the plasmoid which reconnects field lines into the neighbouring arcades, both heating the side arcades and removing field blocking the path of the plasmoid. This run-away process results in the launch of the CME. In this scenario, the CME is triggered by internal reconnection, with breakout reconnection occurring later in the event. Therefore it is expected that signatures of internal reconnection e.g. flaring of the central arcade, a SXR source at the reconnection site etc. will occur before those of the side lobe restructuring. 

\begin{figure}
\centering
\includegraphics[width=\textwidth, trim =0 5 00 0, clip = true]{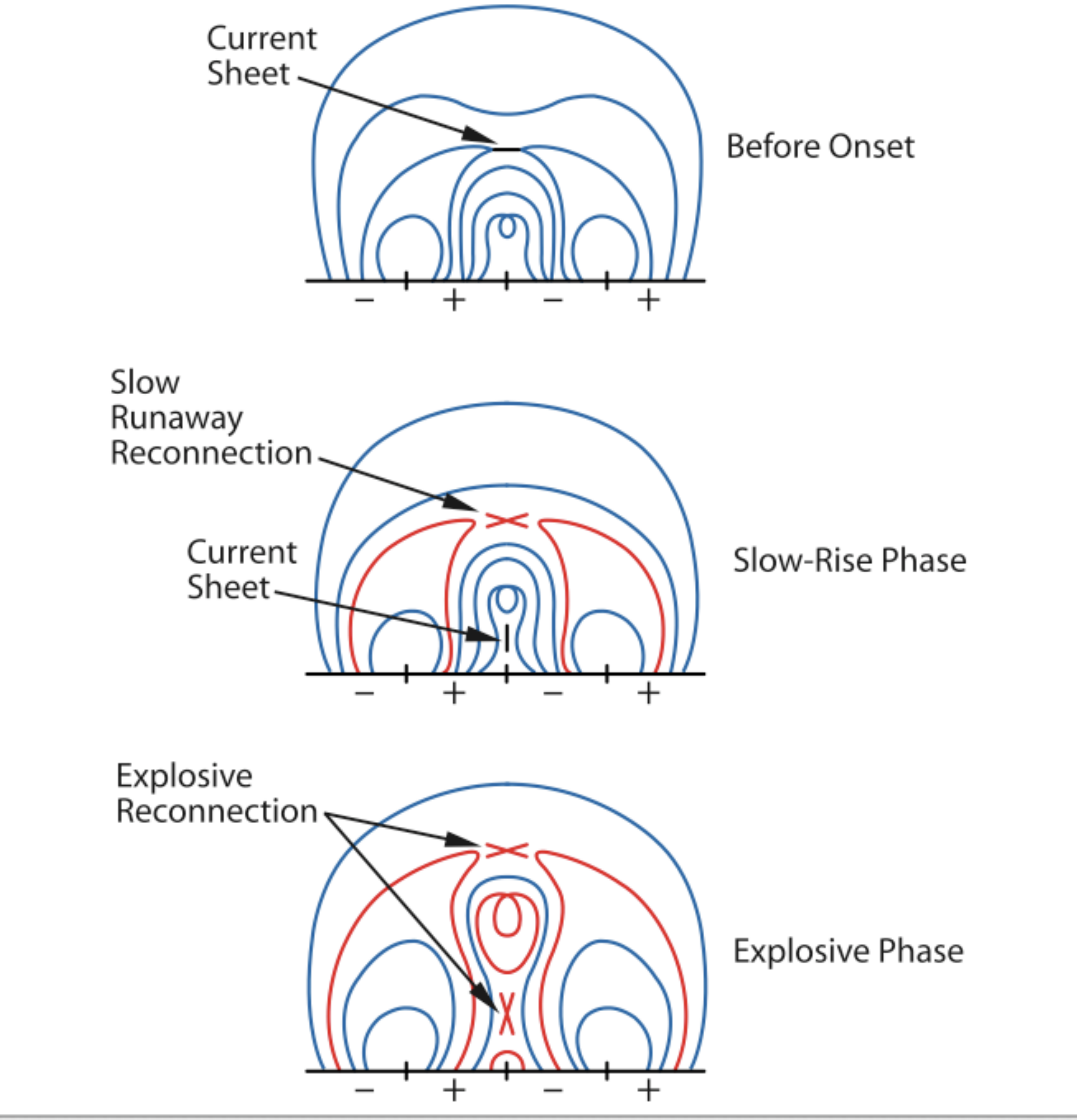} 
\caption{Schematic diagram describing external tether cutting, from \citet{Moore06}, as in Figure \ref{fig:moore_int}.}
\label{fig:moore_ext}
\end{figure}

External tether-cutting, or the ``breakout'' model by \citet{Antiochos1998}, begins with a quadrupolar magnetic topology: a central arcade between two side lobe arcades (Figure \ref{fig:moore_ext}). The structure inside the central arcade is such that no current sheet exists between the arcade legs (or if it does it is too thick to support reconnection) but a current sheet does exist between the top of the arcade and the overlying field, as shown in the top panel of Figure \ref{fig:moore_ext}. This can result from e.g. further emergence of the central arcade, which works to compress the null point between the central arcade and the overlying field without the generation of a current sheet between the arcade legs. Reconnection above the arcade shifts the force balance so that the central arcade begins to rise, stretching the field and drawing the legs of the central arcade together to create a second current sheet beneath the filament (Figure \ref{fig:moore_ext} middle panel). This results in run-away tether-cutting reconnection as in the internal tether-cutting case. Unlike the internal tether-cutting case, breakout reconnection begins first. Therefore evidence of heating or reconnection in the side lobes would be expected before evidence of the same in the central arcade. 

\begin{figure}
\centering
\includegraphics[width=\textwidth, trim =25 00 -10 0, clip = true]{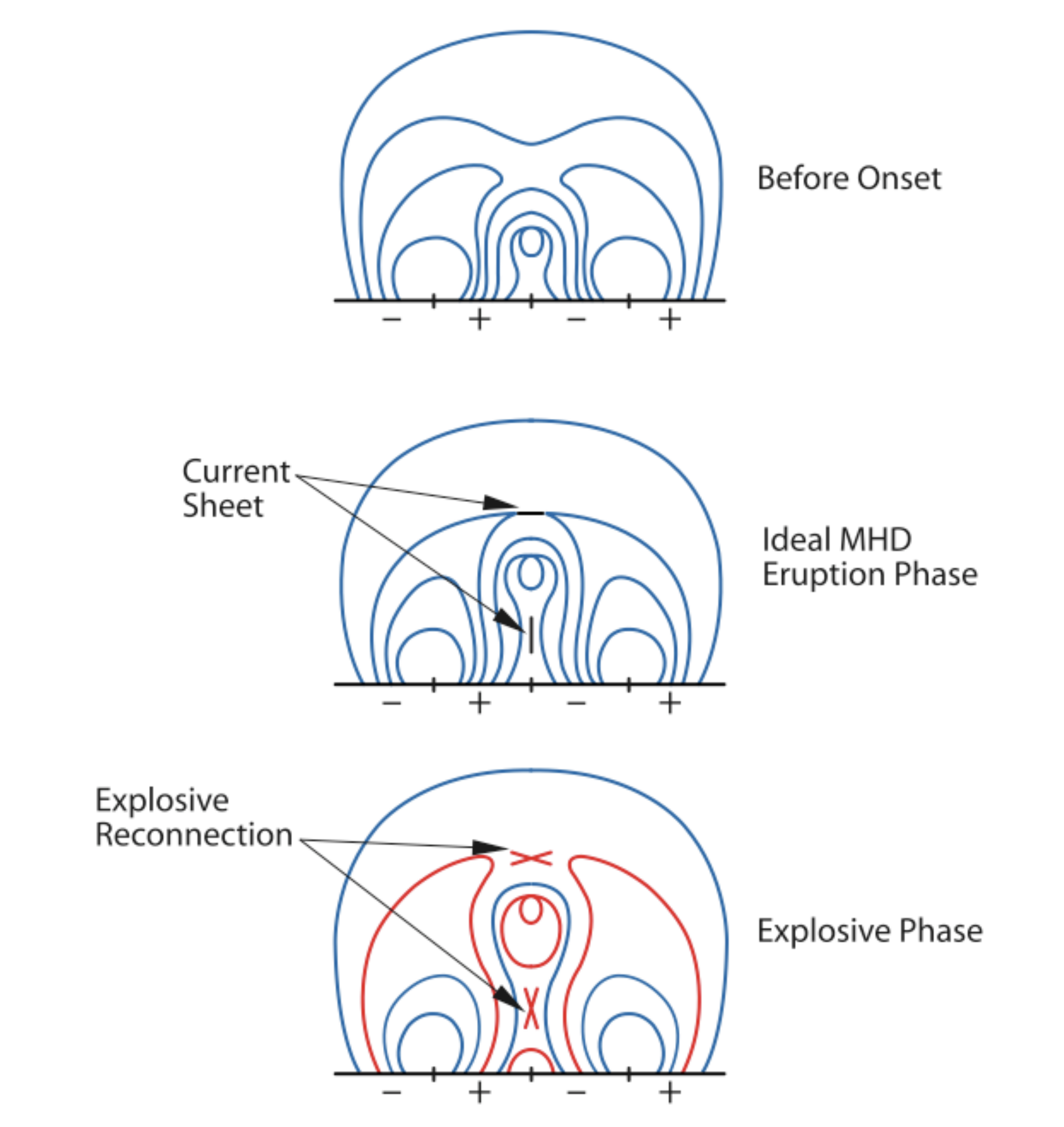} 
\caption{Schematic diagram describing a CME initiated by an MHD instability, from \citet{Moore06}, as in Figure \ref{fig:moore_int}.}
\label{fig:moore_ideal}
\end{figure}
The third case we consider is the catastrophe model \citep[e.g.][]{Forbes_Isen1991, Forbes_Priest1995, Isen93}. This model differs from the first two in that it is not triggered by magnetic reconnection. Instead, the continued shearing and twisting of the central arcade gradually evolves the field until it is forced out of force-free magnetostatic equilibrium. The field seeks a new equilibrium by erupting upwards, generating two current sheets, one between the stretched fields of the arcade legs and a second between the top of the arcade and the overlying field (middle panel Figure \ref{fig:moore_ideal}). It has been shown that this can occur without the use of magnetic reconnection \citep[e.g.][]{Isen93, Chen2000, Roussev03}. Following the formation of the current sheets, magnetic reconnection can take place and run-away tether-cutting drives the launch of the CME, as before (Figure \ref{fig:moore_ideal} bottom panel). In this case, one would expect to see a rising flux rope before any indication that magnetic reconnection had taken place. 

While current theoretical models have made significant progress from the ideas of Carmichael, Sturrock, Hirayama, Kopp and Pneuman there remains significant work to be done in this field. The development of new and improved technology and instruments, e.g. the SWAP instrument on board Proba-2 with its extended field of view (see \S \ref{sect:swap} for further details) continuously pushes the envelope on the performance of current theories.

\section{Outline of thesis}
The work presented in this thesis attempts to improve the understanding of the connection between solar flares and coronal mass ejections. To date, the relationship between these two phenomena has been fraught with complications and competing theories. This thesis attempts to better understand the behaviour of flares, beginning with the evolution of a confined flare. Extending this study to investigate the evolution of an eruptive flare attempts to categorise both the similarities and differences between these two flare categories. 

Chapter \ref{chapter:atom} presents the atomic processes involved in producing the emissions observed from the Sun which are employed when discussing the methods of observation in Chapter \ref{chapter:instrumentation}. As an extension to the discussion of instrumentation, a study of the calibration of EUV imaging telescopes is presented in Chapter \ref{chapter:SWAP}. As part of the Proba-2 team, the analysis of the SWAP instrument's sensitivity to temperature was investigated for coronal hole, quiet sun, active regions and flares. This was extended to investigate the corresponding responses of other EUV imagers, namely TRACE, SOHO/EIT, STEREO A/EUVI, STEREO B/EUVI and SDO/AIA.

Chapter \ref{chapter:mar26} explores the evolution of a confined flare. The physical mechanisms involved in the flaring process are investigated by comparing observations from a wide range of spacecraft to a 0-D hydrodynamic model, EBTEL. The combination of observations and theory allow for a more extensive investigation than otherwise possible. The investigation into the physics of flares is then expanded in Chapter \ref{chapter:CME} to include an ``eruptive'' flare. Using the knowledge and techniques developed in Chapter \ref{chapter:mar26}, the relationship between flares and CMEs are investigated. This is done through a study of the hydrodynamic evolution of the flare and the kinematic evolution of both the flare and the CME. 

Chapter \ref{chapter:concs} discusses the implications of the work presented in this thesis both from observational and theoretical perspectives and suggests improvements and developments for future work.

\chapter{Atomic physics}
\label{chapter:atom}
\ifpdf
    \graphicspath{{1_introduction/figures/PNG/}{1_introduction/figures/PDF/}{1_introduction/figures/}}
\else
    \graphicspath{{1_introduction/figures/EPS/}{1_introduction/figures/}}
\fi

\hrule height 1mm
\vspace{0.5mm}
\hrule height 0.4mm 
\noindent 
\\ {\it The basis of all solar observations, from radio to $\gamma$-ray emission, requires a detailed understanding of how that particular emission is generated. This thesis focusses on the EUV and X-ray regimes. To better understand the production of EUV and X-ray line and continuum emission, the use of an atomic physics package, CHIANTI, is used. In this chapter, we describe how CHIANTI models emission line and continuum radiation in the range of interest, the assumptions made in the calculations and the capabilities of the package. 
}
\\ 
\hrule height 0.4mm
\vspace{0.5mm}
\hrule height 1mm 

\newpage

\section{Introduction}
\begin{figure}[!t]
\centering
\includegraphics[width=0.8\textwidth, trim =40 80 40 80, clip = true]{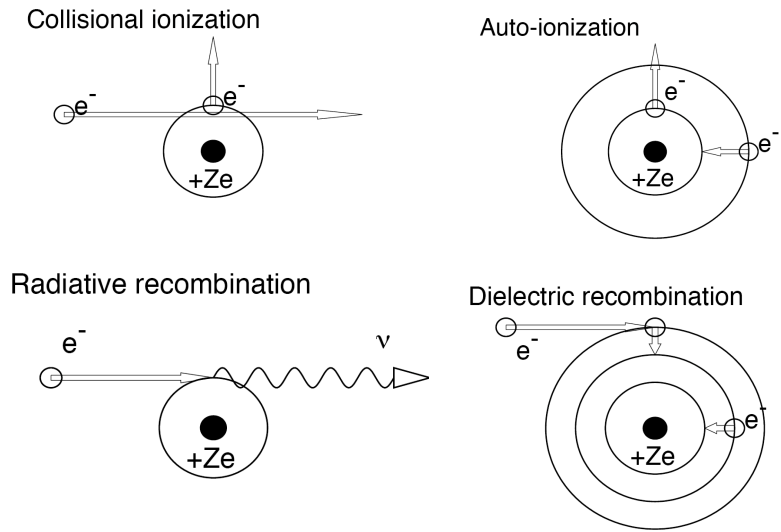} 
\caption{Schematic diagram showing the main ionisation and recombination processes in the solar atmosphere. Adapted from \citet{Aschwanden_book}.  }
\label{fig:ioniz_recomb}
\end{figure}

\begin{figure}[!t]
\centering
\includegraphics[width=\textwidth, trim =90 40 60 40, clip = true]{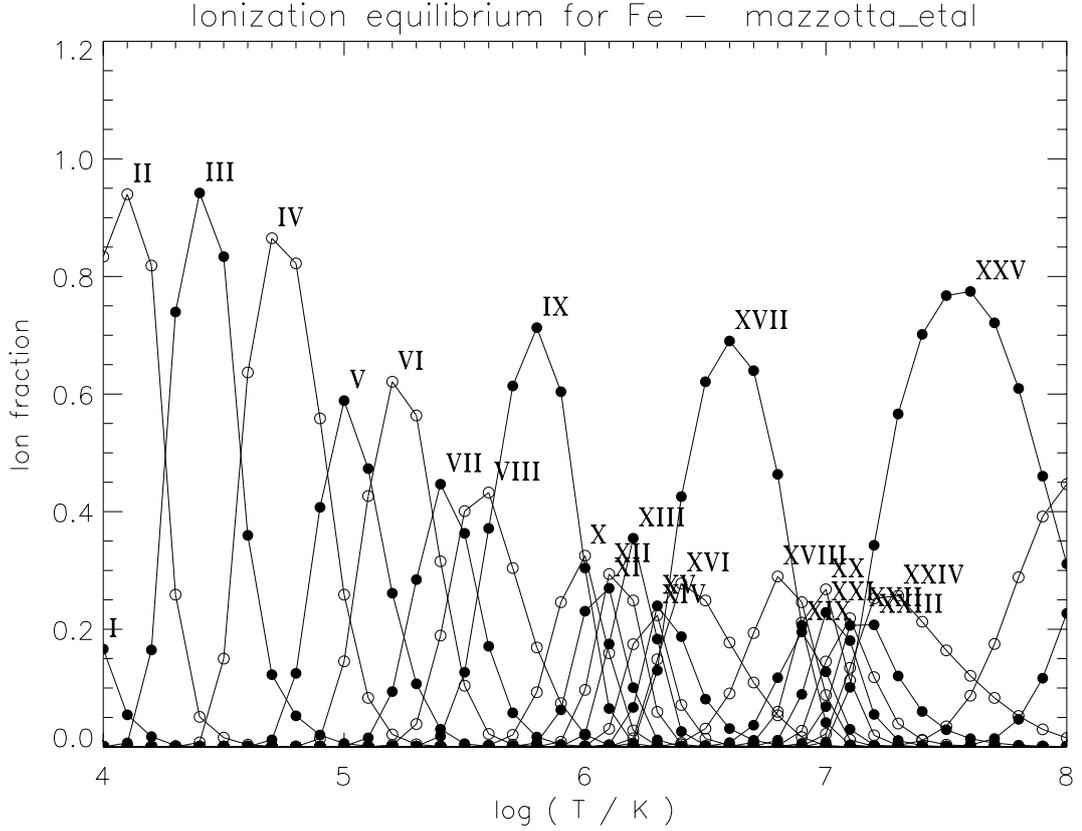} 
\caption{Ionisation fraction of iron calculated using CHIANTI and data from \citet{Mazzotta98}.  }
\label{fig:fe}
\end{figure}

The outer atmosphere of the Sun is a hot, tenuous plasma. Light elements such as hydrogen and helium are completely ionised  while heavier elements are at least partially ionised, depending on the temperature. The main processes involved in the ionisation of atoms, shown in Figure \ref{fig:ioniz_recomb}, are collisional ionisation and excitation autoionisation while atoms recombine by radiative recombination and dielectronic recombination. In equilibrium, the \emph{ionisation fraction} is the number density of a particular ion relative to the number density of a particular element, determined from the balance between the ionisation and recombination processes of that particular element. The ionisation fraction for Fe is shown in Figure \ref{fig:fe} calculated from the ionisation fractions of \citet{Mazzotta98}.

The solar corona emits strongly in the EUV and X-ray part of the spectrum through both emission lines and continuum. In this regime, the spectrum contains a multitude of emission lines, two-photon, free-free and free-bound continuum. To fully understand the origin and the conditions of the plasma emitting such radiation, it is necessary to model the spectrum in this regime. This is not a trivial task. In order to do so, parameters such as the energy levels, transitions, radiative transfer probabilities and excitation rates must be well understood for each individual line. In this thesis, we make use of the CHIANTI\footnote{CHIANTI is a collaborative project involving NRL (USA), RAL (UK), and the following Universities: College London (UK), of Cambridge (UK), George Mason (USA), and of Florence (Italy). } atomic physics package \citep{Dere97}. The CHIANTI database is used extensively by the astrophysical and solar communities to analyse emission line spectra from astrophysical sources. It was established in 1996 has, over the years, been maintained by Ken Dere, Helen Mason, Brunella Monsignori-Fossi, Enrico Landi, Massimo Landini, Peter Young, Giulio Del Zanna. It contains the parameters listed above for the known emission lines and includes sample differential emission measure functions for use in the simulation of astrophysical spectra.

\section{Emission lines}
\label{sect:atom_emis}

\begin{figure}[!t]
\centering
\includegraphics[width=0.5\textwidth, trim =40 20 400 450, clip = true]{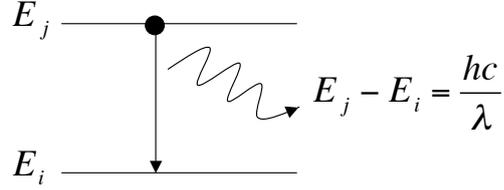} 
\caption{Two level diagram showing an electron decaying from an upper level with energy $E_j$ to a lower level of energy $E_i$ and emitting a photon of energy $\Delta E$ in the process. }
\label{fig:2lev}
\end{figure}

CHIANTI calculates the flux of an emission line by representing the flux in terms of a series measurable and theoretical parameters. Following \citet{Mariska_book} and \citet{Dere97}, an electron in an excited state can spontaneously decay through a bound-bound process with probability $A_{j,i}$. In a species $X$ of ionisation state $m$, an electron that transitions between an upper state $j$ and a lower state $i$ will produce a photon of energy $E_{\gamma} = \Delta E =  E_j - E_i = h\nu = hc/\lambda$ (Figure \ref{fig:2lev}). 
\begin{equation}
X^{+m}_j \rightarrow X^{+m}_i + h\nu
\end{equation}
The volume emissivity, ($\epsilon\ji$) of a plasma with upper level population, $n\subj$, and lower level population density, $n\subi$, is given by:
\begin{equation}
\epsilon_{j,i} = h\nu_{j,i}A_{j,i}n_{j}   \hspace{2cm}  [ergs~cm^{-3} ~s^{-1}]
\label{eqn:emis}
\end{equation}
where $A\ji$ is the Einstein coefficient for spontaneous radiative emission giving the probability per unit time that the electron in the excited state will spontaneously decay to the lower state. 
For a volume of optically thin plasma, $\Delta V$, which is a good approximation for the outer layers of the solar atmosphere, the flux observed at Earth at a distance $R$ from the Sun is proportional to the number of emitting ions in the line of sight to the observer and the fraction of those ions that are in a given energy state producing the emission line. We can begin by writing the flux as:
\begin{eqnarray}
\label{eqn:flux}
F_{j,i} & = & \frac{1}{4\pi R^2} \int_{\Delta V} \epsilon \ji dV    \nonumber \\
  &  = & \frac{1}{4\pi R^2} \int_{\Delta V}  h\nu_{j,i}A_{j,i}n_{j} dV \hspace{2cm}  [ergs~cm^{-2} ~s^{-1}]
 \end{eqnarray}
Practically, the volume element $dV$ is defined by the spatial resolution of the instrument. Assuming the lines observed are optically thin, emission from all material along $R$ will be accounted for. 

The number density of the upper level can be determined from:
\begin{equation}
n \subj = \frac{n \subj}{n_{ion} } \frac{n_{ion}}{n_{el}} \frac{n_{el}}{n_{H}} \frac{n_{H}}{n_e} n_e	
\end{equation}
where $n_j/n_{ion}$ is the population of the upper level relative to the total number density of the ion and is a function of temperature and density, $n_{ion}/n_{el}$ is the relative abundance of the ion and is a function of temperature, $n_{el}/n_H = A_{el}$ is the element abundance relative to hydrogen and $n_H/n_e$ is the hydrogen abundance relative to the electron number density. We can therefore rewrite flux of a line at Earth as:
\begin{equation}
F \ji = \frac{h\nu \ji A\ji}{4\pi R^2} \int_{\Delta V} \frac{n \subj}{n_{ion}} \frac{n_{ion}}{n_{el}} \frac{n_{el}}{n_H} \frac{n_H}{n_e} n_e dV
\label{eqn:f_init}
\end{equation}
The number density of the excited state, $n\subj$ must be populated by balancing the excitation processes with de-excitation processes. The energy state of an emitting ion can be changed by a range of different process. For example, the energy of the ion will change when an electron is excited into a higher level by e.g. a collision. The energy will again change if that electron decays back to its original state. These processes are generally faster than the processes that effect the ionisation state, as shown in Table \ref{table:rate} \citep[][p 18]{Mariska_book}. This makes separating excitation/de-excitation and ionisation/recombination calculations possible. The excitation and de-excitation processes are shown in Figure \ref{fig:excit}. Each  diagram in this figure corresponds to a term in Equation \ref{eqn:excit_balance}. An electron can be excited into a higher energy state by one of four processes: a collision with a free electron, a collision with a free proton or simulation by radiation. The electron can decay by one of four processes: electron and proton collision, stimulation by a photon or by spontaneous radiative decay. This is represented mathematically in Equation \ref{eqn:excit_balance} and schematically in Figure \ref{fig:excit}:
\begin{figure}[!t]
\centering
\includegraphics[width=\textwidth, trim =40 160 40 40, clip = true]{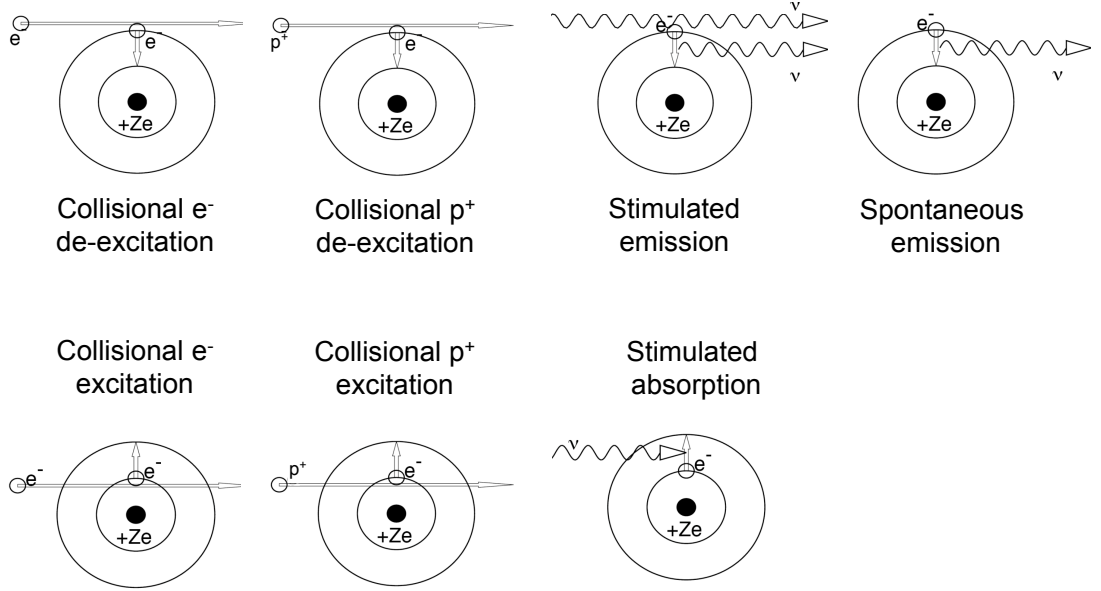} 
\caption{Schematic diagrams of the excitation and decay processes with each figure corresponding to a term in Equation \ref{eqn:excit_balance}. Adapted from \citep{Aschwanden_book}. }
\label{fig:excit}
\end{figure}
\begin{center}
\begin{tabular}{c c c c c c c c}
$n\subj n\sube \sum_iC^e_{j,i}$ &+& $n\subj n_p \sum_i C^p_{j,i}$ &+& $n_{j} \sum_{i>j}R\ji$ &+& $n\subj \sum_{i<j}A\ji$ &= \\
$\sum_i n\subi n\sube C^e_{i,j}$ &+ &$\sum_i n_p C^p_{i,j}$& +&$\sum_{i<j} n\subi R_{i,j}$&  & & \\
e$^-$ collision & & p$^+$ collision & & stimulated & & spontaneous
\end{tabular}
\end{center} 	
\begin{equation}
\label{eqn:excit_balance}
\end{equation}
The $C^{a}_{b, c}$ terms, given in $cm^3~s^{-1}$, are the collision rate coefficients for electrons and protons (where $a = e$ and $p$ respectively) for the $b^{th}$ - $c^{th}$ transition. $R_{b, c}$ in $s^{-1}$ refers to the stimulated absorption coefficient for the $b^{th}$ - $c^{th}$ transition. The radiative transfer probabilities, $A_{b,c}$ for the $b^{th}$ - $c^{th}$ transition is measured in $s^{-1}$. These parameters act to populate and depopulate the excited level $j$. Initially, CHIANTI did not consider the proton excitation rates. However, they were later introduced to take account of the fine structure transitions in highly ionised systems \citep{Young03}. 

\begin{table}[!t] 
\caption{Transition rates of important atomic processes calculated for \ion{C}{iv} at a temperature of 10$^5$~K and a density of 10$^{10}$~cm$^{-3}$ \citep[from][]{Mariska_book}. } 
\centering 
\begin{tabular}{l l l} 
\hline\hline
Process						 & Rate 				&	Characteristic time \\
							& 	[$cm^{-3}~s^{-1}$]  	&		[s]			\\
\hline
(De-)excitation processes			&					&					\\
\hline
Collisional excitation				&	$n_in_eC_{i,j}$		&	$2\times10^{-3}$				\\ 
Collisional de-excitation			&	$n_jn_eC_{j,i}$		& 	$2\times10^{-3}$				\\ 
Spontaneous radiative decay		&	$n_jA_{j,i}$		&	$4\times10^{-9}$				\\ 
\hline
ionisation/recombination processes	&	&									\\
\hline
Collisional ionisation				&	$n_en_{ion}q_{coll}$	&	107				\\ 
Autoionisation					&	$n_en_{ion}q_{auto}$&					\\ 
Total ionisation rate				&	$n_en_{ion}q_{tot}$	&	107				\\ 
Radiative recombination			&	$n_en_{ion}\alpha_{rad}$&	88			\\
Dielectric recombination			&	$n_en_{ion}\alpha_{diel}$&				\\ 
Total recombination rate			&	$n_en_{ion}\alpha_{tot}$&		88			\\ 
 \hline
\hline
     \end{tabular} 
\label{table:rate} 
\end{table}

In what is known as the \emph{coronal approximation}, it is assumed that the population of excited states occurs primarily by collisional excitation by electrons from the ground state and the de-population of excited states occurs primarily by spontaneous emission. Since the majority of electrons are in the lower (ground) state, we can say $\sum_i n_i = n_{ion}$. Thus, Equation \ref{eqn:excit_balance} can be approximated as:
\begin{equation}
n_{ion} n_e C^e_{i, j} = n_j \sum_k A_{j, k} \hspace{1cm} k<j
\label{eqn:coronal_approx}
\end{equation}
With $n_j/n_{ion} = n_eC_{i,j}^e/\sum_k A_{j, k}$, the emissivity can now be written as:
\begin{equation}
\epsilon_{j,i} = h\nu_{j,i}A_{j,i} \frac{n_{ion}}{n_{el}} \frac{n_{el}}{n_H} \frac{n_H}{n_e} \frac{n_e C_{i, j}^e}{\sum_k A_{j, k}} n_e 	
\end{equation}

The electron collision rate coefficient $C_{i,j}$ for Maxwellian distributed electron velocities is given by:
\begin{equation}
C_{i,j} = \int_{v_0}^\infty \sigma_{i,j}(v)f(v)v dv
\label{eqn:C_init}
\end{equation}
where
\begin{equation}
f(v) = 4\pi \left(\frac{m}{2\pi k_B T}\right)^{3/2} v^2 exp\left(\frac{1/2 mv^2}{k_BT}\right)
\label{eqn:max_b}
\end{equation}

$\sigma_{i,j}$ is the electron excitation cross section by collisions and is commonly expressed in terms of the collision strength $\Omega_{i,j}(E)$, the incident electron energy $E=1/2 mv^2$ (measured in Rydbergs), the Bohr radius $a_0$ and the statistical weight of the $i^{th}$ level, $\omega_i$:
\begin{equation}
\sigma_{i,j} = \frac{\pi a_0^2 \Omega_{i,j}(E)}{\omega_i E}
\label{eqn:sigma}
\end{equation}

$\omega_i$ is introduced to adhere to the principle of detailed balance, ensuring the net exchange between any two levels will be balanced i.e. the number of excitations caused by electrons in range $dE_1$ is balanced by collisional de-excitations by electrons in range $dE_2$ so $E_1 = E_2 + \Delta E_{i,j}$. This is only valid for energy levels in thermal equilibrium and is not valid for the coronal approximation. However, CHIANTI does not make the assumption of Equation \ref{eqn:coronal_approx} and so the principle of detailed balance is used for the calculation of collisional excitation rates as:
\begin{equation}
C_{j,i} = \frac{\omega_i}{\omega_j} C_{i,j}exp\left(\frac{\Delta E_{i,j}}{k_BT}\right)
\end{equation}

Combining Equations \ref{eqn:C_init}, \ref{eqn:max_b} and \ref{eqn:sigma}, the collision rate coefficient can be written as:
\begin{equation}
C_{i,j} = \frac{8.63 \times 10^{-6}}{\omega_i k_B T^{3/2}}\int_{\Delta E_{i,j}}^\infty \Omega_{i,j}(E) exp\left(\frac{-\Delta E}{k_B T}\right)dE
\end{equation}

Equation \ref{eqn:f_init} can now be written as 

\begin{equation}
F(\lambda) = \frac{h\nu}{4 \pi R^2}\frac{8.63 \times 10^{-6} \Omega_{i,j}}{\omega_i}\frac{n_H}{n_e}\frac{n_{el}}{n_H}\int_{\Delta V} \frac{n_e^2}{\sqrt{T}} \frac{n_{ion}}{n_{el}} exp\left(\frac{-\Delta E}{kT}\right) dV
\label{eqn:final_flux}
\end{equation}

In order to solve this equation for a given emission line, CHIANTI acquires the parameters from a number of sources. Energy level information is obtained, where possible, from the National Institute of Standards and Technology (NIST) database of observed energy levels \citep{Martin_nist}. This has been supplemented by theoretical estimates where the energy levels are not known \citep{Dere97}. These have been calculated using the UCL SSTRUCT program \citep{Eissner74}. Einstein coefficients for radiative transitions are, for the most part, obtained from literature or calculated using the SSTRUCT code. Collision rate coefficients ($C_{i,j}$) are scaled according to \citet{Burgess92} and the de-excitation rates are obtained from the principle of detailed balance. The elemental abundance and ionisation fraction of the plasma are user defined. Coronal abundances and the ionisation fraction of \citet{Mazzotta98} are used throughout this thesis.

\subsection{Contribution functions and emission measures}
 \label{sect:atom_em}
\begin{figure}[!t]
\centerline{
 \includegraphics[width=\textwidth, trim =25 400 25 70, clip = true]{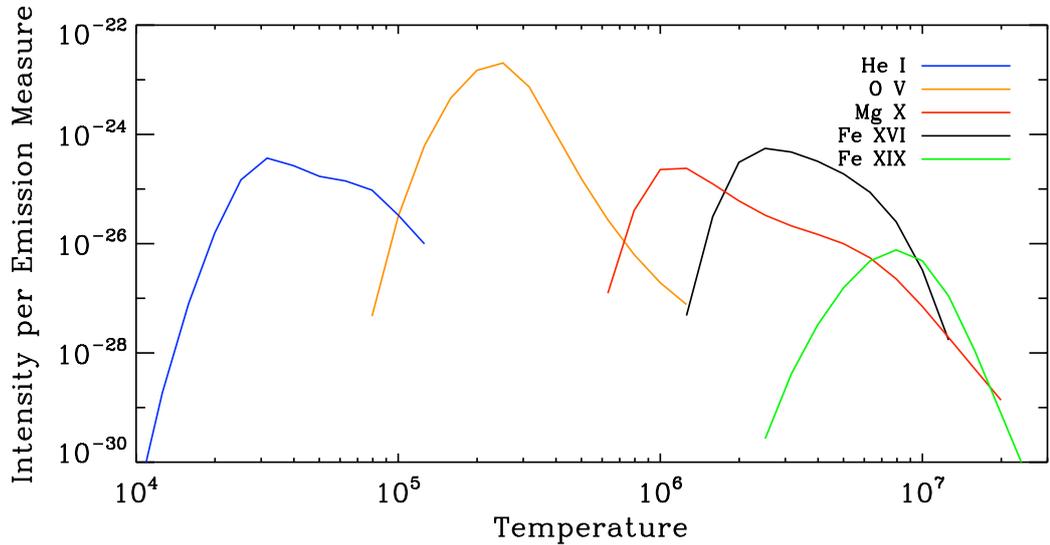} 
}
 \caption{Contribution functions $G(n, T)$, obtained from CHIANTI for \ion{He}{i} (584.33~\AA), \ion{O}{v} (629.73~\AA), \ion{Mg}{x} (524.94~\AA), \ion{Fe}{xvi} (360.75~\AA) and \ion{Fe}{xix} (592.23~\AA). These were calculated using a density of $n_e = 5\times10^9$~cm$^{-3}$, coronal elemental abundances and ionisation fractions from \citet{Mazzotta98}. 
  \label{fig:goft_fns} 
  }

\end{figure}

The temperature sensitive components of Equation \ref{eqn:final_flux} can be extracted in what is known as the contribution function, $G(n, T)$, samples of which can be found in Figure \ref{fig:goft_fns}. The contribution function is given by:
\begin{equation}
G(T) = \frac{n_{ion}}{n_{el}}T^{-1/2}exp\left(\frac{-h\nu}{kT}\right)
\end{equation}
and defining the emission measure (EM) to be the amount of emitting plasma in a given volume $dV$
\begin{equation}
EM = \int n_e^2 dV
\label{eqn:emission_measure}
\end{equation}
or column depth $dh$
\begin{equation}
EM = \int n_e^2 dh
\label{eqn:h_emission_measure}
\end{equation}
the line flux can then be written as:
\begin{equation}
F(\lambda) = c \times EM \times G(T)
\label{eqn:flux_simple}
\end{equation}
where $c$ takes into account the physical constants, the hydrogen abundance ($n_H/n_e$) and the elemental abundance ($n_{el}/n_H$). The contribution function provides information regarding the formation temperature of a given line. In this thesis, we utilise the $G(T)$ of emission lines to convert between the intensity of a line and the EM of the line at a given time and temperature \citep[Chapter \ref{chapter:mar26},][]{Raftery09}. This is done by inversion of Equation \ref{eqn:flux_simple}. Estimation of density can be made from:
\begin{equation}
n_e \approx \sqrt{\frac{EM}{V}}
\label{eqn:dens}
\end{equation}
This inversion requires the estimation of $V$, the volume of the emitting source. This of course can be a significant problem. Even if the source can be imaged using e.g. a spectrometer raster, the two dimensional nature of the observation immediately places uncertainty on the volume. However, with little or no other options, this is sometimes the only method by which the density can be calculated. 

\begin{figure} [!t]
 \includegraphics[width=\textwidth, trim =30 40 25 30, clip = true]{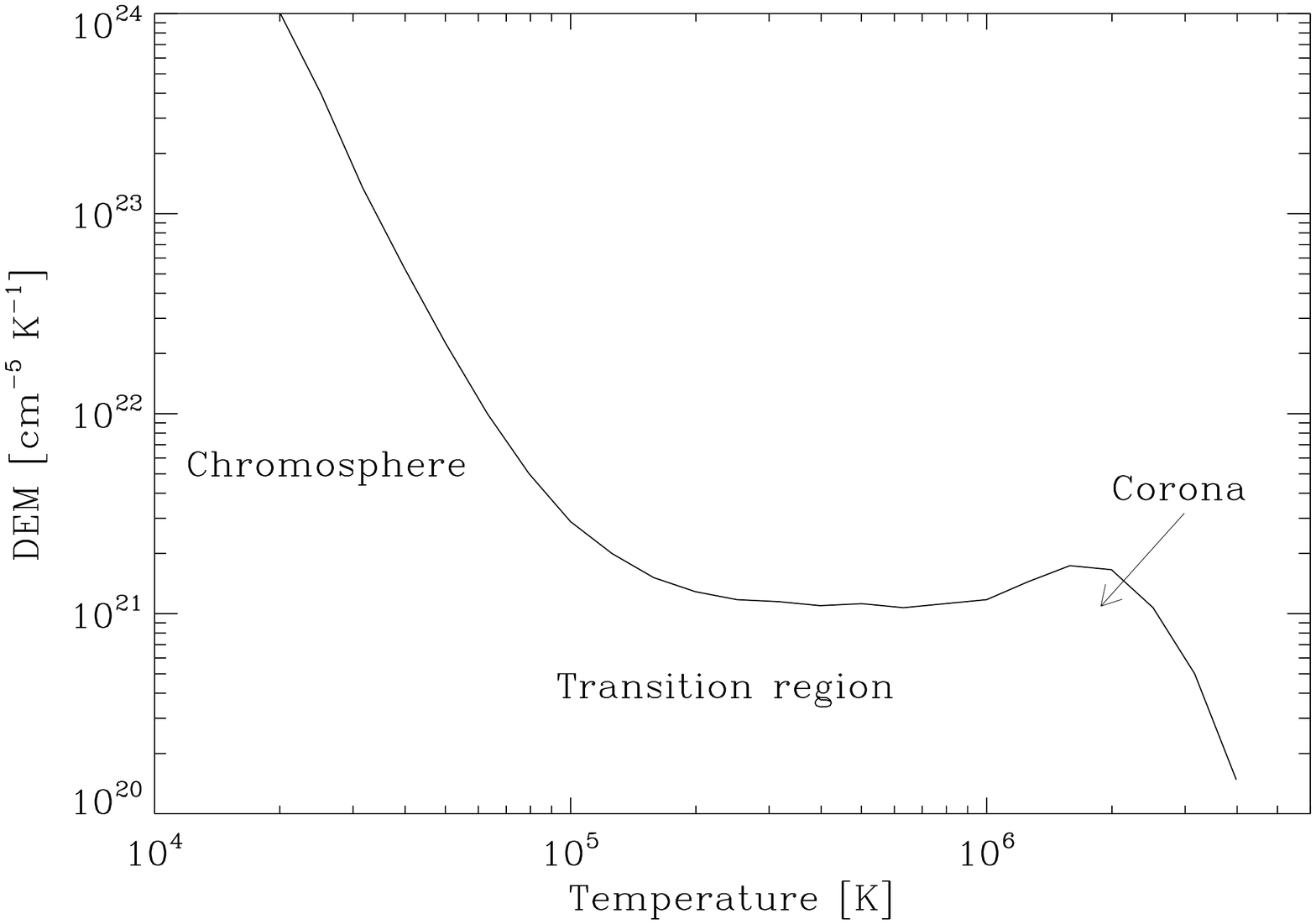} 
 \caption{Quiet Sun DEM function of the solar atmosphere calculated using values from CHIANTI. 
  \label{fig:dem_demo} }
\end{figure}

The emission measure of a plasma is suitable, assuming that the spectral line is emitted over a homogeneous, isothermal volume. For cases of multi-thermal plasma it is useful to define the differential emission measure (DEM) for a volume as
\begin{equation}
DEM(T) =  n_e^2\left(dV/dT\right)
\end{equation}
or for a column of depth as
\begin{equation}
DEM(T) =  n_e^2\left(dT/dh\right)^{-1}
\label{eqn:dem_col}
\end{equation}
The DEM relates the amount of material in the temperature interval $dT$ in a volume and so can give information about the structure of the atmosphere. For example, the temperature gradient $dT/dh$ in the column DEM can be used to determine the temperature gradient between the transition region and the corona. It also takes account of the large volume that is the chromosphere and the temperatures found within it, as shown in Figure \ref{fig:dem_demo}. Likewise, the thinness of the transition region results in a reduced DEM over transition region temperatures. The large volume of the corona means the DEM is increased from transition region values, although the low densities found in the corona suppress the DEM somewhat. During a solar flare however, the coronal temperatures and densities are raised considerably resulting in a DEM peak at high temperatures. CHIANTI supplies sample differential emission measure files. These have been calculated from a combination of the work by Pottasch (1964) and UV/EUV spectra. CHIANTI does not currently include volume emission measures and so column emission measure must be considered throughout while using the database.

\section{Continuum emission}
\begin{figure} [!t]
 \includegraphics[width=\textwidth, trim =0 400 0 30, clip = true]{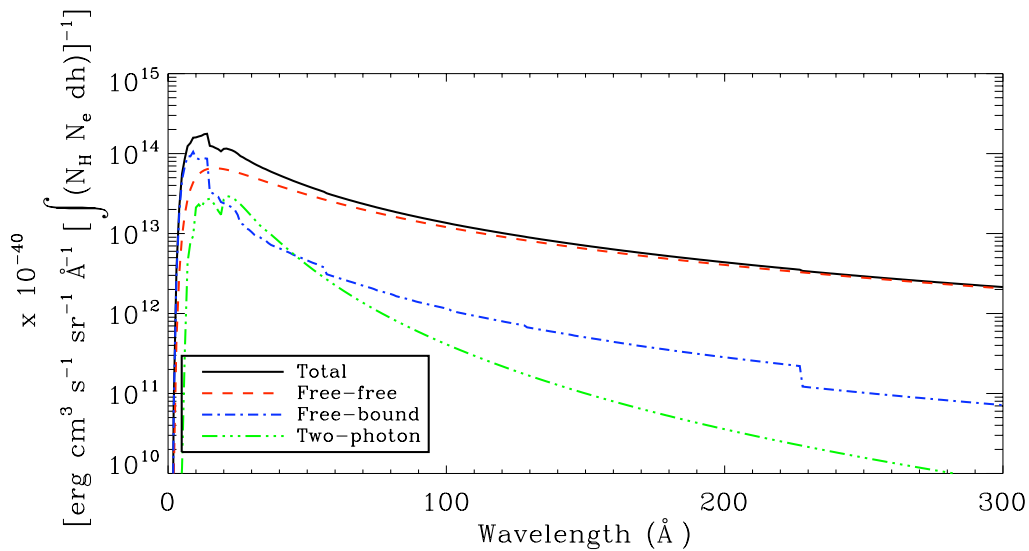} 
 \caption{Continuum emission from free-free, free-bound and two-photon processes in the wavelength range 1 - 500~\AA\ calculated using CHIANTI.   
  \label{fig:ff_fb_tp} }
\end{figure}

\label{sect:continuum}
There are three primary mechanisms by which continuum emission is formed: free-free, free-bound and two-photon. The relative intensity of these processes between 1 and 500~\AA\ are shown in Figure \ref{fig:ff_fb_tp}. In the SXR range, free-bound continuum has a significant contribution, however it is clear that free-free continuum dominates throughout the spectrum. Therefore, we shall focus on free-free continuum emission here.  

\subsection{Free-free continuum}

\begin{figure}[!t]
\centering
\includegraphics[width=0.7\textwidth, trim =0 160 40 40, clip = true]{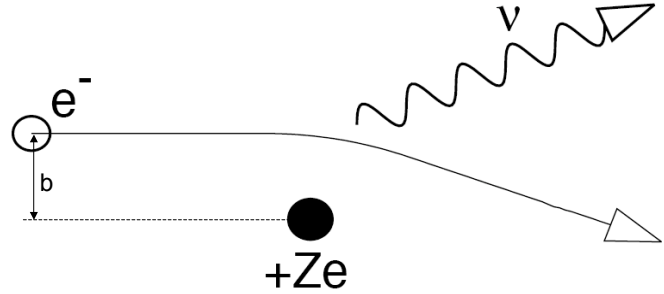} 
\caption{Schematic diagram of how free-free Bremsstrahlung continuum emission is formed: as the path of an electron is bent by interaction with the electric field of a stationary particle, it emits a photon. \citep{Aschwanden_book}. }
\label{fig:brem}
\end{figure}

In a hot coronal plasma, free electrons and ions can suffer multiple interactions. The most frequent of these is when a free electron is scattered in the Coulomb field of an ion ($Z$). The scattering is such that the electron remains free after the interaction. This is what gives rise to the name ``free-free'' continuum, otherwise known as Bremsstrahlung or \emph{braking} radiation. During the scattering process, the electron loses some of its energy which is released as a photon with energy $E_{\gamma} = E_{e_{incid}} - E_{e_{scat}}$. Thick target Bremsstrahlung occurs when electrons are accelerated to high energies in a collisionless plasma and become collisionally stopped when they interact with a thermal plasma such as the chromosphere. 

\begin{figure}[!t]
 \includegraphics[width=\textwidth, trim =0 0 0 450, clip = true]{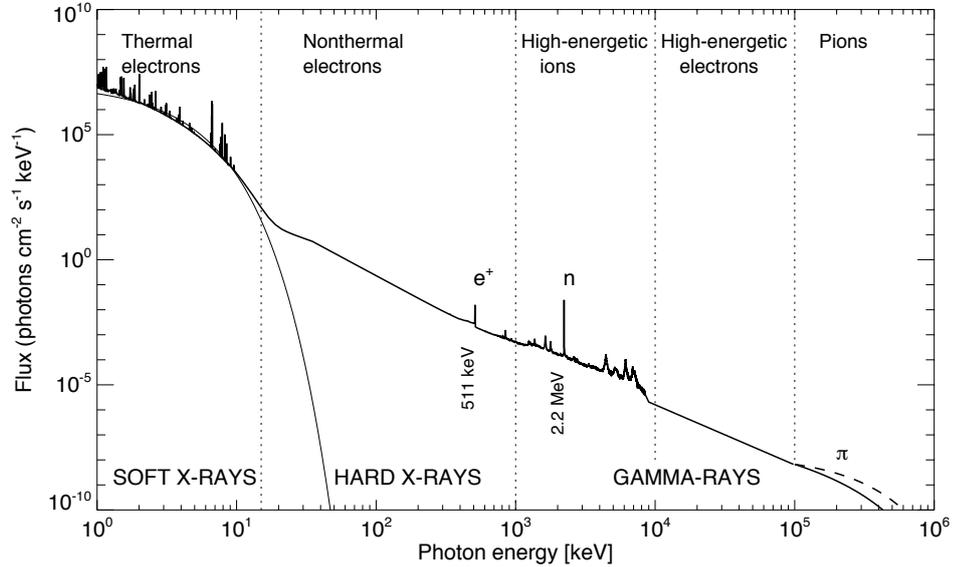} 
 \caption{Photon spectrum of a large solar flare extending from SXR out to $\gamma$-rays. The spectrum shows the different processes that dominate the different regimes. The SXR spectrum is dominated by thermal electrons, HXR (Bremsstrahlung) radiation by non-thermal electrons and $\gamma$-rays by energetic electrons, ions and pions. The electron-positron annihilation line is shown at 511~keV along with the neutron capture line at 2.2~MeV. \citep{Aschwanden_book} .
     \label{fig:en_spec_ex} }
\end{figure}

Following \citet{Aschwanden_book}, the Bremsstrahlung total power $P_i(v, \nu)$ radiated by a single electron in a plasma of $n_i$ ions in a frequency range $d\nu$ is
\begin{equation}
P_i(v, \nu)d\nu = n_i v \sigma_r(v, \nu)d\nu
\label{eqn:brem_p}
\end{equation}
where $\sigma_r$ is the radiating cross section in $cm^2~ergs~Hz^{-1}$ and is given by
\begin{equation}
\sigma_r(v, \nu) = \frac{16}{3} \frac{Z^2e^6}{m^2c^3v^2}\int_{b_{min}}^{b_{max}} \frac{db}{b} = \frac{\pi}{\sqrt{3}} g(\nu, T)
\label{eqn:brem_x}
\end{equation}
where $b$, the distance from the electron to the atom is shown schematically in Figure \ref{fig:brem} and $g(\nu,T)$ is the \emph{Gaunt factor} which is approximately unity in the corona. The volume emissivity for $n_e$ electrons can be written as 
\begin{equation}
\epsilon_{\nu} = \frac{n_{\nu}n_e}{4\pi} \int P_i(v, \nu) f(v) dv
\label{eqn:brem_e}
\end{equation}
where the Maxwell-Boltzmann distribution $f(v)$ is given by 
\begin{equation}
f(v) = 4\pi \left(\frac{m}{2\pi k_B T}\right)^{3/2}v^2 exp \left( \frac{-mv^2}{2k_B T}\right)
\label{eqn:brem_f}
\end{equation}
and the flux of emission at Earth can be written as
\begin{equation}
F \propto \int \epsilon_{\nu} dV
\label{eqn:brem_flux1}
\end{equation}
In the corona, $g(\nu, T) \approx 1$, $n = n_i = n_e$, and since we are predominantly dealing with protons, $Z\approx1$. Therefore, we can combine Equations \ref{eqn:brem_p} to \ref{eqn:brem_flux1} to write the Bremsstrahlung flux in terms of the photon energy $\epsilon_{ph}=h\nu$
\begin{equation}
F(\epsilon_{ph}) \propto \int_V exp\left(\frac{-\epsilon_{ph}}{k_B T}\right)T^{-1/2} n^2 dV
\label{eqn:brem_flux}
\end{equation}
Remembering that emission measure $EM = \int n^2 dV$ we can write Equation \ref{eqn:brem_flux} as
\begin{equation}
F(\epsilon_{ph}) \propto exp\left(\frac{-\epsilon_{ph}}{k_B T}\right)T^{-1/2} EM
\label{eqn:brem_flux_em}
\end{equation}
This function demonstrates the sensitivity of the Bremsstrahlung spectrum to both temperature and emission measure. These are the two significant factors taken into account when fitting the thermal component of the RHESSI and SAX spectra in Chapters \ref{chapter:mar26} and \ref{chapter:CME}. Figure \ref{fig:en_spec_ex} shows an example of a solar energy spectrum where the thermal component is dominated by free-free continuum. The amplitude of the thermal spectrum will change with emission measure while the width of the spectrum will change with temperature. 

The CHIANTI code is extremely useful in identifying the emission lines and continuum found in the solar spectrum in the EUV and X-ray range. A comparison of theoretical line intensities to observed line intensities can reveal the physical conditions required for their production, information that would otherwise be unattainable. This can be especially helpful when determining the contributions from blended lines observed by an instrument that cannot resolve them, such as those in Chapter \ref{chapter:mar26}. The database has been constructed in such a way as to facilitate the easy distribution and updating of files. It is freely accessible through the internet\footnote{http://www.chianti.rl.ac.uk/} and through a package in the Interactive Data Language and SolarSoftware package \citep[\textsc{SSWIDL};][]{Freeland_ssw}.



\chapter{Instrumentation} 
\label{chapter:instrumentation}

\ifpdf
    \graphicspath{{2/figures/PNG/}{2/figures/PDF/}{2/figures/}}
\else
    \graphicspath{{2/figures/EPS/}{2/figures/}}
\fi

\hrule height 1mm
\vspace{0.5mm}
\hrule height 0.4mm 
\noindent 
\\ {\it This chapter presents details of the instruments used in this thesis and how the observations are obtained, prepared and analysed. Beginning with CDS, a discussion of how data are obtained and analysed is presented. Following this, EUV imaging telescopes are discussed, in particular SWAP, TRACE and EUVI. Extending our interest into the heliosphere, the techniques employed to analyse the data from STEREO Cor1 and Cor2 coronagraphs are presented. Going from the EUV and white light emission into X-rays, we begin with the GOES Satellites, focussing on the interpretation of observations for diagnostic purposes. Finally, the capabilities of RHESSI and the Mercury MESSENGER Solar array for X-rays are laid out. The imaging techniques used for RHESSI and the spectroscopic capabilities of both instruments are reviewed, focussing in particular on their performance in the thermal energy range.  
 }
\\ 
\hrule height 0.4mm
\vspace{0.5mm}
\hrule height 1mm 

\newpage

\section{The Coronal Diagnostic Spectrometer}
The Solar and Heliospheric Observatory (\emph{SOHO}) was launched in 1995 as a joint endeavour between ESA and NASA and has been providing uninterrupted observations of the Sun for 13 years (with the exception of a 3 month period, during which communication with the space craft was lost). Orbiting the first Lagrange point -- L1 -- at 1.5 million km from the Earth, it is always on the day-side of the Earth. This thesis takes advantage of observations from one SOHO instrument in particular - the Coronal Diagnostic Spectrometer \citep[CDS;][]{Harrison95}. CDS is an EUV spectrometer designed to study the solar corona. Initially, the instrument was not used for studying solar flares for fear of damaging the detectors. It has however become an integral part of flare studies. This thesis utilises five emission lines at a wide range of temperatures, from \ion{He}{i} ($\sim$0.03~MK) to \ion{Fe}{xix} at $\sim$8~MK. These data can provide users with temperature and emission measure diagnostics, along with Doppler velocities. 

\begin{figure}[!t]
\centering
\includegraphics[width=\textwidth, trim =30 00 30 80, clip = true]{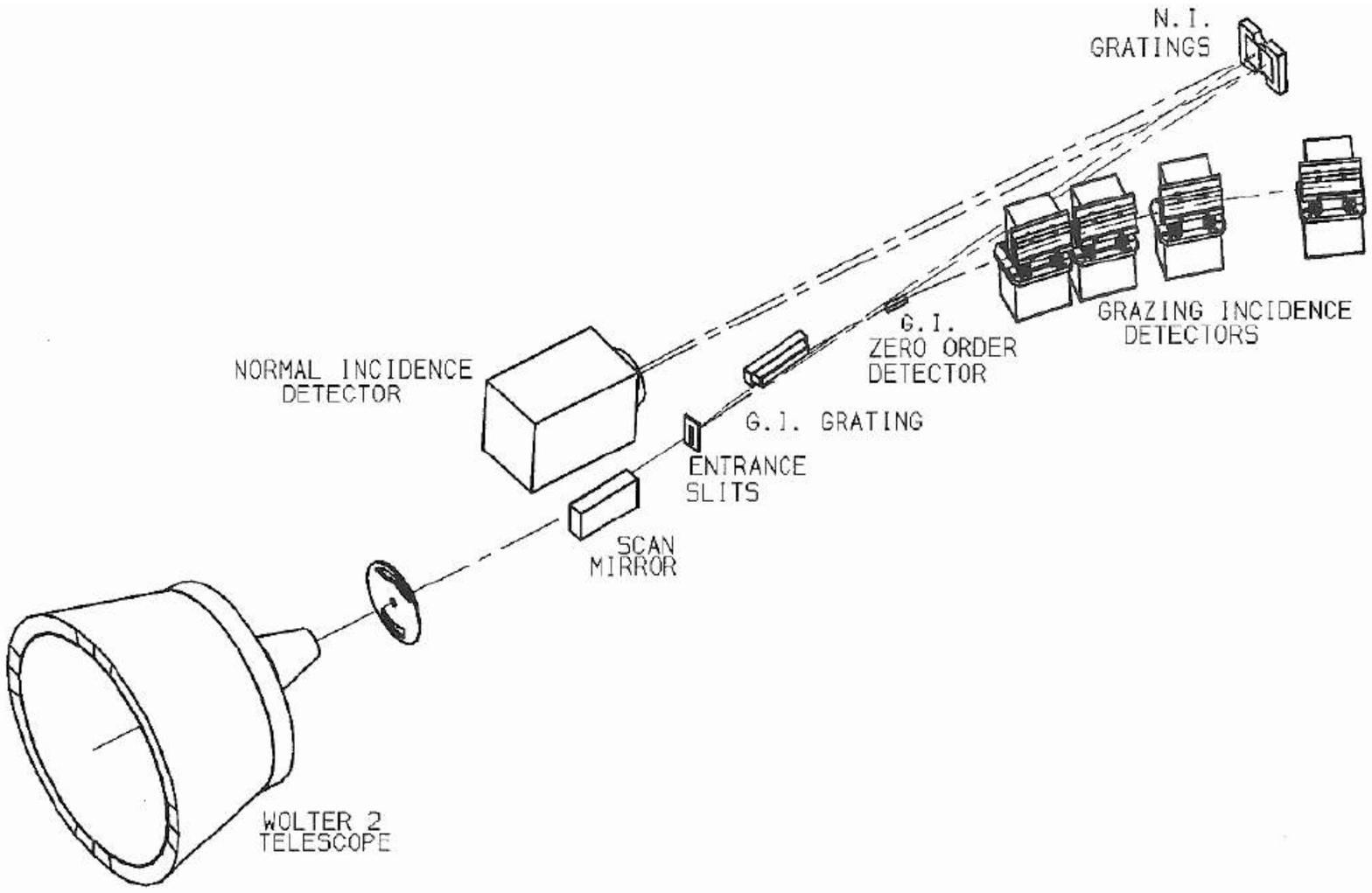} 
\caption{CDS optical layout \citep{Foley03}.}
\label{fig:cds}
\end{figure}
\begin{figure}[!t]
\centering
\includegraphics[width=\textwidth, trim =20 180 50 0, clip = true]{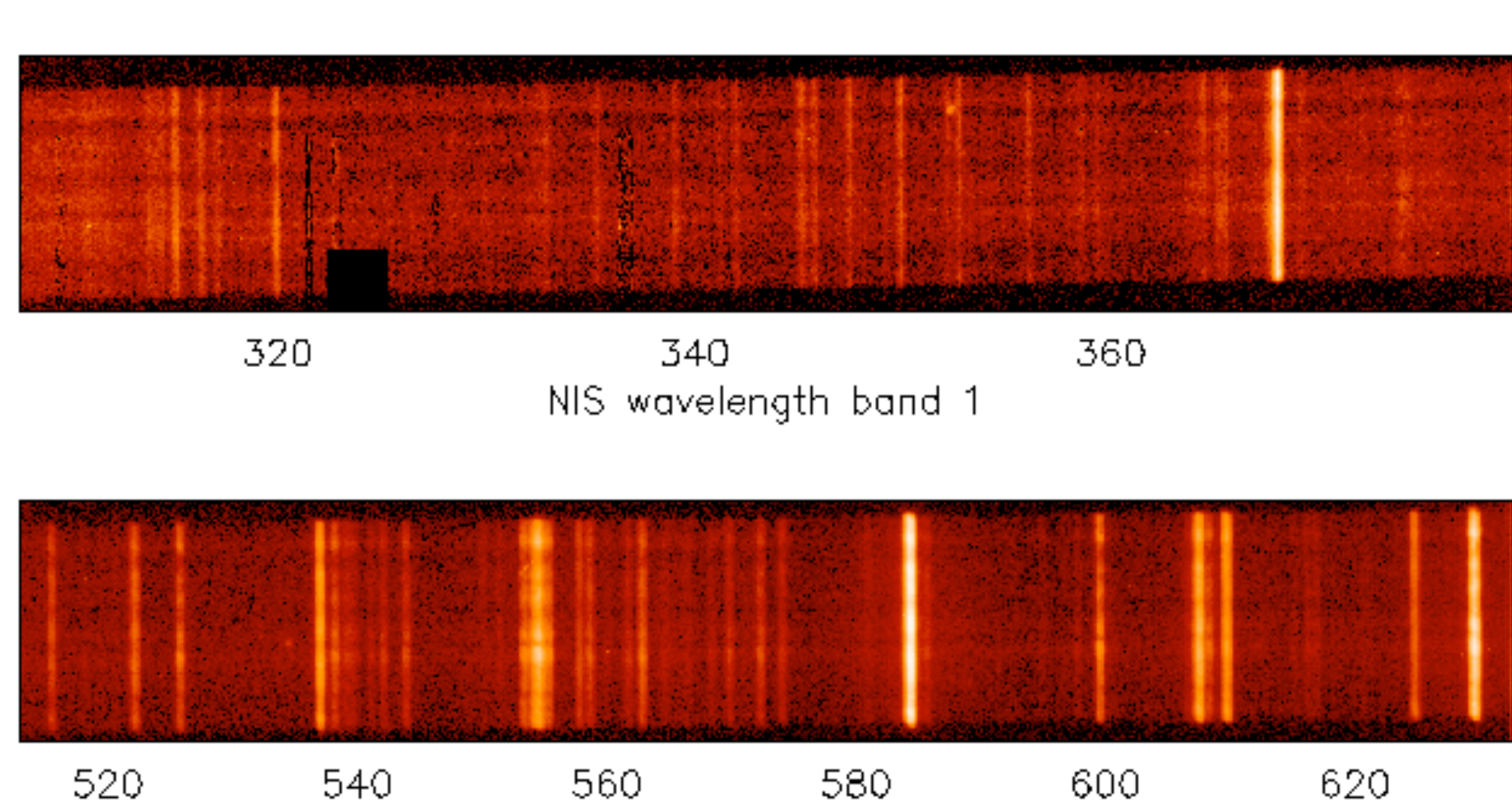} 
\caption{A quiet Sun spectrum as recorded by CDS by passing light through two gratings (NIS1 above and NIS2 below) and imaging the dispersed spectrum onto a detector.}
\label{fig:disp}
\end{figure}

\begin{figure}[!t]
\centering
\includegraphics[width=\textwidth, trim =0 0 0 0, clip = true]{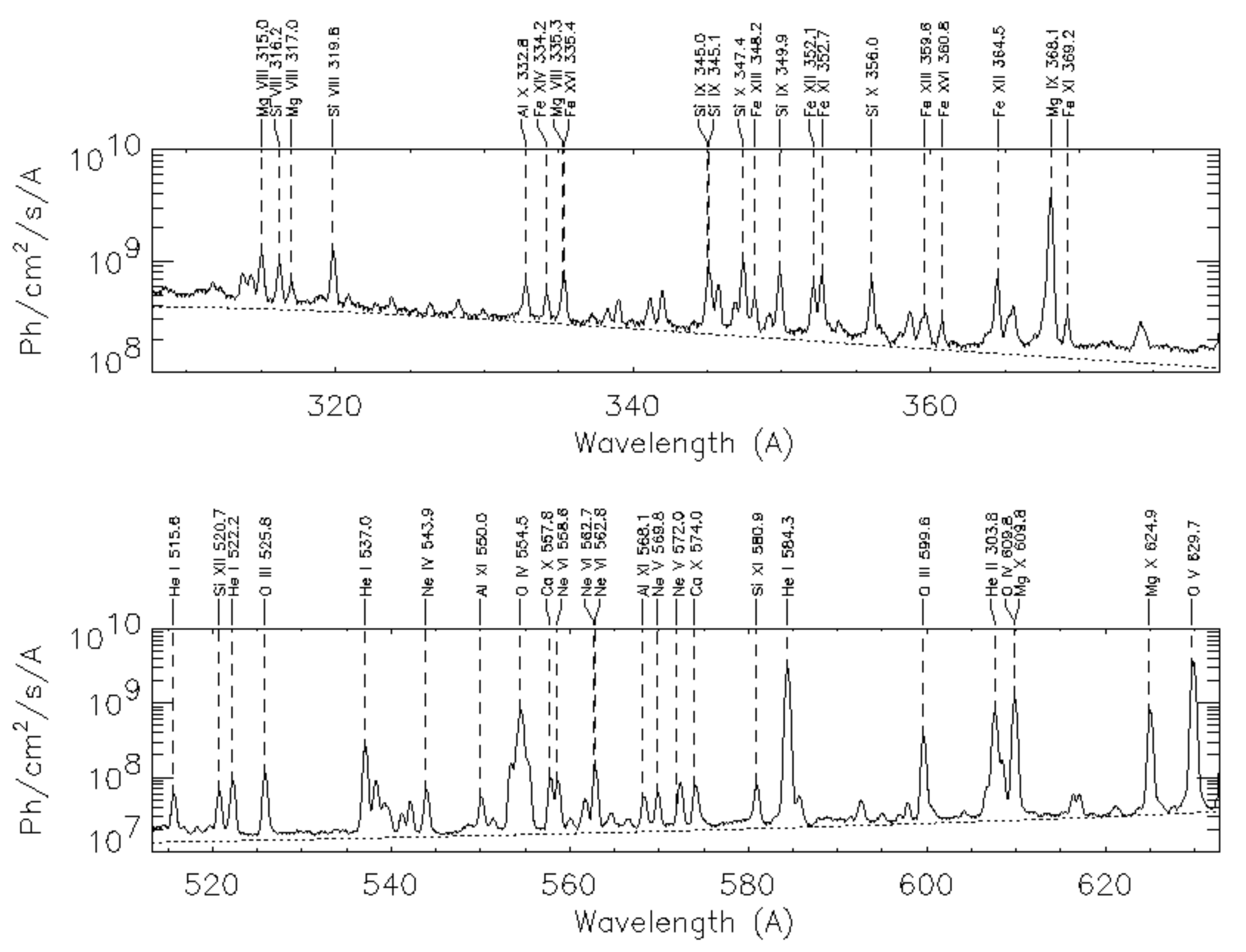} 
\caption{A ``Sun as a star'' spectrum observed with NIS1 and NIS2 across the full spectral range of both gratings. The brightest lines are identified within both spectral ranges. }
\label{fig:cds_spec}
\end{figure}

CDS is a Wolter II grazing incidence telescope whose optical layout is shown in Figure \ref{fig:cds}. Light passing through a scan mirror is focussed onto one of six slits ($2\arcsec \times2\arcsec, 4\arcsec \times4\arcsec, 8\arcsec \times50\arcsec, 2\arcsec \times240\arcsec, 4\arcsec \times240\arcsec, 90\arcsec \times240\arcsec$). The beam is passed into one of two telescope apertures by reflection at grazing incidence. Light entering the Grazing Incidence Spectrometer (GIS) hits a grating in grazing incidence and the spectrum is dispersed onto one of four 1-D detectors. Light entering the the Normal Incidence Spectrometer (NIS) passes through two toroidal gratings in normal incidence. A small out-of-plane tilt in each of the diffraction gratings (mounted side by side) means that the spectra formed by both gratings can be dispersed onto a single 2-D CCD detector as shown in Figure \ref{fig:disp} (note in this figure the spectra are displaced. In practice the spectra are dispersed in the central regions of the CCD). Only the central portion of the CCD is used for observations. In each of the four corners, a 10~pixel $\times$10~pixel box remains unexposed. These regions are used for debiasing and will be explained later (see \S \ref{sect:cds_cal}). 
\begin{figure}[!t]
\centering
\includegraphics[width=\textwidth, trim =60 180 50 50, clip = true]{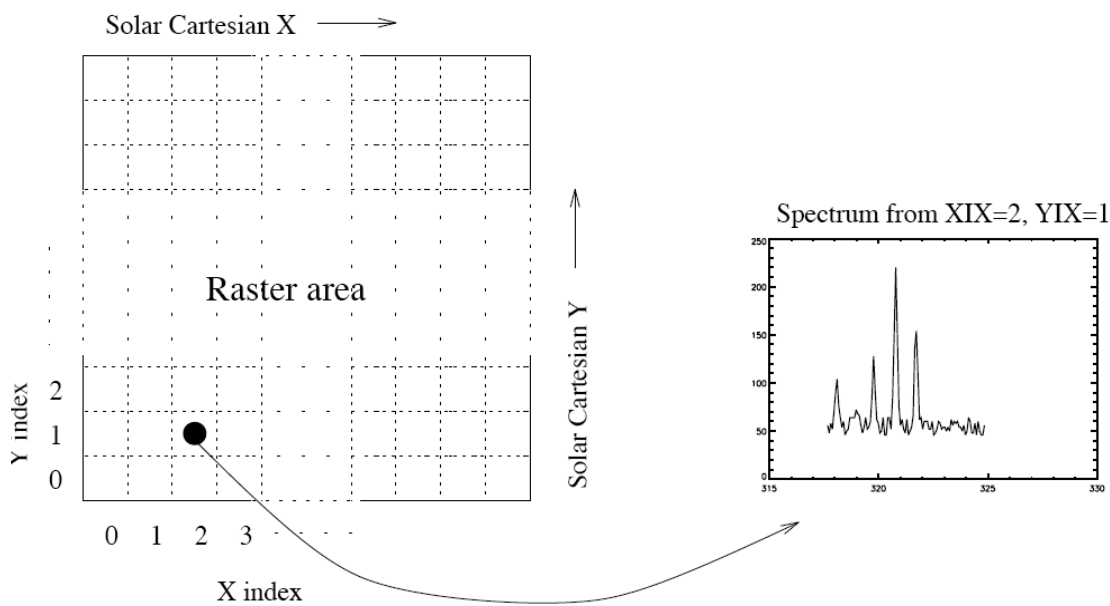} 
\caption{Schematic diagram of an CDS raster. The slit is located along solar Y and stepped from right to left along solar X. At each $\Delta$X$\Delta$Y, the spectrum is recorded onto a CCD \citep{Haugen96}. }
\label{fig:nis}
\end{figure}
NIS is a rastering spectrometer that builds images by stepping a slit from right to left using a scan mirror. A schematic diagram of such an image is shown in Figure \ref{fig:nis}. Here, each column represents the row of pixels along the slit from which a spectrum can be obtained. As the slit moves from right to left, a 3-D structure is built in solar \textsc{x, y} and $\lambda$. This enables the observer to extract the spectrum for an individual, or group of pixels. This type of image is achieved at the expense of temporal resolution. For CDS, the effective cadence of a single image (i.e. the time it takes to step through the \textsc{x} dimension) is $\sim$11~minutes. Alternatively, if timing is an important aspect of the investigation, the slit position can remain fixed and the \textsc{x} dimension can be exchanged for temporal resolution. The instrument has two gratings with wavelength ranges of $310-380$~\AA\ for NIS1 and $520-630$~\AA\ for NIS2, shown in Figure \ref{fig:cds_spec} with the brightest lines identified. Due to telemetry limitations, only a portion of the whole spectrum can be down-linked. The selection of lines to study, the slit used to generate the spectrum and the exposure times are three of the main components that make up a CDS study. 

In this thesis, we utilise the CDS observing study \textsc{flare\_ar}. This focusses on five emission lines spanning a broad range of temperatures. The \ion{He}{i} line at 548.5~\AA is an optically thick chromospheric line and has a formation temperature of $3\times10^4$~K. A 0.25~MK \ion{O}{v} line is observed at 629.8~\AA. Coronal lines \ion{Mg}{x} (625~\AA) and \ion{Fe}{xvi} (350.89~\AA) are formed at 1.2 and 2.5~MK respectively. Finally, an \ion{Fe}{xix} flare line that has a formation temperature of 8~MK is observed at 592.3~\AA. Each raster, taken with the 4$\arcsec$ NIS slit utilising only 180$\arcsec$ in the Y direction, consists of 45 slit positions, each 15~s long. This give an effective cadence of $\sim$11~minutes and a 180$\arcsec \times$180$\arcsec$ field of view.

\subsection{CDS data analysis}
Each CDS file contains spectral and spatial data of a single raster, along with engineering information. This information is stored in a \emph{FITS} file \citep[Flexible Image Transport System;][]{fits}. FITS is designed specifically for scientific data and is widely used for storing photometric and spatial calibration information, together with image origin metadata. In order to retrieve spatial and spectral data from the fits structure, routines from \textsc{ssw} are used. \textsc{sswidl} consists of a ``tree'' of routines developed by the solar physics community, with ``branches'' for the preparation and analysis of different instruments. 

\begin{figure}[!t]
\centering
\includegraphics[width=0.7\textwidth, trim =0 0 0 0,clip = true]{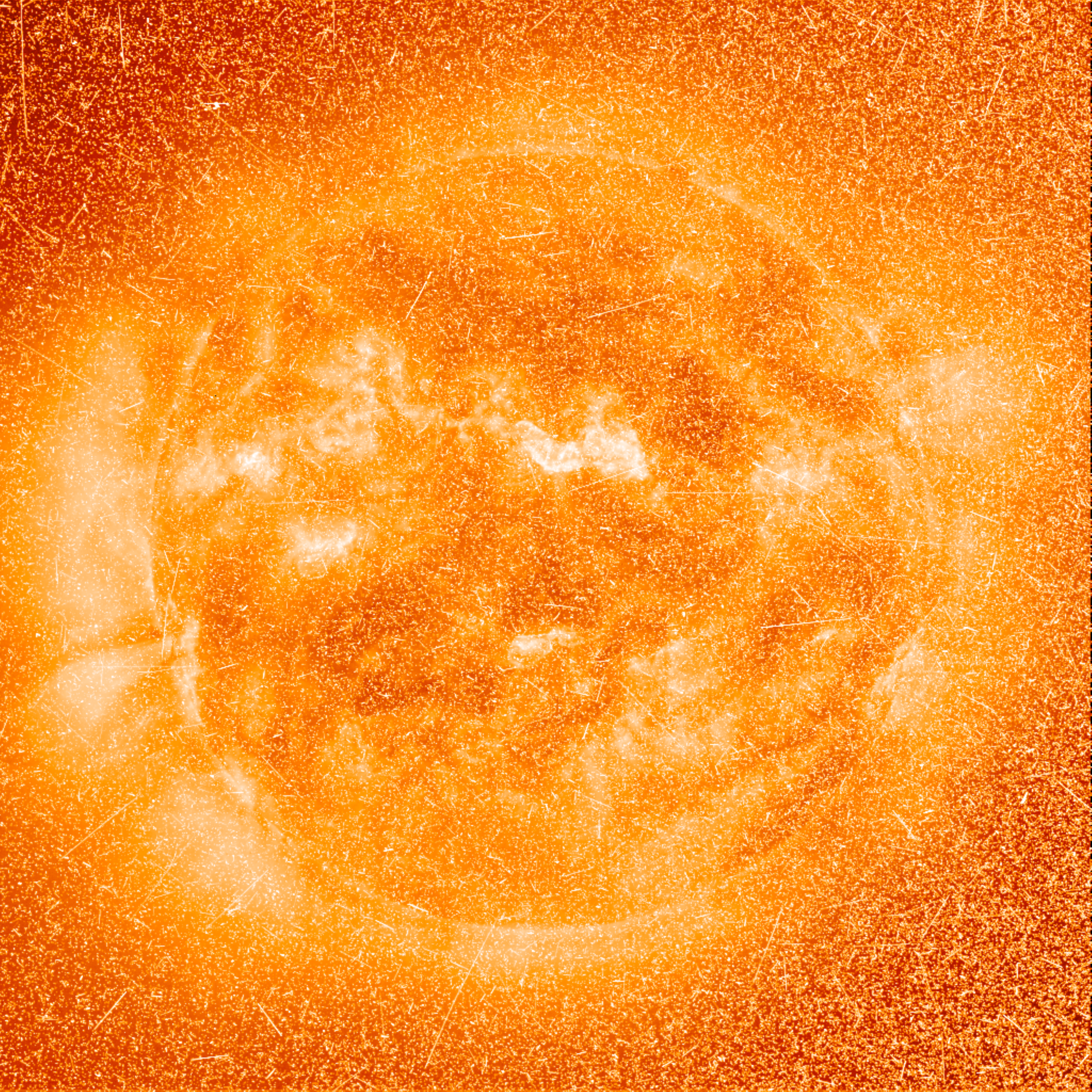} 
\caption{The ``salt and pepper'' effect caused by cosmic rays on a SOHO/EIT image.  }
\label{fig:cosmic}
\end{figure}

\subsubsection{Data preparation}
\label{sect:cds_cal} 
The CDS branch of the \textsc{ssw} tree consists of a series of routines to retrieve and prepare data. The following steps were applied to many of the other instruments used for this thesis, especially the EUV imagers. The preparation steps included:
\begin{itemize}
\item \textbf{Cosmic ray cleaning:} NIS, and all space-based CCD detectors are subject to contamination by cosmic rays passing through the instrument. When this happens, the cosmic ray can leave a high intensity streak across images. Figure \ref{fig:cosmic} shows the ``salt and pepper'' damage these high energy particles can have on an image. Using \textsc{cds\_clean}, any pixels above the $3\sigma$ level are replaced using bilinear interpolation of its neighbours. 
\item \textbf{Flat fielding:} Each CCD pixel has a different response to incoming photons. The difference in gain and dark current can be removed by dividing data by  pre-flight flat-field exposures. This should result in a uniform sensitivity across the pixel. For NIS, the flat fielding is done automatically using \textsc{vds\_calib} when \textsc{nis\_calib} is called. 
\item \textbf{Debiasing:} A DC bias potential is applied to the CCD before it is read out. This ensures that the readout is always positive. This bias is removed using \textsc{vds\_debias} (again called automatically by \textsc{nis\_calib}). In each of the four corners of the CCD, a small (10~pix$\times$10~pix) region remains unexposed. Subtracting the average value of each of these dark ``frames'' from the respective quadrant removes the DC bias.
\item \textbf{Calibration:} The conversion of raw data numbers into physical units requires consideration of the reflectivity of the primary, secondary and scan mirrors, the efficiency of the grating, the quantum efficiency of the detector and the wavelength calibration. For CDS, these effects are considered in \textsc{nis\_calib} along with the flat-fielding and debias techniques mentioned above. This technique, applied to EUV imagers, is described in detail in Chapter \ref{chapter:SWAP}. 
\item \textbf{Slant and tilt correction:} The CDS wavelength axis is not perfectly aligned with the detector axis. This is as a result in an offset between the gratings and detector. This results in the dispersed spectrum appearing slanted relative to the detector (this effect can be seen clearly in Figure \ref{fig:disp}). This tilt, along with an error along the Y axis (due to an error in the slit assembly) is removed using the \textsc{nis\_rotate} routine.
\end{itemize}

\subsection{CDS line fitting}
\begin{figure}[!t]
\centering
\includegraphics[width=1\textwidth, trim =75 40 180 40,clip = true]{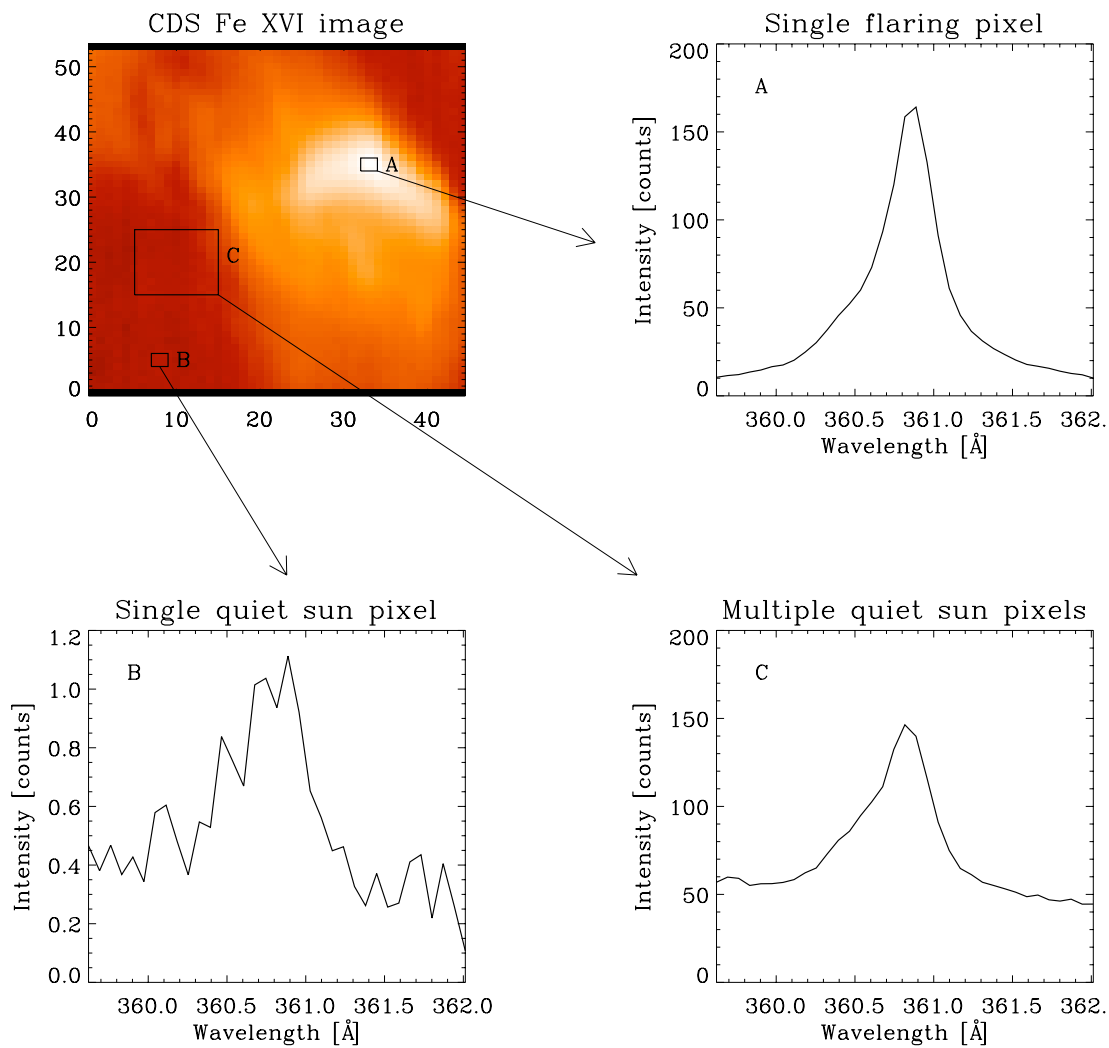} 
\caption{CDS \ion{Fe}{xvi} image (top left) and spectra for a single flaring pixel (spectrum A), single quiet Sun pixel (spectrum B) and multiple quiet Sun pixels (spectrum C). }
\label{fig:indiv_pix}
\end{figure}

Once the data has been prepared and cleaned, an investigation of individual lines can take place. As mentioned before, spectral data is available for individual pixels. However, in many cases (especially for non-flaring conditions), the intensity of a line in a single pixel is too low to be accurately modelled. Therefore, it is often necessary to sum the emission from multiple pixels in order to achieve a well defined spectral line, as shown in Figure \ref{fig:indiv_pix}.

Before the three month loss of contact, the CDS emission lines were modelled using Gaussian distributions. However, damage to the instrument resulted line profiles with extended wings (see comparison in Figure \ref{fig:post_rec}). In order to analyse these lines, a single Gaussian was insufficient. \citet{Thompson99} modelled post recovery line profiles, approximated by one or more broadened Gaussian functions:
\begin{figure}[!t]
\centering
\includegraphics[width=\textwidth, trim =20 100 50 35, clip = true]{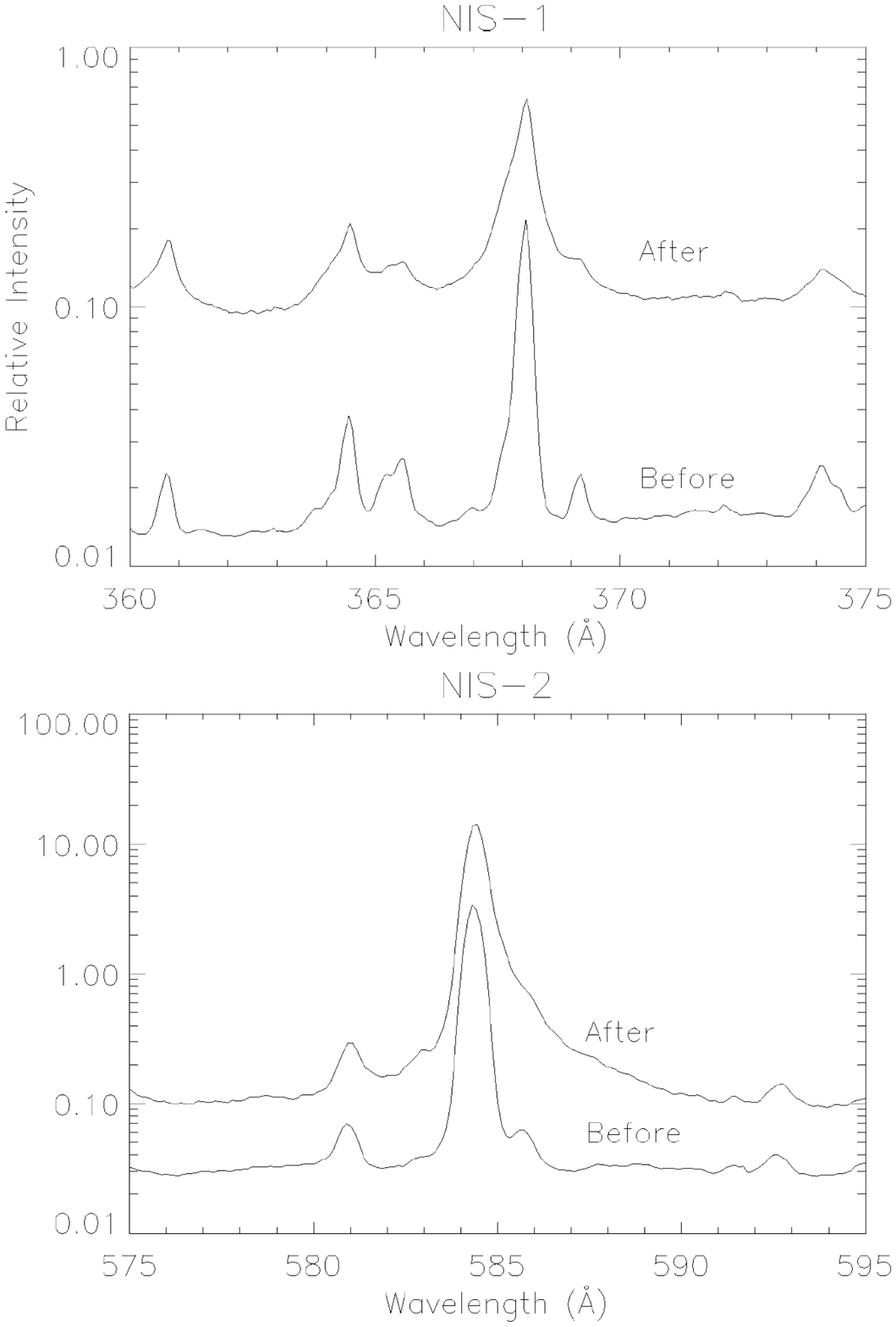} 
\caption{Comparison of line profiles from before and after loss-of-contact for NIS1 and NIS2. The faint lines illustrate the lines profiles prior to loss-of-contact on 25 June 1998. The bolder lines show the broadened profiles after recovery in October 1998 \citep{Thompson99}.}
\label{fig:post_rec}
\end{figure}

\begin{equation}
\label{eqn:post_rec_lines_g}
    G(\lambda) =  \emph{exp}\left[-\frac{1}{2}\left( \frac{\lambda - \lambda_{0}}{\sigma} \right)^{2} \right]
\end{equation}
\begin{equation}
\label{eqn:post_rec_lines_w}
    W(\lambda) =  \frac{1}{  \left(\frac{\lambda - \lambda_{0}}{\sigma '}\right)^{2} +1   }
\end{equation}
where $G$ represents the Gaussian term, $W$ represents the wings, $\lambda$ is the wavelength, $\lambda_{0}$ is the rest wavelength, $\sigma$ is the Gaussian width and $\sigma '$ is the FWHM. The combined profile can be expressed as 
\begin{equation}
\label{eqn:post_rec_lines_b}
    B(\lambda) = I\left[\left(1 - a\right)G(\lambda) + aW(\lambda)\right]
\end{equation}
where $I$ is the amplitude of the line profile and $a$ is the amplitude of the red or blue wing. The best fit of a particular function to a line was established by $\chi^2$ minimisation. 
\begin{equation}
\chi^2 = \sum_i \frac{(I_{i(obs)} - I_{i(mod)})^2}{\sigma_i^2}
\end{equation}
where $\sigma_i^2$ is the RMS error in $I_{obs}$.

\begin{figure}[!t]
\centering
\includegraphics[width=1\textwidth, trim =50 00 50 250,clip = true]{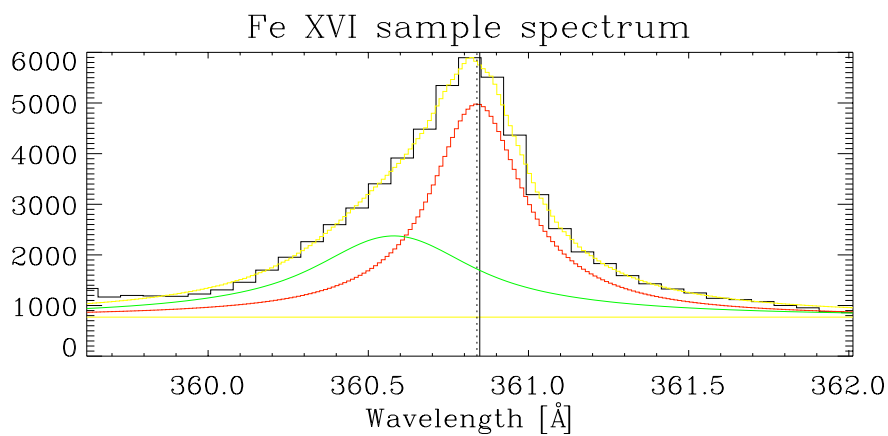} 
\caption{An example of a blended \ion{Fe}{xvi} line, where red corresponds to the \ion{Fe}{xvi} line, the green line is a blend, the yellow histogram is the total fit and the yellow horizontal line is the background level. The black histogram corresponds to observed data and the vertical lines are the centroid position. }
\label{fig:blend_spec}
\end{figure}

If a spectral window is known to contain more than one line, i.e. the line of interest is blended or contaminated, a second, and sometimes third broadened Gaussian function is required to account for and remove all line blends (e.g. Figure \ref{fig:blend_spec}). The relative intensities of the blended lines changes throughout the course of a flare. As the temperature rises, ``flare'' lines, i.e. those formed at high temperatures, become more dominant. Often it is the case that at the peak of a flare, the lines of interest completely dominate the spectral window. However, as the flare cools, the blends become more significant and can influence the parameters such as width and centroid of the line of interest.

\subsection{CDS Doppler shifts}
\label{sect:dop_shift}
\begin{figure}[!t]
\centering
\includegraphics[width=1\textwidth, trim =100 300 100 20, clip = true]{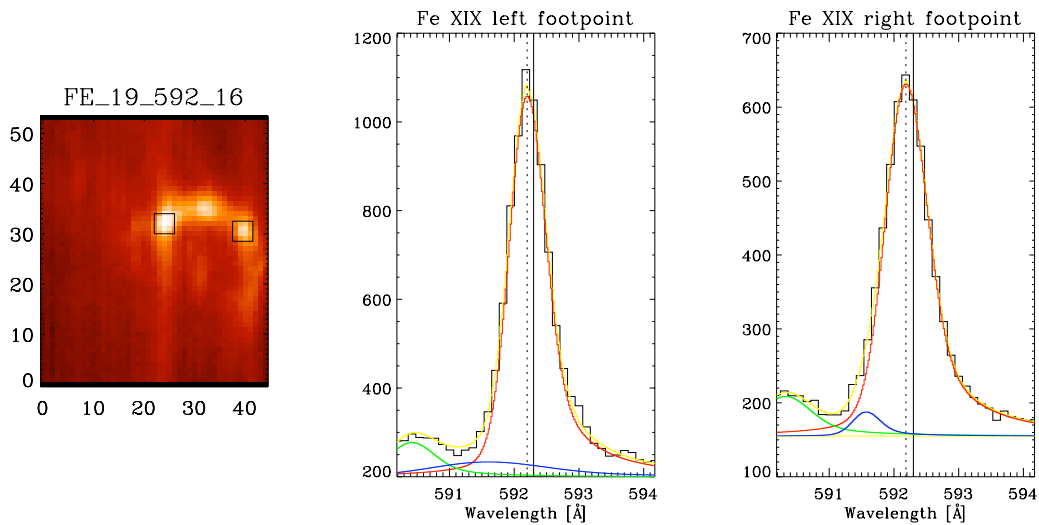} 
\caption{\ion{Fe}{xix} map with footpoints highlighted by black boxes. The integrated spectrum of each footpoint are shown with the centroid and rest wavelengths marked by dotted and solid lines respectively. The red component shows the fit of the \ion{Fe}{xix} lines, the green and blue components are blended lines' fits, the yellow horizontal line is the background and the yellow histogram shows the overall fit. }
\label{fig:dop_shift}
\end{figure}

A useful diagnostic tool that emission line spectroscopy provides is the ability to analyse the velocity of the plasma. Moving plasma will emit Doppler shifted radiation. Measuring the shift in wavelength can give the magnitude of the velocity and measuring the direction of the shift i.e. towards the blue or red end of the spectrum, will give the direction. Caution must be taken when analysing Doppler shifts as only the line of sight component of the velocity will register any shift. If the plasma motion is not parallel to the line of sight, then a correction factor is required. The Doppler shift is given by:

\begin{equation}
\label{eqn:dop_vel_val}
\frac{v}{c} = \frac{\lambda-\lambda_{0}}{\lambda_{0}}
\end{equation}
where $v$ is the plasma velocity, $c$ is the speed of light, $\lambda_0$ is the rest wavelength and $\lambda$ is the shifted wavelength. Figure \ref{fig:dop_shift} gives a clear example of Doppler shifts for the footpoints of a loop. The deviation from the rest wavelength (solid line) is shown, along with the Gaussian profiles and the shifted wavelength (dotted line). CDS has been used to resolve velocities to an uncertainty of 4.7~km~s$^{-1}$ \citep{Gallagher99} for \ion{He}{i}, although a more conservative uncertainty of 10~km~s$^{-1}$ is often used \citep[e.g.][]{Brekke97, Raftery09}.

\subsubsection{Rest wavelength calculation}
\label{sect:rest_wv}
The calculation of the rest wavelength for individual emission lines is essential for the accurate measurement of the Doppler shifts. For the cooler lines (\ion{He}{i}, \ion{O}{v}, \ion{Mg}{x} \ion{Fe}{xvi}) the rest wavelength was obtained from a region of quiet Sun, since it contains plasma emitting at the temperatures at which the above lines are formed. The rest wavelengths for CDS chromospheric and transition regions lines are known to be redshifted \citep[e.g.][]{Doschek76, Feldman82, Brekke97}. \citet{Brekke97} measured the redshift of the \ion{O}{v} 629~\AA\ line using the Solar Ultraviolet Measurements of Emitted Radiation \citep[SUMER;][]{Wilhelm95}. They found a maximum shift of 16$\pm$3~km~s$^{-1}$ for the \ion{O}{v} line and 6$\pm$1.5~km~s$^{-1}$ for \ion{Mg}{x}. Since no absolute Doppler shifts can be obtained with CDS, we adopt the upper limit of 16~km~s$^{-1}$ as the error on our observations. Obtaining a rest wavelength for the \ion{Fe}{xix} line was more difficult since no \ion{Fe}{xix} exists in the quiet sun. The orientation of the loop was such that any plasma flowing along the loop would have little or no line of sight velocity component. Therefore, it was assumed that the net velocity of plasma at the top of the loop would be zero as the components flowing upwards from each leg will cancel out, resulting in the ``rest'' wavelength for \ion{Fe}{xix}.

\section{EUV imagers}

\subsection{Sun Watcher using Active Pixel System detector and Image Processing (SWAP)}
\label{sect:swap}
The launch of the ESA Proba-2 mission in November 2009 has brought with it the commissioning of the Sun Watcher using Active Pixel System detector and Image Processing \citep[SWAP;][]{swap}: a full disk EUV imager with a single filter centred on the 171~\AA\ \ion{Fe}{ix/x} line. With a plate scale of 3.1$\arcsec$ per pixel, the spatial resolution of the instrument is comparable to SOHO/EIT. However, the extra wide field-of-view ($54' \times 54'$) makes this unique instrument ideal for tracking coronal mass ejections from the low corona out to more than 3$R_\odot$. 
\begin{figure}[!t]
\centering
\includegraphics[width=\textwidth, trim = 50  160 40  180, clip = true]{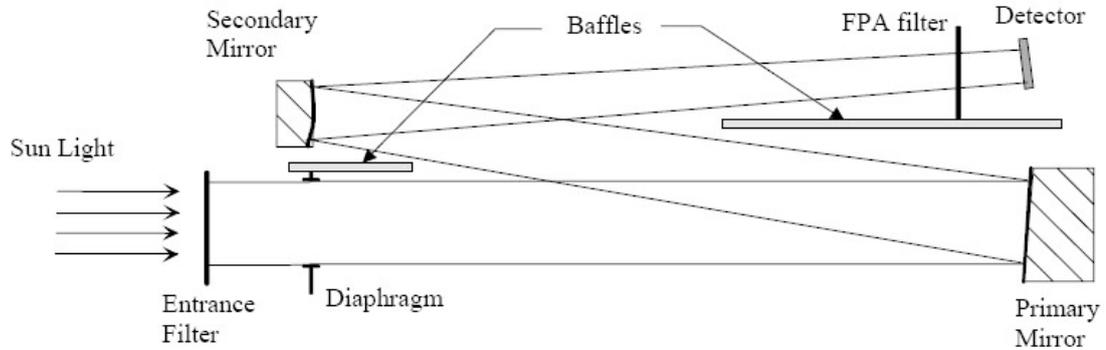} 
\caption{A schematic diagram of SWAP's optical components. }
\label{fig:swap}
\end{figure}

SWAP's optical layout, shown in Figure \ref{fig:swap}, includes a set of aluminium foil filters. The front filter is supported by a polyimide membrane and the rear filter by a nickel grid. The filters, combined with EUV multilayer coatings on the mirrors tune the sensitivity of the instrument to between 171~\AA\ and $\sim$183~\AA. With first light not expected until January 2010, we cannot include details on image processing here, however the telescope was used in a calibration effort for EUV imagers described in Chapter \ref{chapter:SWAP}. 

\subsection{Transition Region and Coronal Explorer (TRACE)}
The Transition Region and Coronal Explorer \citep[TRACE;][]{Handy99} is a space based imaging telescope, designed to take high cadence, high resolution images of the different layers of the solar atmosphere. The large aperture, small field of view (8.5$'\times$8.5$'$) and small plate scale (0.5$\arcsec$ pixels) make the resolution of TRACE the highest of any current EUV imagers, though at the cost of coverage area. Similar to SWAP, TRACE's multilayer mirror coating and filters tune the sensitivity of the instrument to one of eight passbands in the UV and EUV. TRACE has seven filters, three of which are in the EUV range (171, 195 and 284~\AA), four in the UV (1216, 1550, 1600, 1700~\AA) and a broad whitelight passband. TRACE data is available online and, like CDS, can be cleaned and prepared using \textsc{sswidl} routines. The basic techniques are similar to those described in \S \ref{sect:cds_cal}.

\subsection{The Extreme Ultraviolet Imagers (EUVI)}
The Extreme Ultraviolet Imagers \citep[EUVI;][]{Howard08} on board the twin Solar Terrestrial Relations Observatories \citep[STEREO;][]{Kaiser08}, Ahead (A) and Behind (B) have provided the solar community with two full disk EUV imagers whose spatial resolution are almost 40\% better than EIT. EUVI has four passbands; 304~\AA, 171~\AA, 195~\AA, and 284~\AA. Again, the multilayer coatings and filters tune the instrument sensitivity to these passbands. The cadence of EUVI is significantly better than EIT (2 mins vs. 10 mins) and the 43'$\times$43' field of view makes the EUVI instruments optimal for studying large scale coronal disturbances. EUVI (along with the coronagraphs listed in \S \ref{sect:coronagraphs}) are prepared and cleaned using the \textsc{secchi\_prep} routine, in a similar manner to \S \ref{sect:cds_cal}.

\section{Coronagraphs}
\label{sect:coronagraphs}
The Cor coronagraphs from the Sun Earth Connection Coronal and Heliospheric Investigation \citep[SECCHI;][]{Howard08} suite of instruments on board STEREO B were used for work in this thesis. A coronagraph is an instrument that simulates a solar eclipse by blocking radiation from the photosphere. This allows the faint emission from the extended corona to be integrated over the line of sight through the optically thin corona. The instrument is designed not only to eliminate photospheric emissions, but also to reduce stray light from aperture diffraction. 

\subsection{Cor1}
\begin{figure}
\centering
\includegraphics[width=\textwidth, trim = 40  40 50  110, clip = true]{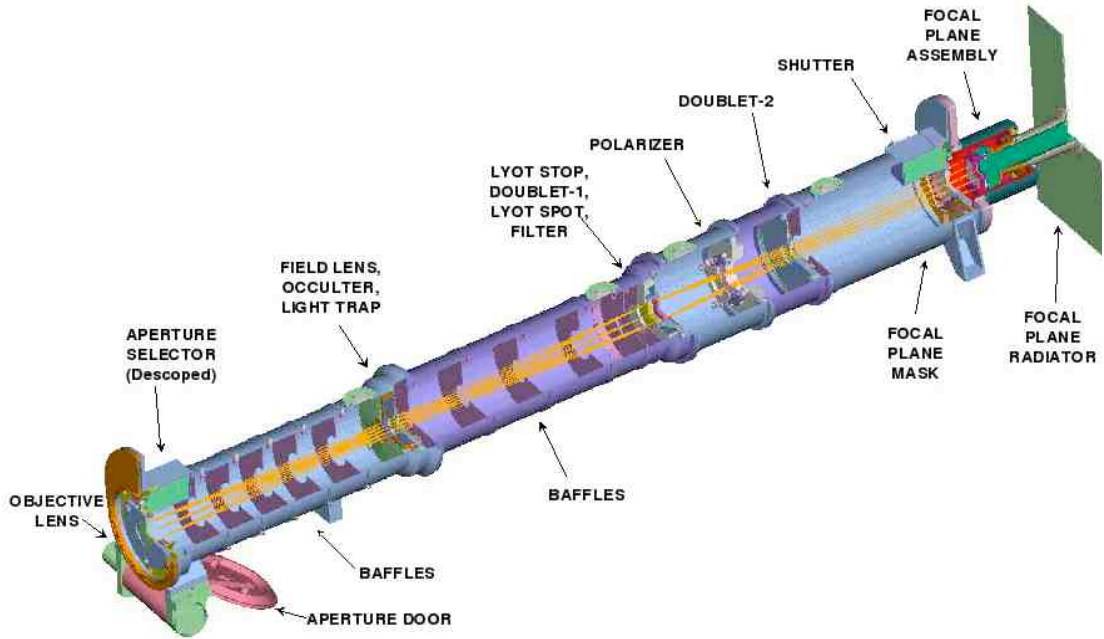} 
\caption{A schematic diagram of the Cor1 coronagraph. }
\label{fig:coronagraph}
\end{figure}

The Cor1 telescope \citep{Howard08} is an internally-occulted coronagraph based on the classic design by \citet{Lyot39} that has a field of view from 1.4 - 4 $R_{\odot}$. A schematic diagram of the telescope's optics are shown in Figure \ref{fig:coronagraph}. Light enters the telescope and is focussed by the objective lens onto the occulting spot. This spot absorbs light from the center of the field of view, removing much of the emission from the central source, in this case, the solar disk. This has the effect of forming diffraction rings around the occulting spot. A Lyot stop is used to remove these rings. Additional stray light is removed by placing baffles at various points between the front aperture and the Lyot stop. A linear polariser placed between the Lyot stop and the final lens is used to suppress scattered light (F corona emission) and to extract the polarised brightness signal from coronal emissions.  

\subsection{Cor2}
Cor2 is an externally occulted coronagraph with a field of view of 2.5 to 15 $R_{\odot}$. Unlike Cor1, the solar disk is occulted at the entrance aperture of the telescope. This has a vignetting effect on the image formed at the objective lens (this is why an internal occulter was chosen for Cor1). The external occulter is cantilevered with a single pylon to reduce diffraction. Internal baffles between the occulter and the objective lens help to reject stray light. Further to this, a heat rejection mirror guides any remaining photospheric light out of the light box, preventing the reflection off the back end of the occulter. An internal occulter rejects the bright fringe of diffracted light that forms around the primary occulter. The light path is then guided through a Lyot stop and a filter wheel, onto the detector. See \citet{Howard08} for further details.

\section{Geostationary Operational Environmental Satellites (GOES)}
\label{sect:instr_goes}

\begin{figure} [!t]
\centering
 \includegraphics[width=\textwidth, trim =0 0 0 0, clip = true]{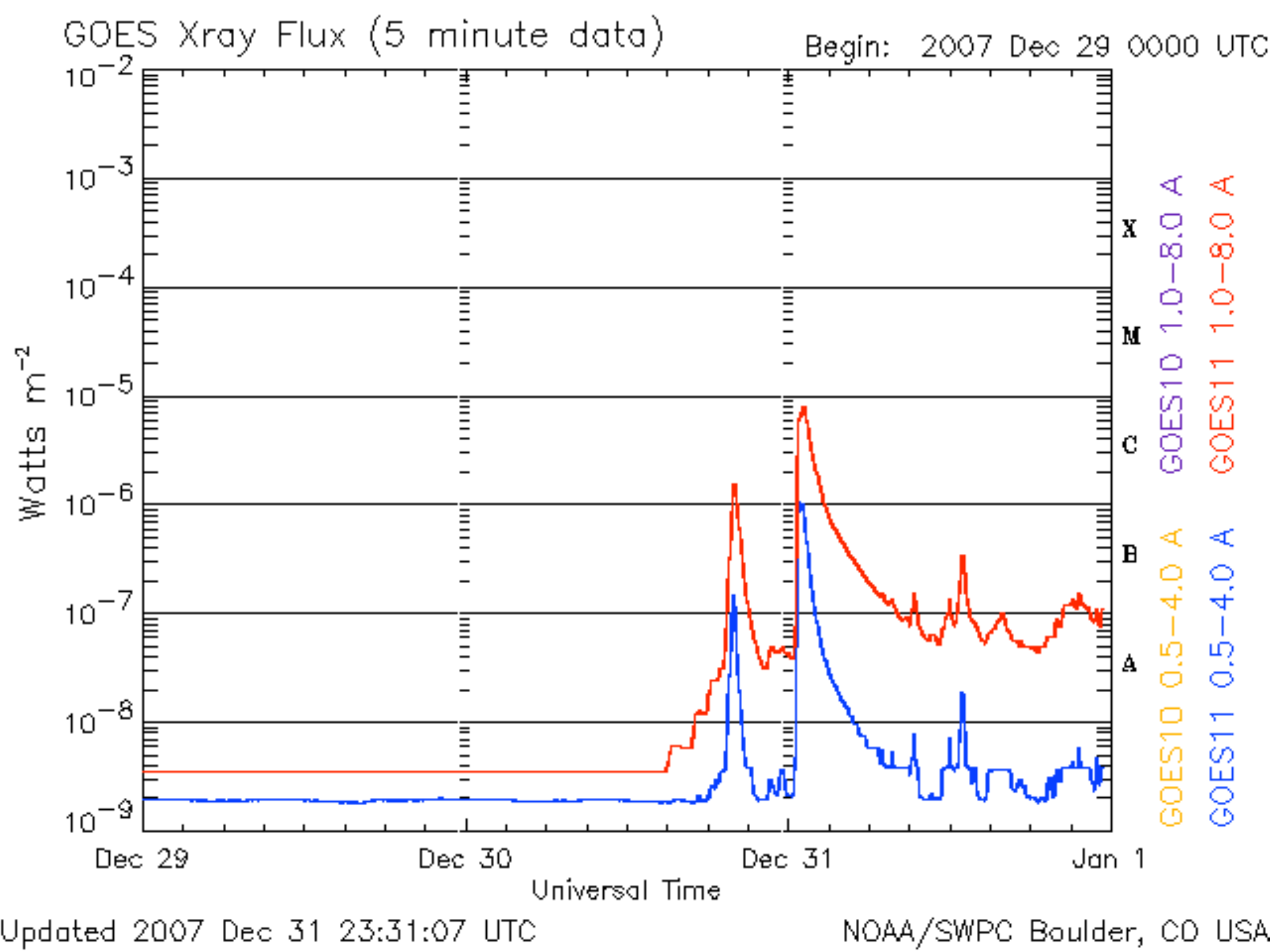} 
 \caption{GOES 0.5-4~\AA\ and 1-8~\AA \ flux for the 29-Dec-2007 to 1-Jan-2008. Note the extremely quiet conditions on the 29-Dec-2007 are suddenly disturbed by the rotation of a flare-producing active region onto the solar limb. Note the large C8.0 class flare at 00:45~UT on 31-Dec-2009 is the focus of Chapter \ref{chapter:CME}. (Image taken from www.solarmonitor.org) }. 
  \label{fig:goes_xray} 
\end{figure}

\begin{figure} [!t]
\centering
 \includegraphics[width=0.8\textwidth, trim =40 0 40 0, clip = true]{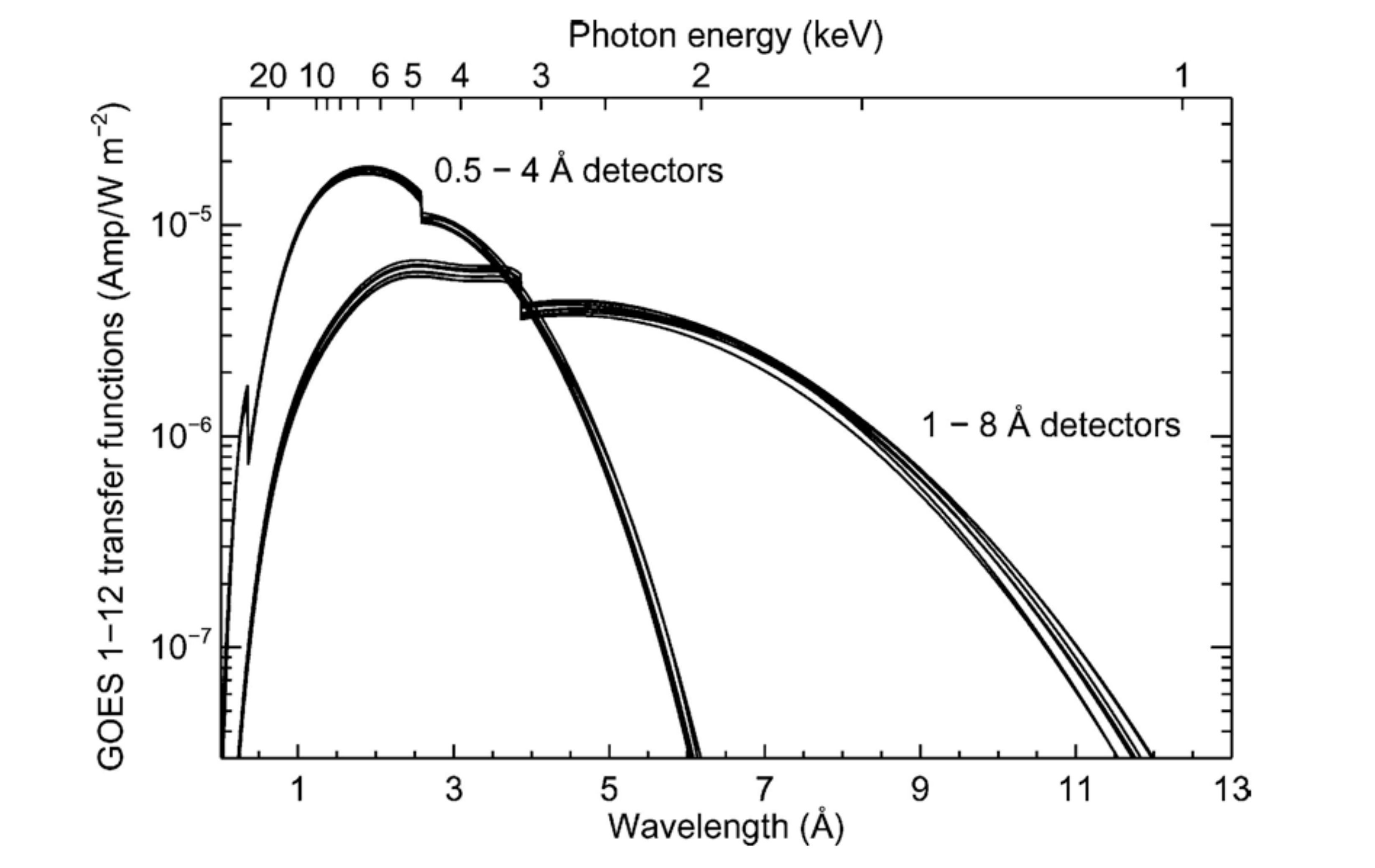} 
 \caption{GOES 0.5-4~\AA\ and 1-8~\AA \ transfer functions \citep{White05}}. 
  \label{fig:gtran} 
\end{figure}

The GOES fleet consist of a series of satellites, the first of which was launched on 16-October-1975. This study makes use of data from the twelfth of the series. These spacecraft are used for ``now-casting'', providing data for severe storm evaluation, information on cloud cover, winds, ocean currents, fog distribution, storm circulation and snow melt on Earth. The instrument also takes continuous observations of the integrated soft X-ray flux of the Sun in the wavelength ranges 0.5-4~\AA \ and 1-8~\AA\ (Figure \ref{fig:goes_xray}). This data is frequently used for Space Weather forecasting. The classification of a flare is obtained from the magnitude of soft X-ray flux observed by GOES (see \S \ref{sect:flares}). This satellite is integral to the past, present and future of solar physics, as it is the only instrument that has been studying the Sun for such an extensive period of time.

Observations from GOES take the form of a flux measurement, taken every three seconds. Two channels are used: 0.5-4~\AA \ and 1-8~\AA. Following \citet{Thomas85}, the flux of a given channel $i$ can be expressed as: 

\begin{equation}
\label{equation:goes1}
F_{i} =  EM \int G(\lambda)_{i}d\lambda /\bar{G} = EM \times b_{i}
\end{equation}
where 
\begin{equation}
\label{equation:goes2}
b_{i} =  \int G(\lambda)_{i}d\lambda /\bar{G}
\end{equation}
contains the temperature dependent components of flux -- the transfer function, $G({\lambda})$, shown in Figure \ref{fig:gtran} and the wavelength range, $d\lambda$ -- and is normalised by the wavelength averaged transfer function, $\bar{G}$.

\begin{figure}[!t]
\centering
 \includegraphics[width=0.8\textwidth, trim =0 00 00 0, clip = true]{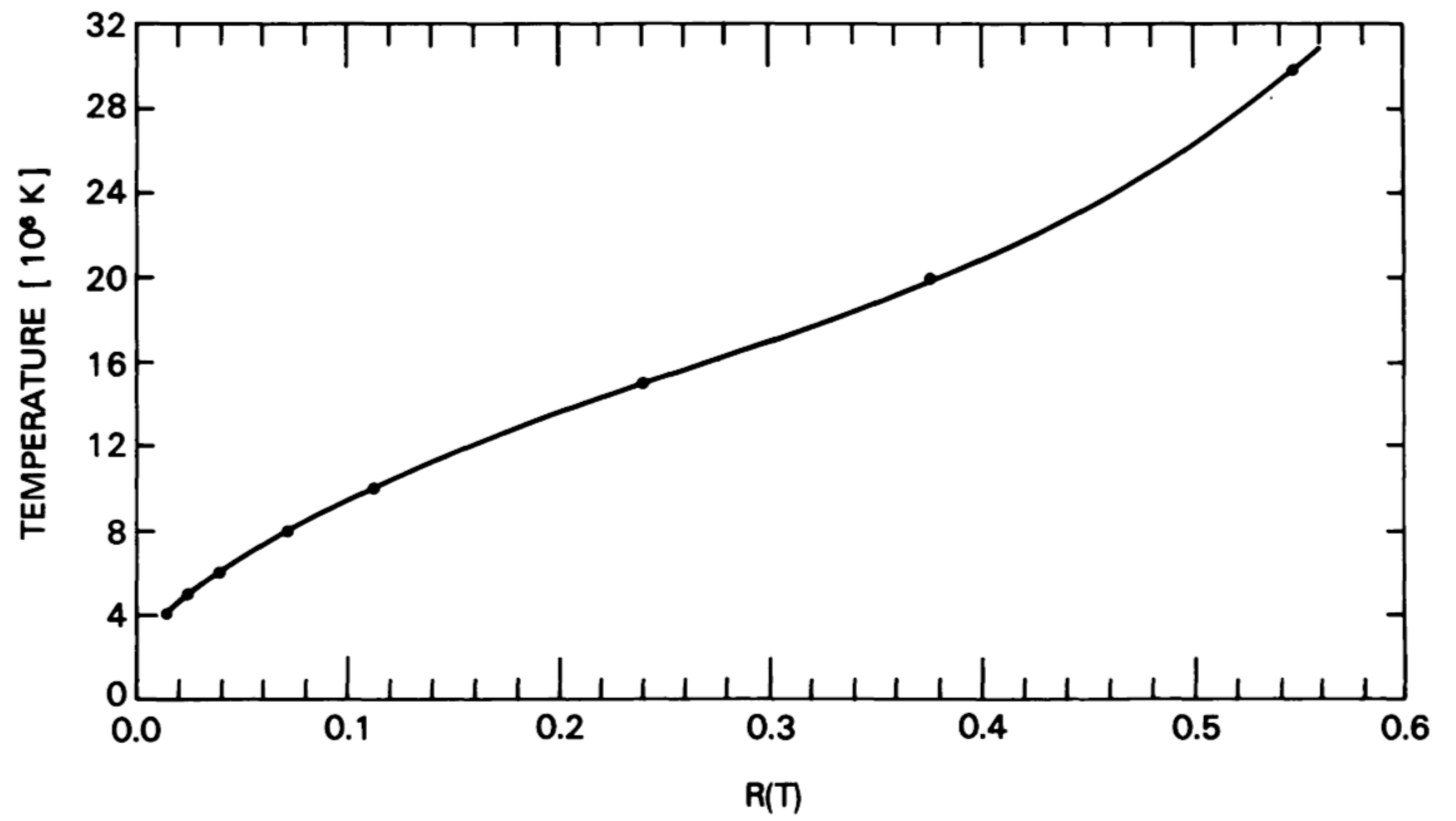} 
 \caption{Flux ratio, $R(T)$ response as a function of temperature \citep{Thomas85}}. 
  \label{fig:goes_resp} 
\end{figure}

The ratio of two channels -- 0.5-4~\AA \ (subscript 4) and 1-8~\AA \ (subscript 8) -- gives the temperature dependent value $R(T)$, where

\begin{equation}
\label{equation:goes3}
R(T) = \frac{F_{4}}{F_{8}} = \frac{b_{4}(T)}{b_{8}(T)}
\end{equation}
and the response of $R(T)$ is shown in Figure \ref{fig:goes_resp}. Once the effective temperature has been found from the ratio of the fluxes, $b_{8}$ can be determined and the emission measure can be calculated by inverting Eqn. \ref{equation:goes1}.


\section{Reuven Ramaty High Energy Solar Spectroscopic Imager (RHESSI)}
\label{sect:rhessi}

RHESSI \citep{Lin02} is a NASA Small Explorer Mission designed to investigate particle acceleration and energy release in solar flares through imaging and spectroscopy of high energy continuum and lines. The instrument consists of nine bi-grid rotating modulation collimators (RMCs) in front of a spectrometer and nine cryogenically cooled germanium detectors, each one behind an RMC. The sensitivity ranges from 3/6~keV up to 17~MeV depending on the attenuator state (see Table \ref{table:attenuator}), covering soft X-rays to $\gamma$-ray emission.

\begin{table}[!t] 
\caption{RHESSI attenuator states and the lowest corresponding energies that can be observed. } 
\begin{center}
\begin{tabular}{l c c c }
\hline
\hline
Attenuator state		&	Attenuator thickness	&	Lowest reliable energy \\ \hline
A0				&	None			&	3~keV\\
A1				&	Thin				&	3~keV\\
A2				&	Thick			&	6~keV\\
A3				&	Thick + thin		&	6~keV\\
\hline
\end{tabular}
\label{table:attenuator}
\end{center} 	
\end{table}

The spacecraft, which rotates once every 4 seconds, is pointed and oriented by the Solar Aspect System (SAS) and the Roll Angle System (RAS). The SAS uses white light images of the solar disk to ensure spacecraft pointing is correct to within 1.5$\arcsec$. This is done through accurate measurement of pitch and yaw angles. The RAS uses background stars to position the spacecraft once every rotation, defining the spacecraft orientation in relation to the star field. 

 \begin{figure} 
\centering
 \includegraphics[width=0.54\textwidth, trim =50 50 70 160, clip = true]{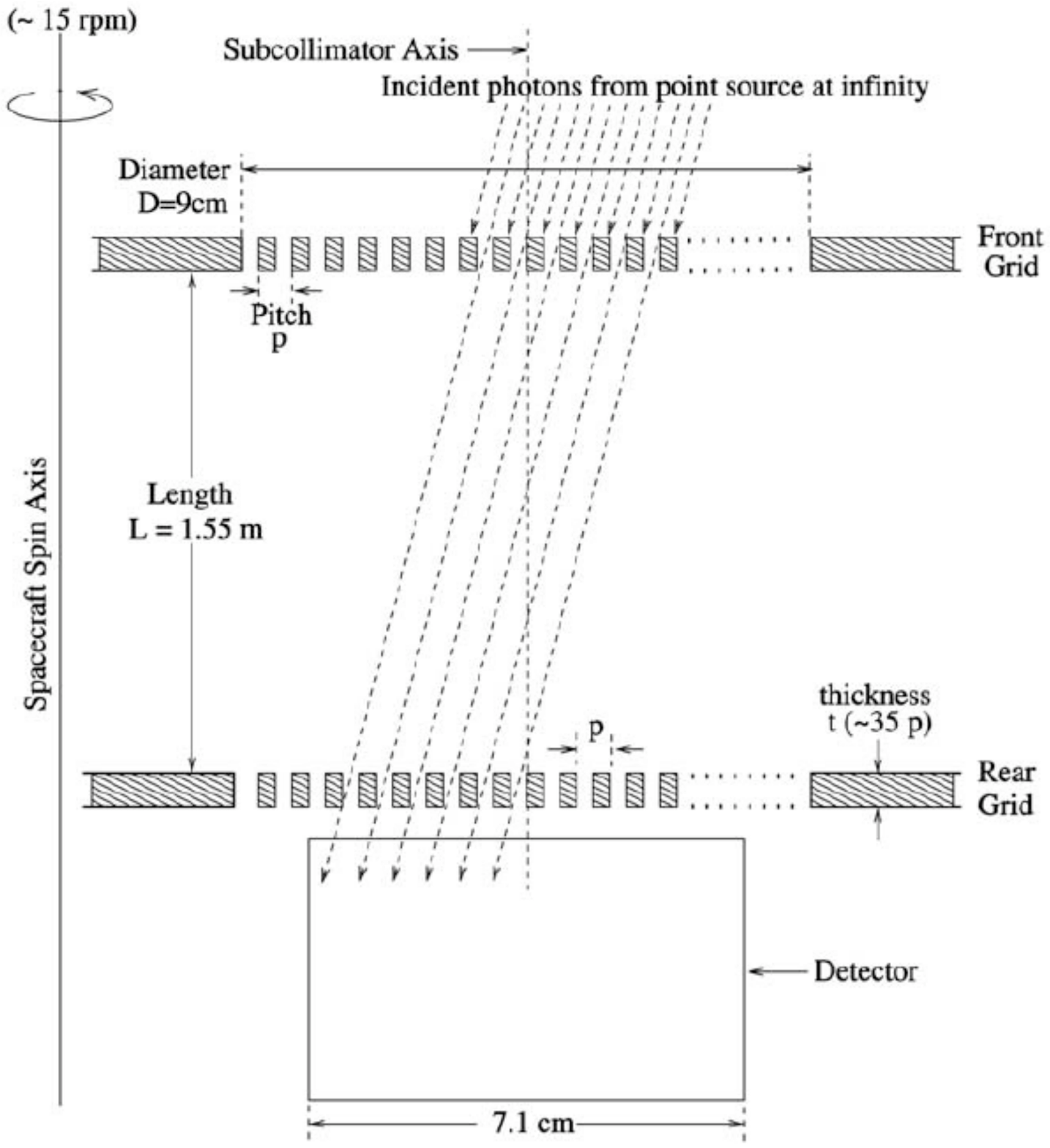} 
 \caption{Schematic of one RHESSI RMC setup. \citep{Hurford02}. } 
  \label{fig:rmc} 
\end{figure}

RHESSI has dual analysis capabilities: imaging and spectroscopy. The details of each are very different and equally complex. They are detailed below.

\subsection{RHESSI imaging}
Imaging high energy photons is very challenging. This is because the wavelength of the light is shorter than the spacing between atoms used in detectors. As such, a more complex method of imaging must be employed to successfully capture the high energy emission from solar flares. This problem is overcome by using the fact that radiation creates a shadow when an object is obstructing the light path. This technique, \emph{collimator-based Fourier-transform imaging} is detailed in \citet{Hurford02}. 

At a given instant in time, an off-axis source will either be occulted by the grid or not. As the spacecraft rotates, the grids will pass infront of the source, modulating the detected signal observed by the detectors as the grids cast a shadow on the detector. The modulation pattern will vary depending on location, intensity and size of the source. The detectors, which have no spatial resolution, record the arrival time and energy of the incident photons, recording the count flux as a function of rotation angle. 

The nine grids each have a different angular resolution, given by:

\begin{equation}
\theta = tan^{-1}\left(\frac{P}{2L}\right)
\end{equation}
The pitch $P$ is the distance between the edge of one slat and the corresponding edge of the neighbouring slat and L is the length of the collimator (see Figure \ref{fig:rmc} for clarification). The grid pitches range from 34~$\mu$m to 2.75~mm in steps of $\sqrt{3}$. Thus, the grids provide resolution from 2.3$\arcsec$ to 180$\arcsec$. 

\begin{figure} [!t]
\centering
 \includegraphics[width=\textwidth, trim =30 50 30 100, clip = true]{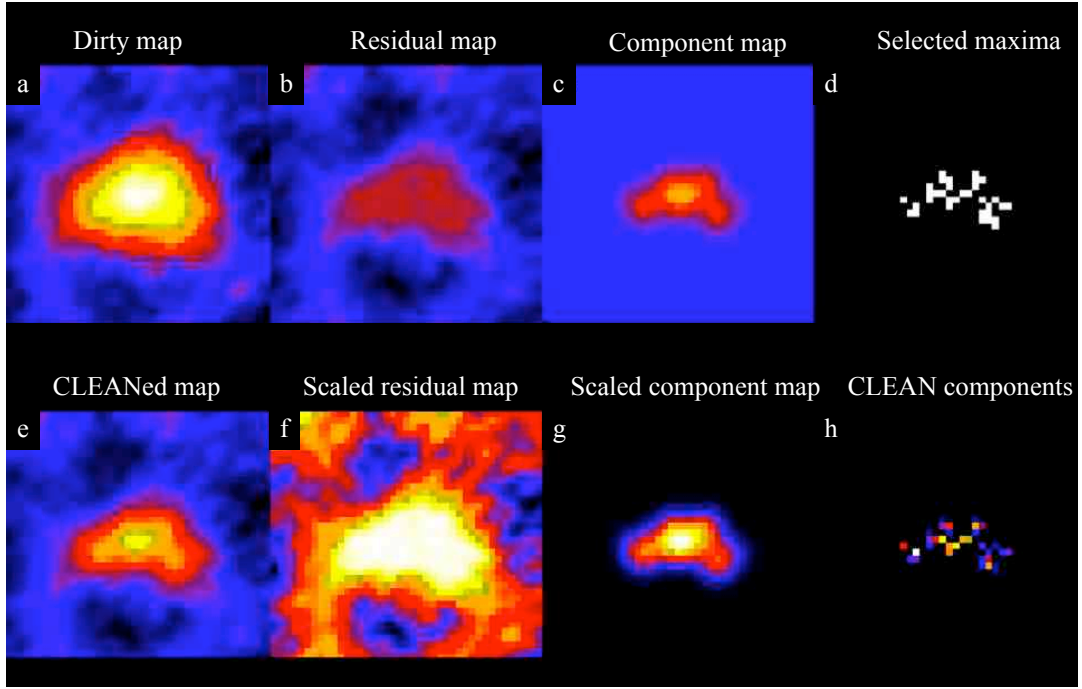} 
 \caption{Steps involved in creating a CLEAN image. (a) shows a ``dirty'' or pre-processed map. (b) shows the residuals following the removal of a CLEAN component (c) which is the convolution of the maximum intensity pixel in the map (d) with the instrument point spread function. The resulting CLEANed map is shown in (e) along with scaled residual and component maps (f, g) and the scaled CLEAN components (h).   } 
  \label{fig:clean} 
\end{figure}

In order to reconstruct an image using RHESSI data, one must select an image reconstruction algorithm. Back projection is the most basic routine. A ``probability'' (or dirty) map is created of the origin of each photon and the probability map for all photons are then summed. In order to remove any artefacts, such as side lobes, a more complex algorithm should be used. The simplest and most efficient of these is the ``Clean'' algorithm, which has been employed for all RHESSI imaging in this thesis. Clean assumes a source can be represented by the superposition of multiple point sources. Figure \ref{fig:clean} shows the various steps in the processing of a ``Clean'' image. Clean removes the sources in the dirty map $(I(x, y))$ which it believes are not real, leaving only the real sources $(O(X, Y))$ convolved with the instrument point spread function $(P(x, y))$.

 \begin{equation}
I(x, y) = \int_{-\infty}^{+\infty} O(X, Y) P(x-X, y-Y) dX dY
\end{equation}

The algorithm begins by first computing the dirty map  $I_n$ (or residual map for $n \ne 0$) as with the back projection method (Figure \ref{fig:clean}A). The maximum value of the dirty map, $I_{n_{max}}$ is ascertained (Figure \ref{fig:clean}D). The point spread function is applied to the point source (Figure \ref{fig:clean}C) and removed from $I_{n}$, leaving the residual map $I_{n+1}$ (Figure \ref{fig:clean}B). This process is repeated until the residual map is at the noise level or a negative maximum is reached. The final Clean map is the convolution of the maxima with the point spread function, added to the final residual map (Figure \ref{fig:clean}E).

\subsection{RHESSI spectroscopy}
\label{sect:instr_hsi_spec}
The second major function of the RHESSI instrument is for spectroscopy of high energy solar events. RHESSI has the capability to observe continuum from the soft X-ray regime to $\gamma$-rays, along with some strong line features (such as the Fe/Ni complex in the thermal range, the electron/proton annihilation line at 511~keV and $\gamma$-ray lines in the 1-10~MeV range). 

Photons entering the Ge detectors lose their energy via photoelectric absorption, Compton scattering and electron/hole pair production. The high voltage across the detector attracts the electron or hole to the relevant electrode, creating a pulse. However, not every photon behaves in exactly this manner. Some effects that may modify the spectrum are:

\begin{itemize}
\item Absorption by the mylar blankets, cryostat windows and grids.
\item Compton scattering out of detectors.
\item Compton scattering off the Earth's atmosphere.
\item Noise in the electronics.
\item Detector degradation due to radiation damage. 
\item Low energy cut off due to electronics. 
\end{itemize} 
In order to accurately interpret the spectrum, it is necessary to understand how detectors interpret photons of different energies. This is known as the detector response matrix (DRM).  The diagonal components represent the efficiency of the instrument for detecting photons at their proper energy and the off diagonal elements represent the changing of photon energy from their true energy to a different energy. 

The detected count spectrum can then be converted to a photon spectrum using
\begin{equation}
CT = BG + DRM*PH
\label{eqn:drm}
\end{equation}
where $CT$ is the count rate, $BG$ is the background count rate and $PH$ is the photon rate. 

The photon spectrum can then be modelled using the Object SPectral EXecutive package \citep[\textsc{OSPEX};][]{Schwartz02} using a variety of photon spectra components. Alternatively, and more applicable to the studies carried out in this thesis, Equation \ref{eqn:drm} can be inverted and the model photon components can be converted to count flux and compared to the raw count flux observations. Since this is highly dependent on temperature and emission measure, the changes in these parameters can be observed more clearly in count space than in photon space

\section{MESSENGER Solar Array for X-Rays}

The Mercury Surface, Space, Environment, Geochemistry and Ranging Instrument \citep[MESSENGER;][]{Santo81} is a satellite that is designed to study the planet Mercury. On board is the Solar array for X-rays \citep[SAX;][]{Schlemm2007}. SAX, a Silicon-PIN detector, is a component of a larger instrument designed to measure X-rays on the surface of Mercury. SAX is a calibration device that measures the incident SXR flux from the Sun. The high spectral resolution in the range 1-10~keV makes it ideal for measuring the temperature and emission measure of flares. 

The \textsc{OSPEX} routines in \textsc{SSW} have been adapted to accommodate SAX spectra. Therefore, the analysis tools used for RHESSI can also be applied to SAX spectra. As with RHESSI, the count spectrum is compared to photon models that have been convolved with the SAX DRM file (Equation \ref{eqn:drm}). 

The SAX instrument is sensitive to only thermal emission (i.e. it observes no hard X-rays). For isothermal plasma, Maxwell Boltzmann distributions are the only available functions with which to fit the spectra. For multi-thermal plasma, there are a number of options. All of these are based on differential emission measure functions. \textsc{OSPEX} has two built in functions: an exponential DEM fit and a power law DEM fit. Both fit a function (either exponential or power law) between the DEM peak at low temperatures (photosphere/chromosphere) and at high temperatures (corona). These models ignore the dip that occurs in the middle of the DEM curve at the transition region. This approach is reasonable for RHESSI observations in which high temperature emission dominates. However, for SAX data, there is also a significant contribution from low temperature plasma. Therefore, the accurate measurement of the relative contributions of both the high and low temperature plasma is important. As such, the observed plasma is approximated by two isothermal functions: one corresponding to the low temperature component of the DEM and one to the high temperature component, thus modelling both the low and high temperature plasma observed by SAX simultaneously.


\chapter{Temperature response of EUV imagers} 
\label{chapter:SWAP}

\ifpdf
    \graphicspath{{6/figures/PNG/}{6/figures/PDF/}{6/figures/}}
\else
    \graphicspath{{6/figures/.pdf/}{6/figures/}}
\fi
\hrule height 1mm
\vspace{0.5mm}
\hrule height 0.4mm 
\noindent 
\\ {\it This chapter discusses the temperature response of six extreme ultraviolet imaging telescopes in the 171~\AA\ passband. The intent is to investigate the response of the instruments to various types of plasma, including coronal holes, quiet sun, active regions and flaring emission. Beginning with a differential emission measure distribution for each type of emission, synthetic spectra at single temperatures were generated in the range $10^4$ to $10^8$~K. The product of these spectra with the instruments' wavelength response functions resulted in a series of isothermal transmitted spectra. The instruments' response to temperature was then obtained by integrating the transmitted spectra over wavelength to give the expected counts per second per pixel at a single temperature and emission measure. Despite variations in the instruments' wavelength responses, the resulting temperature responses were found to agree very well. The most significant deviation was the offset between temperature responses. This is a direct result of the efficiency and size of each instrument. However, comparing the responses for different types of emission revealed significant differences, especially for the coronal hole and flaring scenarios. The peak temperature of the passband when observing coronal holes is $7.5\times10^{5}$~K versus $1\times10^6$~K for quiet sun and active regions and the response to flaring plasma proved to be multi-peaked as a result of the significant contribution from free-free continuum at high temperatures. This method will enable scientists to directly compare the sensitivity of a range of instruments to different solar conditions. This work has been submitted to the Astronomy \& Astrophysics journal \emph{(Raftery, Bloomfield, Gallagher, Seaton \& Berghmans}, 2010, submitted). Considerable assistance was given to the author by Dr. Shaun Bloomfield and Dr. Peter Gallagher. Access to BESSY data was facilitated by Dr. David Berghmans and knowledge of the Proba-2 spacecraft was imparted by both Dr. Berghmans and Dr. Daniel Seaton. \\  }
\hrule height 1mm
\vspace{0.5mm}
\hrule height 0.4mm 

\newpage

\section{Introduction}
\label{section:intro}
The upper layers of the atmosphere emit strongly in the EUV portion of the spectrum, from the chromosphere through to the corona. This part of the solar atmosphere contains features across many different size scales, from oscillations of coronal loops \citep[on the order of arc seconds;][]{DeMoortel02} to EIT ``waves'' \citep[on the order of R$_{\odot}$;][]{Thompson98}. Active regions are the strongest features visible in EUV emission on the solar disk. As the origin of solar flares and CMEs, the accurate interpretation of EUV emission from active regions is of particular interest to the author. The EUV part of the spectrum is dominated by emission lines formed across a wide range of temperatures from $\sim$10$^5$ to more than 10$^{7}$~K. With the strong dependence of temperature on height through the atmosphere, this suggests that EUV emission can originate from anywhere between the upper chromosphere and the corona. Density is also known to scale with height in the atmosphere (Figure \ref{fig:solar_mod}). Since emission measure (EM) scales with density (Equation \ref{eqn:emission_measure}) it is clear that not only does the temperature of a feature give some indication of the height at which it is being observed, but we can also draw conclusions regarding the EM and density of the region, as Chapters \ref{chapter:mar26} and \ref{chapter:CME} showed \citep{Raftery09, Raftery10_cme}. 

Imaging of the Sun using full disk EUV images is a very useful method of studying solar features across many different size scales at a wide range of temperatures. Normal incidence telescopes make use of constructive interference to amplify the reflectivity of their mirrors. The use of varying layers of Molybdenum and Silicon tune the wavelength sensitivity to different portions of the EUV spectrum. At $\sim$10~\AA\ wide, the wavelength sensitivity is broader than that of a spectrometer but generally sufficiently narrow to focus on one or two emission lines under ``typical'' solar conditions. However, the solar spectrum can vary significantly when studying various features, even within the narrow band of sensitivity. For example, the emission coming from a coronal hole is significantly cooler then that coming from a solar flare, or even an active region. Coronal holes have low temperatures and densities ($\sim10^5$~K, $\sim10^{7}$~cm$^{-3}$ respectively) and appear as dark regions in EUV images. A solar flare, on the other hand has emission at very high temperatures and high electron densities ($\sim10^7$~K, $\sim10^{11}$~cm$^{-3}$ respectively). The appearance of a cool line in a coronal hole spectrum or a hot line in a flare spectrum can have a remarkable effect on the temperature of the emission detected by the EUV instrument. This can lead to the misinterpretation of ``bright'' features in EUV images. 

In an attempt to avoid such confusion, investigations into the characterisation of EUV imagers' temperature responses have been carried out. A characterisation of each instrument has been carried out by their respective team and published in the relevant instrument paper (see \S \ref{sect:wav} for papers). Typically this is attained for a single type of solar feature (usually quiet-sun) using a constant emission measure (normally $10^{44}$~cm$^{-3}$). This clearly does not account for the order of magnitude difference in EM between e.g. the transition region and the corona. 

With the level of understanding of atomic processes ever evolving, the responses presented in instrument papers require continuous re-analysis. \citet{DelZanna03}, for example, highlighted the importance of including \ion{Fe}{viii} in the calculation of the TRACE 171 and 195~\AA\ passbands. The \ion{Fe}{viii} line had not been included in the calculation by \citet{Handy99} and had an significant effect on the 195~\AA\ channel in particular. This, however was only carried out for quiet-sun emission at a constant pressure. \citet{Phillips05} continued the development of the TRACE temperature response by incorporating the high temperature contribution to the spectrum by free-free continuum, as predicted by \citet{Feldman99}. Although they expanded their study to include emission from quiet-sun, active regions and flares, the response curves published by \citet{Phillips05} have been normalised, making it difficult to compare to other curves. \citet{Brooks06} again improved the temperature response of the TRACE and EIT EUV filters using spectroscopic observations taken with CDS. Recalculating the ionisation fractions using the CDS data facilitated a more accurate calculation of the instruments' temperature response curves compared to those in the instrument papers. Again, however, this was only completed for quiet-sun emission. It is clear that the studies carried out in the past do not facilitate the \emph{direct} comparison of the temperature response of multiple instruments to many different types of plasma. 

In this chapter, version 5.2 of the CHIANTI atomic physics package is used to investigate the temperature response of \emph{six} solar instruments in the commonly termed ``171~\AA'' wavelength band. These responses are calculated for four classes of solar plasmas: coronal holes, quiet sun, active regions and solar flares. The variation of EM with height is taken into account through the use of a differential emission measure curve (DEM). The calculation of isothermal spectra is detailed in \S \ref{sect:method}. The instruments under investigation and their corresponding wavelength responses are described in \S \ref{sect:wav}. The nature of the transmitted spectra are presented in \S \ref{sect:trans}. The resulting temperature response curves for the four solar conditions and for the six instruments are described in detail in \S \ref{sect:res}. Finally, the conclusion and implications of this study are discussed in \S \ref{sect:concs}.

\begin{figure}[!t]
\centering
\includegraphics[width=0.7\textwidth, trim =40 40 230 40, clip = true, angle = 0]{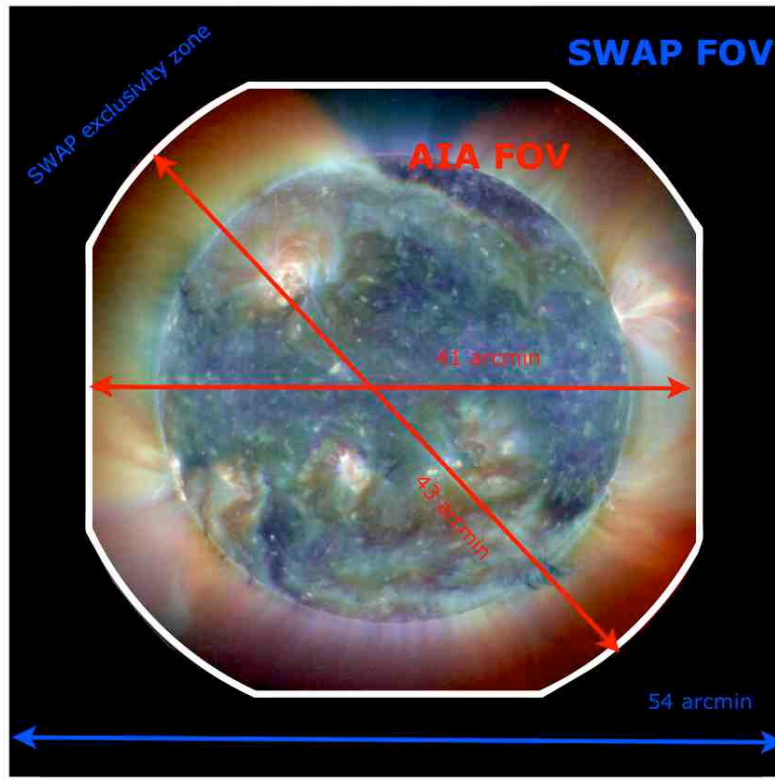} 
\caption{The expected field of view of SWAP compared to AIA. The image shown is an EIT composite of multiple wavelength bands.   }
\label{fig:swap_fov}
\end{figure}

\section{Method}
\label{sect:method}

\subsection{Instruments and wavelength responses}
\label{sect:wav}
There are six instruments under investigation in this chapter. EIT on board SOHO, TRACE, the twin STEREO/EUVI imagers, the soon to be launched Atmospheric Imaging Assembly on board the Solar Dynamics Observatory \citep[SDO/AIA;][]{aia} and the recently launched SWAP instrument on the Proba-2 ESA satellite. As a member of the SWAP instrument team, this telescope will be the focus of my investigation. 

SWAP is a single filter imaging telescope with the unique capability to off-point out to 3~$R_{\odot}$. It was launched successfully in November 2009 with first light expected on 26 January 2010. At 54$'\times54'$, its field of view already surpasses that of all the other instruments (see Figure \ref{fig:swap_fov} and Table \ref{table:cor} for comparison to instruments). With its remarkable ability to move the field-of-view to exclude the solar disk, this technology demonstration will enable users to image the solar corona in EUV to greater distances then ever before using the first active pixel sensors (APS) flown on a satellite. The high cadence of the instrument, at 1 image per minute, will allow investigations of fast moving phenomena such as CMEs and EIT ``waves'' to be carried out in great detail. This will be especially important in the coming years as the STEREO spacecraft move further from the Earth. As their angle increases, their cadence will be reduced to due telemetry limitations. In addition, their field of view will be close to $180^{\circ}$ from each other, making co-ordinated studies with each other and Earth orbiting instruments extremely difficult. SWAP, with its similar capabilities to EUVI, will be an excellent replacement for these instruments. 
\begin{table}[!t]
\begin{center}
\begin{tabular}{l c c c c}
\hline\hline
Instrument	&	Plate scale 	&	Nominal cadence	&	 FOV		&	Aperture area						\\
			& 	[arcsec/pixel]	& 	[Minutes] 			& 	[arcmin]	&	[cm$^{2}]$							\\ \hline
SWAP		&		3.1		& 		1			&	54$^{2}$	&	8.55												\\ 
TRACE		&		0.5		& 		$<$1			&	8.5$^{2}$	&	706				\\ 
EIT			&		2.6		& 		10			&	45$^{2}$	&	13				\\ 
EUVI	 \	&		1.6		&		2			&	43$^{2}$	&	3.01							\\ 
AIA			&		0.6		& 		$<$1			&	41$^{2}$	&	113.1							\\ \hline
\end{tabular}
\label{table:cor}
\caption{Specifications for the EUV instruments used in this investigation.} 
\end{center} 	
\end{table}

Launched in 2006, the EUVI instruments on STEREO A and B are also full disk imagers, though with a slightly smaller field of view compared to SWAP. These twin spacecraft have image cadence of $\sim$2 minutes and have extremely high resolution with a plate scale of 1.6$''$/pixel. The instruments have four EUV filters (171, 195, 284 and 304~\AA) and have provided users with the first EUV stereo images of the Sun. 

EIT, launched in 1995 is often used as a ``third eye'' for the EUVI instruments. With similar field of view and the same filters as EUVI, it can often provide a complementary or comparative view of an event. For example, in Chapter \ref{chapter:CME} \citep{Raftery10_cme}, the EUVI instrument was primarily used for diagnostic purposes as the flare was mostly occulted in the EIT field-of-view. However, the EIT lightcurve provided a very useful tool for proving that a ``double peak'' observed in the GOES lightcurve was as a result of occultation and not a secondary event. EIT has significantly reduced cadence and resolution compared to the newer instruments. However, as a pioneer in the field of solar EUV imaging, it was once revolutionary in its capabilities. 

TRACE is another instrument that transformed solar physics using its seven passbands, 3 of which are in the EUV range (171, 195 and 284~\AA). Unlike the other EUV instruments, TRACE does not have a full disk field-of-view. Instead, it has extremely high resolution. With a cadence of less than one minute and a plate scale of 0.5$\arcsec$/pixel, this instrument revealed small scale features that had never been seen before. TRACE also deviated in the type of detector used. Unlike the other instruments listed here, TRACE used a front-illuminated CCD. The structures in front of the photosensitive portion of the front-illuminated device act as a deadlayer, reducing its sensitivity. Therefore, the overall efficiency of the TRACE instrument is expected to be lower than other instruments, despite its large aperture. 
\begin{figure}[!t]
\centering
\includegraphics[width=\textwidth, trim =40 40 40 40, clip = true, angle = 0]{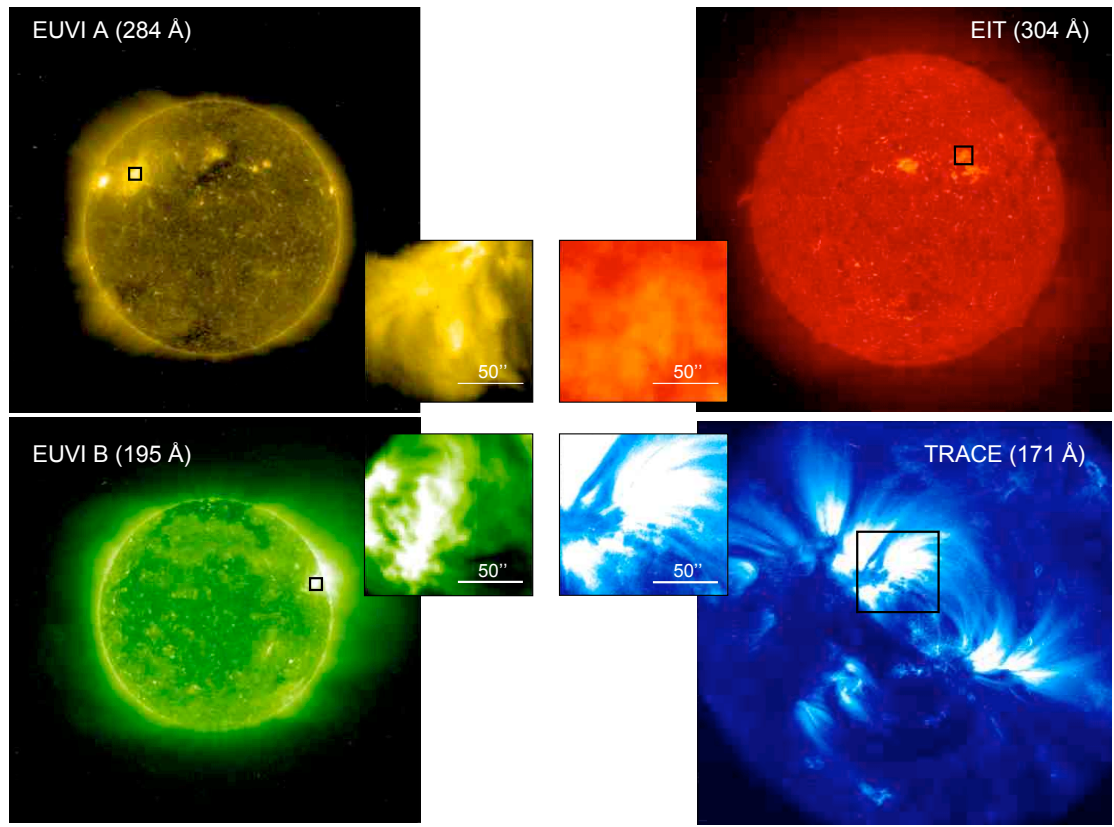} 
\caption{Images of the Sun in all four instruments currently taking measurements (i.e. excluding AIA and SWAP). Top left shows EUVI A in 284~\AA. Bottom left shows EUVI B in 195~\AA. Top right shows EIT in 304~\AA\ and bottom right shows TRACE in 171~\AA. In the centre there are four images corresponding to the 100$\arcsec \times$100$\arcsec$ regions marked on the main images.   }
\label{fig:multi_sun}
\end{figure}

The AIA instrument has significant heritage from TRACE, although it has reverted to the use of a full disk, back-illuminated structure. This instrument, due for launch in early 2010, will be another in a long line of revolutionary telescopes. With resolution comparable to TRACE, nine EUV passbands (94, 131, 171, 193, 211, 304, 335~\AA), full disk resolution and a cadence of less than a minute, AIA will hopefully reveal even more new science of the solar corona. A sample TRACE image is shown in the bottom right of Figure \ref{fig:multi_sun}. AIA will have similar resolution to this but for the entire disk. 

Figure \ref{fig:multi_sun} highlights the different resolutions of the instruments in question. This figure shows images of a recent period of activity on 22-Nov-2009 from the four telescopes currently taking data. In the center is a small 100$\arcsec \times$100$\arcsec$ region to more clearly show the resolution differences. In addition, the different wavelength bands in which these images are taken emphasises the differences in solar emission at different temperatures. At 2~MK, the 284~\AA\ image is the hottest image and it is clear that the structures are significantly more diffuse that the other images. The 195~\AA\ image is showing more structure as is the somewhat cooler, with a peak temperature of just above 1~MK. Slightly cooler again, the 171~\AA\ image displays a similar level of structure to 195~\AA, although the details are far superior thanks to TRACE's high resolution. Finally, the EIT 304~\AA\ at chromospheric temperatures shows a very different Sun. With little extended emission and showing prominences on the limb, it is clear that the resolution of EIT is quite poor. The structure one would expect to observe in the chromosphere is completely blurred out.

The wavelength responses \rlam\ of five of the imagers (all but SWAP) were obtained from \textsc{ssw} routines or from the instrument teams. The SWAP response was calculated from observations taken at the Berliner Elektronenspeicherring-Gesellschaft fŸr Synchrotronstrahlung (BESSY; Berghmans et al. private communication). This experiment consisted of passing a calibrated synchrotron beam through the telescope's optics at wavelengths between 165~\AA\ and 195~\AA. Comparing the transmitted signal $(I_{trans})$ to the known input signal $(I_{beam})$, the number of counts, or data numbers (DN) were obtained for every input photon. 
\begin{equation}
\label{eqn:cal}
R(\lambda) = \frac{I_{trans}}{I_{beam}}  \hspace{2.2cm}   [DN\ phot^{-1}].
\end{equation} 
To ensure the response was compatible with the synthetic spectra generated in the following section, the wavelength responses of the six instruments were, where necessary, corrected for the telescope aperture size (cm$^{2}$) and for the solid angle sky coverage of one pixel (str~pix$^{-1}$). 

\begin{figure}[!t]
\centering
\includegraphics[width=\textwidth, trim =110 90 230 180, clip = true, angle = 0]{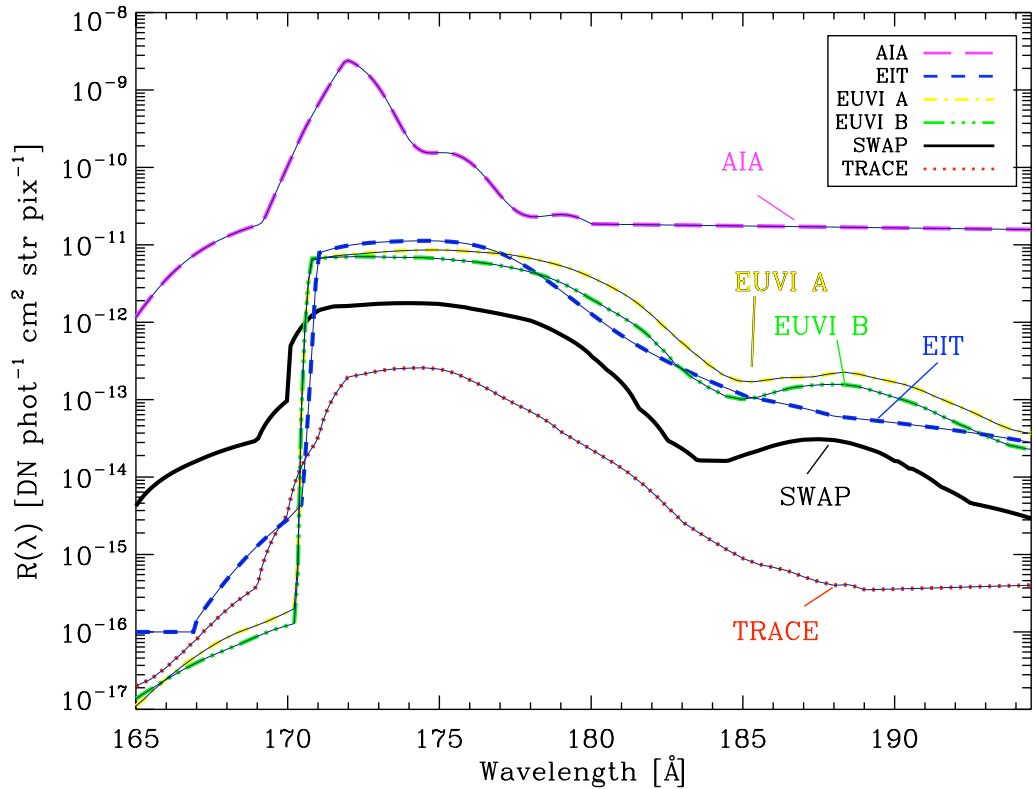} 
\caption{Wavelength response of AIA, EIT, EUVI A, EUVI B, SWAP and TRACE.    }
\label{fig:all_resps}
\end{figure}

The wavelength responses are shown in Figure \ref{fig:all_resps}. The most striking feature is the offset between the instruments' response magnitude. As expected, the TRACE response is the lowest of all, mainly due to the low efficiency of its detector. Despite being built at the same location, AIA has the highest response. The shape of the AIA response is also significantly different to the others. This is a testament to both its large aperture size and the developments in technology since TRACE was built. The EUVI and EIT instruments have very similar responses despite their significantly different aperture sizes. This again, can be attributed to the developments in the efficiency of modern detectors. SWAP falls between TRACE and EUVI/EIT. A combination of its small aperture size and the nature of an APS detector reduces its overall sensitivity. Since an APS chip requires individual electronics for each pixel, the overall effective area of the detector is reduced. 

Moving through increasing wavelength, it is interesting to note that all instruments have a gradual increase in their \rlam\ over the first 5$\pm$1~\AA\ of the function. It is not surprising that AIA and TRACE have similar slopes in this region, as do SWAP and EUVI (and EIT, to a certain extent), since these pairs of instruments were developed by the same institutes. From 169~\AA\ the instruments' responses begin to deviate. SWAP, EUVI B, EUVI A and EIT have a very sharp rise between 170 and 171~\AA, peaking at (1.2, 7.0, 8.5, 11.3)$\times10^{-12}$~DN~phot$^{-1}$~cm$^2$~str~pix$^{-1}$ respectively. Each of these four responses ``plateau'' as they remain at values close the peak 171~\AA\ value for 6$\pm$2~\AA. The TRACE and AIA responses, however, have a significantly different shape. These responses have a much more gradual rise from 169~\AA\ to 172~\AA, peaking at 2.5$\times10^{-13}$ and 2.5$\times10^{-9}$~DN~phot$^{-1}$~cm$^2$~str~pix$^{-1}$ respectively. Neither TRACE nor AIA have as broad a plateau as the other four instruments. The response of AIA falls off at wavelengths immediately after its peak at 172~\AA. It does however, experience a 1~\AA\ wide step at 175~\AA\ before decreasing to an almost flat response above 180~\AA. The TRACE wavelength response remains close to the peak value for $\sim$3~\AA\  and then decreases slowly to an almost constant response above 188~\AA. The decay of the EIT wavelength response follows a similar rate to that of TRACE. EUVI and SWAP however, experience a secondary increase in their respective response functions between $\sim$184 to 192~\AA.

\subsection{Synthetic spectrum calculation}
\label{sect:trans}
The method presented here obtains the temperature response of an EUV imager between the temperatures of $2\times10^{4}$ and $10^7$~K for a range of solar features. In order to accommodate both the change in temperature (T) and EM with height for these different features, the T and EM values were taken from a DEM curve. The amount of emitting plasma in a column $dh$ can be written as:
\begin{equation}
\label{eqn:dem}
EM =  \int n_{e}^2 dh,					
\end{equation}
As discussed in \S \ref{sect:atom_em}, DEM can be written as $DEM = \int n_e^2(dT/dh)^{-1}$. Therefore we can redefine the EM to be:
\begin{equation}
\label{eqn:dem}
EM =  \int_{\Delta T} DEM dT 			
\end{equation}
The EM can therefore be extracted from the DEM curve for a temperature bin log$_{10}\Delta T = 0.1$ wide. 
\begin{figure}[!t]
\centering
\includegraphics[width=1.2\textwidth, trim =130 150 50 80, clip = true, angle = 0]{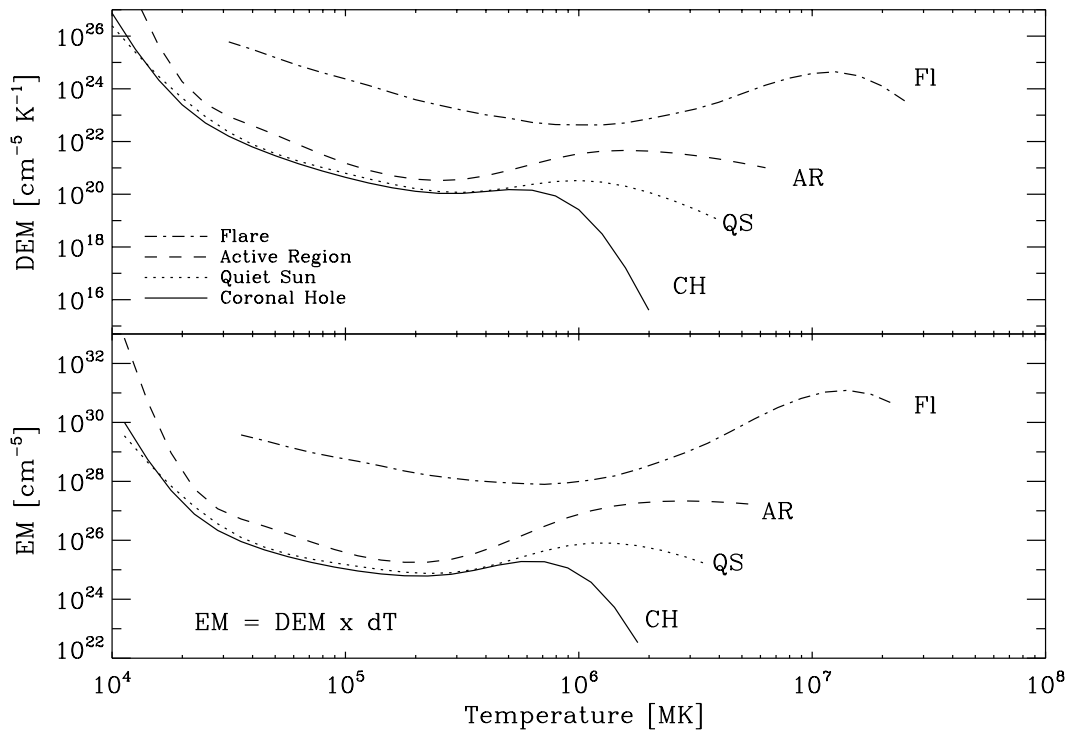} 
\caption{Differential emission measure and emission measure as a function of temperature for coronal hole, quiet sun, active region and flare.   }
\label{fig:DEM_cvs}
\end{figure}

Figure \ref{fig:DEM_cvs} shows  the DEM (top) and corresponding EM (bottom) curves used in this chapter that have been calculated using CHIANTI. The coronal hole, quiet sun and active region curves for both EM and DEM are very similar up to a temperature of $\sim$6$\times10^5$~K. This roughly corresponds to the temperature at which the coronal hole DEM curve peaks. Following the split, the coronal hole curves start to decrease. This means that there is little or no plasma above $\sim$1$\times$10$^6$~K present in a coronal hole. The quiet sun curves continue to rise to a peak at $\sim$1.6$\times10^6$~K, at which temperature, they turn over. While the active region curves follow the same trend as the coronal hole and active region at temperatures less than $\sim$1~MK, it is clear that the active region is approximately two orders of magnitude greater. At temperatures greater than 1~MK, the active region curves remain approximately constant to almost 10~MK. In all of the coronal hole, quiet sun and active region curves, a dip exists at approximately 1.5$\times$10$^5$~K. The small height of the transition region (at this temperature) has a significant effect on the emission levels that originate here. Notice however, that the dip is shifted to closer to 1~MK in the flare EM and DEM curves. This is around the temperature of the active region curves' peak. This suggests that during a solar flare, the plasma within an active region loop is heated to very high temperatures ($\sim$10$^7$~K), thus leaving the active region loops devoid of 1~MK plasma.

At each temperature ($T$) in the range of interest, an EM value was extracted from the appropriate EM curve in Figure \ref{fig:DEM_cvs}. Along with the unique T and its corresponding EM value, a generic density value was supplied. The intensity of an emission line depends on the population of the upper levels of the atomic transition, which itself depends on the plasma density. Understandably, this density will change depending on the feature being studied. However, it can be taken as a constant within the study of an individual feature. The densities used were:
\begin{itemize}
\item Coronal hole: $10^{7}$~cm$^{-3}$ \citep{Wilhelm06}
\item Quiet sun: $6\times10^{8}$~cm$^{-3}$ \citep{Young05}
\item Active region: $5\times10^{9}$~cm$^{-3} $ \citep{Gallagher01}
\item Flare: $10^{11}$~cm$^{-3}$ \citep{Raftery09}
\end{itemize}
The abundances were taken to be those of \citet{Feldman_abundance} and the ionisation fractions of \citet{Mazzotta98} were used. 

\begin{figure}[!t]
\centering
\includegraphics[width=0.8\textwidth, trim =70 50 80 80, clip = true]{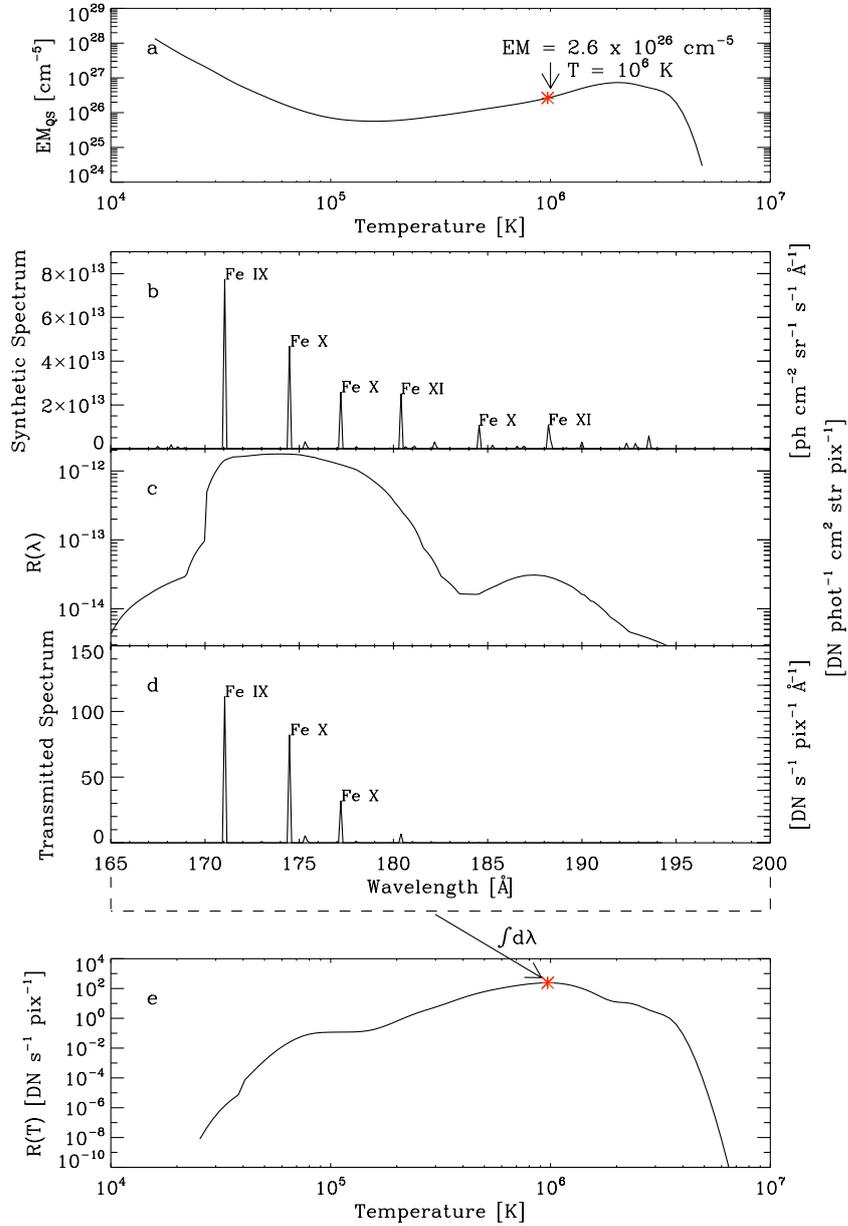} 
\caption{(a) Quiet sun EM curve with a single point highlighted at T = $10^6$~K and EM = $2.6\times10^{26}$~cm$^{-5}$. (b) Synthetic isothermal spectrum calculated for the conditions selected in panel (a). (c) The SWAP \rlam\ function. (d) Transmitted spectrum from the product of (b) and (c). (e) The SWAP temperature response, calculated by integrating the transmitted spectrum over wavelength. The asterisk corresponds to the conditions set in panel (a).  }
\label{fig:schematic}
\end{figure}

\begin{figure}[!t]
\centering
\includegraphics[width=0.85\textwidth, trim =20 60 50 88, clip = true]{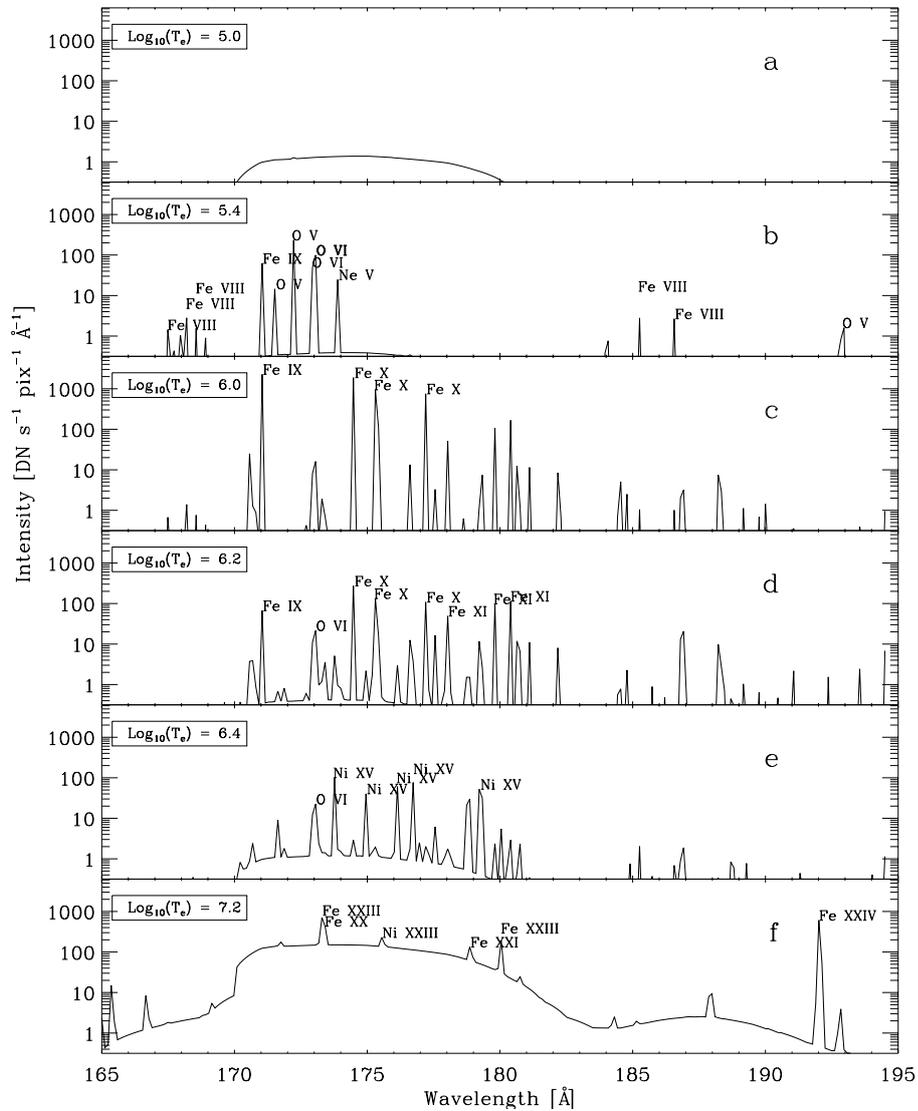} 
\caption{Sample transmitted spectra over the temperature range under investigation calculated using a flare DEM and the SWAP wavelength response function for log$_{10}(T_e)$ = 5.0, 5.4, 6.0, 6.2, 6.4 and 7.2. The ion names of the strongest lines are also listed. }
\label{fig:trans}
\end{figure}

Utilising the parameters listed above, along with a unique T and corresponding EM (Figure \ref{fig:schematic}a, asterisk), an isothermal spectrum was calculated using the CHIANTI routines \textsc{ch\_synthetic} and \textsc{make\_chianti\_spec} (Figure \ref{fig:schematic}b). Each isothermal spectrum was then multiplied by the wavelength response of each instrument (Figure \ref{fig:schematic}c) to give the expected throughput of the instrument: the transmitted spectrum (Figure \ref{fig:schematic}d). The transmitted spectrum, is the number of DNs a single pixel is expected to detect in one second at a given wavelength. A sample of transmitted spectra are shown in Figure \ref{fig:trans}, taken from throughout the temperature range for a flare DEM. Figure \ref{fig:trans}a shows the transmitted spectrum at a temperature of log$_{10}(T_e)$ = 5. It is clear that with no emission lines, the continuum is the dominant source of emission at this temperature. Figure \ref{fig:trans}b shows the transmitted spectrum at log$_{10}(T_e)$ = 5.4 where the spectrum is dominated by an \ion{O}{v} line at 172.2~\AA. The \ion{Fe}{viii} lines between 167 and 169~\AA\ along with the O and Fe lines at wavelengths above 184~\AA\ are all suppressed by the instrument response function. Moving up to log$_{10}(T_e)$ = 6 in Figure \ref{fig:trans}c, the spectrum is now dominated by the characteristic \ion{Fe}{ix} line at 171~\AA\ and has significant contribution from a number of \ion{Fe}{x} lines at slightly higher wavelengths. Notice that at this temperature, emission lines completely dominate the spectrum and the continuum, now an order of magnitude less than in Figure \ref{fig:trans}a, has a negligible contribution to the spectrum. At a higher temperature, log$_{10}(T_e)$ = 6.2 in Figure \ref{fig:trans}d, the contribution of the \ion{Fe}{ix} line is not as significant as the contribution from the series of \ion{Fe}{x} and \ion{Fe}{xi} lines, particularly at 174.5~\AA. Increasing the temperature further to log$_{10}(T_e)$ = 6.4, as in Figure \ref{fig:trans}e, the continuum level begins to rise again, although the line emission still accounts for $\sim$93\% of the total emission. \ion{Ni}{xv} lines between 174 and 188~\AA\ dominate the spectrum at this temperature. Figure \ref{fig:trans}f shows the spectrum at log$_{10}(T_e)$ = 7.2 where the continuum accounts for 41.6\% of the emission. Emission lines at this temperature consist mainly of highly-ionised species of Fe, specifically \ion{Fe}{xxiii} at 173~\AA\ and \ion{Fe}{xxiv} at 192~\AA.

It is clear from Figure \ref{fig:trans} that the relative intensity of the spectrum changes significantly with temperature. Therefore integrating the transmitted spectrum over all wavelengths results in the expected number of counts a single pixel is expected to detect per second at a single temperature, or the instrument's temperature response, $R(T)$ (Figure \ref{fig:schematic}e, asterisk). Repeating this process for all temperatures in the range of interest results in the temperature response curve for a single instrument and solar feature.

\section{Results}
\label{sect:res}
\subsection{Inter-instrument comparison}
\label{sect:instr_comp}

\begin{figure}[!t]
\centering
\includegraphics[width=\textwidth, trim =60 50 120 70, clip = true, angle = 0]{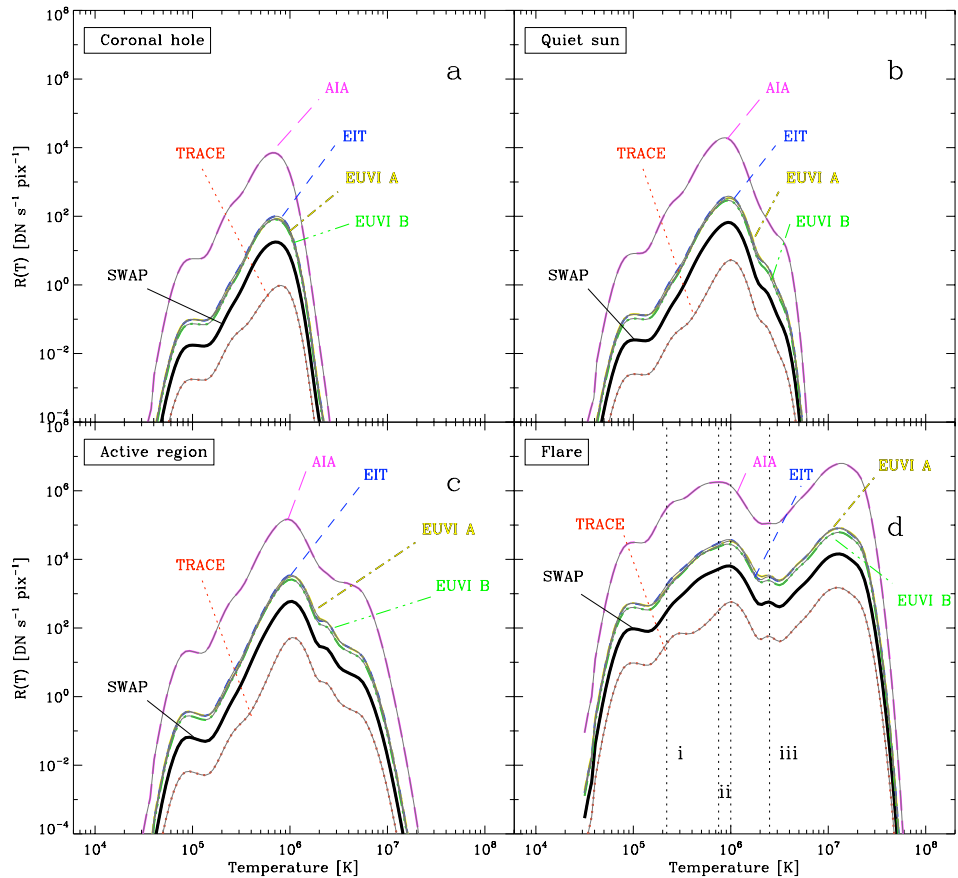} 
\caption{Instrument response to temperature for 6 EUV imagers in the 171 \AA \ passband for (a) coronal hole, (b) quiet sun, (c) active region and (d) flare. The numbers $i, ii, iii$ in the bottom panel refer to the discussion points in \S \ref{sect:instr_comp}. }
\label{fig:all_temps}
\end{figure}

\begin{figure}
\centering
\includegraphics[width=\textwidth, trim =20 400 0 50, clip = true, angle = 0]{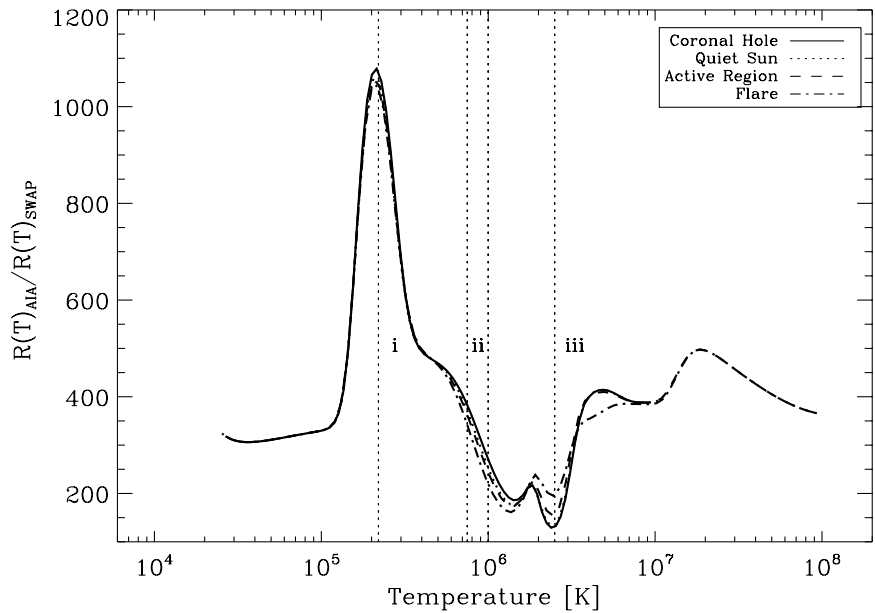} 
\caption{The ratio of the AIA and SWAP temperature responses for coronal hole (solid), quiet sun (dotted), active region (dashed) and flare (dot-dashed). $i$, $ii$ and $iii$ refer to the discussion point in \S \ref{sect:instr_comp}.  }
\label{fig:ratio}
\end{figure}

The temperature responses for each of the four solar conditions are shown in Figure \ref{fig:all_temps} for all six instruments. At first, it appears that aside from the offset in their magnitude, the responses of all six instruments are in good agreement. Considering the significant deviations noted in the respective wavelength responses in \S \ref{sect:wav}, the correlation between the various temperature responses are remarkable. There are some discrepancies worth noting, marked in Figure \ref{fig:all_temps}d and Figure \ref{fig:ratio} with dashed lines. Figure \ref{fig:ratio} shows the ratio of the AIA and SWAP temperature responses since all of the discrepancies involve the AIA temperature response. 

Below the peak temperature, between (2-4)$\times10^5$~K, there is an increase in the relative sensitivity of AIA and TRACE compared to SWAP, EIT and EUVI, marked as $i$ on Figures \ref{fig:all_temps}d and \ref{fig:ratio}. Since the peak of both AIA and TRACE's wavelength responses occur from $\sim$172~\AA, they are more sensitive to the \ion{O}{v} lines that are formed at 172~\AA\ between 0.2 and 0.5~MK (see Figure \ref{fig:trans}b). This also has the effect of shifting the peak response of AIA to 7.5$\times10^5$~K compared to 10$^6$~K for all other instruments (marked $ii$ on Figures \ref{fig:all_temps}d and \ref{fig:ratio}). The other instruments peak at 1~MK because the spectrum is dominated by \ion{Fe}{ix} at 171~\AA\ and \ion{Fe}{x} at 174~\AA, both with a formation temperature of $\sim$1~MK (Figure \ref{fig:trans}c). Since the relative wavelength response of EIT, EUVI and SWAP is high at these wavelengths these instruments are sensitive to plasma at 1~MK. TRACE's wavelength response does not peak until 172~\AA. Therefore it is not sensitive to the 171~\AA\ line. Unlike AIA however, TRACE's wavelength is not shifted. This is because of the 3~\AA\ wide plateau around its peak wavelength response. This increases the relative sensitivity of TRACE to \ion{Fe}{x} lines formed at 174~\AA\ that AIA does not see due to its steep fall off. 

The third significant deviation is marked number $iii$ on Figures \ref{fig:all_temps}d and \ref{fig:ratio}. At this temperature, 2.5$\times10^6$~K, there is significant contribution to the spectrum from \ion{Ni}{xv} lines formed between 172 and 177~\AA (Figure \ref{fig:trans}e). The steep drop off of the AIA wavelength response between these wavelengths significantly reduces the sensitivity of the instrument to these hot lines. Therefore, AIA's relative sensitivity to emission at this temperature is reduced compared to the other instruments.

\subsection{Solar conditions comparison}
\label{sect:sol_condts}
Figure \ref{fig:all_temps}a shows the temperature response to imaging a coronal hole, Figure \ref{fig:all_temps}b to quiet sun plasma, Figure \ref{fig:all_temps}c to an active region and Figure \ref{fig:all_temps}d to a solar flare. It is apparent that the responses of the instruments to these different features are significantly different, as expected. The response to imaging a coronal hole has a primary peak at 7.5$\times10^5$~K, a temperature only slightly higher than the EM peak in Figure \ref{fig:DEM_cvs}. This is not surprising as a comparison between Figure \ref{fig:trans}b and c shows. Since the EM in the corona never rises, the EM at the temperature of formation of \ion{O}{v} dominates over the EM at the formation temperature of \ion{Fe}{ix/x}. Therefore the relative contribution of the O lines is greater than Fe. Thus, the \ion{O}{v} lines will dominate the transmission spectrum and lower the overall temperature. The step that appears at $\sim$10$^5$~K is due to emission from continuum at this temperature (Figure \ref{fig:trans}a). A comparison to the quiet sun temperature response shows that the magnitude of the response to a coronal hole is smaller. Therefore, longer exposures are required to image these regions. Not only has the quiet sun response an overall higher magnitude but the emission at 3~MK is almost eight orders of magnitude greater than that of a coronal hole. This compares to a factor of 10 at their peaks. This is even more exaggerated in the active region case. The ``shoulder'' that appears at 3~MK in the quiet sun is barely noticeable in the active region response as it is dominated by contribution from emission at $\sim$7~MK. Despite this however, the main peak at 1~MK is more than two orders of magnitude greater than the high temperature peak. Therefore, it can still be assumed that the majority of emission observed in an active region is being emitted at 1~MK.  The active region peak response is 10 times greater then the quiet sun peak. This is not surprising given the distribution of plasma in the active region EM curve. The continuum at these temperatures has a considerable effect on the emission levels. Finally, the temperature response to a flare is the most dramatic of all. A second peak that occurs at 10$^7$~K is now the dominant feature of the response and is more than an order of magnitude greater than the response at 10$^6$~K. This is due to a combination of strong, highly ionised emission lines (\ion{Fe}{xxiii}, \ion{Fe}{xxiv}) forming at $\gtrsim$10~MK and the rising levels of continuum at these temperatures (Figure \ref{fig:trans}f).  

\subsection{Comparison to instrument teams' $R(T)$ functions}
\begin{figure}[!t]
\centering
\includegraphics[width=\textwidth, trim =60 40 130 70, clip = true, angle = 0]{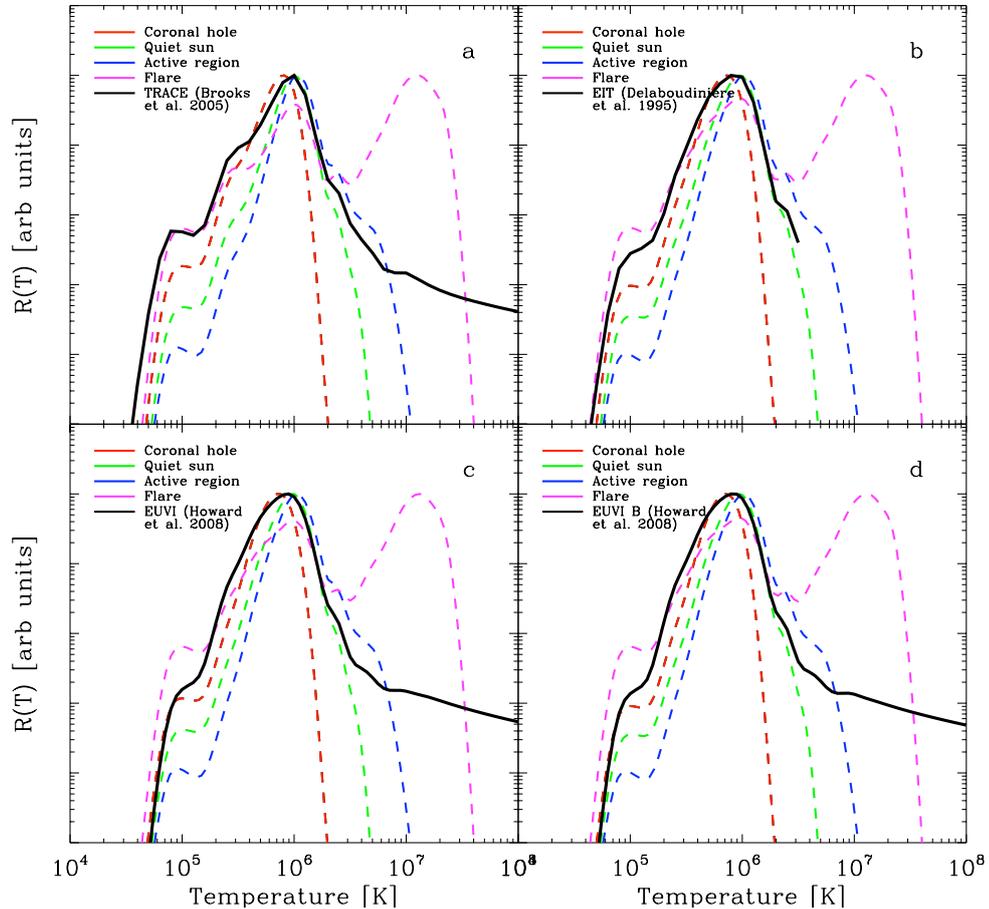} 
\caption{Normalised $R(T)$ functions calculated by the instrument response teams (black lines) for TRACE (a), EIT (b), EUVI A (c) and EUVI B (d). These are compared to the $R(T)$ functions calculated in this chapter for each instrument for coronal hole (pink lines), quiet sun (red lines), active region (green lines) and flares (blue lines).  }
\label{fig:team_resp}
\end{figure}

As discussed in \S \ref{section:intro}, the temperature response functions for these instruments have previously been examined by the instrument teams. A comparison to these response functions was carried out and is presented in Figure \ref{fig:team_resp}. Each panel contains the $R(T)$ from the instrument team (black line) and the $R(T)$ curves calculated in this chapter for the four solar conditions. Beginning with the TRACE plots in panel (a), it is clear that the $R(T)$ calculated by \citet{Brooks06} is in reasonable agreement with our flare $R(T)$ at temperatures less than $\sim$10$^5$~K. At the peak temperature of 1~MK, the Brooks response is most similar to our active region response, although it appears to be shifted slightly towards cooler temperatures in conjunction with our coronal hole $R(T)$. At high temperatures, the correlation breaks down, with the Brooks response almost three orders of magnitude less than our flare response. This is a direct result of the explicit treatment of the four differing plasma types. The Brooks $R(T)$ attempts to incorporate all types of conditions into one single response function and in doing so, is removing the significance of the high temperature peak that may be produced during a solar flare. 

The EIT teams temperature response function (panel b) shows that the response at 10$^5$~K is less than that of a flare, as is the case with TRACE. This is most likely to do with the fact that when this analysis was conducted, the continuum levels were not as well understood \citep{Feldman99}. Like TRACE, there appears to be a broadening of the response across our coronal hole and active region peaks at 1~MK. The EIT response to temperatures above 2$\times$10$^6$~K is not available. 

The responses of the two EUVI instruments shown in panels (c) and (d) are almost identical and so we will discuss them simultaneously. Like EIT, the EUVI $R(T)$ function at 10$^5$~K calculated by the instrument team is not as large as that of our flare and the response at 1~MK is again, broadened across our coronal hole and active region curves. Similar to TRACE, the high temperature continuum is greatly underestimated due to combining the responses of multiple solar conditions into one $R(T)$ function.

\section{Conclusions}
\label{sect:concs}
This chapter studies the temperature response of the SWAP EUV imager, along with five similar instruments in the 171~\AA\ passband. The temperature responses were calculated for four different solar features: coronal holes, quiet sun, active regions and flares. Besides an expected offset in the magnitudes of the temperature responses, the six instruments were found to agree remarkably well, despite significant deviations in their wavelength response functions. Any differences that occurred in the temperature responses were as a result of the shape of the wavelength responses. 

The ability to compare the responses of individual instruments is important to understanding both similarities and differences in co-ordinated observations. Presently there exist comparisons between the TRACE and EIT instruments. This is understandable since these two instruments have been at the forefront of EUV imaging until very recently. However, with the introduction of four new imagers in almost as many years there is a clear need to expand the instrument comparison. While the individual response functions are available in their respective instrument papers, different units, methods and initial conditions in their calculation make comparison difficult. In addition, the instruments' temperature response functions are, in general, presented for only one solar feature - usually quiet sun. This chapter has presented a solution to this problem by combining an in depth study of multiple instruments and different solar conditions. A comparison to the existing temperature response functions reveals that our results are in very good agreement with the response curves of the instrument teams at temperatures below $\sim$3$\times$10$^{6}$~K. Above this temperature, the high temperature continuum contribution that is found, in particular during a solar flare, is highlighted though our method of isolating the individual plasma types.

The facility to analyse the temperature response to different types of plasma is as important as the inter-comparison of the instruments themselves. For quiet sun and active regions, the interpretation of emitting plasma is reasonably straightforward and well known in the community. Coronal hole and flare emission however, cannot be interpreted in the same way. The dominance of continuum and cool O emission lines in coronal holes creates a shift in the maximum temperature sensitivity towards lower temperatures. Likewise, the significant free-free continuum and hot Fe lines in solar flares results in a multi-peaked temperature response, with emission above 10~MK dominating.

This chapter presents a new, more appropriate method of calculating the temperature response of EUV imagers to different types of solar conditions. This method takes account of the variation in EM with temperature for coronal holes, quiet sun, active regions and flaring conditions. It can be applied to multiple instruments, thus facilitating their direct comparison. In addition, the method can be easily adapted to investigate other passbands besides 171~\AA\ and new instruments in the future. This work will facilitate a more precise understanding and interpretation of EUV images.


\chapter{Multi-wavelength observations and modelling of a solar flare}
\label{chapter:mar26}

\ifpdf
    \graphicspath{{4/figures/PNG/}{4/figures/PDF/}{4/figures/}}
\else
    \graphicspath{{4/figures/EPS/}{4/figures/}}
\fi

 \hrule height 1mm
\vspace{0.5mm}
\hrule height 0.4mm 
\noindent 
\\ {\it This chapter discusses observations of a confined, C-class solar flare in light of the standard flare model. The evolution of temperature, emission measure, energy loss and footpoint velocity were investigated from both observational and theoretical perspectives. These properties were derived by analysing the systematic cooling of emission through the passbands of RHESSI ($>$5~MK), GOES (5-30~MK), TRACE 171~\AA\ (1~MK), and CDS ($\sim$0.03-8~MK). Comparing these observations to a 0-D hydrodynamic model, EBTEL, enabled the author to diagnose gentle and explosive chromospheric evaporation and analyse the heating and cooling mechanisms involved in the event. This is the first extensive study of the evolution of a canonical solar flare using both spectroscopic and broad-band instruments, in conjunction with a 0-D hydrodynamic model. While our results are in broad agreement with the standard flare model, the simulations suggest that both conductive and non-thermal beam heating play important roles during the impulsive phase of at least this event. This work was published by \emph{Raftery, Gallagher, Milligan \& Klimchuk}, Astronomy \& Astrophysics, 494, 1127, 2009. In the analysis of this event, significant assistance was given by Dr. Ryan Milliagan for the preparation and interpretation of the CDS and RHESSI observations. The theoretical model, EBTEL, was written by Dr. James Klimchuk and considerable assistance in the use of this model was offered by him. Guidance and advice on this project was given by Dr. Peter Gallagher. 
\\}

\hrule height 1mm
\vspace{0.5mm}
\hrule height 0.4mm 

\newpage

\section{Introduction}
The temporal evolution of most solar flares can be divided into two distinct phases. During the impulsive phase, temperatures rise to $\gtrsim$10~MK via direct heating below the reconnection site in the corona and/or chromospheric evaporation due to accelerated particles \citep{Kopp76}. Chromospheric evaporation, discussed in detail in \S \ref{sect:flares}, can be classified in one of two ways: explosive or gentle \citep{Fisher85, Milligan_explosive, Milligan_gentle}. Explosive evaporation occurs when the flux of non-thermal particles impacting the chromosphere is greater than a critical value of approximately 3$\times10^{10}$~ergs~cm$^{-2}$~s$^{-1}$. In this case, the chromosphere cannot dissipate the absorbed energy efficiently enough. The plasma is forced to expand into the corona as hot upflows of hundreds of~km~s$^{-1}$ and simultaneously into the chromosphere as cooler downflows of tens of~km~s$^{-1}$. Beam driven gentle evaporation occurs when the non-thermal flux is less than $\sim$10$^{10}$~ergs~cm$^{-2}$~s$^{-1}$. Under these circumstances, the chromosphere is efficient in radiating the absorbed energy and plasma expands slowly (tens of~km~s$^{-1}$) upwards into the loop. Gentle evaporation can also be driven by a downward heat flux from the corona in what is known as conduction driven gentle evaporation: conduction fronts that propagate from the hot looptop to the footpoints heat the chromospheric plasma and drive the gradual expansion of plasma upwards into the loop. 

Once the energy release has ceased, the hot plasma returns to its equilibrium state during the decay phase. The cooling process begins with thermal conduction as the dominant energy loss mechanism due to the high temperatures present (see Equation \ref{eqn:spitzer}). As the temperature decreases and the radiative loss function begins to increase (Figure \ref{fig:rtv}) and radiative cooling becomes the primary energy loss mechanism \citep{Culhane1970}. Finally, the ``evaporated'' material drains back towards the solar surface, returning the system to equilibrium.

There have been a wealth of studies that focus on the hydrodynamic modelling of these heating and cooling mechanisms (e.g. \citeauthor{Antiochos78},  \citeyear{Antiochos78}; \citeauthor{Fisher85}, \citeyear{Fisher85}; \citeauthor{Doschek83}, \citeyear{Doschek83}; \citeauthor{Cargill93}, \citeyear{Cargill93}; \citeauthor{Klimchuk01}, \citeyear{Klimchuk01}; \citeauthor{Reeves02}, \citeyear{Reeves02}; \citeauthor{Bradshaw05}, \citeyear{Bradshaw05}; \citeauthor{Klimchuk06}, \citeyear{Klimchuk06}; \citeauthor{Warren07}, \citeyear{Warren07}; \citeauthor{Sarkar08}, \citeyear{Sarkar08}). For example, \citet{Reale07} conducted an analysis of the details of stellar flares using the Palermo-Harvard theoretical model \citep{Peres82, Betta97}. This paper fully describes the cooling timescales and plasma parameters of flares in terms of their phases, including an investigation of the thermal heating function. However, these, and most other theoretical results were not compared to observations. The majority of investigations that make this comparison concentrate on broad-band instruments and utilise very simple models. For example \citet{Culhane1994} compared Yohkoh observations to an over-simplified power-law cooling curve. \citet{Aschwanden2001} compared broad-band observations to a model similar to that of \citet{Cargill94} that considers a purely conductive cooling phase followed by a purely radiative cooling phase. \citet{Vrsnak06} conducted a similar study, again concentrating on broad-band observations and a simple, independent cooling mechanism model. \citet{Teriaca06} conducted multi-wavelength analysis of a C-class flare, incorporating observations from CDS, RHESSI, TRACE and ground based detectors. The cooling timescales were obtained using the simple, independent-cooling Cargill model. 

In this chapter, we attempt to improve upon previous studies by comparing high resolution observations over a wide range of temperatures to a theoretical model. Observations of a GOES C-class solar flare were made with several instruments, including TRACE, RHESSI, GOES-12 and CDS. There are many advantages to using spectroscopic data in conjunction with broad-band observations. The identification of emission lines are for the most part, well documented and therefore individual lines can be isolated for analysis. Also, material as cool as $\sim$30,000~K can be observed simultaneously with emission at $\sim$8~MK. Furthermore, it is possible to carry out velocity, temperature and emission measure diagnostics over a wide range of temperatures for the duration of the flare, significantly improving the scope of the analysis undertaken. These observations were compared to a highly efficient 0-D hydrodynamic model, EBTEL. The combination of this extensive data set and the new modelling techniques enabled the author to carry out a comprehensive investigation of the heating and cooling of a confined solar flare. 

\section{Observations and data analysis}
\label{section:observations}

This investigation concentrates on a GOES C3.0 flare that occurred in active region NOAA 9878 on 26-March-2002. Located close to disk centre ($-92\arcsec, 297\arcsec$) the event began at $\sim$15:00~UT. The CDS observing study that was used (\textsc{flare\_ar}) focused on five emission lines spanning a broad range of temperatures. The rest wavelengths and peak temperatures of the lines are given in Table \ref{table:lam0}, where the quoted temperatures refer to the maximum of the contribution function. The contribution functions for the five CDS lines are shown in Figure \ref{fig:goft_fns}. Each CDS raster consists of 45 slit positions, each 15~s long, resulting in an effective cadence of $\sim$11~minutes. The slit itself is 4$\arcsec \times$180$\arcsec$, resulting in a 180$\arcsec \times$180$\arcsec$ field of view.

\begin{table}[!t] 
\caption{Rest wavelengths and temperature of emission lines and bandpasses used in this study.} 
\centering 
\begin{tabular}{c c c} 
\hline\hline
Ion& $\lambda_{0}$~[\AA] & Temperature~[MK]\\
\hline
 \ion{He}{i} & 584.45 & 0.03 \\
 \ion{O}{v} & 629.80 & 0.25 \\
 \ion{Mg}{x} & 625.00 & 1.2 \\
 \ion{Fe}{xvi} & 360.89 & 2.5 \\
 \ion{Fe}{xix} & 592.30 &  8.0\\
 TRACE & 171 & 1.0 \\
 \hline
\hline
Instrument & Range & Temperature~[MK]\\
 \hline
 GOES & 0.5-4\AA \ \& 1-8\AA & 5 - 30 \\
 RHESSI & 3~keV-17~MeV &  $\gtrsim$5 \\
\hline
     \end{tabular} 

\label{table:lam0} 
\end{table}

 \begin{figure}
      \includegraphics[width=\textwidth, trim =50 85 50 80, clip = true]{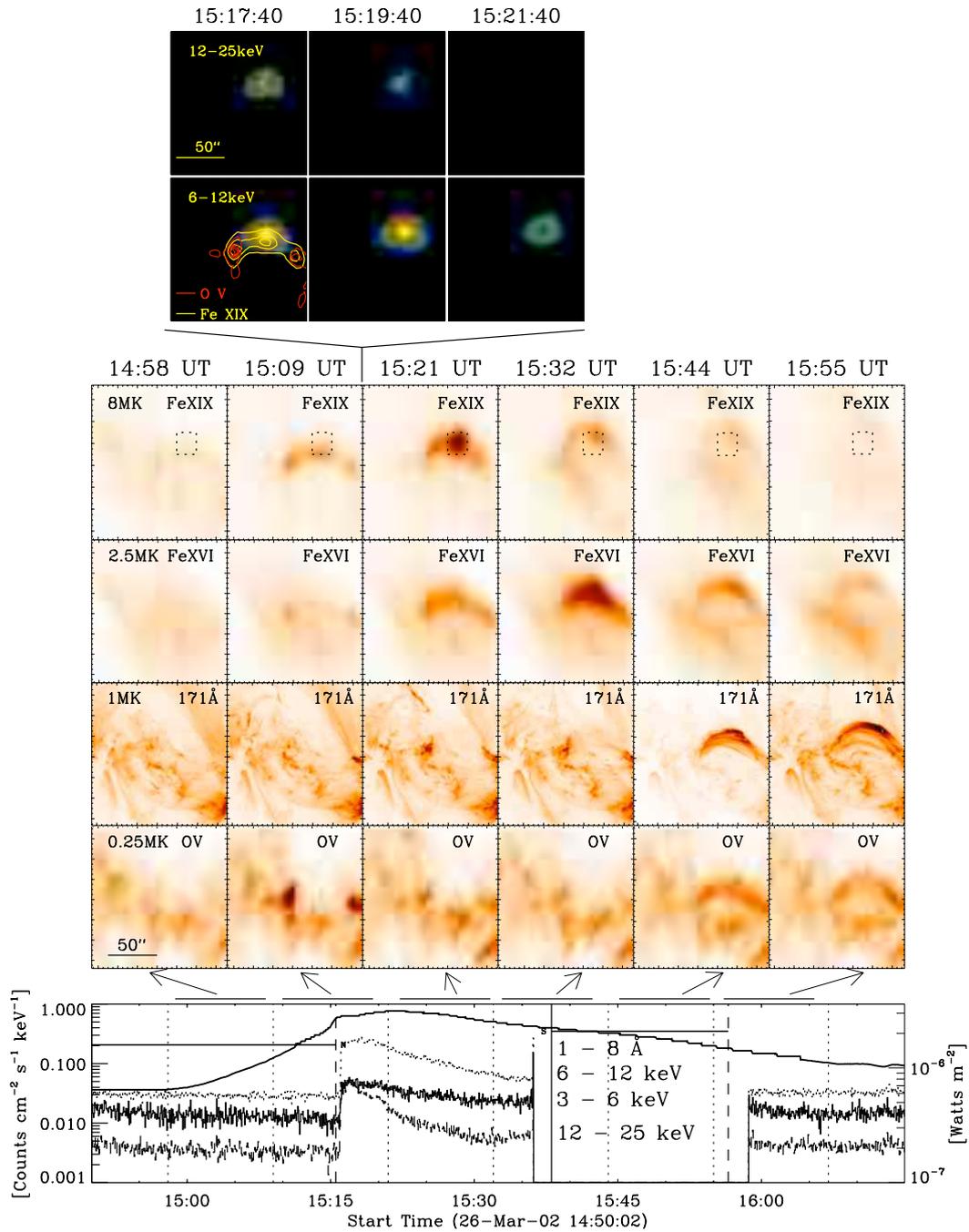}
\caption{RHESSI 6-12 and 12-25~keV images are shown in the top two panels with \ion{O}{v} and \ion{Fe}{xix} contours overplotted in red and yellow respectively. \ion{Fe}{xix} (8MK), \ion{Fe}{xvi} (2.5~MK), TRACE/171~\AA \ (1.0~MK) and \ion{O}{v} (0.25~MK) images are shown in the next four rows. The dotted box represents the loop ``apex'' used for comparison to EBTEL. The bottom panel shows the GOES 1-8~\AA, RHESSI 3-6, 6-12 and 12-25~keV lightcurves. RHESSI was in eclipse until 15:15~UT and passed through the South Atlantic Anomaly between 15:35~UT and 15:58~UT. The vertical dotted lines (and corresponding arrows) on the GOES plot represent the start and end times of the CDS rasters above. 
       }
   \label{fig:F-4panels}
\end{figure}

Figure \ref{fig:F-4panels} shows the evolution of the flare in multiple wavelengths. The top two rows show the looptop source observed in RHESSI 6-12 and 12-25~keV energy bands, with the \ion{O}{v} and \ion{Fe}{xix} (15:09~UT) contours overplotted. The next four rows of this figure show emission observed in \ion{Fe}{xix}, \ion{Fe}{xvi}, TRACE 171~\AA \ and \ion{O}{v}. At $\sim$15:00~UT, before the main impulsive phase of the flare began, evidence of low level \ion{Fe}{xix} loop emission was observed (first \ion{Fe}{xix} image in Figure \ref{fig:F-4panels}). By $\sim$15:09~UT, the footpoints were seen in \ion{O}{v} while the \ion{Fe}{xix} loop top emission continued to brighten. At 15:16:40~UT, when RHESSI emerged from eclipse, a thermal looptop source was observed in both 6-12 and 12-25~keV energy bands. By 15:21~UT the loop was emitting predominantly at $\sim$8~MK and a bright ``knot'' was seen at the top of the loop. The area of the knot is marked with a dotted box on the \ion{Fe}{xix} images. Such features have been observed in the past, although they have not been readily explained (e.g. \citeauthor{Doschek05}, \citeyear{Doschek05}). By $\sim$15:32~UT the loop was emitting mainly in \ion{Fe}{xvi} at 2.5~MK and had cooled to $\le$1~MK, into the TRACE and \ion{O}{v} passbands by $\sim$15:44~UT. 

The lightcurves plotted at the bottom of Figure \ref{fig:F-4panels} show the GOES 1-8~\AA \ and RHESSI 3-6, 6-12 and 12-25~keV lightcurves. Although RHESSI was in eclipse for the majority of the impulsive phase (up to $\sim$15:15~UT), the observed continued rise of the 6-12~keV lightcurve after emergence from night implies that the peak of the soft X-rays was observed. However, while a hard X-ray component (12-25~keV) was observed,  we believe that the HXR peak occurred before this time.


\subsection{Temperature and emission measure}
\label{section:temperature}

\begin{figure}[!t]
\centering
 \includegraphics[width=0.7\textwidth, trim =50 200 30 50, clip = true]{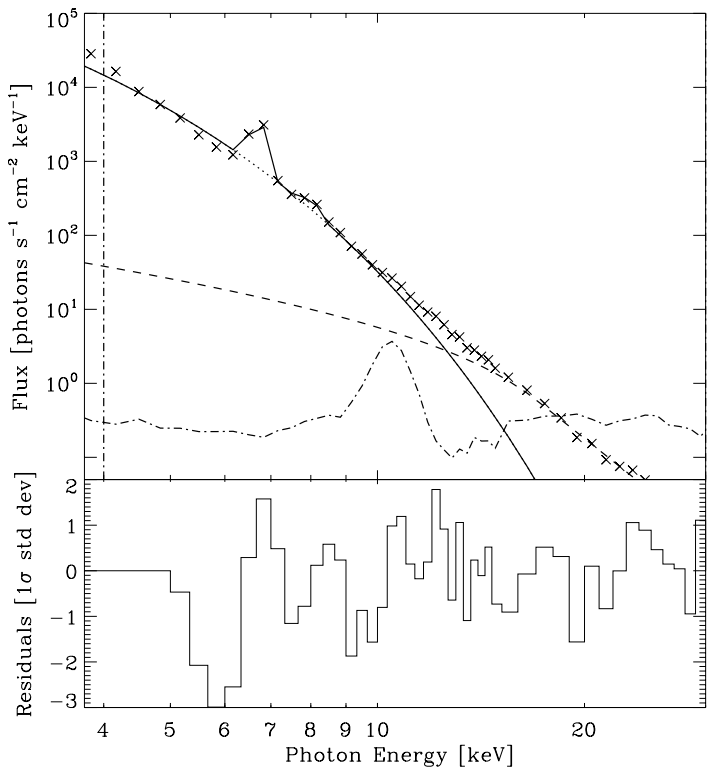} 
 \caption{The top panel shows the RHESSI photon spectrum between 15:16:30~UT and 15:17:30~UT. The data points (crosses) were fit with a thermal Maxwell distribution (dotted line) and a thick target, non-thermal model (dashed line). Combining these model parameters with the background function (dot-dash line) resulted in the overall fit to the spectrum (solid line). The residuals of the spectrum fit are shown in the bottom panel.   } 
 \label{fig:hsi} 
\end{figure}

The RHESSI spectrum, shown in Figure \ref{fig:hsi}, was analysed for one minute between 15:16:30~UT and 15:17:30~UT. Since it is believed that this was after the the HXR peak, this may mean that the maximum temperature was not in-fact observed. Following previous studies (e.g. \citeauthor{Saint-Hiliare02}, \citeyear{Saint-Hiliare02}), the data were fitted with an isothermal model at low energies and a thick-target model up to $\sim$30~keV (see \S \ref{sect:instr_hsi_spec} and Equation \ref{eqn:brem_flux}). The thick target component yielded a low energy cutoff of 17~keV and a power law index of 8.2. The isothermal fit to lower energies resulted in temperature and emission measure values of $\sim$13~MK and  $\sim$1$\times$10$^{48}$~cm$^{-3}$ respectively. A non-thermal electron flux of $\sim$7$\times$10$^{9}$~ergs~cm$^{-2}$~s$^{-1}$ was also calculated by approximating the footpoint area from \ion{He}{i} and \ion{O}{v} observations. Since it is probable that the HXR peak occurred before this time, this non-thermal electron flux must be assumed to be a lower limit to the peak flux. The filter ratio of the two GOES passbands produced the temperature and emission measure of the event as described in \S \ref{sect:instr_goes}, giving a peak temperature of 10~MK and an emission measure of 4$\times$10$^{48}$~cm$^{-3}$. 
 
\begin{figure}[!t]
 \includegraphics[width=\textwidth, trim =45 140 35 30, clip = true]{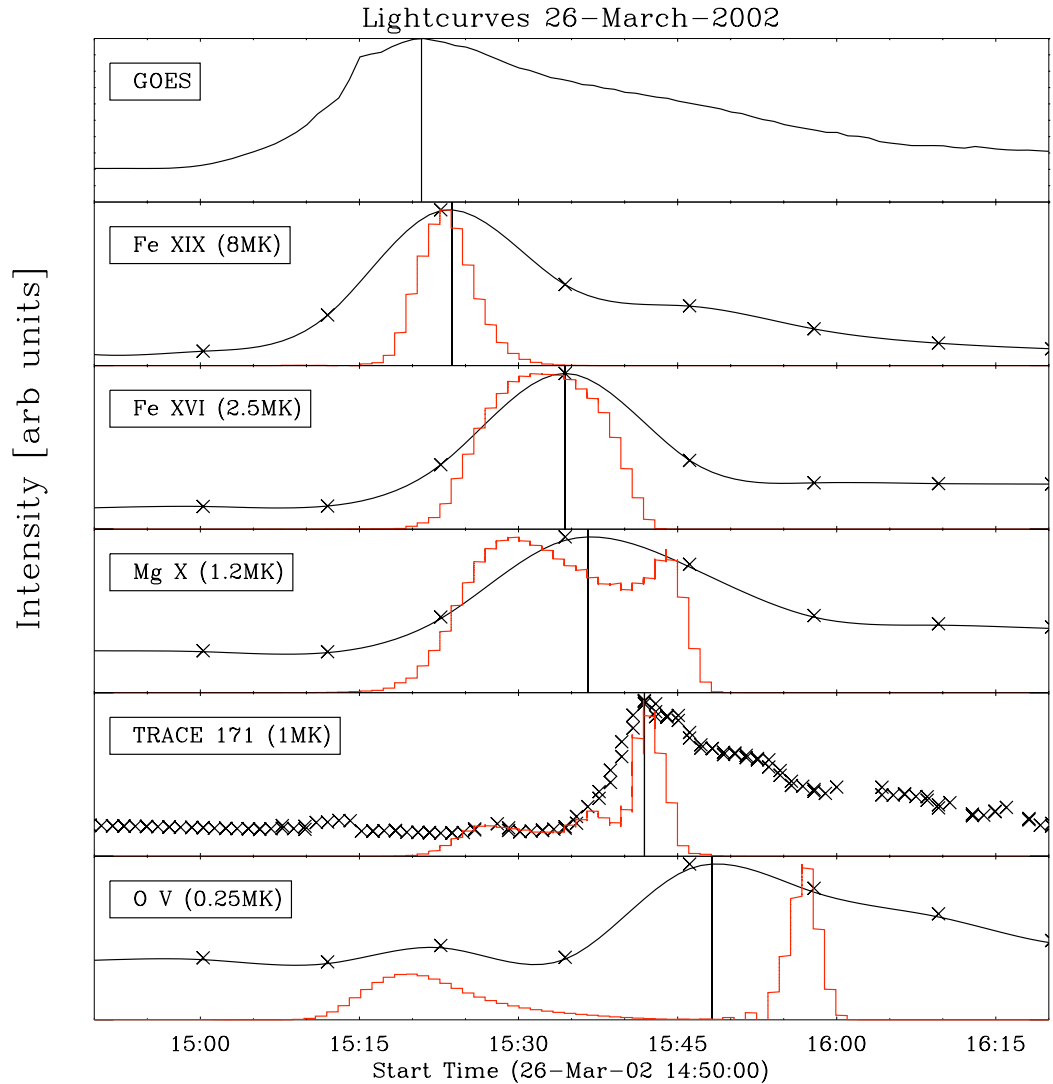} 
 \caption{Lightcurves from GOES, \ion{Fe}{xix}, \ion{Fe}{xvi}, \ion{Mg}{x}, TRACE 171~\AA\ and \ion{O}{v}. Overplotted on the data points are spline interpolations (black lines) and the lightcurves predicted by Equation \ref{equation:lgtcv} (red lines). The vertical lines represent the times adopted as the peak of the lightcurves.   } 
 \label{fig:lgtcv} 
\end{figure}

Four of the CDS emission lines (excluding \ion{He}{i} as it is optically thick) and TRACE~171~\AA\ images were integrated over the area described by the bright ``knot'' mentioned above. Although the loop is believed to consist of multiple magnetic strands, it was assumed that the majority of strands within this region were heated almost simultaneously during the impulsive burst. A small number of strands can be heated before or after this time, producing a multi-thermal plasma, however from studying Figure \ref{fig:F-4panels}, this small region was assumed to be approximately isothermal at any one time. The lightcurves of this region are shown in Figure \ref{fig:lgtcv}. In order to ascertain the thermal evolution of the flare, a temperature was assigned to each lightcurve. This was achieved through the fitting of a spline interpolation to the data points and comparing them to forward modelling results of various emission lines and of the TRACE 171~\AA\ passband (Figure \ref{fig:lgtcv}). The emission measure and temperature evolution of the event as predicted by EBTEL (top two panels of Figure \ref{fig:full}) were used to calculate the value of the contribution functions $G(T_i)_{ion}$ shown in Figure \ref{fig:goft_fns} and emission measure $EM_i$ at each timestep $i$. Their product resulted in the predicted lightcurve for a particular ion at a given time, as in Equation \ref{equation:lgtcv}:
\begin{equation}
\label{equation:lgtcv}
I_i = G{(T_i)}_{ion} \times EM_i
\end{equation}

As Figure \ref{fig:lgtcv} shows, the peak of the predicted lightcurves agree very well with the observed lightcurves. The predicted lightcurves, in general, reach their maximum intensity at the time ($t$) the contribution function is maximised (i.e. $t(G(T)_{max}) = t(I_{max})$). Therefore, the temperature of the peak of the contribution functions were were assigned to the time of the maximum intensity. The poor cadence of the CDS instrument was such that the time of maximum intensity could only be estimated by a spline fit to the datapoints (black lines, Figure \ref{fig:lgtcv}).

The predicted lightcurves for \ion{Fe}{xix} and \ion{Fe}{xvi} are in very good agreement with the observed lightcurves. This is most likely due to the high intensity of these lines during a flare. The predicted lightcurve for \ion{Mg}{x} however, requires some explanation. It is clear that there are two peaks in intensity for \ion{Mg}{x}, one each side of the spline peak. By 15:30~UT, the time of the first predicted peak, the temperature (shown in the top panel of Figure \ref{fig:full}) had already reached its maximum and had begun to pass back into the band of sensitivity of the \ion{Mg}{x} contribution function. At this time, the levels of $EM$ were raised to a significantly higher level compared to during the impulsive phase, when the temperature initially rose through the \ion{Mg}{x} $G(T)$. The combination of the increasing $G(T)$ and elevated $EM$ resulted in a broad peak in the predicted lightcurve around 15:30~UT. A dip is then seen in the predicted lightcurve at $\sim$15:40~UT. This is as a result of the falling $EM$ at this time and the leveling off of the contribution function at $\sim$2$\times$10$^6$~K. As the contribution function rises again to its peak at 1~MK, the lightcurve too rises to a sharp peak. Although we have ascertained the theoretical time of the 1~MK peak, the poor cadence of the instrument restricts us from being as exact with the observation. 

Similar to \ion{Mg}{x}, the \ion{O}{v} lightcurve peak was difficult to establish. This may be due to the high levels of \ion{O}{v} in the surrounding areas, along with the low cadence of the instrument. In addition, the $T$ and $EM$ values used for predicting the lightcurves  were obtained from Figure \ref{fig:full}. It can be seen that the EBTEL temperature is not in good agreement with the \ion{O}{v} data point, predicting that the loop will reach this temperature a short time later.

The instrument response function used in the forward modelling of TRACE were those calculated in Chapter \ref{chapter:SWAP}. The first small peak at 15:13~UT was reproduced using the \emph{flare} response function. However, in that case, the main peak was not predicted and so we find it inappropriate to use the flare response to predict the temperature of TRACE during the decay phase. The active region response curve (Figure \ref{fig:all_temps} panel \emph{c}) however, predicts very nicely the main late phase intensity peak along with a smaller peak at 15:28~UT (which also corresponds to the broad peak seen in \ion{Mg}{x} formed at a similar temperature) as shown in the fifth panel of Figure \ref{fig:lgtcv}. In \S \ref{sect:sol_condts}, it was established that when imaging an active region, TRACE is most sensitive to plasma at 1~MK. For this event, by the time the TRACE intensity begins to increase, the main phase of the flare has passed and the system is similar to a bright active region loop. Therefore, assigning a temperature of 1~MK to the time of the TRACE lightcurve maximum is justifiable. 

The emission measure was calculated by inverting Equation \ref{equation:lgtcv} and assuming an isothermal plasma at any given time. We have not included TRACE in the analysis of EM due to its broad and complex response function (see Chapter \ref{chapter:SWAP} for further details). For the uncertainty in the CDS EM, there are a number of factors to consider. These include uncertainties in the intensity of the line, the contribution function, the volume element $dV$ assumed for Equation \ref{equation:lgtcv} and the CDS calibration. While the uncertainty in measuring the line intensity is small for strong lines such as those used in our study, \citep[typically $\sim$10\%;][]{DelZanna01}, a consideration of the contribution functionÕs FWHM yields an uncertainty in the EM of up to 30\%. In addition, the CDS calibration is known to be good to within 15-20\% \citep{Brekke00}. Considering these factors, the combined photometric error was taken to be 50\%. 

 \begin{figure}[!t]
                 \includegraphics[width=0.5\textwidth,trim =50 350 70 0, clip = true]{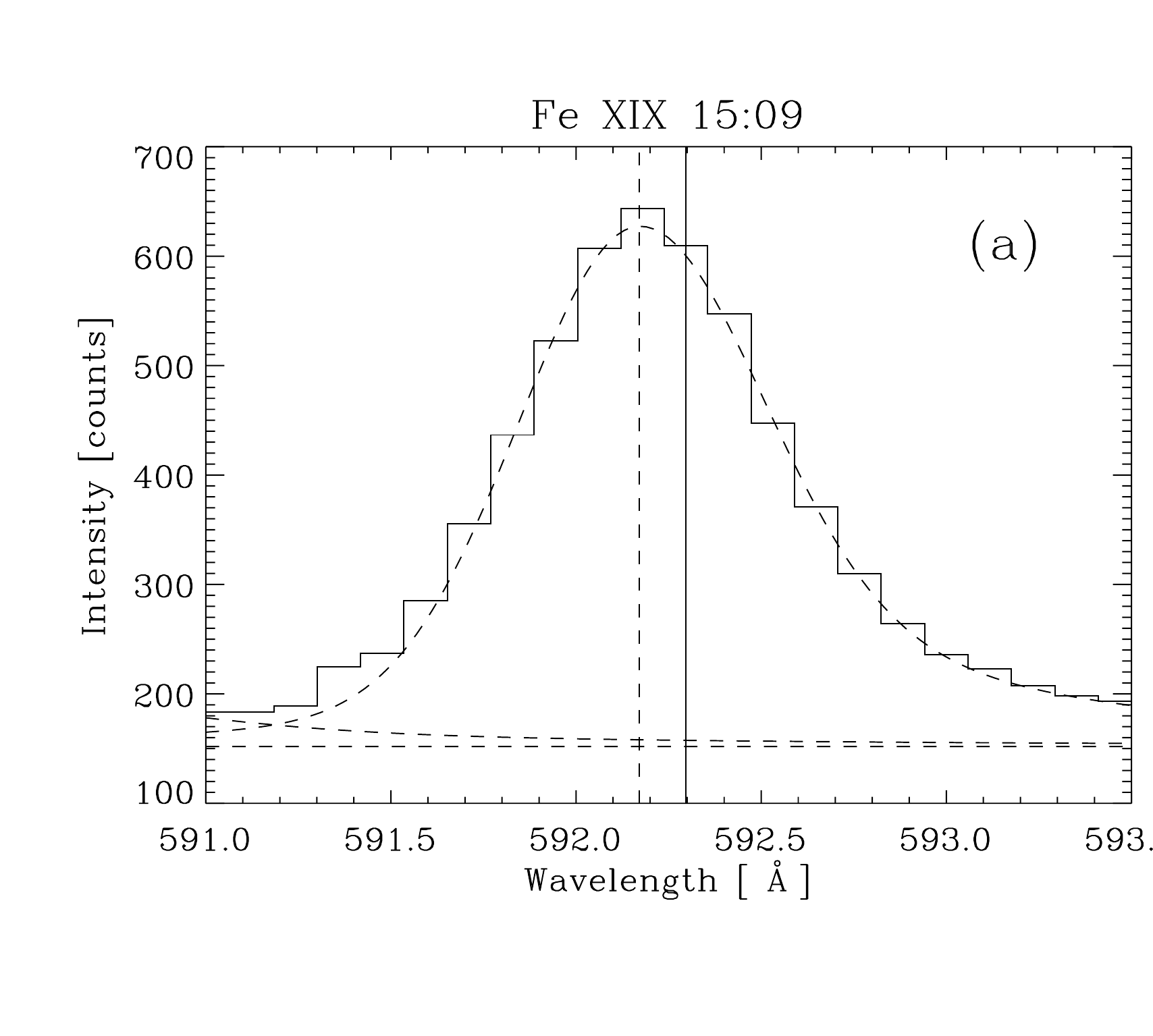}
               \includegraphics[width=0.5\textwidth,trim =50 350 70 0, clip = true]{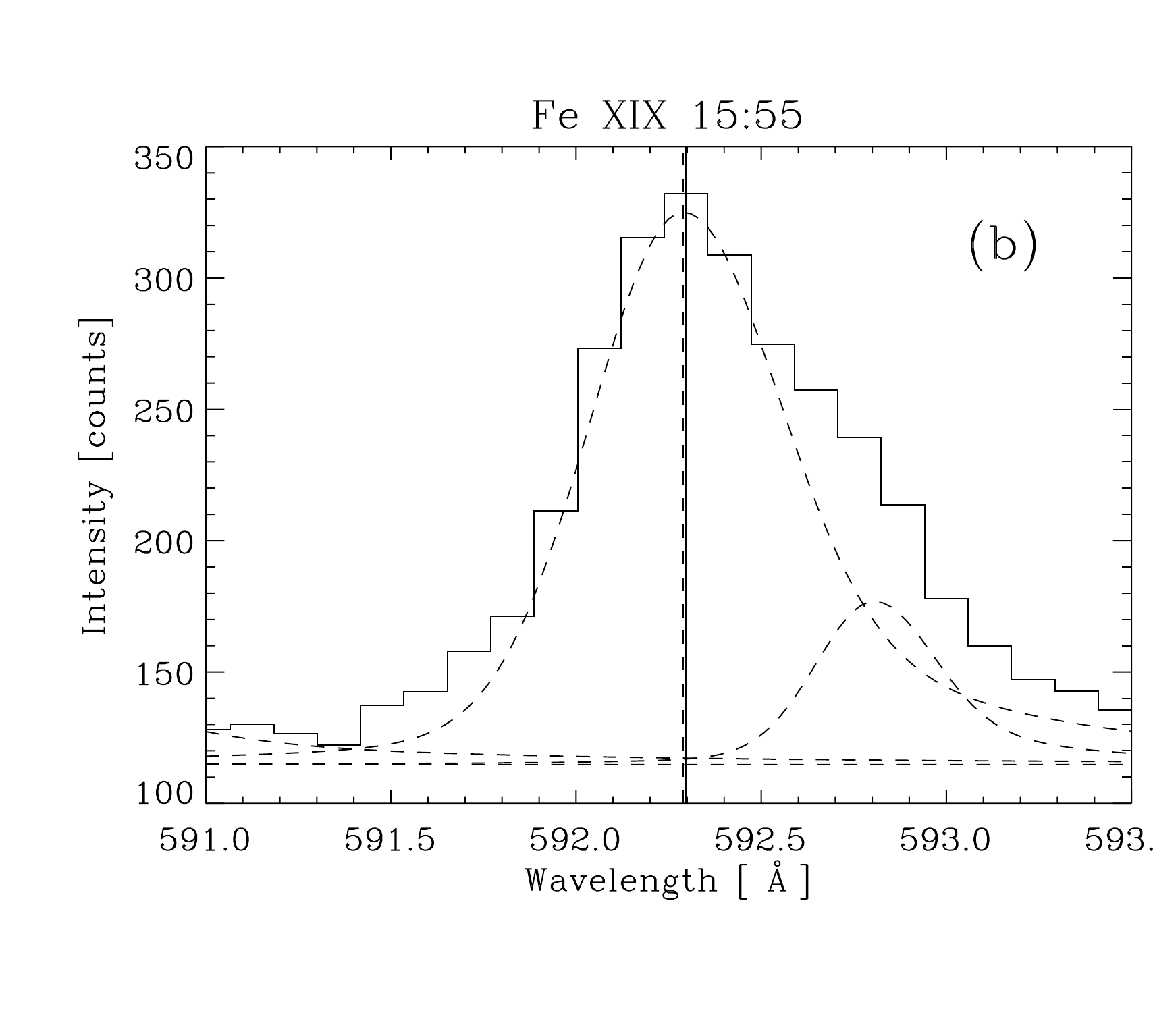}
             \caption{The line profiles of the \ion{Fe}{xix} emission line for the right footpoint during the impulsive phase (a) and the decay phase (b). Note the increasing intensity of the \ion{Fe}{xii} blend at $\sim$592.8~\AA \ \citep{DelZanna05} during the decay phase. The solid vertical line represents the rest wavelength while the dashed vertical line is its centroid.    }
   \label{fig:line_prof}
   \end{figure}

\subsection{Velocity}
\label{section:velocity}
The relative Doppler shifts at both footpoints were calculated for the duration of the flare using the five CDS emission lines. The centroids were calculated following \S \ref{sect:rest_wv} and were corrected for both heliocentric angle and an average inclination of 44$^{\circ}$ to obtain the rest wavelengths listed in Table \ref{table:lam0}. As mentioned previously, it is not possible to obtain absolute velocities using CDS and so these should be interpreted as relative velocities. Figure \ref{fig:line_prof} shows the \ion{Fe}{xix} line profiles for the right footpoint during (a) the impulsive phase and (b) the decay phase.

\subsection{Modelling}
\label{section:modelling}
The EBTEL model simulates the evolution of the average temperature, density, and pressure along a strand at any given time (\S \ref{sect:ebtel}). The flare was modelled as a single, monolithic loop. This was justifiable since the observations suggest that most of the strands are heated in approximately the same way and at the same time. The standard EBTEL pre-flare conditions included a temperature of 0.3~MK, an initial density of 5$\times$10$^{7}$~cm$^{-3}$ and an emission measure of 4$\times$10$^{43}$~cm$^{-3}$. The observations described in this chapter were, where possible, used to constrain the model's input parameters. The ranges of acceptable parameters studied, along with the observed values are shown in Table \ref{table:params}. The method presented here does not use any rigorous statistical fitting methods. However, \citet{Adamakis09} utilised a Bayesian fitting technique to better identify the best-fit model parameters. The results of their study are also shown in Table \ref{table:params} for comparison. In general, the statistically robust parameters are in good agreement with those published in \citet{Raftery09} with the exception of the $direct/non-thermal$ ratio which is discussed below. The loop length was determined from magnetic field extrapolations of the region (P. A. Conlon, private communication). Since it was assumed that the heating function is associated with the HXR burst, the majority of which was not observed, the shape of the heating function was inferred both from previous observations of HXR bursts and the slow rise of the GOES SXR lightcurve. Thus, the most appropriate heating function was deemed to be Gaussian in shape. This choice was later justified by \citet{Adamakis09} who proved statistically that a Gaussian distribution is is in better agreement with observations than a half-Gaussian distribution. The amplitude of the non-thermal electron flux was constrained by the lower limit calculated from RHESSI observations and the width was inferred from the derivative of the SXR flux \citep[\S \ref{sect:flares};][]{Neupert68, Zarro93}. While the direct heating rate was not constrained by observations, it was assumed to have the same width as the non-thermal heating flux and to occur at the same time. The range of parameter values shown in Table \ref{table:params} correspond to the maximum and minimum values that produce an acceptable fit to data. The ratio of the heating components (i.e. direct to non-thermal) is also shown for the best fit parameters. Note the discrepancy between the \citet{Adamakis09} heating ratio and the value used in this chapter result from the exclusion of background heating levels in the ratio calculated by Adamakis and the inclusion of the background heating here.

	\begin{table}[!t]
\caption{Input parameters used for EBTEL simulation for work shown in this Chapter and work carried out by \citet{Adamakis09}. The parameters were constrained by data when possible and the ranges of parameters investigated are shown. } 

\centering 
\begin{tabular}{ l  c  c  c  c c } 
\hline\hline
	
Parameter 				 								& Observed 			&		EBTEL			& Adamakis		\\
\hline
 Loop half-length [cm] 										& 3$\times$10$^{9}$		&		$(3 \pm 0.2)\times10^{9}$	&	$(3.5 \pm 0.4)\times10^{9}$	\\
 Non-thermal flux  											&					& 							&	\\
 					- Amplitude [ergs~cm$^{-2}$~s$^{-1}$]		& 7$\times$10$^{9}$		&		$5 \times 10^{8\pm1}$	&	$(3.2\pm0.4) \times 10^{8}$	\\
					- Width [sec]							& $\sim$100					&		$100\pm50$	&	$92\pm20$		\\
					- Total  [ergs~cm$^{-2}$]					& $\sim$1.7$\times$10$^{12}$		&	$2.5\times10^{10\pm1}$	& $3\pm4\times10^9$	\\
 Direct  heating rate										 	& 							&					&			\\	
 					- Amplitude [ergs~cm$^{-3}$~s$^{-1}$]		& -							&		$0.7\pm 0.3$	&	-		\\
					- Width [sec]							& -							&		$100\pm 50$	&	-		\\
					- Background [ergs~cm$^{-3}$~s$^{-1}$]		& -							&		$\le1\times10^{-6}$&	-	 	\\
					- Total  [ergs~cm$^{-3}$]					& -							&		$175\pm150$	&	-		\\					 
Direct/non-thermal heating									&		(best fit)		&		$\sim4$		&	23 $\pm$ 16	\\

 \hline
     \end{tabular} 

\label{table:params} 
\end{table}

In order to highlight the importance of the effect of cooling by conduction and radiation simultaneously, the data were also compared to the Cargill model (\S \ref{sect:cargill}). Following \citet{Antiochos76}, \citet{Cargill93, Cargill94} presented a model that considered a flare that is cooling purely by conduction for a time $\tau_{c}$, followed by purely radiative cooling for a time $\tau_{r}$, as given in \S \ref{sect:cargill}. The cooling times, along with $\tau_{*}$ and $T_{*}$, the time and temperature at which the cooling mechanisms change were calculated for the flare and compared to the results of the 0-D EBTEL model.

\section{Results}
\label{section:results}

Combining the observations from the different instruments used for this study with results from EBTEL, the heating and cooling phases of this flare can be comprehensively described. These results are presented in Figs. \ref{fig:full} and \ref{fig:flare_life}. These figures show  the evolution of the flare through the dependence of temperature, emission measure, energy losses and velocity.

\subsection{Comparison of model to data}

The EBTEL parameters were allowed to vary within the limits listed in Table \ref{table:params} until a good fit to the cooling phase data points was found. The cadence of CDS was such that it did not observe any significant intensity deviations during the impulsive phase and since RHESSI was eclipsed during the impulsive phase, the only reliable observations available were for the decay phase. Therefore, no observations could be used to constrain the model during the rise phase of the flare. 

\begin{figure}
\centerline{\hspace*{0.0\textwidth}
 \includegraphics[width=\textwidth, trim =15 90 20 50, clip = true]{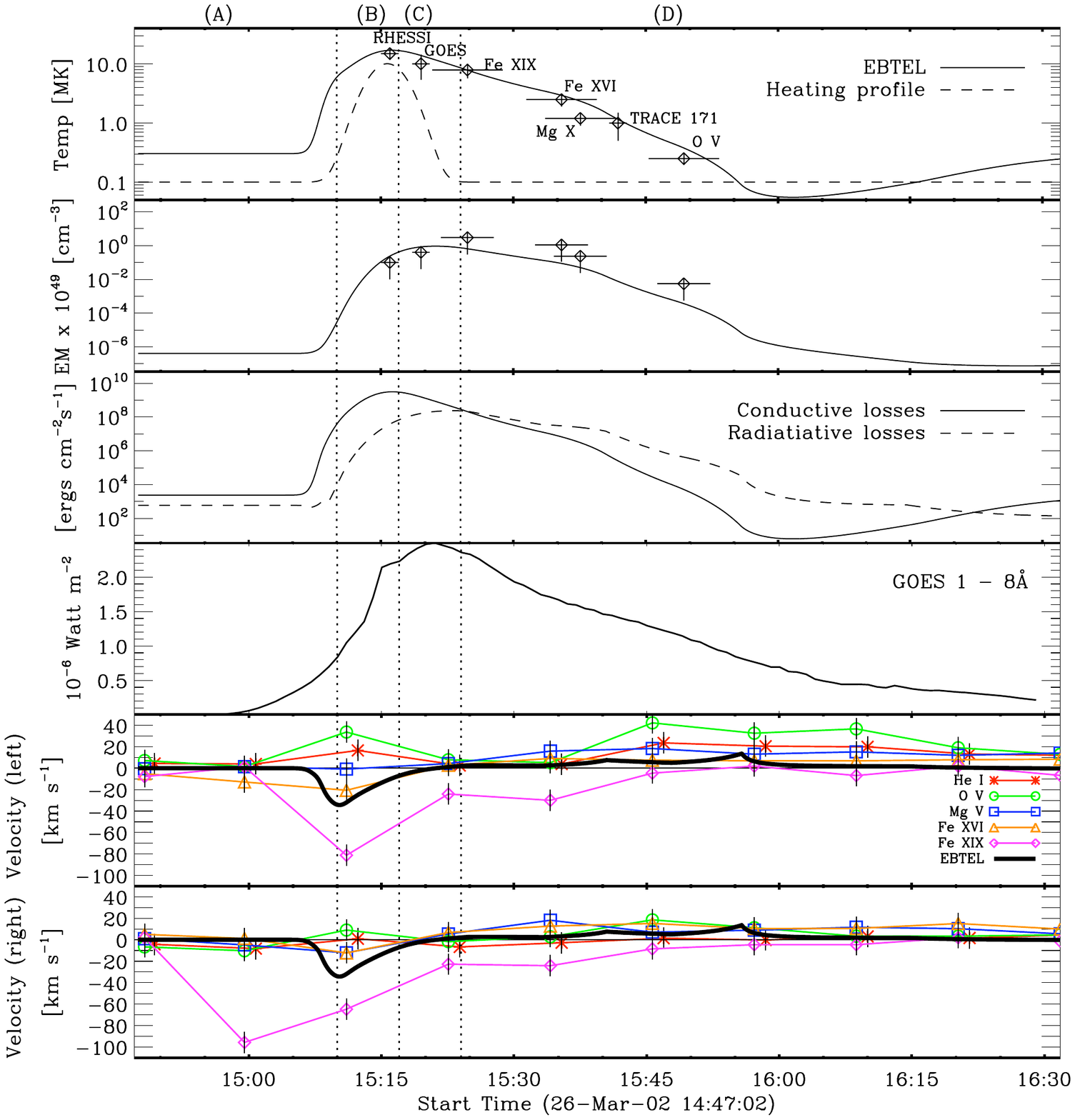} 
}
 \caption{Comparison of observations (data points) to EBTEL model (solid lines, except panel 4). From top to bottom, the panels describe the temperature, emission measure, loss rate, GOES SXR flux and velocity profiles of spectral lines of the left and right footpoints of the flare as a function of time. The adopted heating function for EBTEL is shown overplotted on the first panel. The data points in the second panel correspond in time to the ones above in the temperature panel. The EBTEL modelled velocity (thick black line in the bottom panel) is for \ion{Mg}{x}. The dotted vertical lines correspond to the flare phases (A)--(D) explained in \S \ref{section:phases}.} 
 \label{fig:full} 
 \vspace{-0.1cm}
\end{figure}

The top two panels of Figure \ref{fig:full} describe the evolution of the flare temperature and emission measure from both observations (data points) and model (solid line). The conductive and radiative loss curves generated by EBTEL for the flare are shown in the third panel of Figure \ref{fig:full}. Conduction was found to dominate initially, with radiation becoming prevalent for the remainder of the decay phase. This is consistent with previous observations \citep[e.g.][]{Culhane1994, Aschwanden2001}. Both the Cargill and EBTEL simulations found conduction to dominate for the first 200 - 400 s of the decay phase, with radiation dominating for the remaining $\sim$4000 s, referring to $\tau_{c}$ and $\tau_{r}$ respectively. The time $\tau_{*}$ at which $\tau_{c} \approx \tau_{r}$ is $\sim$15:24~UT in both cases. However, the temperature at which this occurs, $T_{*} = T(\tau{_*})$, was found to be $\sim$12~MK and $\sim$8~MK according to Cargill and EBTEL respectively. This is not surprising since EBTEL uses both radiation and conduction to remove heat from the corona while Cargill uses only one of these mechanisms at any given time. Therefore the overall efficiency of heat removal will be greater for EBTEL than for Cargill. The fourth panel of \ref{fig:full} shows the GOES 1-8~\AA \ lightcurve for context. The last panel shows the velocities at the loop footpoints, calculated following the analysis in \S \ref{section:velocity} where negative velocities correspond to upflowing (blueshifted) plasma. The flow velocity at both left and right footpoints are shown for the coolest and hottest lines -- \ion{He}{i} and \ion{Fe}{xix} respectively while, for clarity, only the right footpoints for the remaining three lines are shown. The \ion{Mg}{x} velocity simulated by EBTEL is represented by the thick black line. The simulations are in reasonable agreement with the observations. Upflows are of course predicted during the evaporation phase and downflows are predicted during the draining phase. However, the magnitudes are generally larger than those observed in the \ion{Mg}{x} line. This likely to be as a result of the treatment of the HXR beam by EBTEL. By assuming all of the beam energy is transferred into upflowing plasma, the velocity attained will be greater than the velocity observed which, in reality, will be driven by only a portion of the beam energy. The small peaks seen at later times in the simulated velocity curve are a result of the piece-wise continuous form used for the radiative loss function. 

\begin{figure} 
\centerline{\hspace*{0.0\textwidth}
 \includegraphics[width=0.8\textwidth, trim =60 400 60 50, clip = true]{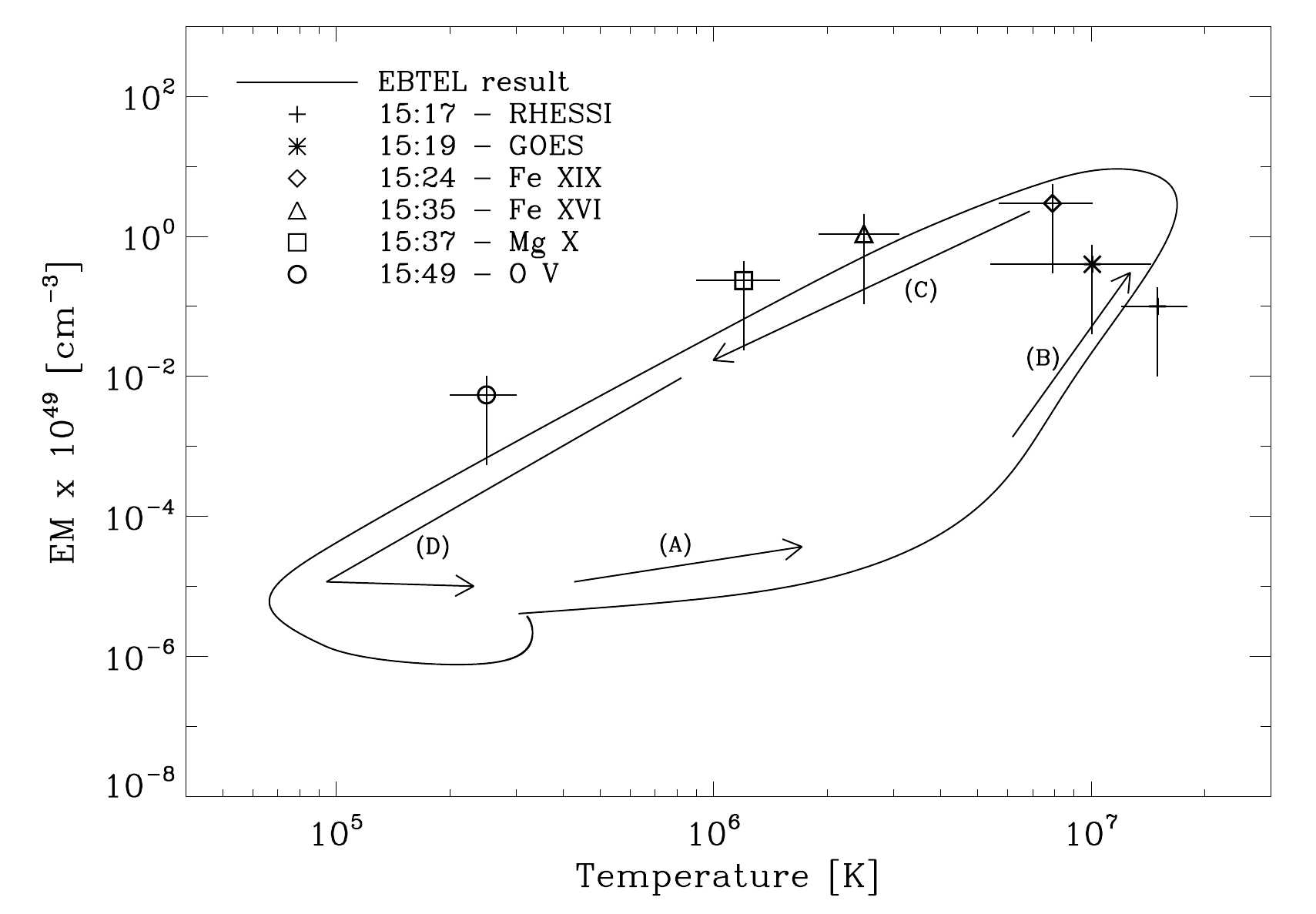} 
}
 \caption{This shows the dependence of emission measure on temperature for both model and data. The different phases of the flare are marked (A)--(D). Over-plotted are the emission measure data-points as a function of their temperature.  } 
 \label{fig:flare_life} 
 \vspace{-0.1cm}
\end{figure}

Figure \ref{fig:flare_life} shows the evolution of the flare through the interdependence of emission measure and temperature. The data points obtained during the analysis described in \S \ref{section:temperature} were computed at the same time for any one emission line or bandpass. Figures \ref{fig:full} and \ref{fig:flare_life} show four phases of the flare evolution: the pre-flare heating of the plasma (A) followed by evaporation of hot plasma (B), cooling (C) and draining (D).

\subsection{Flare phases}
\label{section:phases}

\begin{inparaenum}
\item[\itshape A\upshape)]
\emph{14:45--15:10; Pre-flare phase:} For the majority of this phase, the EBTEL parameters remained at quiet Sun values, as  phase (A) of Figure \ref{fig:full} shows (i.e. between 14:45 and 15:10~UT). At 15:07~UT the EBTEL temperature and emission measure began to rise. Figure \ref{fig:flare_life} shows the steep temperature gradient and the initial gradual rise in emission measure. However, as the fourth panel in Figure \ref{fig:full} shows, the GOES soft X-rays began to rise slowly before this. Since the impulsive phase of the flare was not constrained, this pre-flare heating was not modelled. At $\sim$15:00~UT, a small amount of \ion{Fe}{xix} emission was seen in the loop (Figure \ref{fig:F-4panels}, first \ion{Fe}{xix} image). In addition, velocities of 90$\pm$16~km~s$^{-1}$ observed in \ion{Fe}{xix} can be seen in the bottom panel of Figure \ref{fig:full} while all of the cooler lines remain at rest. This could be evidence for pre-flare gentle chromospheric evaporation in a small number of strands before the HXR burst.

\item[\itshape B\upshape)]
\emph{15:10--15:17; Impulsive phase:} During the impulsive phase of a flare, the standard model predicts the propagation of non-thermal electrons to the chromosphere where they heat the ambient plasma, causing it to rise and fill the loop. Upflows of 81$\pm$16~km~s$^{-1}$ in \ion{Fe}{xix} and simultaneous cool downflows of 16 and 33$\pm$16~km~s$^{-1}$ \ion{He}{i} and \ion{O}{v} respectively were observed and shown in the bottom panel of Figure \ref{fig:full}. A non-thermal electron flux of $\sim$7$\times10^{9}$~ergs~cm$^{-2}$~s$^{-1}$ was determined between 15:16:30 and 15:17:30~UT from RHESSI observations. This is slightly lower than the 3$\times10^{10}$~ergs~cm$^{-2}$~s$^{-1}$ required to drive explosive chromospheric evaporation \citep{Fisher85, Milligan_explosive}. This may be due to the uncertainty in estimating the area of the HXR beam, although it is most likely due to RHESSI being in eclipse during the main release of hard X-rays. It is assumed that 7$\times10^{9}$~ergs~cm$^{-2}$~s$^{-1}$ is a lower limit to the peak HXR flux for this event. 

\item[\itshape C\upshape)]
\emph{15:17--15:24; Soft X-ray peak:} The top panel of Figure \ref{fig:full} shows the temperature has peaked and begun to fall and that the emission measure and SXRs were at a maximum in this phase. As Figure \ref{fig:F-4panels} shows, the non-thermal flux levels are significantly reduced at a value of $\sim$9$\times10^7$~ergs~cm$^{-2}$~s$^{-1}$. Conduction however, was highly efficient at this time (third panel of Figure \ref{fig:full}). The blueshifts observed in \ion{Fe}{xix} (25$\pm16$~km~s$^{-1}$) during this phase may be as a result of conduction driven gentle chromospheric evaporation. The lack of any significant downflowing plasma at this time corroborates this theory.

\item[\itshape D\upshape)]
\emph{15:25--16:30; Decay phase:} This phase is dominated by radiative cooling, as seen in the third panel of  Figure \ref{fig:full}. Velocities in \ion{Fe}{xix} were returning to quiet Sun values. If the upflows observed at 15:32~UT in \ion{Fe}{xix} is not as a result of continued conduction driven gentle evaporation, they may be explained by the line profile containing components from evaporating strands that were heated after the main loop bundle. Between approximately 15:45 and 16:20~UT \ion{Mg}{x}, \ion{O}{v} and \ion{He}{i} showed downflows of up to 42$\pm$16~km~s$^{-1}$. This implies loop draining was occurring \citep{Brosius03}. By the end of the simulation, all of the parameters had returned to quiet sun values. 

\end{inparaenum}

\section{Momentum Balance}
Following the approach taken by \citet{Teriaca06}, the momentum, $p=mv$, of the upflowing and downflowing plasma was calculated. Mass density $\rho$ can be expressed as the product of the mean molecular mass, the proton mass and the total particle density:
\begin{equation}
\rho = \mu m_p n_{tot}
\end{equation}
where $m_p = 1.67\times10^{-24} g$. $\mu = 1.27$ for a neutral plasma and 0.61 for a fully ionized plasma and $n_{tot}$ can be expressed in terms of the chromospheric and coronal densities. Since $m = \rho  V$, the mass can now be expressed as 
\begin{equation}
m = \mu m_p n_{tot} V = \mu m_p n_{tot} A \Delta h
\label{eqn:mom_den}
\end{equation}
where $A$ is the cross sectional area of the loop and $\Delta h$ is the height through which the plasma can flow. The footpoint area was estimated from \ion{O}{v} images taken at the peak of the flare and was found to be $\sim10^{18}$~cm$^2$. This is consistent with a typical footpoint area \citep[e.g.][]{Canfield_nat}. Following Equation \ref{eqn:mom_den}, the momentum can be expressed as:
\begin{equation}
p = \mu m_p n_{tot} Av \Delta h 
\label{eqn:mom}
\end{equation}
for average upflow or downflow velocity $v$. 

For chromospheric evaporation into the chromosphere, we assume the total density can be taken to be 1.1 times the typical chromospheric density where $n_{ch} \approx 10^{13}$~cm$^{-3}$ (see e.g. Figure \ref{fig:solar_mod}). From Figure \ref{fig:fisher} it is clear that the high density found in the chromosphere restricts the depth to which downflowing plasma can reach. This leads to an estimate of $\Delta h \approx$ 200~km \citep{Teriaca06, Abbett99}. With average downflow velocities of $\sim$20~km~s$^{-2}$ during the impulsive phase of this flare, the momentum of the downflowing plasma was calculated to be $9\times 10^{20\pm1}$~g~cm~s$^{-1}$. 

Chromospheric evaporation upwards into the corona can occur along approximately the entire loop length. Therefore, in the coronal calculation, we take $\Delta h = 3 \times 10^9$~cm. The coronal density can be approximated as $n_{tot} = 1.91\times n_e$ where $n_e \approx \sqrt(EM/A \times \Delta h) = 1 \times 10^{10}$~cm$^{-3}$. Considering an average upflow velocity of $\sim$80~km~s$^{-1}$, the momentum of upflowing plasma was calculated to be $5\times 10^{20\pm1}$~g~cm~s$^{-1}$. These values are consistent with momentum calculations carried out for flares of similar sizes in e.g. \citet{Teriaca06, Canfield_nat}.

\section{Conclusions and discussion}
\label{section:concs}
This chapter compares a flare observed with CDS, TRACE, GOES and RHESSI. Early in the impulsive phase of the flare, evidence of 8~MK emission and 95$\pm$16~km~s$^{-1}$ upflows suggest the pre-flare heating of the loop. During the impulsive phase, hot upflowing plasma at velocities of 81$\pm$16~km~s$^{-1}$ and cool downflows of up to 33$\pm$16~km~s$^{-1}$ imply explosive chromospheric evaporation, in accordance with previous observations such as \citet{Milligan_explosive, Teriaca06}. Upflowing plasma at velocities of 24$\pm$16~km~s$^{-1}$ were observed around the time of the SXR peak in \ion{Fe}{xix}, along with negligible (6$\pm$16~km~s$^{-1}$) \ion{He}{i} upflows. Since conduction was found to be highly efficient at this time, this may be evidence for conduction driven gentle chromospheric evaporation.

As expected, once the temperatures within the loop had fallen and temperature gradients were significantly reduced, the conductive loss rates fell considerably. Thus, the late decay phase of the flare was dominated by radiative cooling. Late in the flare (post 15:45~UT), the draining of plasma from the looptop was observed in \ion{He}{i}, \ion{O}{v} and \ion{Mg}{x}. 

During the explosive chromospheric evaporation phase, the momentum of the moving plasma was calculated. The upflowing plasma momentum was found to be $4 \times 10^{20\pm1}$~g~cm~s$^{-1}$ while for downflowing plasma it was $9 \times 10^{20\pm1}$~g~cm~s$^{-1}$. Not only do these results show that the momentum was conserved within the loop, they are also consistent with previous studies of flares of similar GOES class. \citet{Canfield_nat} for example found the upflow momentum of a small C2 class flare to be $\sim$2$\times 10^{20}$~g~cm~s$^{-1}$ and the corresponding downflow momentum to be $\sim$6$\times 10^{20}$~g~cm~s$^{-1}$ while \citet{Teriaca06} found the momentum of a GOES C2.3 flare to be $(1\pm5)\times 10^{20}$~g~cm~s$^{-1}$ in both directions. The balance between these values means momentum is conserved within the system during chromospheric evaporation. Since the momentum of the material appears to be consistent (at least for C-class flares), the momentum of the plasma may lead to furthering the understanding of the force applied to the plasma due to the electron beam.

The observations of the flare were compared to the 0-D hydrodynamic model, EBTEL. The comparison of the simulations and observations were confined to the decay phase of the event, due to data limitations. It is believed that for future analysis, the investigation into different direct heating models (e.g. the width of the function and the time at which it peaks) along with the inclusion of low level beam heating early in the flare, may have the capability to reproduce the pre-flare heating that is observed in the GOES lightcurve. For this work however, the behaviour of the flare was tracked as it cooled through the response functions of the many instruments and emission lines that were available. This provided details such as the ratio of the heating functions, which cannot easily be obtained from observations. It was found that the data were best reproduced when the plasma was heated approximately equally by direct and non-thermal mechanisms. This implies that both of these processes are vital during the flaring process and that flares may not be energised solely by non-thermal particles, as previously believed \citep[e.g.][]{Brown71}. This is in agreement with recent results found by \citet{Milligan08}. There it was shown that a non-thermal electron beam is not necessarily required to obtain the high-temperature, high-density material we see in flares. However, it should be noted that the EBTEL value of the flux of non-thermal electrons required for equal heating is below the critical value for explosive evaporation hypothesised by \citet{Fisher85}. This can be explained by the over-simplified treatment of the non-thermal beam by EBTEL. This requires caution for a flare of this nature, where it is evident that non-thermal particles play an important role. The model assumes that all non-thermal energy is used for evaporating plasma upward into the loop. However, it is well known that should the beam flux be sufficient, it will drive plasma downwards into the dense chromosphere \citep{Woods04, Allred05}. Despite the approximations made by EBTEL, such as the homogenous nature of the loop or the disregard for the location of energy deposition, the temperature and emission measure curves reproduce observations very well. It is computationally efficient, running a complete simulation in a matter of seconds. 

The cooling timescales modelled by \citet{Cargill93, Cargill94} were calculated to be within a factor of 2 of EBTEL timescales. The critical time, $\tau_*$ was found to be $\sim$15:24~UT for both models, however the temperature $T_*(\tau_*)$ did not agree. The temperature at $\tau_*$ was found to be 8~MK for EBTEL and 12~MK for Cargill. This discrepancy is as a result of the way in which the cooling mechanisms are treated by the models. Since EBTEL takes account of the energy loss by conduction and by radiation at all times during the event, heat is removed more quickly than by the Cargill model, which considers only a single energy loss mechanism at any one time. This highlights the importance of considering the simultaneous cooling by both conduction and radiation. 

This chapter has established a method that will be applied to the analysis of future events. For this case, the data was manually compared to theoretical models. However, the fitting of the parameters together with model comparison techniques has been investigated using a Bayesian technique for simulating values from the posterior distributions of the parameters \citep{Adamakis09}. The results of the statistically robust approach have been found to be within the errors of the parameters presented by \citet{Raftery09}. The main discrepancy is regarding the ratio of the thermal to non-thermal heating. The value found by \citet{Adamakis09} is significantly larger than that found by the author ($\sim23$ vs $\sim$4). This can be explained since in the calculation of this parameter, \citet{Raftery09} includes the background heating while \citet{Adamakis09} does not. However in each case it was found that $F_{thermal}/F_{non-thermal} > 1$, meaning the thermal (direct) heating of the plasma must have at least as significant an impact on the heating of the plasma as beam heating. Statistically, the use of a Gaussian profile for the heating function was found to be in better agreement with observations than a half-Gaussian, which corroborates the assumption that the heating function can be obtained from the derivative of the soft X-ray flux. The Bayesian technique will be used when comparing theoretical models to future data sets. The authors intend to carry out an investigation of flare hydrodynamics using the improved cadence and extensive spectral range of the Extreme Ultraviolet Imaging Spectrometer (EIS) on board Hinode. Combining these data with RHESSI spectral fits will vastly improve observations and allow for even more accurate modelling.


\chapter{The flare-CME connection} 
\label{chapter:CME}

\ifpdf
    \graphicspath{{5/figures/PNG/}{5/figures/PDF/}{5/figures/}}
\else
    \graphicspath{{5/figures/EPS/}{5/figures/}}
\fi

\begin{figure}[!t]
\centering
\includegraphics[width=0.7\textwidth, trim =30 50 200 0, clip = true, angle = 0]{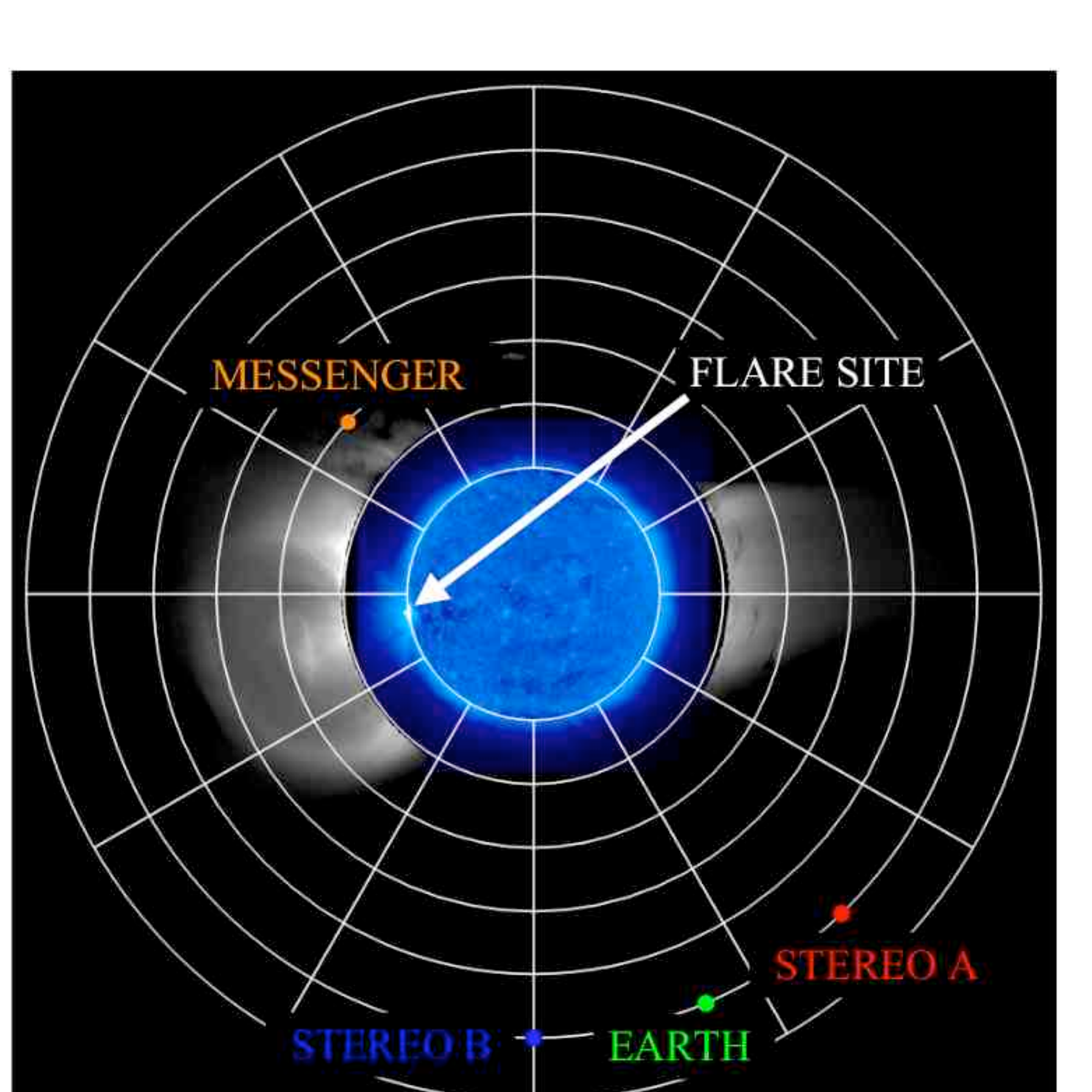} 
\caption{Position of the spacecraft used to observe the flare-CME event on 2007 December 31. RHESSI and GOES are orbiting Earth. This is a top-down grid overlaid on a side-on image taken with STEREO B/EUVI (blue) and Cor1 (black and white).}
\label{fig:instr_loc}
\end{figure}

\hrule height 1mm
\vspace{0.5mm}
\hrule height 0.4mm 
\noindent 
\\ {\it The connection between solar flares and CMEs has long been a topic of debate. We investigate this relationship using a well observed flare-CME event that occurred on the east limb of the Sun on 31-December-2007. The event was analysed using X-ray observations from MESSENGER and RHESSI, EUV observations from STEREO B/EUVI and white light observations from STEREO B/Cor1 and Cor2. The hydrodynamic evolution of the flare was analysed along with the kinematic evolution of both the flare and the CME. Evidence for pre-eruption magnetic reconnection was found to occur at least 2 minutes before the CME was launched. This was evidenced by a soft X-ray source at the top of the pre-flare arcade and temperature and emission measure values significantly higher than equilibrium. At the time of eruption, a HXR source (30-50~keV) was observed above the thermal sources and the RHESSI HXR flux reached a maximum. Evidence of breakout reconnection was not observed until 8 minutes later when a second set of HXR fluctuations were observed in conjunction with the restructuring of neighbouring magnetic field. Since magnetic reconnection occurred before the CME erupted and breakout reconnection was not observed until minutes after, it is evident that this event was not triggered by an ideal MHD instability or by external tether-cutting mechanisms. The pre-flare SXR source and the HXR source observed at the time of the eruption were both found to be located above the central arcade. This provides evidence that this CME was in fact triggered by internal tether-cutting reconnection. This work is currently in review for publication in the Astrophysical Journal (\emph{Raftery, Gallagher, McAteer, Lin \& Delahunt}, 2010, in review). Considerable intellectual conversation was conducted with Dr. Peter Gallagher, Dr. Chia-Hsien Lin and Dr. James McAteer. In addition, Dr. McAteer constructed Figure \ref{fig:instr_loc} and Gareth Delahunt conducted the SXR time-height analysis with the assistance of the author and Dr. Peter Gallagher. }
\\ 
\hrule height 0.4mm
\vspace{0.5mm}
\hrule height 1mm 

\newpage

\section{Introduction}
\label{section:intro}

The behaviour of CMEs and eruptive flares have been investigated on two fronts - theoretically and observationally. Unfortunately, the comparison between theory and observations is difficult and in most cases is a qualitative one. The mechanism involved in the initiation of CMEs is still a topic of hot debate with many competing theories, three of which are internal tether-cutting, external tether-cutting and ideal MHD instability. These models are discussed in detail in \S \ref{sect:cme_models}. 

In the case of internal tether-cutting (Figure \ref{fig:moore_int}), magnetic reconnection between the legs of a sheared arcade (i.e. ``internal'' reconnection) trigger the eruption. Following this, breakout reconnection above the freed ``plasmoid'' removes flux from its path and allows it to erupt as a CME. For external tether-cutting (Figure \ref{fig:moore_ext}) or breakout reconnection \citep{Antiochos1998, Lynch04}, these processes are reversed. The eruption is triggered by breakout reconnection above the central arcade (i.e. ``external'' magnetic reconnection). As the central arcade rises, the magnetic field in the legs is stretched and drawn together beneath the reconnection region, generating a second current sheet. Internal tether-cutting along this second current sheet frees the plasmoid, enabling it to erupt upwards. The final scenario involves the CME being initiated by an ideal MHD process (Figure \ref{fig:moore_ideal}). Should the central arcade be forced out of equilibrium by e.g. flux emergence or continued shearing of the field, it rises upwards to seek a new equilibrium. As this happens, it creates current sheets between its legs and above it with the overlying field, as before. Magnetic reconnection then takes place along these current sheets, resulting in the eruption of an accelerated CME. 

Observationally, complete analyses of flare-CME systems have been hindered by the lack of suitable data. Considering the broad range of temperatures ($\sim$8,000~K to $\sim$20~MK), energies (few eV to MeV), and distances (1 to $>$30 R$_{\odot}$), it is clear that a multi-instrument approach is necessary. In the past, the kinematics of CMEs (e.g. velocity, acceleration) have been well studied. \citet{Gallagher02} presented the first observations of a rising SXR source that occurred in conjunction with a coronal mass ejection. The kinematics of the SXR loops were analysed in detail and the thermal emission was found to originate from successively higher altitudes as the flare progressed, agreeing with the standard flaring picture. The CME associated with this flare was studied in detail by \citet{Gallagher2003}. The CME's acceleration was found to be best represented by a double exponential function. \citet{Temmer08} furthered the investigation into the flare-CME connection and found that the CME acceleration occurs simultaneously with the hard X-ray (HXR) burst of the corresponding flares. This lends further support to the proposal by \citet{Zhang2001} who suggested that CMEs and flares are driven by the same mechanism but do not have a cause and effect relationship. The importance of the flare-CME onset has been well documented by \citet{Harrison07}. Using the evolution of the pre-flare arcades though a series of EUV spectroscopic observations, they established the importance of the pre-eruption activity. Often, the temperature and density (emission measure) evolution of a system is conducted using spectroscopic data, such as that used in \citet{Harrison07} and \citet{Raftery09}. However, since these instruments are most effective close to disk centre, observations of limb flares are not as readily available. An alternative method of studying flare hydrodynamics is to use the spectroscopic capabilities of RHESSI. However, since RHESSI is primarily designed to study high energy emission, caution must be taken when analysing the lower energy end of the spectrum, especially when the attenuators are in use. As such, while high temperature emission can be modelled accurately with RHESSI, lower temperature emission ($\lesssim$5~MK while in the A1 state) is harder to observe. In comparison, CMEs originating on the limb are better observed than those originating on disk. For limb events, the components of the CME can be easily observed against the sky and avoid contamination by disk emissions. As a result of these obstacles, the hydrodynamic evolution of eruptive flares with well observed CMEs are rare.
 In this Chapter, the hydrodynamic evolution of a CME-associated solar flare is examined in a unique way, using MESSENGER/SAX. The temperature and emission measure of the flare is investigated in conjunction with the kinematic evolution of both the CME and the post-flare loop system. The availability of this unique data set has made this extensive study possible. The instruments used were located throughout the heliosphere and provided excellent coverage of the event, both temporally and spectrally. Along with MESSENGER/SAX the instruments used included RHESSI, GOES-12, EUVI, Cor1 and Cor2. The observations and data analysis techniques are discussed in \S \ref{section:obs}. \S \ref{section:res} describes the main results of this investigation which are discussed in light of current theory in \S \ref{section:concs}. 
 
\section{Observations and data analysis}
\label{section:obs}

This event occurred on the east limb of the Sun on 31-December-2007. The SXR flux began to rise from approximately 00:30~UT. The CME was launched at 00:48~UT and the SXR flux was above background levels for more than 4 hours. As observed from Earth, the footpoints in the low corona were occulted. Therefore, Earth orbiting satellites (RHESSI and GOES) only observed looptop emission from the event. As a result, it is likely that that GOES classification of C8.3 is an underestimation of the total flux. STEREO B and MESSENGER however, had a clear view of the entire system, as Figure \ref{fig:instr_loc} shows.

\subsection{X-ray spectroscopy}

\begin{figure}[!t]
\centering
\includegraphics[width=\textwidth, trim =0 10 100 60, clip = true, angle = 0]{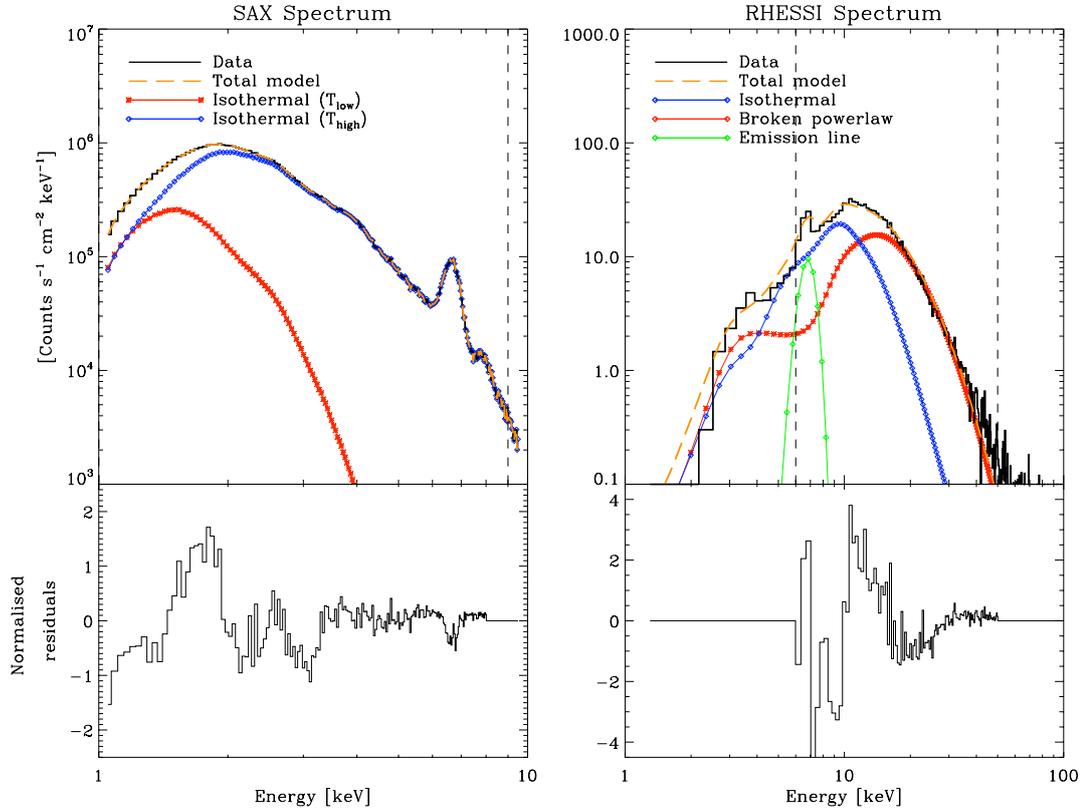} 
\caption{Top left panel shows the MESSENGER/SAX spectrum from the peak of the flare between 00:47~UT and 00:52~UT (solid black line). The isothermal fits, shown as red asterisk (low temperature) and blue diamonds (high temperature) were applied between 1-9~keV. Top right panel shows the RHESSI spectrum taken between 00:47~UT and 00:48~UT (solid black line) and its fit components constrained between 6-30~keV: an isothermal function (blue), a broken power-law (red) and a single emission line (green). The normalised residuals are shown in the bottom panels. }
  \label{fig:msgr} 
\end{figure}

The RHESSI A1 attenuators were in place for the duration of this event. As such, the RHESSI spectrum was only analysed above 6~keV (see Table \ref{table:attenuator} and Figure \ref{fig:msgr}). A sample spectrum taken at the peak of the flare is shown in Figure \ref{fig:msgr} (right) showing the observed flux (black) along with the fit components. The thermal continuum was well modelled using a Maxwell-Boltzmann distribution, as discussed in \S \ref{sect:continuum} (blue line) and the non-thermal emission was modelled using a broken power-law (red line). A single emission line was also included at 8~keV (green line). By fitting the thermal part of the RHESSI spectrum with a Maxwell-Boltzmann distribution, the temperature and emission measure could be calculated during the early decay phase of the event. As the lower energy range of the spectrum ($<$6~keV) was not analysed (due to the attenuator state), any plasma with a temperature of $\lesssim5$~MK present in the loop was not observed with this instrument. The MESSENGER/SAX instrument was used to account for any low temperature emission. This instrument is designed to measure characteristic X-ray emissions from the surface of Mercury. Incident solar flux is measured using SAX for calibration purposes (when analysing Mercury). SAX is sensitive to the 1-10~keV range, overlapping with the thermal range of the RHESSI spectrum. However, significantly better spectral resolution and sensitivity to lower energies allows both high and low temperature emission to be observed and modelled accurately. Figure \ref{fig:msgr} (left) shows a sample SAX spectrum taken at the peak of the flare. The spectra were fit using the \textsc{ospex} software in the SolarSoft \textsc{spex} package. Two isothermal functions were used to approximate a differential emission measure function to accommodate both high and low temperature plasma present in the flare. 

\subsection{Imaging}

RHESSI images were reconstructed using the CLEAN algorithm and detectors 3, 4, 5, and 6 integrated over 2 minute periods. The source was imaged in the 3-6, 6-12, and 12-25~keV energy bands. The 30-50~keV coronal sources imaged by \citet{Krucker2009} have also been incorporated into this study. The evolution of the flare and the acceleration phase of the CME were imaged using 171~\AA\ EUVI passband on board STEREO B, while the propagation phase of the CME was imaged using Cor1 and Cor2.

\begin{figure}
\centering
\includegraphics[width=0.95\textwidth, trim =70 100 140 80, clip = true, angle = 0]{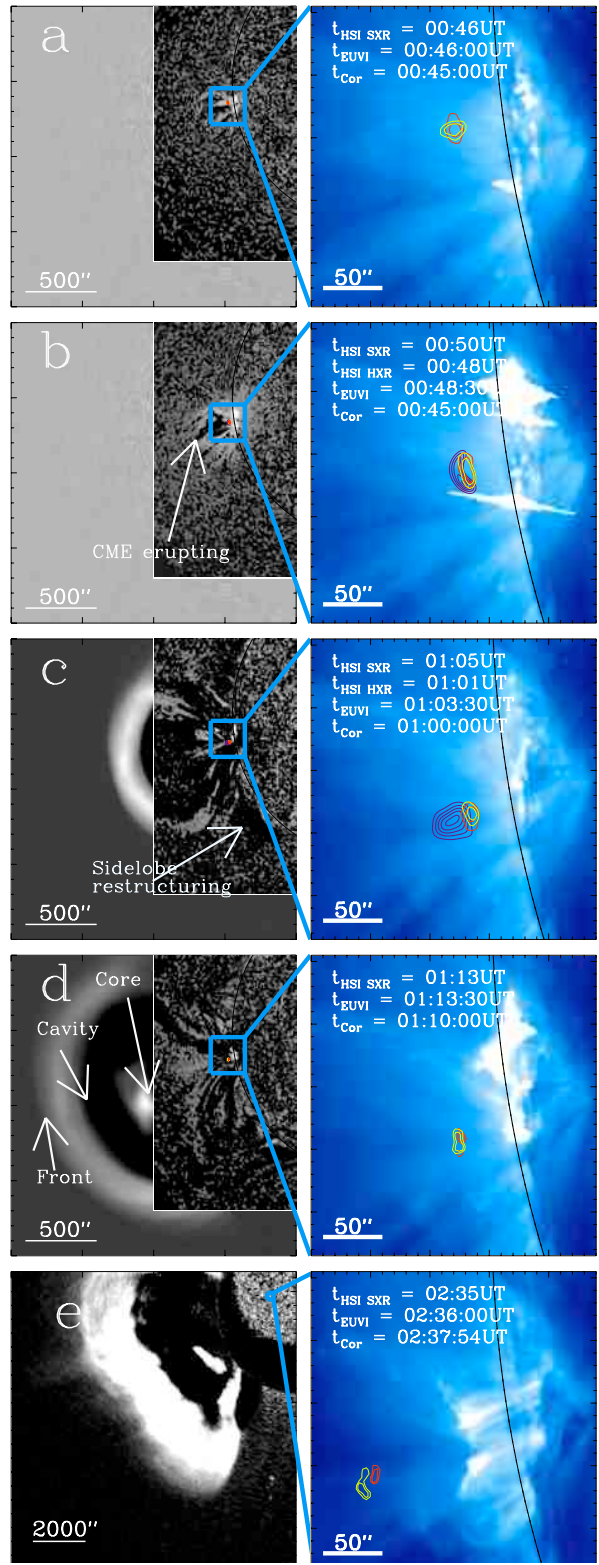} 
\caption{Evolution of the 31-Dec-2007 flare-CME as observed by EUVI, Cor1, Cor2 (panel e only), RHESSI 3-6~keV (yellow), 6-12~keV (red), and 30-50~keV (purple).   }
    \label{fig:multi} 
\end{figure}

Figure \ref{fig:multi} shows the evolution of the event. The left hand panels are interlaced difference images from EUVI and Cor1 (a - d) and Cor2 (e). Overlayed on these are the RHESSI sources, 3-6~keV (red), 6-12~keV (yellow), and 30-50~keV (purple, panels b and c only). The flaring region of the event has been expanded on the right hand side to highlight the evolution of a post-flare loop system and the motion of the soft X-ray sources. 

\section{Results}
\label{section:res}

\begin{figure}[!t]
\centering
\includegraphics[width=\textwidth, trim =10 275 50 100, clip = true, angle = 0]{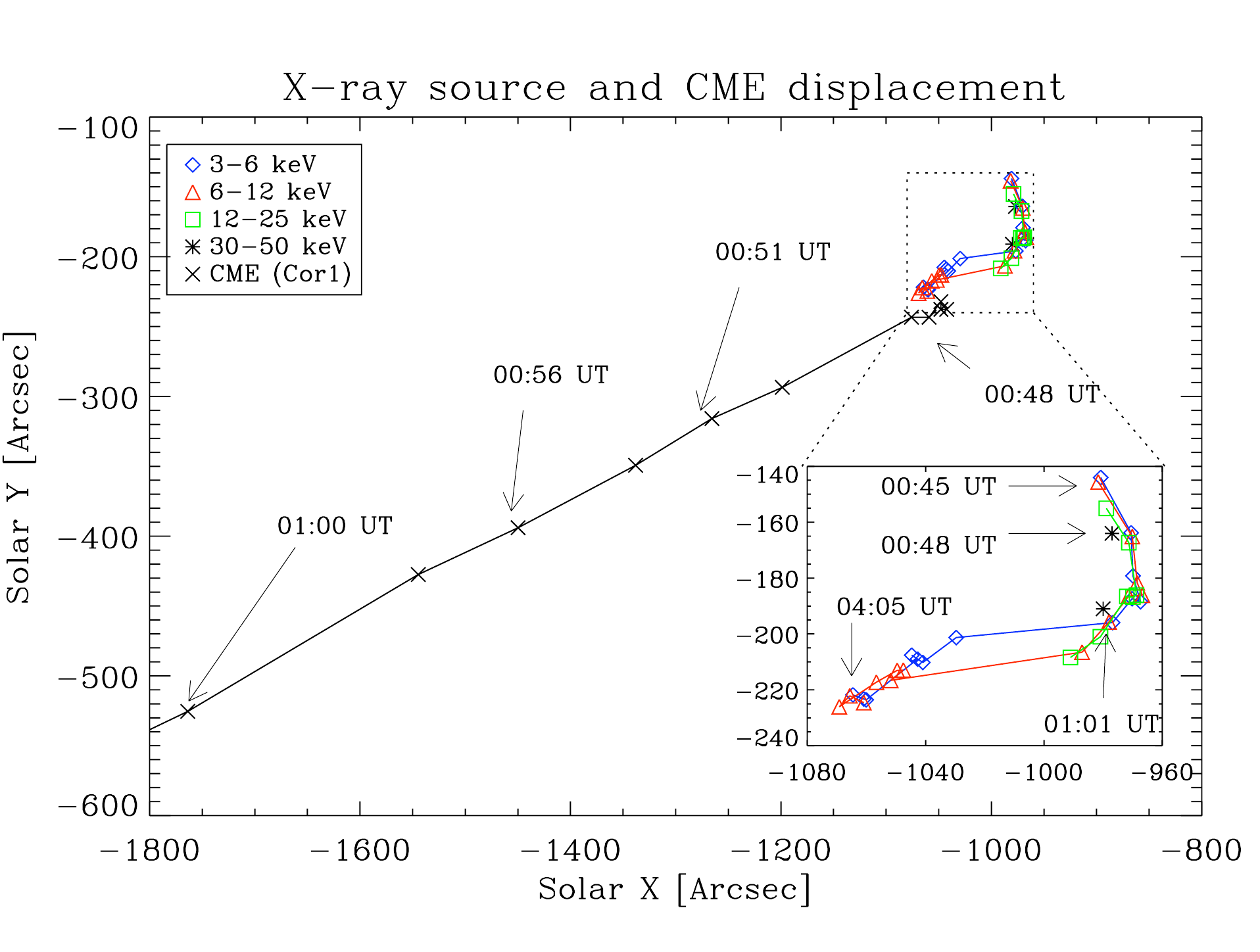} 
\caption{Position of the CME front in the Cor1 field of view (black crosses) along with the centroid of the RHESSI sources: 3-6 (blue diamonds), 6-12 (red triangles), 12-25 (green squares), and 30-50~keV (black asterisk). The inset panel is a magnified view of the flaring region. Time stamps for both the flare and CME are also marked.}
  \label{fig:sxrht}
\end{figure}

This CME-flare system was observed for more than four hours using a multitude of instruments. Figure \ref{fig:multi} shows the evolution of the system and Figure \ref{fig:sxrht} explicitly shows the motion of both the CME and SXR loop top sources (inset). As RHESSI emerged from eclipse at $\sim$00:46~UT, a SXR source was observed to lie between and above two bright EUV ribbons (Figure \ref{fig:multi}a). Between 00:46~UT and 00:48~UT, the SXR sources appeared to move in a southerly direction, decreasing in altitude at the same time (Figure \ref{fig:sxrht} inset). The decreasing altitude of the SXR source may correspond to the collapse of a magnetic X-point in X- to Y-type reconnection \citep{Sui03}. At 00:48~UT (Figure \ref{fig:multi}b), a 30-50~keV coronal source was observed above the now extended thermal sources. Simultaneously, the CME was launched. Figure \ref{fig:multi}b also shows the location of the SXR source to be at the base of the CME, between the flare footpoints. Following the CME eruption, the SXR source began to rise beneath the CME while still moving south along what would become the post-flare arcade. This can be seen in the main frame of Figure \ref{fig:sxrht}: the direction of motion of the post-flare loops follow the same path as the CME. Figure \ref{fig:multi}c shows that the field surrounding the erupting arcade has been disturbed by the launch of the CME, primarily to the south. This disturbance of the neighbouring field is also evident in Figure \ref{fig:multi}d, along with the 3 part structure of the CME - the core, the cavity and the front. The evolving post-flare arcade can also been seen in this figure. The bottom panel of Figure \ref{fig:multi} shows the system approximately 2 hours after eruption. The CME is observed in Cor2, the post-flare loop system has evolved and the X-ray sources are still located above the post-flare EUV loops, at the base of the CME. 
\begin{figure}
\centering
\includegraphics[width=1.07\textwidth, trim =0 57 0 90, clip = true, angle = 0]{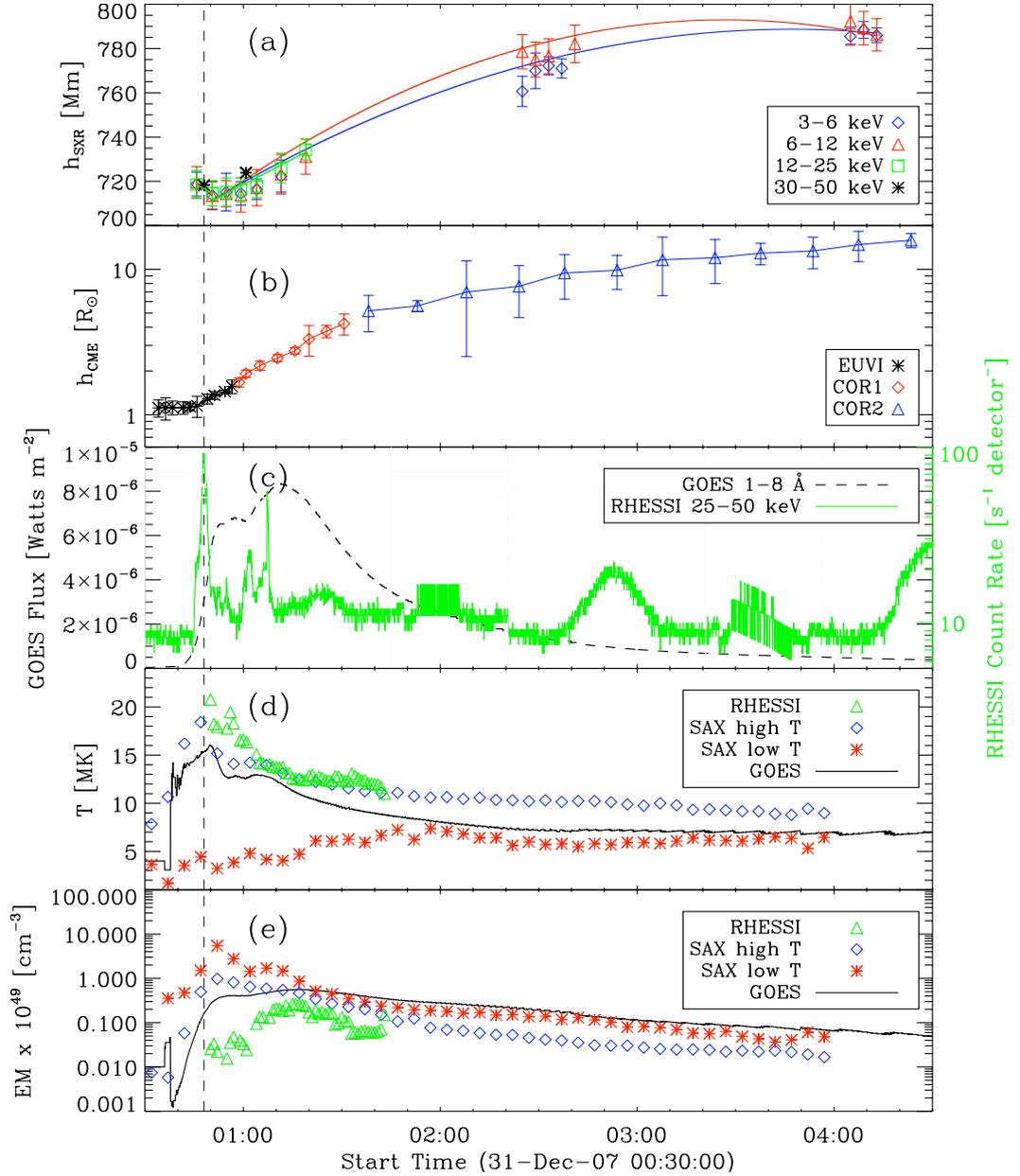} 
\caption{Evolution of the flare and CME kinematics and flare hydrodynamics in time. (a) SXR displacement from Sun centre, together with the fits to the constant acceleration model of the 3-6 (blue diamonds), 6-12 (red triangles), and 12-25~keV (green squares) sources. The displacement of the HXR 30-50~keV source is also shown (black asterisk). (b) CME displacement from Sun centre in the EUVI (black asterisk), Cor1 (red diamonds), and Cor2 (blue diamonds) fields of view. (c) GOES 1-8 \AA \ soft X-ray lightcurve (black dashed line) overplotted on the RHESSI 25-50~keV lightcurve (green solid line). (d) Flare temperature measured by GOES (black solid line), RHESSI (green triangles), and the two components of the SAX fit: high temperature (blue diamonds) and low temperature (red asterisk). (e) Flare emission measure from GOES (black solid line), RHESSI (green triangles), and the two components of the SAX fit: high temperature (blue diamonds) and low temperature (red asterisk). }
  \label{fig:kine} 
\end{figure}

Figure \ref{fig:kine} shows the kinematics of the flare and the CME, along with the hydrodynamic evolution of the flare. The CME liftoff time (00:48~UT; defined as the time the CME is first observed in EUVI, marked in Figure \ref{fig:multi}b) is highlighted in each panel with a dashed vertical line. Figure \ref{fig:kine}a shows the displacement of the RHESSI SXR sources (3-6, 6-12, and 12-25~keV) relative to Sun centre. The 30-50~keV sources are also shown here. The average velocity of the loops as they descend (pre 00:50~UT) is $ 18 \pm 4$~km~s$^{-1}$. The height of the ascending loops (post 00:50~UT) were fit with a constant acceleration model: 
\begin{equation}
h(t) = h_{0} + v_{0}t + \frac{1}{2}at^{2}
\end{equation}
where $h_{0}, v_{0}$, and $a$, averaged across the three energy ranges are $711.1\pm0.5$~Mm, $15\pm2$~km~s$^{-1}$, and $-1.1\pm0.9$~m~s$^{-2}$. These fits are shown in Figure \ref{fig:kine}a.

Figure \ref{fig:kine}b shows the displacement of the CME apex relative to Sun centre measured using STEREO B instruments (EUVI, Cor1 and Cor2). A detailed analysis and comparison to theoretical models of the CME velocity and acceleration can be found in \citet{Lin09}. Panel (c) of Figure \ref{fig:kine} shows the RHESSI 25-50~keV HXR lightcurve (green). The peak in the hard X-rays corresponds exactly to the time the CME is launched. Overplotted on this is the GOES 1-8~\AA \ lightcurve\footnote{The appearance of multiple peaks in the GOES lightcurve is as a result of occultation and not multiple events. This was verified by comparing EIT intensity (partially occulted) to that of EUVI on STEREO B (not occulted).}. Figure \ref{fig:kine}d shows the thermal evolution of the flare using results from RHESSI, SAX, and GOES. The GOES temperature (and later, emission measure) were obtained from the ratio the two flux channels (see \S \ref{sect:instr_goes}). It is clear that the lower temperature component of the SAX spectrum remains approximately constant during the initial and early decay phases of the flare while the higher temperature component rises quickly to a maximum value of 19$\pm$2~MK. The high and low temperature components converge as the flare decays. Figure \ref{fig:kine}e shows the evolution of emission measure of the flare, again from RHESSI, SAX, and GOES. As expected, the level of high temperature emission increases as the flare approaches its peak. Surprisingly however, the level of cooler emission also increases during the flare impulsive phase.

\section{Conclusions and discussion}
\label{section:concs}

This chapter presents observations of a CME and associated solar flare that were observed using a range of spacecraft positioned at various points in the heliosphere. The eruption took place on the east limb of the Sun and the footpoints were occulted as viewed from Earth. Before the eruption, a SXR source was observed in conjunction with increasing flare temperature. At 00:48~UT, the time the CME was launched, a 30-50~keV HXR looptop source was observed. Following the liftoff of the CME, the SXR source was found to rise beneath it as the temperature and emission measure both began to fall. 

\begin{figure}[!t]
\centering
\includegraphics[width=1\textwidth, trim =0 0 150 30, clip = true, angle = 0]{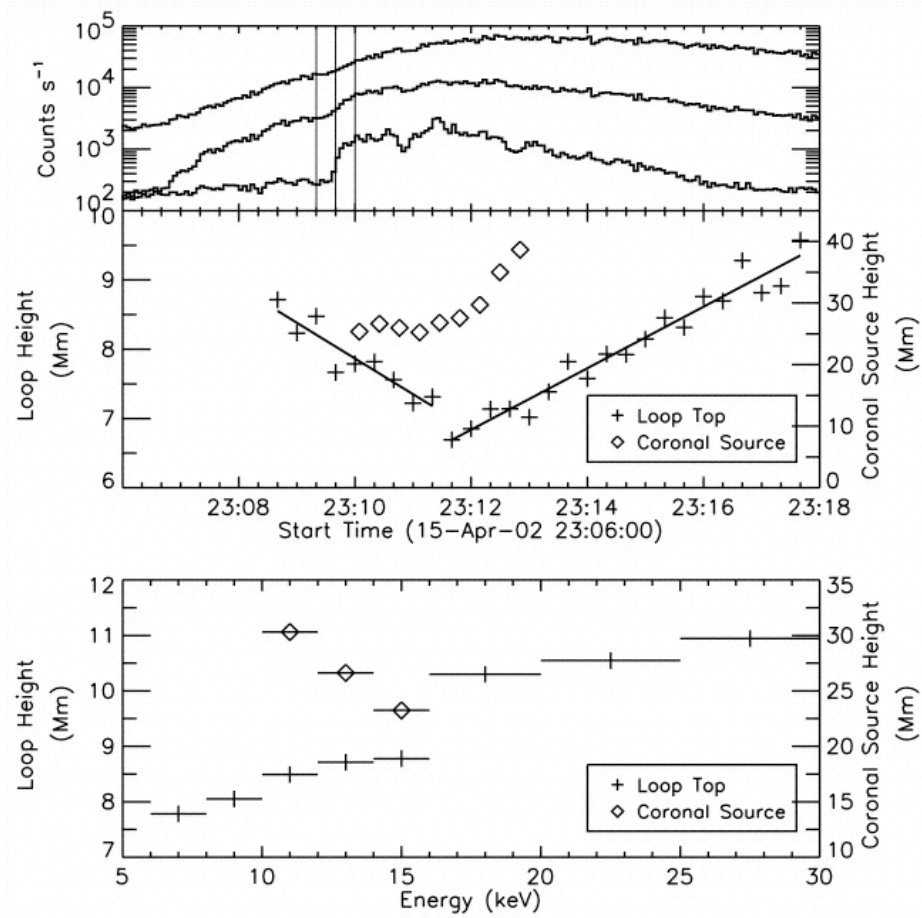} 
\caption{Observations published in \citet{Sui03} that display similar behaviour to the observations presented in this chapter. Top panel: RHESSI light curves in three energy bands (from top to bottom): 3-12, 12-25, and 25-50 keV, scaled by 2.0, 0.5 and 1.0 respectively. Middle panel: Time histories of the loop height (obtained from 10-12 keV images) and the coronal source height (obtained from 10-25 keV images). Bottom panel: Height of the loop and the coronal source at different energies at 23:11:00 UT. The horizontal bars represent the energy bandwidths of the RHESSI images.}
  \label{fig:sui1} 
\end{figure}

\begin{figure}[!t]
\centering
\includegraphics[width=0.5\textwidth, trim =40 40 400 300, clip = true, angle = 0]{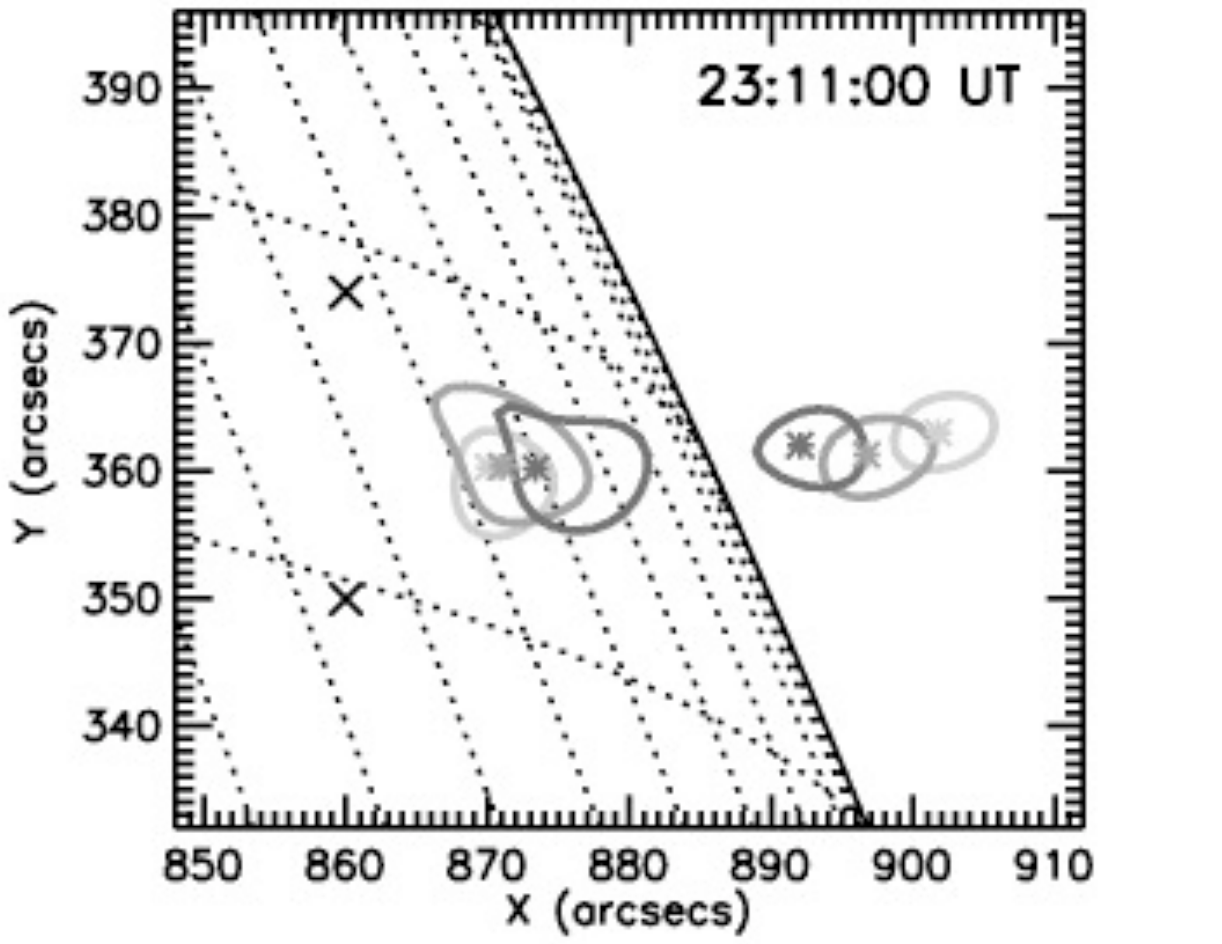} 
\caption{RHESSI images in different energy bands from \citet{Sui03}. The three contours on the solar disk indicate the top of loops in the energy bands (contour line shade from light to dark): 6-8, 10-12, and 16-20 keV. The contours above the limb are (from light to dark) 10-12, 12-14, and 14-16 keV bands. The asterisk marks the centroid of each source. The crosses mark the two footpoints of the X-ray loop.}
  \label{fig:sui_source} 
\end{figure}

The pre-eruption increase in temperature and emission measure, along with the presence of the SXR source early in the event is strong evidence for pre-eruption reconnection. Following the interpretation of \citet{Sui03}, we also believe that this event exhibits signs of X- to Y-type reconnection. \citet{Sui03} observed a looptop source that appeared to shrink during the impulsive phase of a flare (Figure \ref{fig:sui1}). Following the flare peak, the loops were observed to rise. In addition, they too observed a HXR coronal source that was located above the looptop sources (Figure \ref{fig:sui_source}). They hypothesise that a current sheet is located between the looptop sources (on disk in Figure \ref{fig:sui_source}) and the coronal source (off limb).  In this chapter, the coronal source was observed for only two instances in time so a direct comparison is difficult. However, the apparent downward motion of our looptop (SXR) source is in very good agreement with the motion of that of \citet{Sui03}. They explain the stratification in temperature (energy sources) within the flare loops as the heating of material close to a current sheet formed between the looptop and the coronal source. According to \citet{Krucker2009}, this coronal source is the location of particle acceleration in the Dec-31 event. This, therefore is a good explanation for why the 12-25~keV source is located above the 3-6~keV source (see Figure \ref{fig:sxrht} inset for clarification). The temperature distribution can be explained in terms of the energy supply by reconnection at the coronal source heating the plasma closest to the current sheet. This produces the higher energy sources. Subsequent cooling of the inner loops result in the lower energy sources observed at lower altitudes. In addition, the loop shrinkage is believed to be due to a change in the magnetic field configuration as an X-point collapses into a current sheet. As this happens, the looptop source moves downwards, as is observed in this chapter. Since the observations presented here are clearly in good agreement with \citet{Sui03}, we can conclude that we are observing the changing in magnetic topology from an X-point (X-type) to current sheet (Y-type) reconnection region. 

\begin{figure}[!t]
\centering
\includegraphics[width=0.5\textwidth, trim =40 40 40 100, clip = true, angle = 0]{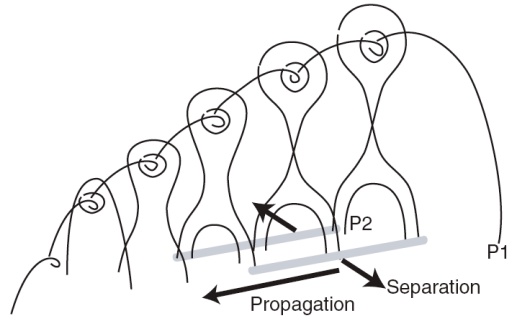} 
\caption{Schematic diagram showing the asymmetric eruption of a CME that fits well with the observations presented in this chapter. \citep{Tripathi06}}
  \label{fig:arcade} 
\end{figure}

The extension of the SXR source across the arcade top may be interpreted as successive reconnection events progressing along the arcade. These reconnections are likely be the source of the heated flare plasma, either by chromospheric evaporation \citep{Brown71, Milligan_explosive, Raftery09} or by a collapsing magnetic trap \citep{Somov97, Karlicky04, Veronig06, Joshi08}. Pre-eruption magnetic reconnection would also explain the apparent motion and elongation of the SXR source: as the event progressed and neighbouring loops within the arcade reconnect, emission from more and more loops contribute to the SXR source, resulting in the apparent elongation of the source along the arcade. This is similar to the asymmetric eruption described in \citet{Tripathi06} and shown in Figure \ref{fig:arcade}. 

The timing in this event is crucial for the interpretation of the initiation mechanism. The three indicators are the initial rising of the CME, the evidence for magnetic reconnection in the central arcade (e.g. rising temperature, X-ray sources) and evidence for reconnection in neighbouring arcades. This event began at$\sim$00:46~UT with the appearance of a SXR source at a height of 22$\pm$1~Mm above the photosphere. This corresponds very closely to the height of the pre-eruption arcade loops. At this time, the temperature of the region has risen from its equilibrium value of 5.3$\pm$2~MK to 17$\pm$2~MK (see Figure \ref{fig:kine}d). The rise in temperature and the SXR source are clear evidence that magnetic reconnection has taken place before the CME is launched. This rules out the ideal MHD trigger as a mechanism for CME initiation. At $\sim$00:48~UT, approximately 2 minutes later, we see the first evidence of a rising CME in EUV difference images (see Figure \ref{fig:multi}b) with the SXR source at the base of the CME. It is not until $\sim$00:56~UT, that any restructuring of side lobes is observed. This would indicate that the breakout reconnection occurred \emph{after} the internal reconnection between the legs of the arcade. It is possible that the limitations of the EUVI's 171~\AA\ passband did not observe the neighbouring loops until after they were heated and were cooling back into the 171~\AA\ passband. However, observations of the 284~\AA\ passband, whose peak sensitivity is at 2~MK did not observe any change until this time either. As well as this, RHESSI observed an increase in HXR fluctuations from 00:57~UT. This can be interpreted as evidence that breakout reconnection occurs \emph{after} the CME is initiated. Therefore, we conclude that this event is most likely triggered by an internal tether-cutting mechanism. This is in good agreement with \citet{Chifor07} who also diagnosed tether-cutting CME initiation from the presence of HXR sources (or ``precursors'') during the early phase of an eruption and observed a similar asymmetric evolution to their eruption. 

The unique multi-spacecraft observations used for this study, specifically the high spectral resolution of SAX, enabled a detailed analysis of the hydrodynamic evolution of an eruptive flare. High resolution data from the STEREO and RHESSI instruments allowed the kinematic evolution of both the flare and the CME to be analysed simultaneously. Our results clearly demonstrate there is a physical connection between the two phenomena. We can also disregard a number of theoretical models that are not applicable to this event and conclude that this CME was initiated by an internal tether-cutting mechanism, with breakout reconnection occurring afterwards.


\chapter{Conclusions and Future work} 
\label{chapter:concs}


\ifpdf
    \graphicspath{{8/figures/PNG/}{8/figures/PDF/}{8/figures/}}
\else
    \graphicspath{{8/figures/EPS/}{8/figures/}}
\fi


\hrule height 1mm
\vspace{0.5mm}
\hrule height 0.4mm 
\noindent 
\\ {\it The goal of my thesis work was to better understand the behaviour of the active Sun. This incorporated a study of a confined solar flare, a study of an eruptive flare/CME system and an investigation into the sensitivity of EUV imaging telescopes. In this chapter, the primary findings of these investigations and their implications are presented. This is followed by a discussion of the problems encountered in this thesis and how they they can be rectified in the future.  \\ }

\hrule height 1mm
\vspace{0.5mm}
\hrule height 0.4mm 

\newpage
\section{Thesis results and comments}

The goal of my thesis was to improve our understanding of the active Sun, specifically the behaviour of solar flares. It was found that the use of multi-wavelength analysis is essential when studying the hydrodynamic properties of solar flares. The broad temperature range that these events cover require at least observations from HXR through SXR and into the EUV regime. Ideally this should be extended even further to include UV, white light and radio emissions \citep[e.g.][]{Fletcher07, Benz_radio09, Benz_radio97, Coyner_uv, Alexander06}. Attempting to understand the full effects of a solar flare using a single passband is not advisable. The temperature of a flare can change very quickly, especially during the impulsive phase and early decay phase \citep{Raftery09, Raftery10_cme}. Therefore a single passband, sensitive to a narrow range of temperatures will only observe the flare for a short period of time. A clear example of this is shown in Chapter \ref{chapter:mar26}: the 171 \AA\ images in Figure \ref{fig:F-4panels} show that the TRACE instrument did not register the occurrence of a solar flare until approximately 30 minutes after it began. Even the use of multiple broadband telescopes for such studies must be done with care. For example, \citet{Aschwanden2001} utilised integrated flux measurements from four imagers -- TRACE, Yohkoh/HXT, SXT and GOES -- to analyse the cooling curve of the Bastille day flare. The cooling curve was obtained by assigning the characteristic temperature of a particular instrument to the time of maximum flux, much like the approach taken by \citet{Raftery09}. Unlike \citet{Raftery09} however, integrating the flux of a source observed with a broadband imager can lead to problems in obtaining the temperature of the emission, especially during the impulsive phase of a flare. As Chapter \ref{chapter:SWAP} \citep{Raftery10_swap} clearly demonstrates, the temperature responses of broadband instruments vary significantly under atypical conditions, such as a flare. Therefore, the nominal temperature assigned to these instruments is not likely to be reliable during the impulsive phase and early decay phase of a flare. 

The question of plasma temperature can be alleviated through the use of spectroscopy. Emission line spectroscopy, specifically the use of rasters, is extremely useful for investigations of solar flares \citep[e.g. Chapter \ref{chapter:mar26},][]{Raftery09, Milligan_gentle, Milligan_explosive, Brosius03, Cirtain07, Schmelz07}. The ability to image a solar flare simultaneously in multiple temperatures can provide a detailed insight into the evolution of the event along with the distribution of plasma temperature and densities in the system. It can also be used for the analysis of plasma velocity through the use of Doppler shifts. In Chapter \ref{chapter:mar26} of this thesis, we utilised the presence of upflowing \ion{Fe}{xix} plasma at $95\pm16$~km~s$^{-1}$ to diagnose pre-flare heating. A combination of hot (\ion{Fe}{xvi} and \ion{Fe}{xix}) upflows at up to $81\pm16$~km~s$^{-1}$ and cool (\ion{He}{i} and \ion{O}{v}) downflows at approximately 30~km~s$^{-1}$ were characteristic of explosive chromospheric evaporation. Other studies have utilised the versatility of the multi-temperature nature of spectroscopic observations. For example, \citet{Milligan_vels} used the broad range of emission lines observed by the EUV Imaging Spectrometer (EIS) on board Hinode to find the cut off temperature at which plasma is evaporated \emph{upwards} into the loop rather than downwards towards the chromosphere. This was achieved by studying the Doppler shifts of 15 Fe lines from 0.05 - 16~MK. A clear cut-off was observed \ion{Fe}{xiv} (2~MK) whereby plasma at temperatures above that ion were evaporated upwards and plasma at temperatures below flowed downwards.

This thesis also showed that soft X-ray spectroscopy is also a very powerful tool when analysing the hydrodynamic properties of plasma \citep[Chapter \ref{chapter:CME},][]{Raftery10_cme}. The sensitivity of thermal continuum to both temperature and density makes it ideal for analysing their variation in time. A comparison of the RHESSI SXR continuum to that of SAX revealed that while RHESSI is not as reliable when analysing plasma at temperatures of less than $\sim$5~MK (in the A1 state), it is ideal for observing emission at high temperatures that cannot be observed with e.g. SAX. In Chapter \ref{chapter:CME} \citep{Raftery10_cme}, the combination of these two instruments enabled the author to accurately analyse a solar flare across its entire evolution, from pre-flare emission at $<$5~MK using SAX to 19~MK using RHESSI. The limitations of the SAX spectrum, sensitive to emission up to $\sim9$~keV constrained the upper fitted temperature to $\sim$17~MK. Therefore, careful analysis of both instruments provides the most complete and accurate results. 

\subsection{Principal results}
The primary aim of this thesis was to better understand the behaviour of solar flares. This was done by examining the differences and similarities between an eruptive and a confined flare through quantitative and qualitative comparison of multi-wavelength observations to theoretical models. From this comparison, we draw the following conclusions regarding the behaviour of flares. 
\begin{itemize}
	\item In the case of the confined flare (CF) the HXR profile consisted of a single burst during the impulsive phase that lasted approximately 3 minutes. The eruptive flare (EF) however was observed to have a single HXR burst during the impulsive phase, lasting $4\pm2$~minutes, followed by HXR bursts of lower magnitude but similar duration during the decay phase. This difference is most likely due to the fact that the CF does not experience continuous reconnection to any great extent during the decay phase. However, as a CME erupts above an EF, it is believed that breakout reconnection takes place to free the CME \citep{Zhang2001, Antiochos1998, Raftery10_cme, Moore06}. This continued reconnection may be the cause of the continued HXR bursts. 
	
\item	The continued HXR flux that occurs in conjunction with the hypothesised breakout reconnection, mentioned above, corresponds in time to the restructuring of the magnetic field surrounding the erupting arcade. This manifests itself as an ``EIT wave'' \citep{Thompson98}. The nature of these disturbances has been debated for many years, with the interpretation varying between various types of waves \citep[e.g.][]{Long08, Patsourakos09} to the chromospheric component of breakout reconnection \citep[e.g.][]{Attrill07}. The association between the HXR flux and this ``wave'' lends support to the CME theories: that the wave is in fact observations of the coronal magnetic field being restructured to allow the release of the CME.

\item The time to heat the flare to its maximum temperature took a comparable amount of time for both events. The CF took approximately 11 minutes to reach its maximum temperature of 15$\pm$2~MK while the EF took approximately 20 minutes to reach a temperature of 19$\pm$2~MK. 

\item The time it took for the flares to cool varied considerably. In the case of the CF, the SXR flux took approximately 1 hour to reach background levels following the peak. The EF flare however, took more than 7 hours to return to equilibrium. This could also be attributed to the extended period of reconnection hypothesised above, leading to the continued heating of the plasma within the flare for a longer time, thus keeping temperatures higher for longer than in the confined case. 

\item The topological configurations of each system were considerably different. The CF existed as an individual loop with discrete footpoints. The flare was an isolated event that did not affect the global magnetic field to any great extent. The EF however occurred in extended arcade whose footpoints were extended ribbons. It was a global event that occurred in conjunction with a CME eruption and an EIT wave that reconfigured the magnetic field on a scale much larger than the flare. 
\end{itemize}
The comparisons and differences between the eruptive and confined events has led the author to the conclusion that while eruptive and confined flares have very different properties and manifest themselves in different ways, the fundamental processes involved are the same in both cases. 

The analysis of the individual events had some interesting conclusions that are important in their own right. The main results and conclusions from each of the events studied in Chapters \ref{chapter:mar26} and \ref{chapter:CME} are discussed in detail in \S \ref{sect:hydro_conc} and \ref{sect:cme_conc}. The use of EUV imagers was key to both the solar flare investigations and so a study into the sensitivity of EUV imagers in the 171~\AA\ passband was carried out to ensure accurate interpretation of EUV images in Chapter \ref{chapter:SWAP}. The results and conclusions of this study are presented in \S \ref{sect:swap_conc}. 

\subsection{Hydrodynamic modelling}
\label{sect:hydro_conc}
\begin{enumerate}

\item The use of a 0-D hydrodynamic model, EBTEL, proved to be extremely useful for the interpretation of observations. Despite averaging temperature, density and pressure through the corona, it reproduced observations accurately and in a very efficient manner. Combining modelling results with observations revealed flare properties that would otherwise be unattainable such as the significance of various heating and cooling mechanisms. 

\item The approximate magnitude and duration of the HXR burst were determined from this comparison and found to be $5\times10^{(8\pm1)}$~ergs~cm$^{-2}$~s$^{-1}$ and $100\pm50$~seconds respectively. These are not exact values as the over-simplified consideration of the HXR heat flux by EBTEL will affect the total flux. In order to accurately determine the total non-thermal flux, a more thorough treatment of the effects of beam heating must be considered. The non-thermal flux determined by EBTEL is most likely underestimating of the actual non-thermal flux since it is approximately an order of magnitude less than the flux required by \citet{Fisher84} for explosive chromospheric evaporation. This is an important result as it will effect how EBTEL should be treated in future investigations.	
	
\item  Despite the issue of the treatment of the non-thermal heat flux, EBTEL enabled the author to get a reasonable approximation for the non-thermal heat flux and for the direct heating rate. Although the values obtained from the model were not absolute, it was however, possible to establish that solar flares are not heated purely by beam heating as hypothesised by e.g. \citet{Brown71}. They also require some kind of direct heating mechanism such as shock heating or particle trapping at the top of the loop \citep[e.g.][]{Joshi08, Karlicky04, Veronig06_wave}. 

\item An investigation into the significance of flare cooling mechanisms proved the importance of simultaneous cooling by both conduction and radiation throughout a solar flare. Conduction can be considered independently at the beginning of a flare decay phase when temperatures are high and conduction is efficient. Likewise, radiation can be considered independently late in the decay phase when temperatures are low and conduction is not efficient. However, for the majority of the flare cooling phase, both conduction and radiation are important mechanisms for the removal of heat from the system. Therefore, following the techniques of e.g. \citet{Cargill94, Cargill95, Aschwanden2001, Culhane1994}, using a single loss mechanism at any one time will significantly reduce the efficiency of heat removal from the system, resulting in longer cooling timescales and higher predicted temperatures, as demonstrated in Chapter \ref{chapter:mar26}.

\item Studying the CDS spectral data and the associated Doppler shifts revealed evidence for pre-flare heating and chromospheric evaporation \citep{Raftery09}. Analysis of the \ion{Fe}{xix} images revealed the loop emitting at 8~MK up to 10 minutes before the peak of the hard X-rays. Analysis of the Doppler shifts of that line resulted in the fastest velocities for the entire event, at $95\pm16$~km~s$^{-1}$. Following the precursor phase, explosive chromospheric evaporation was determined from a non-thermal flux of $5\times10^{(8\pm1)}$~ergs~cm$^{-2}$~s$^{-1}$ combined with 8~MK plasma rising into the loop at $81\pm16$~km~s$^{-1}$ and 0.25~MK plasma being forced towards the chromosphere at $33\pm16$~km~s$^{-1}$. These observations (with the exception of the low non-thermal flux, see above) are characteristic of explosive chromospheric evaporation \citep{Fisher84, Milligan_explosive}. The explosive evaporation is believed to have been followed by a period of conduction driven gentle chromospheric evaporation \citep{Fisher84, Milligan_gentle}. This was evidenced by plasma rising into the loop at much lower velocities: $24\pm16$~km~s$^{-1}$ for \ion{Fe}{xix} and $9\pm16$~km~s$^{-1}$ for \ion{He}{i}. This is believed to be driven by conduction as the flux of non-thermal electrons had almost returned to background levels by this time and, according to EBTEL simulations, conduction is highly efficient. Thus, the most likely explanation is conduction fronts from the hot ($\sim$20~MK) loop top propagate towards the chromosphere where they thermalise the plasma at a much slower rate than a non-thermal flux. The small pressure increase due to heating results in a minimal deviation from equilibrium. Thus plasma is forced upwards into the loop. The pressure is not sufficient to overcome the large densities through the chromosphere and so no downflows are observed. 

\item Despite the lack of a rigorous statistical test when comparing the results of the hydrodynamic model to observations, the input parameters used by the author in the EBTEL modelling of a confined flare and published in \citet{Raftery09} agree very well with those found by \citet{Adamakis09} using Bayesian statistics to optimise the model parameters. 
\end{enumerate}

 \subsection{CME initiation mechanism}
\label{sect:cme_conc}
\begin{enumerate}
\item The initiation mechanism for the launch of a CME was investigated by qualitative comparison of multi-wavelength observations and theoretical models. SXR loops are believed to be a signature of magnetic reconnection \citep[\S \ref{sect:flares};][]{Dennis89}. Therefore, the presence of an SXR source at the top of the arcade before the CME erupts suggests magnetic reconnection took place \emph{within} the arcade system (i.e. not breakout reconnection). 

\item The presence of an X-ray source observed before the CME began to rise
excludes triggering by an ideal MHD instability. According to the catastrophe model, a CME flux-rope forced out of equilibrium should rise to find a new equilibrium. This is believed to take place initially through an ideal MHD process i.e. without magnetic reconnection \citep[e.g.][]{Isen93, Forbes_Isen1991, Forbes_Priest1995}. Therefore, the possible evidence for magnetic reconnection observed before the loops began to rise (mentioned above) suggests that the catastrophe model is not applicable to this event. 

\item Approximately 8 minutes after the main CME lift off, restructuring of the surrounding magnetic field was found to correspond to secondary HXR fluctuations that occurred after the main impulsive burst at the time of launch. This may be representative of breakout reconnection occurring \emph{after} lift-off. Breakout is expected to occur high above the photosphere and so the location is not discernible from current instruments. However, considering the chromospheric field motions along with secondary HXR peaks imply some large-scale field reorganisation which fits with the expectations of \citet{Antiochos1998}. However, unlike the predictions of the breakout model, the CME is not triggered by breakout reconnections, rather they occur in order to allow for the propagation of the CME.  

\item Considering the observational evidence and a qualitative comparison to the models presented in \citet{Moore06}, we can conclude that the CME in question was triggered by internal tether-cutting reconnection, followed by breakout reconnection after the CME had been launched \citep{Raftery10_cme}. This is an interesting result as evidence for internal tether cutting is reasonably scarce \citep[e.g.][]{Chifor07, Sterling01}, compared to observations and simulation of the other two models \citep[see e.g.][]{Lynch04, Roussev03, Chen2000}. 

\item The behaviour of the eruptive flare was found to be in good agreement with the CSHKP model \citep{Carmichael64, Sturrock66, Hirayama74, Kopp76}. The positioning of the HXR source above the pre-existing SXR source at the time of eruption agrees with the general topology of the CSHKP model, as does the motion of the SXR source between the footpoints of the CME. In addition, the connection between the SXR source and the CME topology and the relationship between the launch of the CME and the hydrodynamic behaviour of the flare reaffirms the postulation of \citet{Zhang2001}: flares and CMEs are separate manifestations of the same driving mechanism, and therefore are not independent of each other. 
\end{enumerate}

\subsection{EUV temperature response functions}
\label{sect:swap_conc}
\begin{enumerate}

\item A new method for investigating the temperature response of EUV imagers has been developed and presented in this thesis. This method accounts for the varying emission measure with temperature for four types of solar plasma: coronal holes, quiet sun, active region and flares. The method not only enables the interpretation of different plasmas, it can be applied to any EUV imager whose wavelength response is in units of DN~phot$^{-1}$~cm$^{2}$~str~pix$^{-1}$. This will facilitate the comparison of results from different instruments. So far it has been applied to six instruments in the 171~\AA\ passband, however it can easily be adapted to accommodate any of the typical EUV passbands commonly used in solar physics.  

\item The interpretation of images taken with EUV imagers requires special attention. The typical response curves for the 171~\AA\ passband revealed stark differences between the response of EUV imagers to quiet sun/active regions and the extreme conditions of flares and coronal holes \citep{Raftery10_swap}. The maximum sensitivity of a 171~\AA\ passband imager when observing coronal holes is 7.5$\pm0.3\times10^5$~K. This is cooler than the typical peak temperature of $10^6$~K. During a flare, the instrument will observe not only the ``characteristic'' temperature peak at 1~MK but a second peak above 6~MK. 

\item The temperature response of 6 EUV imagers in the 171~\AA\ passband were studied and found to be in very good agreement. There was an offset between the instruments' responses, primarily as a result of their varying aperture size. However the shape of the AIA wavelength response was dramatically different to all the others and this had a large impact on the overall sensitivity of the instrument. It was found to be almost 300 times more sensitive than the next most sensitive instrument, EUVI on STEREO~A. This suggests that developing instruments to study the spectrum around 171~\AA\ wavelength will produce similar results despite different optical systems or detector technologies. 
\end{enumerate}

\section{Future work}
This thesis incorporates many basic assumptions adopted in the solar physics community. For example, coronal abundances are assumed throughout this thesis. However, it has been shown that the ion abundance in the corona changes during non-relativistic electron events such as solar flares \citep{Reames85, Ramaty80}. The assumption that abundances remain constant throughout an energetic event is into question. While it is likely that this assumption is valid through the decay phase of flares, changing abundances during the impulsive phase will affect the hydrodynamic properties derived from observations and used to diagnose energetic phenomena. In order to fully understand the properties of solar plasmas, this effect should be taken into account.

A second approximation is that of ionization equilibrium. For example, the hydrodynamic models used in Chapter \ref{chapter:mar26} assumes that the heating rates involved in the solar flare are longer than those of the atomic processes. Again, since Equation \ref{eqn:final_flux} is based upon the assumption of ionization equilibrium and is used to calculate the emission measure and temperature of emission lines during solar flares, this calls into question the validity of our results. A very large quantity of energy \citep[$\sim$10$^{25}$~J;][]{Emslie05} is deposited in the chromosphere over a very short period of time ($\sim$minutes) and so it is quite possible that the heating timescales are at least comparable to the ionization timescales \citep[$\sim$100 seconds;][]{Mariska_book}. In addition to this, when modelling the thermal continuum of RHESSI and MESSENGER, the assumption is made that the thermal electrons adhere to a Boltzmann distribution of velocities. This assumption must be made for RHESSI spectroscopy as there is no other way to differentiate the thermal from non-thermal components of the spectrum. Should there be a second population of electrons in, for example a bump-on-tail instability, which is known to occur during the production of type-III radio bursts frequently associated with solar flares \citep{Benz03}, this second population would not be modelled accurately using a Maxwell-Boltzmann distribution. This would alter both the incoming flux of electrons and the temperature of the system. 

This thesis tested the robustness and applicability of the EBTEL hydrodynamic model. The model proved to be highly efficient in producing results and is an ideal candidate for use with statistical methods. It will be very useful for probing the broad parameter space such as the heating functions of solar flares. By narrowing the parameter space, this 0-D model can be used as a ``first estimate'' for more robust 1-D hydrodynamic models, such as those by \citet{Bradshaw05}, \citet{Warren07} and \citet{Antiochos91}. In addition, the EBTEL model will be very useful in the investigation of the multi-thread model of solar flares. This theory involves the magnetic reconnection of individual field lines within a larger loop (like strands within a rope). The number of field lines reconnected at a given time will combine to give you the time dependent heating of the loop \citep{Klimchuk01, Klimchuk_soho06, Patsourakos06}. The short computation time of the EBTEL model will enable investigators to probe the frequency and magnitude of individual reconnection events, thus furthering the understanding of solar flares. 

The instruments available for analysis of the Sun have undergone significant improvement since the days of Skylab. The number of available spacecraft is at an all time high and the quality of the instruments by far surpasses anything that has gone before. For example, the innovative approach to stereoscopic imaging of energetic events used by STEREO has revolutionised our understanding of coronal activity, from the structure of coronal loops \citep{Aschwanden08a, Aschwanden08b, Aschwanden09c} to understanding the propagation of CMEs \citep{Byrne09, Maloney09}. 

The launch of Proba-2 has brought the commissioning of a new EUV imaging telescope, SWAP. The extra-wide field of view of this instrument will hopefully facilitate a more accurate analysis of the acceleration phase of CMEs. CMEs are believed to be accelerated close to the surface of the Sun, in the vicinity of the edge of EUV images \citep[Chapter \ref{chapter:CME};][]{Raftery10_cme, Lin09, Gallagher02, Gallagher2003}. SWAP however, will allow users to study evolving CMEs out to 3~R$_{\odot}$ which will cover the entire acceleration and early propagation phase. In addition to this, the upcoming launch of SDO will inundate the solar community with more data than ever before - terabytes per day! The extremely high cadence and resolution of its instruments will give investigators a completely new insight into the dynamic nature of the corona. 

The CDS instrument has recently been accompanied by the Extreme ultraviolet Imaging Spectrometer \citep[EIS;][]{EIS} on board the Hinode satellite. This instrument has higher cadence, spatial and spectral resolution than CDS which has already advanced the understanding of coronal physics significantly in its three years since launch \citep[e.g.][]{Milligan_vels, Mcintosh09, Jin09}. 

Despite these significant developments both in theory and observation, there remain many unanswered questions in the field of solar physics. With the proposed commissioning of new and innovative instruments such as Solar Probe, some of these questions may be answered. However, there remains significant work to be done. 

\section{Outstanding questions}
\begin{figure}[!t]
\centering
\includegraphics[width=0.7\textwidth, trim =30 50 200 0, clip = true, angle = 0]{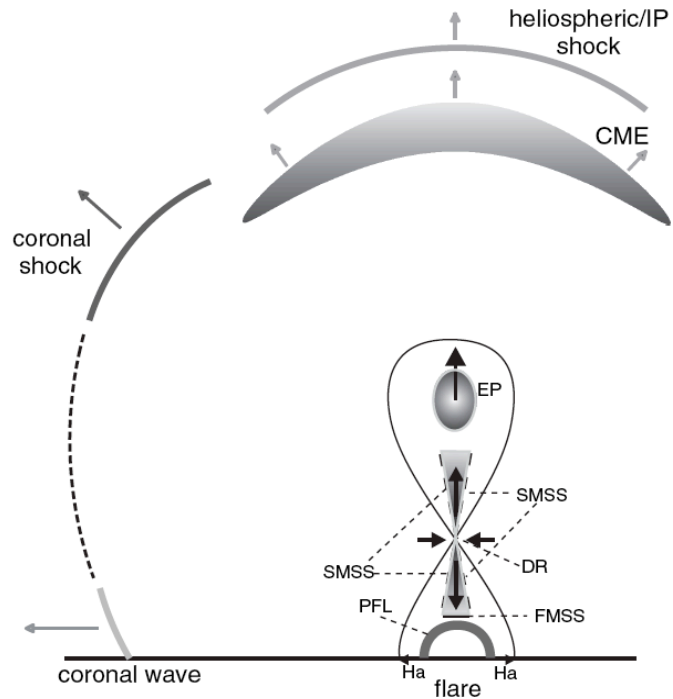} 
\caption{Schematic diagram from \citet{Warmuth07} describing the various coronal disturbances that are known to occur during a solar eruption \citep[adapted from][]{Aurass02}. }
\label{fig:overall}
\end{figure}

In the past three decades, there have been many developments in the area of solar physics, specifically in the field of coronal disturbances. Figure \ref{fig:overall} describes this idea well. Magnetic reconnection is believed to occur within the diffusion region (DR) below an erupting prominence (EP). Slow mode standing shocks (SMSS) occur at the boundary of outflow jets which may generate a fast mode standing shock (FMSS) if the downflow jet is supersonic. As the flux rope evolves into a CME, a heliospheric interplanetary (IP) shock can be driven infront of it, producing metric type-II bursts. Finally, large scale restructuring of the magnetic field can be observed to propagate radially from the eruption site. Despite this apparently concise picture, it is still unclear how these various phenomena are connected and often they are studied as independent events. Here I discuss some outstanding questions regarding the overall future direction of flare physics.

\subsection{What is the role of magnetic reconnection?} 
It has become clear over the past few decades that magnetic reconnection plays a role in solar activity as a mechanism for converting magnetic energy into heat and kinetic energy \citep{Priest_book}. The exact nature of this role, however, remains elusive. The concept of magnetic X-points being the site of particle acceleration and heating, for example, was proposed over 50 years ago by \citet{Giovanelli46} and \citet{Hoyle49}. However, it wasn't until the observations of the Masuda flare \citep{Masuda94} that this could be confirmed observationally. Currently, the standard thick target model for solar flares by \citet{Brown71} hypothesises magnetic reconnection at the top of the flare loop as the particle acceleration mechanism for precipitating electrons. In the CSHKP model, magnetic reconnection is believed to be the method by which a rising plasmoid is untethered from the solar photosphere. Observational evidence for these processes so far consist of secondary effects such as particle acceleration, heating effects and rising loops. Confirmation of the actual reconnection process has not yet been seen. The extremely short length scales over which magnetic reconnection occurs \citep[$\sim$meters;][]{Petschek64} is well below the resolution of even the most technologically advanced instruments. Resolving this process is essential for fully understanding the processes involved in energy release on the Sun. 

\subsection{What is the nature of CME initiation?}
CMEs have been the focus of the solar community since Col. Sabine hypothesised a connection between the Sun and the Earth more than 150 years ago. Since then, significant progress has been made in understanding the nature of these eruptions. Their kinematics have been analysed from many viewpoints over their entire propagation, from source \citep[e.g.][]{Chifor07, Gallagher02, Feynman95, Chifor06, Asai06} to the heliosphere \citep[e.g.][]{Byrne09, Maloney09, Schwenn00, Gopalswamy_catalogue}. Despite this however, much remains unknown regarding CMEs. While many models exist to explain their initiation \citep[e.g.][]{Antiochos1998, Forbes_Isen1991, Moore06}, we are far from isolating one in particular that is accurate every time. Observations of X-ray sources at the base of these features, such as those observed by e.g. \citet{Raftery10_cme} and \citet{Krucker2009, Krucker08_review, Krucker08, Krucker07} may help to shed some light on the nature of their initiation and acceleration. Unfortunately, the dynamic range of RHESSI, the instrument used to image these X-ray sources, is such that any chromospheric emissions will dominate the spectrum, making it impossible to observe coronal emission. Thus, occulted flares are the most reliable. This however, severely limits the number of events suitable for study. 

\subsection{The connection between EIT ``waves'' and CMEs?}
Since their discovery in 1998, so-called EIT ``waves'' have been a topic of hot debate. Initially, they were proclaimed as fast-mode magnetoacoustic waves: the coronal counterpart of chromospheric ``Morton waves'', following \citet{Uchida68}. EIT waves were believed to be fast-mode magnetoacoustic waves as this would explain the EUV brightenings, slow expansion rates \citep{Wang00} and reflection and refraction at coronal hole boundaries \citep{Veronig06, Gopalswamy09, Long08}. Since then, two more theories have been suggested. \citet{Delannee07, Attrill07, Chen02} have suggested various models by which EIT waves are not true waves, but are the propagation of perturbation sources due to a CME lift-off. Finally, a recent suggestion by \citet{Wills07} states that ``EIT waves'' are MHD solitons. 

Recently, an event was studied by \citet{Patsourakos09} using both STEREO spacecraft. The spacecraft were located 90$^{\circ}$ from each other at the time of the event, which occurred on the west limb from STEREO B's point of view and at disk centre from STEREO A. 3D modelling of the event led \citet{Patsourakos09} to conclude that initially, the ``wave'' is associated with opening field, but eventually disassociates from the magnetic field and travels as an independent wave. The analysis however, remains controversial and inconclusive. There is still significant work to be done before a definitive understanding of these events can be established. Continued observations and analysis of quadrature events using the capabilities of RHESSI X-ray imaging is currently the most productive method for establishing this.

\subsection{How do we predict solar storms?}
The prediction of solar storms, both flares and CMEs has not yet been tied down. The statistics for flare prediction are such that if one predicts no flares will be produced in a 24 hour period, more often than not, this will be correct. Therefore, with such poor statistics, it is difficult to make a quantitative prediction for these events. Currently work is being carried out using machine learning techniques that consider the complexity, fractal dimension and magnetic field strength, among other parameters, to identify regions on the cusp of energy release \citep{Gallagher_solmon, Colak09, Conlon08, Qah07, Leka07, Wheatland05}. The use of flares as a mechanism for CME prediction is also important. \citet{Svestka01} pointed out that flares are excellent indicators of CME activity. Using flare monitoring resources such as \emph{www.solarmonitor.org} to gather observations into one central database can help in making the best informed and complete predictions. The process of modelling these events from the Sun out to the Earth is as complex as the initial prediction. Where a CME occurs is an important aspect of its potential impact on Earth. A CME on the west limb of the Sun will not pass close to the Earth. However, a CME that erupts from the east limb has a good chance of interacting with the Earth's magnetic field. This, of course is important for the maintenance of satellites and electricity grids, along with the protection of astronauts \citep{Smith07}. While incremental progress is being achieved in this field, there remains a long way to go before it is sufficiently reliable.






\bibliographystyle{Latex/Classes/jmb} 
\renewcommand{\bibname}{References} 

\bibliography{9_backmatter/references} 








\end{document}